\documentclass[aps,prb,twocolumn,floatfix, superscriptaddress]{revtex4-2}

\usepackage[sort&compress]{natbib}
\usepackage{graphicx,times}
\usepackage[dvipsnames]{xcolor}
\usepackage{amssymb}
\usepackage{subeqnarray}
\usepackage{diagbox}
\usepackage{makecell}
\usepackage{amsmath}
\usepackage{bbm}
\usepackage{mathtools}
\usepackage{empheq}
\usepackage{bbding}
\usepackage{cellspace}

\usepackage[colorlinks=true,citecolor=blue,linkcolor=blue,urlcolor=blue,pdfencoding=auto]{hyperref}
\usepackage{cleveref}
\crefname{equation}{Eq.}{Eqs.}
\crefname{equations}{Eqs.}{Eqs.}
\Crefname{equation}{Equation}{Equations}
\crefname{figure}{Fig.}{Figs.}
\Crefname{figure}{Figure}{Figures}
\crefname{section}{Sec.}{Secs.}
\Crefname{section}{Section}{Sections}
\crefname{appendix}{Appendix}{Apps.}
\Crefname{appendix}{Appendix}{Apps.}
\crefname{paragraph}{Sec.}{Secs.}
\crefname{table}{Table}{Tables}

\usepackage{physics}

\newcommand{\msf}[1]{\mathsf{#1}}

\DeclareSymbolFont{greekletters}{OML}{FiraSans}{m}{n}

\DeclareMathOperator{\msZ}{\msf{Z}}
\DeclareMathOperator{\msY}{\msf{Y}}
\DeclareMathOperator{\msS}{\msf{S}}

\newcommand{\bone}{\mathbbm{1}}
\newcommand{\mzero}{\mathsf{0}}

\newcommand{\Zmat}{\mathsf{Z}}
\newcommand{\Imat}{\mathsf{I}}
\newcommand{\Umat}{\mathsf{U}}
\newcommand{\Omat}{\mathsf{O}}
\newcommand{\Cmat}{\mathsf{C}}
\newcommand{\Emat}{\mathsf{E}}
\newcommand{\Tmat}{\mathsf{T}}
\newcommand{\Smat}{\mathsf{S}}
\newcommand{\Wmat}{\mathsf{\Omega}}
\newcommand{\Rmat}{\mathsf{R}}
\newcommand{\Nmat}{\mathsf{N}}
\newcommand{\Dmat}{\mathsf{D}}
\newcommand{\Amat}{\mathsf{A}}
\newcommand{\Bmat}{\mathsf{B}}
\newcommand{\Gmat}{\mathsf{G}}
\newcommand{\Lmat}{\mathsf{L}}
\newcommand{\Jmat}{\mathsf{J}}
\newcommand{\Kmat}{\mathsf{K}}
\newcommand{\Mmat}{\mathsf{M}}
\newcommand{\Pmat}{\mathsf{P}}
\newcommand{\Vmat}{\mathsf{V}}
\newcommand{\Ymat}{\mathsf{Y}}
\newcommand{\Hmat}{\mathsf{H}}
\newcommand{\cmat}{\mathsf{c}}
\newcommand{\Sigmamat}{\mathsf{\Sigma}}

\newcommand{\Dissipmat}{\mathsf{M}_D}
\newcommand{\Voltmat}{\mathsf{M}_V}
\newcommand{\lambdamat}{\mathsf{\lambda}}
\newcommand{\etamat}{\mathsf{\eta}}
\newcommand{\zetamat}{\mathsf{\zeta}}
\newcommand{\mumat}{\mathsf{\mu}}
\newcommand{\omegamat}{\mathsf{w}}
\newcommand{\mmat}{\mathsf{m}}
\newcommand{\Lambdamat}{\mathsf{\Lambda}}
\newcommand{\deltamat}{\mathsf{\delta}}
\newcommand{\alphamat}{\mathsf{\alpha}}
\newcommand{\boldleft}{\boldsymbol{(}}
\newcommand{\boldright}{\boldsymbol{)}}

\newcommand{\lgr}{\mathcal{L}}
\newcommand{\vphi}{\vb{\Phi}}
\newcommand{\vx}{\vb{x}}
\newcommand{\vp}{\vb{p}}
\newcommand{\vV}{\vb{V}}

\newcommand{\vq}{\vb{q}}
\newcommand{\vQ}{\vb{Q}}
\newcommand{\vPi}{\vb{\Pi}}
\newcommand{\vX}{\vb{X}}
\newcommand{\wphi}{\widetilde{\phi}}
\newcommand{\wq}{\widetilde{q}}
\newcommand{\sDelta}{\scalebox{.5}{$\Delta$}}
\newcommand{\adag}{\hat{a}^{\dagger}}
\newcommand{\alow}{\hat{a}}
\newcommand{\bdag}{\hat{b}^{\dagger}}
\newcommand{\blow}{\hat{b}}

\newcommand{\black}{\color{black}}

\definecolor{adrcolor}{rgb}{0.75,0.1,0.2}
\definecolor{othcolor}{rgb}{0.1,0.1,0.75}
\definecolor{laucolor}{rgb}{0.2,0.5,0.2}
\definecolor{editcolor}{rgb}{0,0,0}

\newcommand{\edit}[1]{{\color{editcolor}{#1}}}

\usepackage[normalem]{ulem}%to cross text
\usepackage{cancel}

\begin{document}
	\title{
    Toolbox for nonreciprocal dispersive models in circuit QED}
	\author{Lautaro Labarca}
    \thanks{These authors contributed equally.}
    \affiliation{Institut Quantique and Département de Physique, Université de Sherbrooke, Sherbrooke, Qu\'ebec J1K 2R1, Canada}
    \author{Othmane Benhayoune-Khadraoui}
    \thanks{These authors contributed equally.}
    \affiliation{Institut Quantique and Département de Physique, Université de Sherbrooke, Sherbrooke, Qu\'ebec J1K 2R1, Canada}
    \author{Alexandre Blais}
    \affiliation{Institut Quantique and Département de Physique, Université de Sherbrooke, Sherbrooke, Qu\'ebec J1K 2R1, Canada}
    \affiliation{Canadian Institute for Advanced Research, Toronto, ON M5G 1M1, Canada}
    \author{Adrian Parra-Rodriguez}
    \email{adrian.parra.rodriguez@gmail.com}
    \affiliation{Institut Quantique and Département de Physique, Université de Sherbrooke, Sherbrooke, Qu\'ebec J1K 2R1, Canada}
    \affiliation{Instituto de Física Fundamental IFF-CSIC, Calle Serrano 113b, 28006 Madrid, Spain}
	\begin{abstract}
    We provide a systematic method for constructing effective dispersive Lindblad master equations to describe weakly anharmonic superconducting circuits coupled by a generic dissipationless nonreciprocal linear system, with effective coupling parameters and decay rates written in terms of the immittance parameters characterizing the coupler. This article extends the foundational
    work of \textcite{Solgun:2019} 
    for linear reciprocal couplers described by an impedance response. Notably,
    we expand the existing toolbox to incorporate nonreciprocal elements, account for direct stray coupling between immittance ports, circumvent potential singularities, and include     collective dissipative effects that arise from interactions with external common environments. We illustrate the use of our results with a circuit of weakly anharmonic Josephson junctions coupled to a multiport nonreciprocal environment and a dissipative port. The results obtained here can be used for the design of complex superconducting quantum processors with nontrivial routing of quantum information, as well as analog quantum simulators of condensed matter systems.
    \end{abstract}
    
    \maketitle
    
    \section{Introduction}
    
    Superconducting circuits~\cite{Devoret:2013} have become one of the most prolific platforms for the design and operation of small-scale quantum processors and quantum simulators, with basic quantum algorithms and error-correction protocols already being implemented~\cite{Ofek:2016,Arute:2019,Campagne-Ibarcq:2020,Krinner:2022,Zhao:2022,GoogleQAI:2023}. Based on Josephson junctions (JJs)~\cite{Josephson:1962}, nonlinear elements with  
    negligible dissipation at cryogenic temperatures, superconducting circuits show macroscopic quantum coherence and are useful platforms 
    for the engineering of light-matter interaction
    giving rise to the field of circuit quantum electrodynamics (cQED)~\cite{Blais:2021}. 
    
    In typical cQED setups, transmons are used as qubits due their simplicity and reduced sensitivity to charge noise, a feature that is linked to their 
    low anharmonicity \cite{Koch:2007}. These qubits are integrated in circuits with lumped or distributed element couplers, readout resonators and control lines \cite{Wallraff:2004}. With these various components operated in the dispersive regime, it is possible to obtain effective models where the qubits are dressed by their electromagnetic environment \cite{Blais:2004}.    
    To address the growing complexity of quantum processors,  
    systematic approaches have been developed to characterize the effective low-energy quantum Hamiltonian of the qubits, accounting for the multimode nature of the distributed circuit elements, while remaining agnostic on the specific circuit design \cite{Nigg:2012,Solgun:2014,Solgun:2015,Solgun:2019,Minev:EPR}. These methods rely on the classical properties of the linearized microwave structure in which the qubits are embedded. This includes knowledge of the immittance response of the microwave circuit, encompassing both admittance~\cite{Nigg:2012} and impedance~\cite{Solgun:2014,Solgun:2015,Solgun:2019} responses, or the computation of eigenmodes and their corresponding fields through three-dimensional electromagnetic simulations~\cite{Minev:EPR}. The nonlinearity arising from the qubits is subsequently introduced, with its effects being systematically computed at any desired level of precision, provided it remains weak. 

    Although general, these methods are not tailored to describing nonreciprocal elements that are characterized by nonsymmetric immittance response~\cite{Newcomb:1966}, and the need for \emph{in situ} physical or synthetic magnetic fields to break time-reversal symmetry. Nonreciprocal components, such as circulators and isolators, are essential for routing signals in and out of quantum processors. They also find application in the simulation of photonic lattices with broken time-reversal symmetry \cite{Koch:2010, Owens:2022}, enabling the exploration of topological phases with superconducting circuits \cite{Anderson:2016}.
    
    In this article, we provide a systematic tool to engineer nonreciprocal Lindbladian in cQED platforms based on Schrieffer-Wolff (SW) transformations, i.e., in the dispersive regime, of general nondissipative linear couplers described by an immittance matrix. 
    This work thus constitutes a generalization of the analysis conducted by \textcite{Solgun:2019} to nonreciprocal circuit elements. 
    Here, the qubit's electromagnetic environment is described not only by their impedance response, but, when applicable, by an admittance response, removing singularities appearing in the characterization of some linear systems. Moreover, due to the mixed couplings between phase-space variables (i.e., flux-charge coupling), appearing in the description of nonreciprocal elements, a generalized symplectic Schrieffer-Wolff transformation is used.
    In addition to this transformation, another perturbation theory is employed to eliminate the dissipative ports (i.e., the input-output lines used to address and measure the qubits) within the Born-Markov and partial secular approximations. This allows us to deduce not only the local decay rates, commonly known as Purcell rates, as discussed in Ref.~\cite{Solgun:2019},
    but also the correlated decay rates resulting from the interaction of the qubit modes with a common bath. 
    The general method we use, which is based on the exact fraction expansion of immittance responses (a.k.a.~black-box modeling~\cite{Nigg:2012,Solgun:2014,Solgun:2015,Hassler:2019,Newcomb:1966}), is applicable to electromagnetic environments containing an infinite number of modes and does not suffer from divergence issues~\cite{Gely:2017,Malekakhlagh:2017,ParraRodriguez:2018,ParraRodriguezPhD:2021}.

   This article is structured as follows. In \cref{Section:Results}, we present our main results, which includes 
    an overview of the derivation of the qubits's Hamiltonian and decay rates, while deferring the technical details to the Appendices.
   This allows us to formulate the master equation for the qubits in its Lindblad form in terms of the circuit's immittance parameters. 
    It is worth emphasizing that our results subsume and extend those derived in Ref. \cite{Solgun:2019}, and are capable of describing a broader range of circuits, including nonreciprocal elements, direct stray couplings both capacitive and inductive, and correlated decay. 
    In \cref{Section:Example}, we consider a simple case involving a nonreciprocal three-port scattering device, showing how our results can be used to achieve unidirectional signal transmission. Finally, in \cref{Section:Application} we give insight into potential applications that can benefit from our work, and we discuss possible directions for future works.

    \section{Effective Lindblad Master Equation}
     \label{Section:Results}

    In this section we present our main results and provide sketches for their derivations, with full details found in the appendices.
    \cref{fig:Intro} (a) schematically illustrates our conceptual framework:  qubits (green and blue) are interacting via an arbitrary linear and lossless electromagnetic environment that can contain nonreciprocal elements, and coupled to drive and dissipative ports (orange and brown). The electromagnetic environment is described by its multiports impedance $\msZ(\omega)$ or admittance $\msY(\omega)$. Moreover, following \textcite{Solgun:2019}, `qubit ports' are defined between the terminals of nonlinear dipole elements described by their flux degree of freedom (such as Josephson junctions), which we decompose into a linear part (represented schematically by the bare inductance $\Tilde{L}_{J_i}$) and a purely nonlinear contribution $U_{\text{nl}} (\phi_i)$ (represented by the spider symbol). The `drive ports' are defined as the terminals at the ends of transmission lines carrying the signals to the chip. We note that, with the port definitions we use here, the response matrices $\Zmat(\omega)$ and $\Ymat(\omega)$ do not include the linear part of the dipole potential, inline with the method of  Refs.~\cite{Solgun:2014,Solgun:2015,Solgun:2019}. This approach contrasts to that used in, e.g.,~Refs.~\cite{Nigg:2012,Minev:EPR}.

    \begin{figure}[t]
    \centering 
    \includegraphics[width=\linewidth]{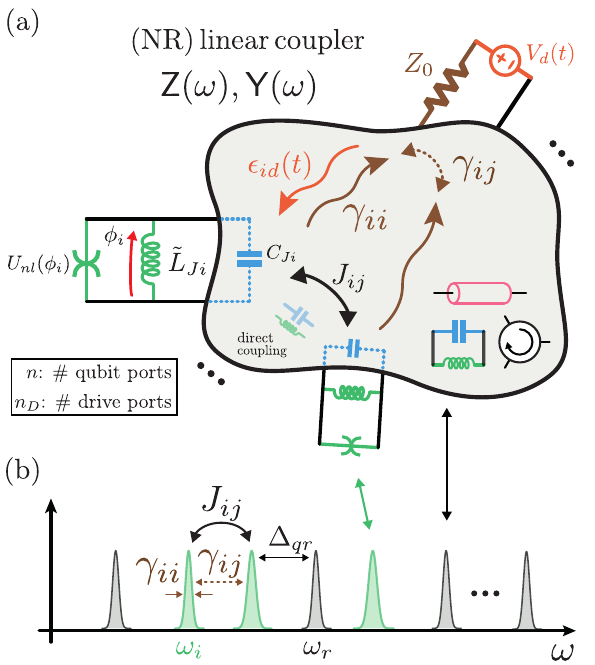}
    \caption{(a) Paradigm of the article: a circuit containing nonlinear elements divided into linear and nonlinear parts, and transmission lines modeled as classical voltage drive ports coupled by a general (nonreciprocal) linear system described by an immittance frequency-dependent matrix, i.e., $\msZ(\omega)$ or $\msY(\omega)$. After dispersive elimination of the black-box inner modes an effective Lindblad master equation including effective couplings $J_{ij}=|J_{ij}|e^{i\theta_{ij}}$with nontrivial phases $\theta_{ij}$, correlated decay rates $\gamma_{ij}$, and drive amplitudes $\epsilon_{id}$ is obtained for the qubit modes. (b) Schematic illustration of the qubit modes (green) and electromagnetic environment modes (black).}
    \label{fig:Intro}
    \end{figure}

    In general, any lossless causal immittance has a canonical lumped representation known as its Cauer circuit that synthesizes its response~\cite{Newcomb:1966}; see \cref{Duality}. Thus, in the circuit of \cref{fig:Intro} we substitute $\msZ$ (or $\msY$) by its Cauer circuit composed of capacitors, inductors, and gyrators coupled to the external ports by ideal transformers (see \cref{fig:Duality} in \cref{Duality}). As shown in \cref{Derivation-main}, using the methods of canonical circuit quantization of Refs.~\cite{ParraRodriguez:2019, Egusquiza:2022}, we construct the exact classical Lagrangian of the 
    circuit.
    From this Lagrangian, we obtain the full classical system-bath Hamiltonian that we decompose in the standard way as
    \begin{equation}
        H = H_{S}+H_I + H_B +H_D(t).
    \end{equation}
    In this expression, $H_S$ is the Hamiltonian of the junctions and inner modes of the circuit (i.e.~resonators and gyrators coupled to the qubit modes), $H_I$ is the interaction Hamiltonian of the coupling between qubit and inner modes with the dissipative ports, and $H_B$ is the bath Hamiltonian corresponding to the dissipative ports. Finally, $H_D(t)$ accounts for 
    classical external voltage sources at the drive ports. 
    
    Having obtained a general classical representation of the circuit, we now assume the circuit's inner modes to be dispersively coupled to the qubit modes; see \cref{fig:Intro} (b). As usual, we first approximately diagonalize the linear sector of the circuit, and then add the qubit nonlinearity. The presence of gyrators, which are the minimal lumped element circuits needed to synthesize a nonreciprocal response, introduces nondynamical modes in the  circuit description~\cite{ParraRodriguez:2019}. Following Ref.~\cite{Egusquiza:2022}, we eliminate these nondynamical modes with a symplectic transformation mixing the flux-charge variables of the inner modes, see \cref{Derivation-main} for details.
    Following this elimination, 
    we move to the dispersive frame 
    by applying 
    a second symplectic perturbation theory akin to the Schrieffer-Wolff transformation in quantum mechanics, see \cref{SSW} for details \edit{and a worked example}. 
    This transformation dresses the qubit and inner modes while preserving the symplectic structure of the Hamiltonian. In contrast to the standard SW transformation approach used in, e.g., Ref.~\cite{Solgun:2019}, the transformation used here preserves the symplectic structure of the Hamiltonian even with flux-charge couplings present in the description of gyrators.  

    The next step is to add back the qubit's nonlinearity. This approach can be applied to different types of low-anharmonicity qubits \edit{operating within a single deep potential well. For such qubits, the charge dispersion is further suppressed when coupled to a linear (non)reciprocal electromagnetic environment \cite{CaldeiraLeggett:1981,Nigg:2012,Leggett:1984} }. Here, we focus on the transmon by keeping the first nonlinear term of the Josephson junction's cosine potential. After quantizing the modes and applying a rotating-wave approximation, we obtain in this way an effective master equation for the qubit modes and their dispersive couplings to inner circuit modes taking the familiar Lindblad form
    \begin{equation} \label{main: master equation}
        \dot{\hat{\rho}} 
        =-i[\hat{H}_q +\hat{H}_\chi +\hat{H}_v(t),\hat{\rho}] 
        + \mathcal L_\gamma\hat\rho. 
    \end{equation}
    In this expression, $\hat{H}_q$ is the qubit Hamiltonian
   
    \begin{equation} \label{main: Hamiltonian}
        \hat{H}_q = \sum_{i} \omega_i\bdag_i\blow_i+\frac{\delta_i}{2}\bdag_i\blow_i(\bdag_i\blow_i-1) + \sum_{i\neq j}J_{ij}\blow_i\bdag_j,
    \end{equation}
    where $\blow_i$ is the annihilation operator for the dressed qubit mode $i$. The first two terms of $\hat{H}_q$ correspond to the free dressed qubit Hamiltonian, and the last term to qubit-qubit coupling mediated by the circuit. As in \textcite{Solgun:2019}, the parameters entering \cref{main: Hamiltonian} can be expressed in terms of the impedance of the full circuit. Generalizing those results, here expressions for these parameters are also obtained in terms of the circuit admittance. Crucially, this allows us to avoid singularities appearing in the characterization of some linear circuits (see \cref{Duality} for details). Moreover, our description  can account for the presence of nonreciprocal elements. To simplify the presentation, the expressions for the parameters entering $\hat H_q$ are provided below for the special case where there is no direct coupling (capacitive, inductive or nonreciprocal) between the 
    qubit ports (see \cref{Derivation-main} for details of the derivation, including the case with direct coupling). 
    
    First, to second-order in perturbation theory, the qubit frequencies in \cref{main: Hamiltonian} are 
    \begin{align}
        \omega_i &= \overline{\omega}_i-\frac{\Im{\msf{Y}^{\text{ac}}_{ii}(\overline{\omega}_i)}}{2C_{J_i}},\\
        \omega_i &= \overline{\omega}_i-\frac{\Im{\msf{Z}^{\text{ac}}_{ii}(\overline{\omega}_i)}}{2\overline{L}_{J_i}},
    \end{align}
    where the last term accounts for the Lamb shift introduced by the inner modes, with $\overline{L}_{J_i} = 1/C_{J_i}\overline{\omega}_i^2$ the dressed junction inductance. For $\Imat = \Ymat, \Zmat$ the ac part of the response is defined as $\msf{I}^{\text{ac}}=\msf{I}-\msf{I}^{dc}$, while the dc part $\msf{I}^{dc}$ is the sum of poles at zero and infinity, i.e., $\Ymat^{\text{dc}}=i\omega\Cmat_Y+\Lmat_Y^{-1}/i\omega+\Emat_{\infty}$, where $\Cmat_Y$ ($\Lmat_Y$) is the capacitive (inductive) matrix extracted from the admittance, and $\Emat_{\infty} = [\Ymat^{\text{dc}}-(\Ymat^{\text{dc}})^T]/2=\Ymat^{\text{dc,NR}}$ is the nonreciprocal part of $\Ymat^{\text{dc}}$ corresponding to direct nonreciprocal coupling between the ports. We emphasize that the dc part of the impedance contains only the capacitive response, $\Zmat^{\text{dc}}=\Cmat_Z^{-1}/i\omega$, as we consider ports shunted by capacitors, see \cref{Duality} for details.
    In the expression for $\omega_i$, the frequencies $\overline{\omega}_i$ take the form
    \begin{equation}
    \label{paper-eq:overline-w-y}
        \overline{\omega}_i = \widetilde\omega_{J_i}\sqrt{1+\zeta_i}\qty(1-\frac{E_{C_i}/ \widetilde\omega_{J_i}}{(1+\zeta_i)^{3/2}-E_{C_i}/\widetilde\omega_{J_i}}),
    \end{equation}
    where $\widetilde\omega_{J_i} = 1/(\widetilde{L}_{J_i}C_{J_i})^{1/2}$ is the plasma frequency of junction $i$, with $\widetilde{L}_{J_i} = \phi_0^2/E_{J_i}$ the bare junction inductance, and $C_{J_i}$ its total shunt capacitance. 
    The frequency $\overline{\omega}_i$ has two types of correction over the bare qubit frequency. First, the presence of the charging energy (bare anharmonicity) $E_{C_i}=e^2/2C_{J_i}$. Second, and in contrast with the similar expression from  Ref.~\cite{Solgun:2019}, the presence of $\zeta_i = \widetilde{L}_{J_i}/L_{s_i}$ which accounts for a possible effective shunting inductance $L_{s_i}=[\lim\nolimits_{s\to 0} s\Ymat_{ii}(s)]^{-1}$ \footnote{This correction is accurate provided the qubit potential is still accurately described by a single-well.}. 
    Importantly, a correction of this form can only be systematically obtained from the admittance response. When working with the impedance representation, such an inductive correction could in principle be obtained from $\lim_{s\to \infty}[\Zmat(s)/s]$. 
    Nonetheless, as we show in \cref{Duality}, for realistic circuits in which the ports are shunted by capacitances (e.g.~transmons), $\lim\nolimits_{s\to \infty}\Zmat(s) = 0$ and the correction vanishes. In short, an inductive energy correction cannot be systematically obtained from the impedance representation. 
    However, in order to use the impedance representation with a shunting inductance, the latter must be taken out of the response and be directly added to the bare junction inductance as part of the dipole self-inductance. 
    
    Moving on to the second term of \cref{main: Hamiltonian}, the qubit anharmonicity takes the form $\delta_i = -E_{Ci}(\omega_{J_i}/\overline{\omega}_i)^2$~\cite{Solgun:2019}. To second-order in perturbation theory, the qubit-qubit coupling $J_{ij}$ in the last term of \cref{main: Hamiltonian} is
    \begin{align}
    \label{paper:eff-J-y}
        J_{ij} &= \frac{i}{4}\sqrt{\frac{\overline{\omega}_i\overline{\omega}_j}{C_{J_i}C_{J_j}}}\qty[ \frac{\Ymat_{ij}(\overline{\omega}_i)}{\overline{\omega}_i}+\frac{\Ymat_{ij}(\overline{\omega}_j)}{\overline{\omega}_j}],\\
    \label{paper:eff-J-z}
        J_{ij} &= \frac{i}{4}\sqrt{\frac{\overline{\omega}_i\overline{\omega}_j}{\overline{L}_{J_i}\overline{L}_{J_j}}}\qty[\frac{\Zmat_{ij}(\overline{\omega}_i)}{\overline{\omega}_i}+\frac{\Zmat_{ij}(\overline{\omega}_j)}{\overline{\omega}_j}],
    \end{align}
    In these expressions, the admittance and impedance account for both the reciprocal (symmetric) and  nonreciprocal (antisymmetric) response, $\Ymat=\Ymat^{\text{R}}+\Ymat^{\text{NR}}$ and $\Zmat= \Zmat^{\text{R}} + \Zmat^{\text{NR}}$.
    
    Crucially, the $J_{ij}$ coupling has a nontrivial phase resulting from the interplay of the reciprocal and nonreciprocal responses. Indeed, writing $J_{ij} = |J_{ij}|e^{i\theta_{ij}}$, the phase is determined by the expression
    \begin{align}
    \label{paper:phaseij-Y}
        \tan{\theta_{ij}} = -\frac{\overline{\omega}_j\Ymat_{ij}^{\text{NR}}(\overline{\omega}_i)+\overline{\omega}_i\Ymat_{ij}^{\text{NR}}(\overline{\omega}_j)}{\Im{\overline{\omega}_j\Ymat_{ij}^{\text{R}}(\overline{\omega}_i)+\overline{\omega}_j\Ymat_{ij}^{\text{R}}(\overline{\omega}_i)}},
    \end{align}
    with an identical expression for the impedance response obtained by substituting $\Ymat(\omega)\to \Zmat(\omega)$. 
    In general, effective quantum models breaking time-reversal symmetry require a nontrivial phase in the hopping between sites \cite{Koch:2010}. Here, we find that this nontrivial phase can be manipulated by adjusting the ratio between reciprocal and nonreciprocal microwave responses between qubit ports at the qubit frequencies. 
    We note that \cref{paper:eff-J-y,paper:eff-J-z,paper:phaseij-Y} can lead to different interaction amplitudes $|J_{ij}|$ and hopping phases $\theta_{ij}$ when obtained from the admittance or the impedance responses. This difference between the two approaches arises from our perturbative derivation, which leads to different final effective frames. Crucially, in the dispersive regime, where these expressions remain valid, any discrepancies between them are negligible.
    
    The second term of the Hamiltonian appearing in \cref{main: master equation} describes the cross-kerr interactions between the qubit modes and the inner circuit modes
    \begin{equation}
        \hat{H}_{\chi} = \sum_{i,\mu} \chi_{i\mu}\bdag_i\blow_i\adag_{\mu}\alow_{\mu},
    \end{equation}
    where $\alow_\mu$ is the annihilation operator for the dressed inner circuit mode $\mu$. To second-order in perturbation theory and sixth-order in the nonlinear terms of the junctions potential, the cross-Kerr coefficients are (see \cref{NL})
    \begin{align}
        \chi_{i\mu} &= 2\delta_i\left(1-\frac{2E_C^{(i)}}{ \overline{\omega}_i}\right)\qty(\frac{\omega_{\mu}}{\omega_{\mu}^2-\overline{\omega}_i^2})^2\boldleft (g_{i\mu}^{qQ})^2+(g_{i\mu}^{q\Pi})^2\boldright,\\
        \chi_{i\mu} &= 2\delta_i\left(1-\frac{2E_C^{(i)}}{ \overline{\omega}_i}\right)\qty(\frac{\omega_{\mu}}{\omega_{\mu}^2-\overline{\omega}_i^2})^2\boldleft (g_{i\mu}^{\phi \Pi})^2+(g_{i\mu}^{\phi Q})^2\boldright,
    \end{align}
    for the admittance and the impedance, respectively. In these equations, $\omega_\mu$ is the frequency of mode $\mu$ obtained from the ac poles of the response. The coefficients $g_{i\mu}^{qQ}$ ($g_{i\mu}^{\phi \Pi}$) and $g_{i\mu}^{q\Pi}$ ($g_{i\mu}^{\phi Q}$) are the bare charge-charge (flux-flux) and charge-flux (flux-charge) couplings between qubit mode $i$ and circuit mode $\mu$ and are proportional to the residue of the admittance (impedance) at frequency $\omega_\mu$. Detailed expressions for these couplings can be found in  \cref{Appendix-NL:dispersive-shifts-Y,Appendix-NL:dispersive-shifts-Z} of \cref{NL}. Notably, the presence of nonreciprocity in the circuit, captured by these charge-flux interactions, modifies the dispersive shifts.
    
    Hamiltonian $\hat H_v(t)$ describes coupling of the qubits to external drive ports and it is given by
    \begin{equation} 
    \hat{H}_v(t)=\sum_{i=1}^{n}\sum_{d=1}^{n_D}\varepsilon_{id}(t)\hat{b}_i+\varepsilon_{id}^\star(t)\hat{b}_i^\dagger, 
    \end{equation}
    where $n$ and $n_D$ are respectively the number of qubit and drive ports. 
    Focusing on the situation where there is no direct coupling between drive ports and assuming a single tone voltage drive $V_d(t)=v_d\sin(\omega_d t)$ at each drive port $d$, the drive amplitudes take the form
    \begin{align} 
    \varepsilon_{id}(t)=&\,\frac{v_d{\abs{\Ymat_{dd}^{\text{drive}}(\omega_d)}}^{-1}}{\sqrt{2\overline{\omega}_iC_{J_i}}Z_0}\left[\left(\Ymat_{id}^{\text{ac}}(\overline{\omega}_i)+\Ymat_{id}^{\text{dc,NR}}(\omega_d)\right)\right.\nonumber\\ 
    &\,\times \sin(\omega_dt-\phi)\left. -i\Ymat_{id}^{\text{dc,R}}(\omega_d)\cos(\omega_dt-\phi)\right],\label{drive_Y_main}
    \end{align} 
   when expressed using the admittance case, and
    \begin{equation} \label{drive_Z_main}
    \begin{split}
         \varepsilon_{id}(t)=&-\frac{v_d{i\abs{\Zmat_{dd}^{\text{drive}}(\omega_d)}}^{-1}}{\sqrt{2\overline{\omega}_i\overline{L}_{J_i}}}\cos(\omega_dt-\phi')\\
         &\,\times \left[\Zmat_{id}^{\text{ac}}(\overline{\omega}_i)+\frac{\omega_d}{\overline{\omega}_i}\Zmat_{id}^{\text{dc}}(\omega_d)\right].
    \end{split}  
    \end{equation}
    expressed using the impedance. In these expressions, $\phi=\arctan [Z_0\Im\Ymat^{dc}_{dd}(\omega_d)]$  and $\phi'=-\arctan[Z_0/\Im\Zmat^{dc}_{dd}(\omega_d)]$ are the phase shift introduced by the external transmission line $Z_0$ at drive port $d$. Moreover, the reciprocal and nonreciprocal components of the dc response take into account possible direct couplings between qubit and drive ports.
    Additionally, $\Ymat^{\text{drive}}$ and $\Zmat^{\text{drive}}$ denote, respectively, the external admittance and impedance 
    filtered by the capacitances and inductances at drive ports. They are obtained by adding the lower $n_D\times n_D$ block of the dc immittance response $\Ymat^{\text{dc}}_D$  ($\Zmat^{\text{dc}}_D$) to the characteristic value of the admittance (impedance) of the external transmission lines located at each drive ports, $\Ymat^{\text{drive}}(\omega)=\Zmat_0^{-1}+\Ymat^{\text{dc}}_D(\omega)$ and $\Zmat^{\text{drive}}(\omega)=\Zmat_0+\Zmat^{\text{dc}}_D(\omega)$. For the sake of simplicity and without loss of generality, we will consider all the external transmission lines to have the same characteristic impedance $Z_0$, that is $\Zmat_0=Z_0 \mathbbm{1}_{n_D}$. 
    More general expressions accounting for arbitrary voltage pulse shapes as well as direct capacitive and inductive couplings between drive ports can be found in \cref{Derivation-main}. 

    The external ports also open decay channels for the qubit, something which we  model as baths of harmonic oscillators following the 
   Caldeira-Leggett approach~\cite{Leggett:1984,CaldeiraLeggett:1981}.
   Going to the dispersive frame as above, we obtain an effective qubits-baths interaction, which in addition to the usual flux-flux and charge-charge couplings include a nonreciprocal flux-charge interaction responsible for breaking time-reversal symmetry.
   Using the Born-Markov and partial secular approximations, which allows us to account for possible qubits quasi-degeneracies (see \cref{masterequation}), we then trace out the baths to obtain the correlated decay rates of the qubits.
   As above, these rates are expressed in terms of the admittance and impedance responses yielding the expressions
    \begin{equation} \label{paper:eff-gamma-y}
        \gamma_{ij} =\!\frac{1}{\sqrt{C_{J_i} C_{J_j}}}\sum_{d,d'=1}^{n_D}\!\!\Re{\Ymat_{dd'}^{\text{drive}^{-1}}(\overline{\omega}_{ij})}\Ymat_{id}(\overline{\omega}_i)\Ymat^\star_{jd'}(\overline{\omega}_j),
    \end{equation}
    and
    \begin{equation} \label{paper:eff-gamma-z}
        \gamma_{ij} =\!\frac{1}{\sqrt{\overline{L}_{J_i} \overline{L}_{J_j}}}\sum_{d,d'=1}^{n_D}\!\!\Re{\Zmat_{dd'}^{\text{drive}^{-1}}(\overline{\omega}_{ij})}\Zmat_{id}(\overline{\omega}_i)\Zmat^\star_{jd'}(\overline{\omega}_j),
    \end{equation}
    where $\overline{\omega}_{ij}=(\overline{\omega}_{i}+\overline{\omega}_{j})/2$ is the average frequency.  
    Notably, the Purcell decay rates of qubit $i$ due to its coupling to the drive ports can be determined from the diagonal elements of the matrix $\gamma_{ij}$. In the absence of direct stray coupling between drive ports, these rates are given by
      \begin{align} \label{Purcell_Y_main}
        \gamma_{i\kappa}
        =&\frac{1}{C_{J_i}}\sum_{d=1}^{n_D}\Re{\Ymat_{dd}^{\text{drive}^{-1}}(\overline{\omega}_i)}\abs{\Ymat_{id}(\overline{\omega}_i)}^2,\\
        \gamma_{i\kappa}
        =&\frac{1} {\overline{L}_{J_i}}\sum_{d=1}^{n_D}\Re{\Zmat_{dd}^{\text{drive}^{-1}}(\overline{\omega}_i)}\abs{\Zmat_{id}(\overline{\omega}_i)}^2,\label{Purcell_Z_main}
      \end{align}
      where $\abs{\mathsf{I}_{id}(\overline{\omega}_i)}^2=\abs{\mathsf{I}^R_{id}(\overline{\omega}_i)}^2+\abs{\mathsf{I}^{NR}_{id}(\overline{\omega}_i)}^2$ with $\mathsf{I}=\Zmat,\: \Ymat$. 
    Here, we have omitted the coherent contribution from the baths which in the dispersive and weak direct coupling regime only leads to a small renormalization of the qubits' Hamiltonian $\hat{H}_q$. Notably, as shown in Ref.~\cite{Correa:2023}, this coherent contribution is exactly cancelled in our regime of interest by second-order terms in the system-bath couplings that enters in the Caldeira-Leggett Hamiltonian (see \cref{masterequation}).
    
    Combining these expressions, we finally have the last term of $\mathcal L_\gamma\hat\rho$ of the master equation of \cref{main: master equation} which takes the form 
    \begin{equation} 
        \mathcal L_\gamma\hat\rho = \sum_{ij}\gamma_{ij}\mathcal{D}(\hat{b}_i,\hat{b}_j)\hat{\rho}.
    \end{equation}
    In this expression, $\mathcal{D}(\hat{b}_i,\hat{b}_j)\hat{\rho}=\hat{b}_j\hat{\rho}\hat{b}_i^\dagger -\frac{1}{2}\{\hat{b}_i^\dagger\hat{b}_j,\hat{\rho}\}$ and the sum is on qubits $i$ and $j$ such that $\abs{\overline{\omega}_i-\overline{\omega}_j}\lesssim \min(\gamma_{i\kappa},\gamma_{j\kappa})$.
    
    In summary, we have shown how the master equation of a nonreciprocal circuit in the dispersive regime is determined by the total linear response exhibited by the microwave structure connecting the qubits and the drive ports. That is $\dot{\hat\rho} = \mathcal{L}\hat\rho$ where the Lindblad superoperator is a function of the circuit's admittance or impedance.
    For completeness, the cases with direct capacitive, inductive, and nonreciprocal coupling between the qubit ports can be found in \cref{Derivation-main} for both admittance and impedance responses.
    
    \section{Example of application} \label{Section:Example}
    
    \begin{figure} 
        \centering
        \includegraphics[scale=1]{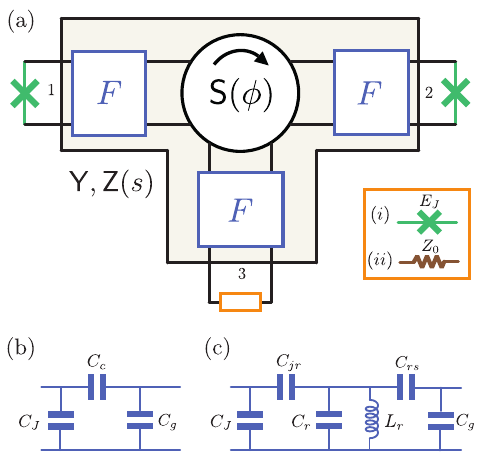}
        \caption{(a) 3-port nonreciprocal linear coupler connecting ($i$) 3 Josephson junctions, or ($ii$) 2 Josephson junctions and an ohmic dissipative element (a resistor). (b) A $\pi$-capacitive filter. (c) One-pole represention of a transmission line filter.
        }
        \label{fig:tuning-nonreciprocity-1}
    \end{figure}

    We now illustrate how the results of the previous section can be used with a simple circuit example that consists on 
    three Josephson junctions coupled via filters to a nonreciprocal scattering element,
    see \cref{fig:tuning-nonreciprocity-1}(a). We take the scattering matrix of the nonreciprocal circuit element to be
    \begin{equation}
    \label{paper:eq-S-definition}
        \Smat(\phi) = \frac{1}{3}\mqty(r(\phi) & t(\phi)-c(\phi) & t(\phi)+c(\phi)\\
        t(\phi)+c(\phi) & r(\phi) & t(\phi)-c(\phi)\\ t(\phi)-c(\phi) & t(\phi)+c(\phi) & r(\phi) ),
    \end{equation}
    with $r(\phi)=1+2\cos(\phi)$, $t(\phi) = 1-\cos(\phi)$ and
    $c(\phi)=\sqrt{3}\sin(\phi)$. Geometrically, this scattering matrix is a rotation around the symmetric axis $\vb{n}=(1,1,1)^T/\sqrt{3}$ and, for $\phi = 2\pi/3$, it corresponds to an ideal circulator.

    \begin{figure}[t]
        \centering
        \includegraphics[width=1\linewidth]{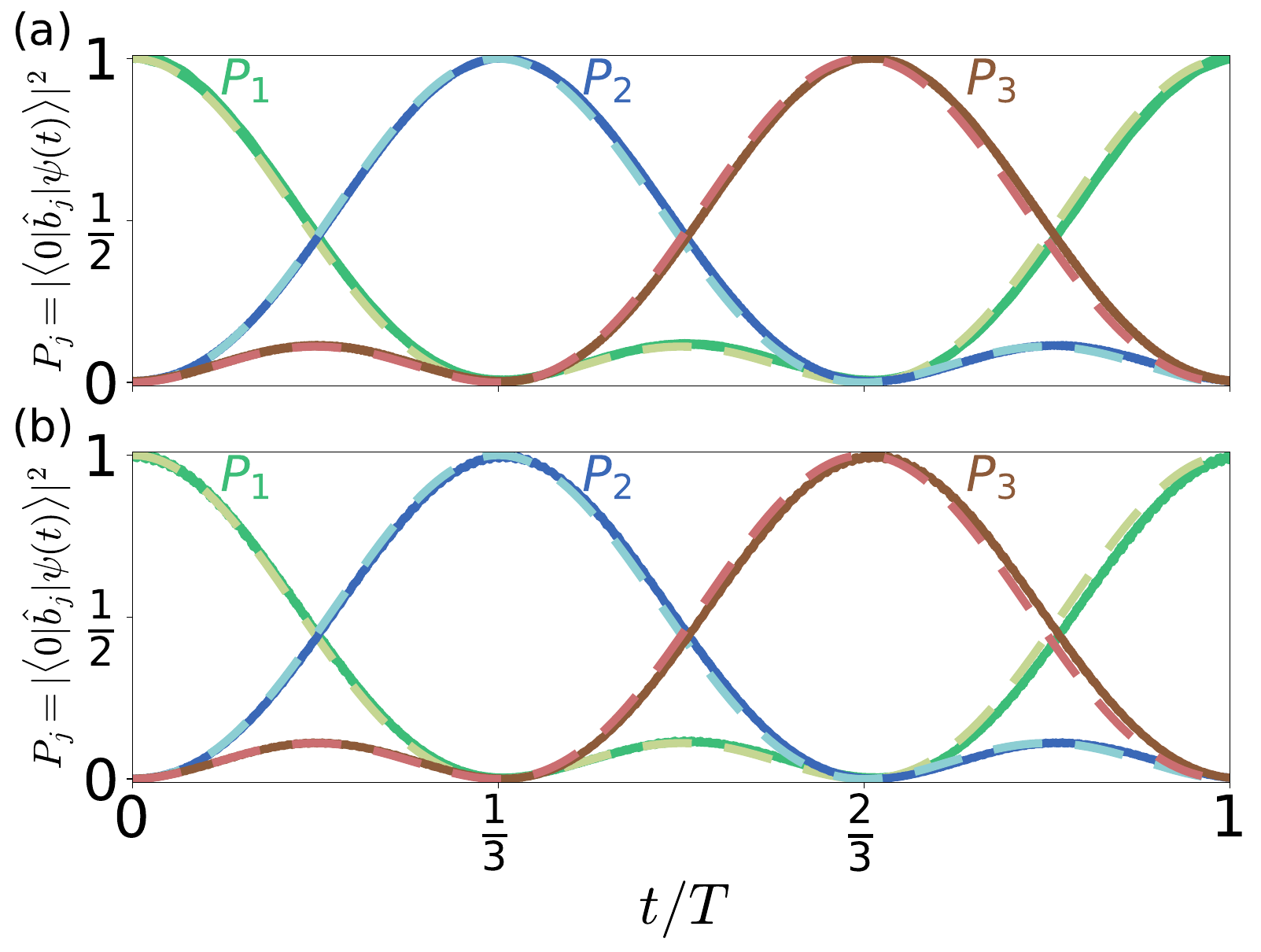}
        \caption{\edit{ Dynamics of qubit populations transfer with an initial state $\ket{\psi(0)}=\ket{100}$ using the circuit in \cref{fig:tuning-nonreciprocity-1} (a,$i$) with (a) the capacitive filter of \cref{fig:tuning-nonreciprocity-1} (b), and (b) with the resonator filter in \cref{fig:tuning-nonreciprocity-1} (c). The inner circulator phase is set to $\phi=\pi/3$. Solid (dashed) lines were obtained from the exact (effective) Hamiltonian of the circuits. The approximate period is $T = 2\pi/\sqrt{3}J(\phi)$.
         The circuit parameters for (a) are $C_J=100$ nF, $C_c = 0.1C_J$, $C_g = 1.5C_J$, $R= 50$ $\Omega$, $E_J/h = 11.37$ GHz, which leads to $E_J/E_C\simeq 64$, $\delta_i \simeq -0.05 \overline{\omega} $ and $\omega_{y}(\pi/3) \simeq 2\pi \times 11.5$ GHz. On the other hand, the parameters for (b) are $C_J = 100$ nF, $C_{jr}=0.1 C_J$, $C_g = 0.1 C_J$, $C_{rs}=C_J$, $1/\sqrt{L_rC_r} = 2\pi \times 7.0$ GHz, $\sqrt{L_r/C_r} = 50$ $\Omega$, $R=50$ $\Omega$, $E_J/h = 14.51$ GHz, which leads to $E_J/E_C\simeq 82$, $\delta_i\simeq -0.04 \overline{\omega}$. }
        } 
        \label{fig:tuning-nonreciprocity-2}
    \end{figure}
    
    Focusing first on the case where the filter is a simple $\pi$-capacitive filter, see \cref{fig:tuning-nonreciprocity-1} (b), we obtain the admittance and impedance response of the circuit in \cref{Appendix main-example (a)}.
    Equipped with these responses, we use \cref{paper:phaseij-Y} to directly obtain the hopping phase $\theta_{ij}$ between qubit modes in the circuit's effective quantum Hamiltonian. We find
    \begin{equation}
        \tan{\theta_{ij}}(\phi)= -\sqrt{3}\overline{\omega}/\omega_{y}(\phi),
        \label{main tan-theta}
    \end{equation}
    where $\overline{\omega} =\tilde{\omega}_J +\delta$ \edit{(see \cref{paper-eq:overline-w-y})}, with $\tilde{\omega}_J = 1/\sqrt{L_J\widetilde{C}_J}$ the junction's plasma frequency, $\delta=-E_C/(1-E_C/ \tilde\omega_J)$ the anharmonicity and $E_C$ the charging energy which we take to be equal for all three junctions. Moreover, $\omega_y(\phi) = \tan(\phi/2)/[R(C_c+C_g)]$ is the ac pole of the admittance response, with $R$ the characteristic impedance of the nonreciprocal scattering element.  
    As noted in Ref.~\cite{Koch:2010}, chiral dynamics with complete population transfer is  
    obtained for $\theta_{ij}(\phi)=\pi/6$, which leads to the condition
    \begin{equation}
        \overline{\omega} = \omega_{y}(\phi)/3, \label{paper:omega/3-condition}
    \end{equation}
    valid for all values of $\phi \neq 0,\pi$. 
    Therefore, we conclude that for the circuit of \cref{fig:tuning-nonreciprocity-1} (a) with the  filter of \cref{fig:tuning-nonreciprocity-1} (b), chiral dynamics can in principle be obtained with any value of $\phi\neq \{0,\pi\}$ , 
    something which only requires a suitable adjustment of the qubits frequencies.
    
    As an illustration of this chiral dynamics, starting with one excitation in the first qubit and all other modes in the vacuum, we plot in \cref{fig:tuning-nonreciprocity-2} (c) the evolution of the qubit population $P_{j=1,2,3}(t) = |\langle 0|\hat b_{j=1,2,3}|\psi(t)\rangle|^2$ for $\phi=\pi/3$.  
    There, we compare the evolution obtained from the exact Hamiltonian of the circuit obtained after the elimination of the nondynamical modes (full lines, see \cref{Elimination-Nondynamical}), with the expectation values obtained using the effective Hamiltonian of the qubit sector constructed using the formulas from the preceding section (dashed lines). The agreement between the two approaches is excellent and shows the expected circulation dynamics.
    
    Other filters can be used instead of the capacitive filter of \cref{fig:tuning-nonreciprocity-1} (b). 
    \edit{For example, consider the LC-resonator filter depicted in \cref{fig:tuning-nonreciprocity-1} (c), which is equivalent to the one studied in Sec. II of Ref.~\cite{Koch:2010}.} \edit{In this case, the interaction between the qubit modes is mediated by four bosonic modes instead of one. }
    %the condition for attaining chiral dynamics with complete population transfer becomes approximately $\overline{\omega}_i(\overline{\omega}_i/\omega_r)^2(1-C_c^2/C_rC_g)\simeq \omega_y(\phi)/3,$ \sout{where} $\omega_r = 1/\sqrt{(C_c+C_r)L_r}$. 
    \edit{However, under %the considered assumptions of 
    dispersive coupling %previously explained, 
    this additional circuit complexity does not substantially increase the difficulty of the application of our formulas. As before, we fix the qubit frequencies such that $\theta_{ij}=\pi/6$, and plot in \cref{fig:tuning-nonreciprocity-2} (b) the resulting dynamics starting with one excitation in the first qubit and all others in the vacuum state of both exact and effective Hamiltonians. Once again, we find excellent agreement between both approaches. Notably, and in contrast with the analysis in Ref. \cite{Koch:2010}, the inner modes do not need to be weakly coupled to achieve the desired chirality. Moreover, as in the previous example, we find that for any value of $\phi \neq {0,\pi}$, the effective phase between qubit modes can be adjusted to any desired value. Details of the simulation %and comparison with the analysis conducted in Ref. \cite{Koch:2010} 
    are provided in \cref{Appendix main-example (b)}.}
    
    %\lc{The next sentence we could eliminate} For more complex circuits lacking analytical \edit{representations}, numerical optimization methods over qubit frequencies and circuit design parameters can be used to obtain the desired effective Hamiltonian.

    As an additional example of application, we now consider the circuit of \cref{fig:tuning-nonreciprocity-1} where we replace the Josephson junction of port 3 by a resistor of impedance $Z_0$, see panel (a, ii). Following the above approach, the circuit parameters can be optimized to obtain partial excitation transfer between the ports. Indeed,  in the absence of a drive, $\Hat{H}_v=0$, the master equation \cref{main: master equation} leads to the following equations of motion for the qubit operators $\hat{b}_i$ \footnote{Here, we assume that the inner circuit mode is always on its ground state. As a result, the cross-Kerr Hamiltonian $\hat{H}_\chi$ is omitted in the master equation.}
    \begin{equation} \label{ME_example}
    \begin{split}
       \frac{d\hat{b}_1}{dt}&=-i\left(\overline{\omega}+\delta\hat{b}_1^\dagger \hat{b}_1-i\frac{\gamma_{11}}{2}\right)\hat{b}_1 -\left(iJ_{21}+\frac{\gamma_{12}}{2}\right)\hat{b}_2,\\
       \frac{d\hat{b}_2}{dt}&=-i\left(\overline{\omega}+\delta\hat{b}_2^\dagger \hat{b}_2-i\frac{\gamma_{22}}{2}\right)\hat{b}_2 -\left(iJ_{12}+\frac{\gamma_{21}}{2}\right)\hat{b}_1.
    \end{split}   
    \end{equation}
    By imposing $iJ_{21}+\gamma_{12}/2=0$, the qubit at port 1 can be isolated from qubit at port 2 \cite{Metelmann:2015}. Using \cref{paper:eff-gamma-z,paper:eff-gamma-y}, we have that this condition is equivalent to 
    \begin{equation} \label{isolation cond}
      \Ymat_{21}(\overline{\omega}) =\frac{Z_0}{1+\overline{\omega}^2\overline{C}_D^2Z_0^2}\Ymat_{13}(\overline{\omega})\Ymat_{23}^\star(\overline{\omega}),  
    \end{equation}
    when expressed in terms of the admittance response, with an analogous expression for the impedance response. Isolation is possible only if 
    $0\leq \Ymat_{21}(\overline{\omega})/\Ymat_{13}(\overline{\omega})\Ymat_{23}^\star(\overline{\omega}) \leq 1/\overline{\omega}\overline{C}_D$, where $\overline{C}_D=C_D+ C_cC_g/(C_c+C_g)$ and $C_D$ corresponds to the shunting capacitance of the dissipative port. For the  $\pi$-capacitive filter, this leads to the following constraints 
    \begin{equation}
    \begin{split}
      \overline{\omega} \gg \omega_y(\phi)&=\frac{\tan\frac{\phi}{2}}{R(C_c+C_g)},\\
      \frac{\sqrt{3}}{\omega_y(\phi)C_c}&=\frac{Z_0}{1+\overline{\omega}^2\overline{C}_D^2Z_0^2}.
    \end{split}
    \end{equation}
    
    \begin{figure}[t]
        \includegraphics[width=1\linewidth]{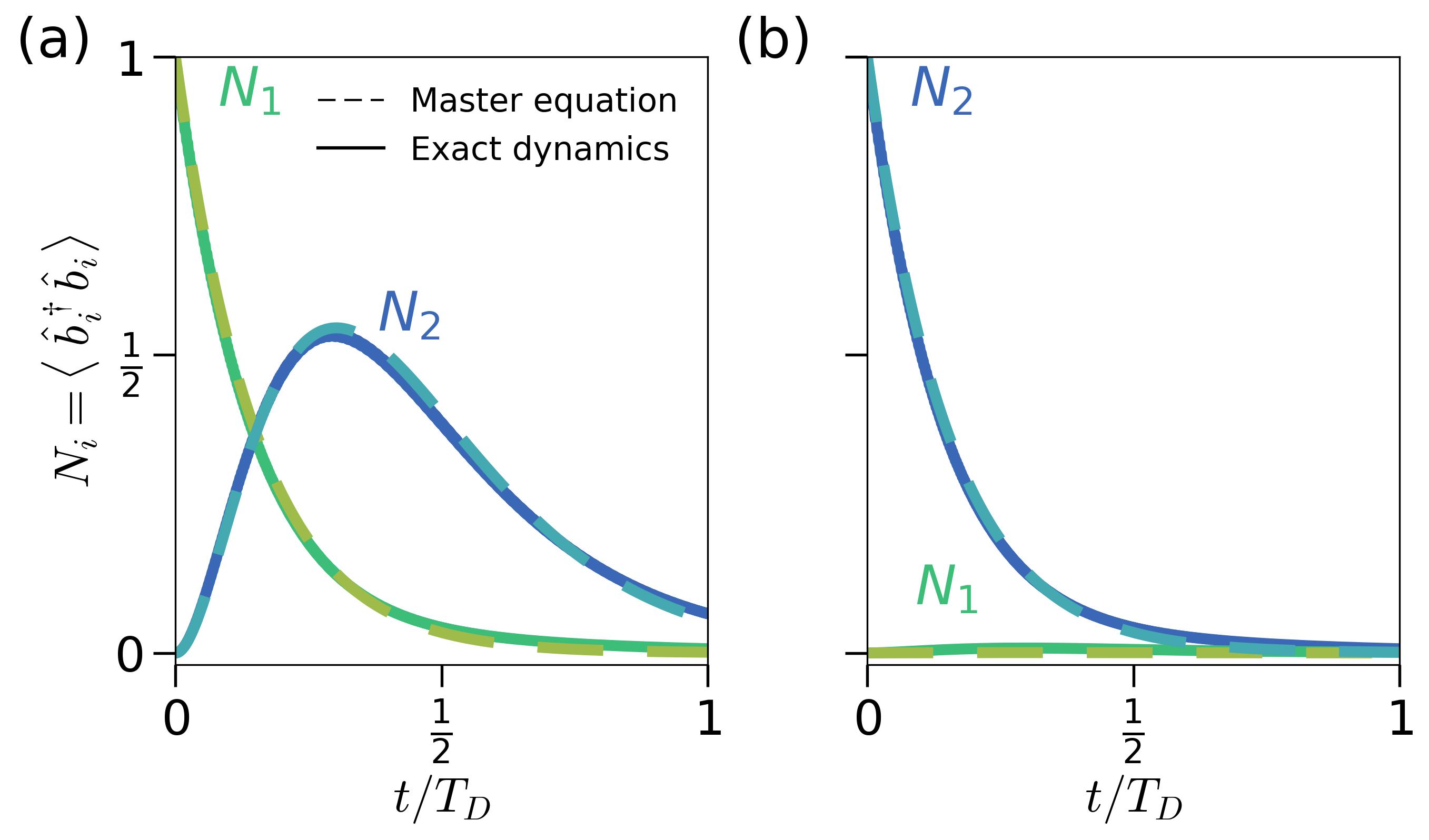}
        \caption{Isolator-like behavior is realized through adequate tuning of circuit parameters. (a) One excitation in mode $\hat{b}_1$ and mode $\hat{b}_2$ in vacuum at $t=0$. The population is transferred to mode $\hat{b}_2$, and then lost to the resistive port. (b) Reversed situation with qubit 2 having the excitation at $t=0$. Because of chirality,  qubit 1 remains underpopulated (less than 1\%) while qubit 2 decays. The dashed lines are obtained from the master equation \cref{ME_example} (dashed lines). Those results are compared to the exact dynamics obtained from classical Kirchhoff equations (solid lines). The circuit parameters are $C_c=0.01C_J$, $C_D=C_g=0.01C_c$, $\phi=\pi/3$, $Z_0=3R$, $\overline{\omega}=10\omega_y(\pi/3)$, $\delta=-0.05 \overline{\omega}$, $T_D=20RC_J$.}
        \label{fig:Isolator_behavior}
    \end{figure}
    
    The first condition results from the filter's high-pass behavior, while the second represents an impedance matching condition for the dissipative port of impedance $Z_0$, and admits a solution only when $C_c/\overline{C}_D \geq  2\sqrt{3}\overline{\omega}/\omega_y(\phi)$. This inequality is satisfied when $C_c \gg C_D \text{, } C_g$. In that case, the matched load should be chosen such that $Z_0\simeq\sqrt{3}R/\tan\phi/2$. For $\phi=2\pi/3$, we recover the usual impedance matching condition $Z_0=R$ of an ideal circulator. Importantly, isolation can also be obtained for $\phi\neq 2\pi/3$ provided a suitable adjustment of the impedance matching condition is made. \cref{fig:Isolator_behavior} shows the evolution of the initial 1-excitation state in (a) qubit 1 or (b) qubit 2 for $\phi=\pi/3$. 
    Panel (a) shows the expected population transfer from qubit 1 to qubit 2, while panel (b) illustrates the chirality of the evolution with the excitation in qubit 2 not reaching qubit 1 but instead being lost to the environment. These results are obtained from integration of the master equation \cref{ME_example} using expressions  \cref{paper:eff-gamma-y,paper:eff-J-y} for $J_{ij}$ and $\gamma_{ij}$, respectively (dashed lines). On the same figure, the full lines are obtained by solving numerically the corresponding classical Kirchhoff equations, yielding an excellent agreement. 

    The three-port network that we have considered here \edit{represents a small nontrivial example of the general results of the previous section.}
    Naturally, extending the application of the above results to an $N$-port network is straightforward. The underlying principles remain unchanged \cite{Aash:2022}, facilitating the use of our formulas to more complex and larger multi-port systems. Our results explicitly show that a \edit{linear element with an arbitrary level of nonreciprocity, e.g.,  the scattering matrix of %representing a symmetric rotation between its ports 
    \cref{paper:eq-S-definition},} is sufficient to reproduce all dynamics achievable with an ideal circulator, provided a suitable adjustments of the circuit parameters is made. In other words, 
    the design of on-chip circulators could in principle be relaxed to engineer the studied class of nonreciprocal synthetic models. We provide further examples with simple circuits in \cref{circuit-examples}.
    
    \section{Summary  and outlook} 
    \label{Section:Application}
    We have shown that knowledge of the classical immittance response together with qubit design parameters (e.g.,~Josephson's energies) is sufficient to fully characterize the dispersive master equation for transmon qubits even in the presence of nonreciprocal elements. While we have focused on transmon qubits, our results can be readily extended to any weakly anharmonic qubit operating in a single well potential, e.g.,~building upon the results of Ref.~\cite{Miano:2023}. 
    Naturally, it can also account for the presence of higher harmonics of the Josephson potential~\cite{Willsch:2024}. Moreover, having obtained expressions for both impedance and admittance responses offers two significant advantages. First, it enables the description of a wide range of circuits, effectively eliminating some singularities that may arise during the characterization of linear systems (examples are provided in \cref{Duality}). 
    Second, this dual approach allows us to use our results for qubits that are dual to the transmons, i.e., for which the charge degree of freedom is a good quantum number. Prominent examples of such qubits include phase-slip junctions \cite{Mooij1:2005,Mooij:2006,Arutyunov:2008}.
    
    Expanding beyond weakly anharmonic qubits, the formalism presented here could potentially be adapted to inductively shunted qubits, such as the fluxonium qubit, following Ref.~\cite{Smith:2016}. In that case, the normal modes of the electromagnetic environment can be extracted from the poles of the impedance response, while the Josephson junction is treated separately outside the response, as we have done in this work. However, becCianiause a truncated Taylor expansion of the Josephson energy is no longer a good approximation to the potential, one should then consider the full cosine operator, whose matrix elements in the normal mode basis can be obtained  
    from the zero-point fluctuations
    of the junction phase \cite{Smith:2016}. Crucially, these zero-point fluctuations can be directly computed from the immittance response, akin to the approach detailed in Ref.~\cite{Nigg:2012}.
    
    More generally, we expect that our work will facilitate the exploration of nonreciprocal models in circuit and cavity QED platforms~\cite{Owens:2022,Anderson:2016,Roushan:2017,WangYingYing:2023,WangYuXin:2023}. It can, for example, be used to systematically explore novel designs for on-chip nonreciprocal elements, building upon the principles established in Refs.~\cite{Koch:2010,Navarathna:2023}.
    Moreover, apart from the dispersive shifts, the parameters of the master equation are determined by the immittance response assessed at the frequencies of the qubits. Hence, only finite-element simulations around this range of frequencies are needed making this approach efficient with respect to energy-participation ratio-based methods \cite{Minev:EPR}. 
    
    More broadly, our work bridges the gap between electrical engineering via immittance design and quantum simulation of nonreciprocal and topological models. This connection relies on the mapping between the phase of the tunneling rate and immittance parameters provided by \cref{paper:phaseij-Y}, highlighting that different nonreciprocal immittance designs yield distinct topological models \cite{Ozawa:2019}. Finally, further work includes extending the treatment of nonlinearities to higher-order corrections, incorporating weakly nonlinear elements within the nonreciprocal response, and investigating tunable couplers in the spirit of Ref.~\cite{Solgun:2022}.  Furthermore, there will be a need for additional work to extend the systematic construction of effective models using the immittance-based paradigm to be applicable to the waveguide QED context, where the black-box structure would encompass a continuous spectrum. 

    All the codes to reproduce the main results are available online \cite{Labarca_Toolbox_for_nonreciprocal_2024}. 
    \begin{acknowledgments}
    We thank I. L. Egusquiza for early-stage helpful discussions on the symplectic Schrieffer-Wolff transformation. This work was undertaken thanks to funding from NSERC, FRQNT, Canada First Research Excellence Fund and the Ministère de l’Économie et de l’Innovation du Québec. This material is partially based upon work supported by the U.S. Department of Energy, Office of Science, National Quantum Information Science Research Centers, Quantum Systems Accelerator (QSA). A. P.-R. is funded by the Juan de la Cierva fellowship FJC2021-047227-I. O.B.K is supported by the UM6P International PhD Program.
    \end{acknowledgments}

    \appendix
    \begin{figure*}[h!]
    \centering 
    \includegraphics[width=1\linewidth]{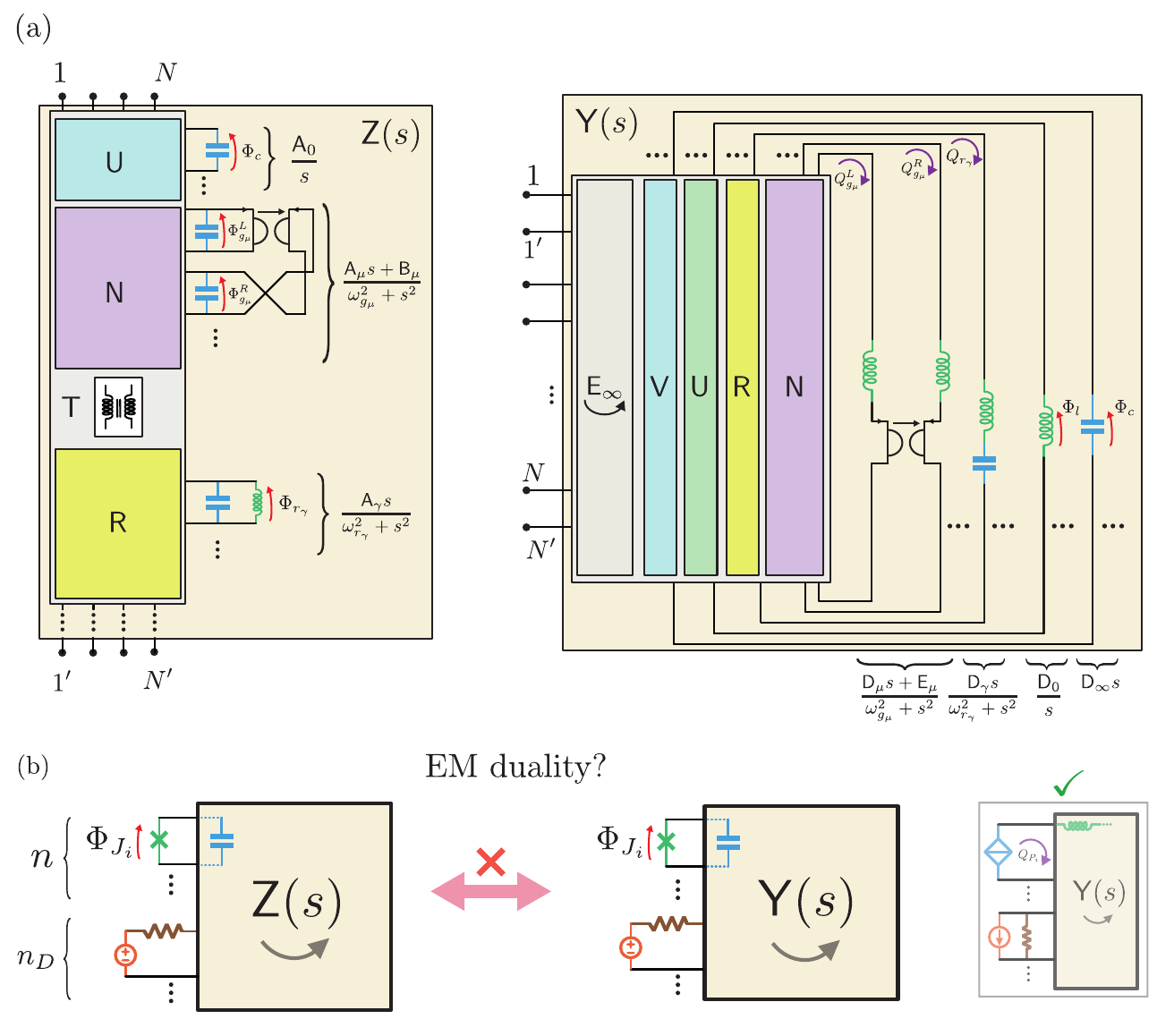}
    \caption{(a): Multiport canonical Cauer circuit for impedance (left) and admittance (right) responses. In this representation, $\Umat$, $\Vmat$, $\Rmat$, and $\Nmat$ correspond respectively to the multiport Belevitch transformer that connects the capacitive, inductive, reciprocal, and nonreciprocal stages to the rest of the circuit \cite{Newcomb:1966}. Additionally, $\Emat_{\infty}$ represents the dc nonreciprocal response. The presence of a non-singular capacitive stage in the admittance picture, reflecting the shunting capacitances of the ports, implies that there is neither an inductive stage $\Vmat$ nor dc nonreciprocal response in the impedance decomposition. (b): Electromagnetic duality between the impedance and admittance pictures, which is broken in the presence of Josephson junctions in the port terminals. However, full duality is recovered when connecting admittance `qubit' ports to a phase-slip flux qubit described by its loop charge degree of freedom \cite{Mooij1:2005,Mooij:2006, Arutyunov:2008} in series with a linear inductor, which is the dual circuit of a capacitive shunted JJ, as well as a current source-admittance representation of the drive ports.}
    \label{fig:Duality}
    \end{figure*}
    
    \section{Immittance response formulas, electromagnetic duality and singular cases}
    \label{Duality}

    In this appendix, we first provide a review of  canonical immittance matrix decompositions for lossless circuits. We then discuss the electromagnetic duality between impedance and admittance responses, and analyze singular cases in both approaches. For more details on linear response synthesis, the reader is referred to Ref.~\cite{Newcomb:1966}. Linear systems can be generically  characterized by a scattering matrix in Laplace space $\msS(s)$ and, in most cases, by an impedance $\msZ=R(\bone-\msS)^{-1}(\bone+\msS)$ matrix, an admittance $\msY=(1/R)(\bone-\Smat)(\bone+\Smat)^{-1}$ matrix, or both provided that the involved matrix inverses exist. Here, $R$ is a characteristic parameter in resistance units. These immittance responses can be decomposed into their multipole expansions 
    \begin{align}
    \label{Appendix-A:fractional expansion}
    \begin{split}
        \Zmat(s) =& \frac{\Amat_0}{s}+\Amat_{\infty}s+\Bmat_{\infty}+\sum_{\beta}\frac{\Amat_{\beta}s+\Bmat_{\beta}}{\omega_{\beta}^2+s^2},\\
        \Ymat(s) =& \frac{\Dmat_0}{s}+\Dmat_{\infty}s+\Emat_{\infty}+\sum_{\beta}\frac{\Dmat_{\beta}s+\Emat_{\beta}}{\omega_{\beta}^2+s^2}.
    \end{split}
    \end{align}
    where $\Amat_\alpha$ and $\Dmat_\alpha$ with $\alpha \in \{0,\infty,\beta\}$ are $N\times N$ real symmetric matrices with $N$ the number of ports. Similarly, $\Bmat_\zeta$ and $\Emat_\zeta$ with $\zeta\in \{\infty,\beta\}$ are $N\times N$ real antisymmetric matrices. These matrices are defined from the residues of the impedance (admittance). Explicitly, we have 
    \begin{equation}
    \label{Appendix-A:residues}
        \begin{split}
            \Amat_0 &= \text{Res}[\Zmat(0)], \\ 
            \Amat_\beta &= 2\Re{\text{Res}[\Zmat(\omega_{\beta})]},\\
            \Bmat_\beta &=-2\omega_{\beta}\Im{\text{Res}[\Zmat(\omega_{\beta})]},\\
            \Amat_\infty &= \lim_{\omega\to \infty}\partial\Zmat(\omega)/\partial\omega = \lim_{\omega\to \infty}\frac{\Zmat(\omega)}{i\omega},\\
            \Bmat_\infty &= \lim_{\omega\to\infty}[ \Zmat(\omega)-i\omega\Amat_\infty ],
        \end{split}
    \end{equation} 
    where the residue is given by $\text{Res}[\Zmat(\omega_0)]=\lim_{s\to i\omega_{0}}(s-i\omega_{0})\Zmat(s)$, and $\Zmat(\omega)\equiv \lim_{s\to i\omega} \Zmat(s)$. The same expressions hold for $\Dmat_\alpha$ and $\Emat_\zeta$ exchanging $\Zmat$ with $\Ymat$. The synthesis of these responses can be done using their canonical Cauer circuit, see \cref{fig:Duality} (a). We now turn to important facts on how this synthesis is done.
    
    First, we note that the expansions in \cref{Appendix-A:fractional expansion} are in general valid for any lossless causal linear response \cite{Newcomb:1966}. However, we consider all of the ports to be capacitively shunted, a physically realistic assumption also taken in Ref.~\cite{Solgun:2019}. Hence, from the definition of the impedance matrix $\Zmat_{ij}(s)=v_i(s)/i_j(s)|_{i_k=0}=v^c_i(s)/i_j(s)|_{i_k=0}$ with $v_i(s)$ the voltage at port $i$, $i_j(s)$ the current at port $j$, and $v^c_i=i^c_i/sC_i$ the voltage of the capacitor shunting port $i$, it follows directly that $\lim_{s\to \infty}\Zmat_{ij}(s)=0$. As such, the impedance responses we will be considering have no poles at infinity and $\Amat_\infty=\Bmat_\infty=0$. Moreover, we define the reciprocal (nonreciprocal) poles with frequencies $\omega_{r_\gamma}$ ($\omega_{g_\mu}$) where $1\leq \gamma \leq m$ ($1\leq \mu \leq l$), as those with zero (nonzero) imaginary part in their residue, see \cref{Appendix-A:residues}. Doing so, the expansions of $ \msZ$ and $\Ymat$ read 
    \begin{align}
    \label{appendix-A:reduced fractional expansion}
    \begin{split}
        \Zmat(s)&= \frac{\Amat_0}{s}+\sum_{\gamma=1}^m\frac{\Amat_{\gamma}s}{\omega_{r_\gamma}^2+s^2}+\sum_{\mu=1}^l\frac{\Amat_{\mu}s+\Bmat_{\mu}}{\omega_{g_\mu}^2+s^2},\\
        \begin{split}
        \Ymat(s)&= \frac{\Dmat_0}{s}+\Dmat_{\infty}s+\Emat_{\infty}\\
        &+\sum_{\gamma=1}^m\frac{\Dmat_{\gamma}s}{\omega_{r_\gamma}^2+s^2}+\sum_{\mu=1}^l\frac{\Dmat_{\mu}s+\Emat_{\mu}}{\omega_{g_\mu}^2+s^2}.
        \end{split}
    \end{split}
    \end{align}
    The Cauer decomposition of these responses rests on two key facts. First, with an appropriate choice of reactive elements, the circuit poles will be at the same frequencies as the response poles. Importantly, the minimal number of reactive elements necessary is directly given by the rank of the response residues. Second, the residue matrices $\Amat_\alpha$ and $\Bmat_\zeta$ ($\Dmat_\alpha$ and $\Emat_\zeta$) encode the topology of the circuit. This topology is realized in the circuit with the use of ideal transformer matrices ($\Umat,\Vmat,\Rmat,\Nmat$ in \cref{fig:Duality} (a)) and gyrators \cite{Newcomb:1966}. An ideal transformer matrix $\Tmat_{b\times a}$ is a constraint between its $a$ primary and $b$ secondary ports voltages ($\vb{v}_a$, $\vb{v}_b$) and currents ($\vb{i}_a$, $\vb{i}_b$), such that $\vb{v}_{a} = \Tmat^T \vb{v}_b$ and $\vb{i}_{b} = -\Tmat \vb{i}_a$. An ideal gyrator is a two-port nonreciprocal constraint relating its left and right port voltages ($v_l$, $v_r$) and currents ($i_l$, $i_r$), such that 
    \begin{equation}
        \mqty(v_l \\ v_r) = R\mqty(0 & -1 \\ 1 & 0)\mqty(i_l \\ i_r),
    \end{equation}
    where $R$ is the gyration resistance of the gyrator. In the nondegenerate case, the rank of each residue at frequencies $\omega_{r_\gamma}$ and $\omega_{g_\mu}$ is one. We elucidate the process of obtaining explicit expressions for the transformer matrices, gyration ratios, and reactive element parameters in the subsequent paragraphs.

  We focus first on the impedance response synthesis, see \cref{fig:Duality} (a). The transformer matrix $\Umat$ is obtained from the orthogonal decomposition of $\Amat_0 = \Umat^T \overline{\Cmat}^{-1} \Umat$. The capacitances of the purely capacitive stage are given by the inverse of the eigenvalues of $\Amat_0$ (the entries of the diagonal matrix $\overline{\Cmat}$). The transformer matrix $\Rmat$ is given by $\Rmat^T = \mqty(\vb{r}_{1},\dots, \vb{r}_m)$, where the transformer ratios $\vb{r}_{\gamma}$ are $(N\times 1)$ column vectors coupling the external ports with the inner mode resonators of frequency $\omega_{r_\gamma}=1/\sqrt{C_{r_\gamma}L_{r_\gamma}}$. Such ratios are the electrical engineer's equivalent~\cite{Ciani:2023} to the energy-participation ratios and signs used in Ref.~\cite{Minev:EPR}. These transformer ratios are obtained from the residue matrices $\Amat_\gamma = \vb{r}_\gamma\vb{r}_\gamma^T/C_{r_\gamma}$. As there is a degree of freedom in the choice of $C_{r_\gamma}$, following the standard convention we set $C_{r_\gamma}=1$ and $L_{r_\gamma} = 1/\omega_{r_\gamma}^2$. Therefore, the transformer ratios are directly given by the normalized eigenvector with nonzero eigenvalue ( $\boldsymbol{\lambda}_\gamma$, $\lambda_{\gamma}$) of $\Amat_\gamma$ as $\vb{r}_\gamma = \sqrt{\lambda_{\gamma}}\boldsymbol{\lambda}_\gamma$, and hence have units of $[C]^{-1/2}$. The transformer matrix $\Nmat$ is given by $\Nmat^T = \mqty(\vb{n}_1^L,\vb{n}_1^R,\dots,\vb{n}_l^L,\vb{n}_l^R)$, where the transformer ratios $\vb{n}_\mu^{L,R}$ are $(N\times 1)$ column vectors coupling the external ports with the left and right branches of the gyrators, with gyration ratio $R_\mu$ capacitively shunted at each branch with $C_{g_\mu}$, synthetizing the nonreciprocal resonators of frequency $\omega_{g_\mu} = 1/R_\mu C_{g_\mu}$. These transformer ratios are obtained from both $\Amat_\mu = \boldleft \vb{n}_\mu^L(\vb{n}_\mu^L)^T+\vb{n}_\mu^T(\vb{n}_\mu^R)^T\boldright/C_{g_\mu}$ and $\Bmat_\mu =R_\mu \omega_{g_\mu}^2\boldleft \vb{n}_\mu^R(\vb{n}_\mu^L)^T-\vb{n}_\mu^T(\vb{n}_\mu^R)^T\boldright $. As before, following standard convention we set $C_{g_\mu}=1$ and $R_\mu = 1/\omega_{g_\mu}$. 
  One can show that in the nondegenerate case, the transformer ratios can be given by $\vb{n}_\mu^{L,R} = \sqrt{\lambda_\mu^{L,R}}\boldsymbol{\lambda}_\mu^{L,R}$, where $\boldsymbol{\lambda}_\mu^{L,R}$ and $\lambda_\mu^{L,R}$ are the normalized eigenvectors with nonzero eigenvalue of $\Amat_\mu$. In summary, we have
    \begin{equation}
    \label{Appendix-A:Transformer-eqs-impedance}
        \begin{split}
            \Amat_0 &= \Umat^T\overline{\Cmat}^{-1}\Umat,\\
            \Amat_{\gamma} &= \vb{r}_\gamma\vb{r}_\gamma^T, \\
            \Amat_\mu &= \vb{n}_\mu^L(\vb{n}_\mu^L)^T+\vb{n}_\mu^R(\vb{n}_\mu^R)^T,\\
            \Bmat_\mu &=\omega_{g_\mu}\boldleft \vb{n}_\mu^R(\vb{n}_\mu^L)^T-\vb{n}_\mu^L(\vb{n}_\mu^R)^T\boldright,\\
            \Rmat^T &= \mqty(\vb{r}_1,\dots, \vb{r}_m),\\
            \Nmat^T &= \mqty(\vb{n}_1^L,\vb{n}_1^R,\dots,\vb{n}_l^L,\vb{n}_l^R).
        \end{split}
    \end{equation}

    For the admittance [see right panel of \cref{fig:Duality} (a)], the synthesis procedure is very similar and we directly provide the summary 
    \begin{equation}
        \label{Appendix-A:Transformers-eqs}
        \begin{split}
            \Dmat_\infty &= \Vmat\overline{\Cmat}\Vmat^T,\\
            \Dmat_0 &= \Umat\overline{\Lmat}^{-1}\Umat^T,\\
            \Dmat_{\gamma} &= \vb{r}_{\gamma}\vb{r}_{\gamma}^T,\\
            \Dmat_{\mu} &= \vb{n}_{\mu}^L(\vb{n}_{\mu}^L)^T+\vb{n}_{\mu}^R(\vb{n}_{\mu}^R)^T,\\
            \Emat_{\mu} &= \omega_{g_\mu}\boldleft\vb{n}_{\mu}^L(\vb{n}_{\mu}^R)^T-\vb{n}_{\mu}^R(\vb{n}_{\mu}^L)^T\boldright,\\
            \Emat_\infty &= \Nmat_{\infty}\Ymat^G_{\infty}\Nmat_{\infty}^T,\\
            \Rmat &= \mqty(\vb{r}_1,\dots, \vb{r}_m),\\
            \Nmat &= \mqty(\vb{n}_1^L,\vb{n}_1^R,\dots,\vb{n}_l^L,\vb{n}_l^R).
        \end{split}
    \end{equation}
    Here, we have used the convention $L_{r_\gamma}=L_{g_\mu}=1$, $C_{r_\gamma}=1/\omega_{r_\gamma}^2$ and $R_{\mu} =\omega_{g_\mu}$. The capacitive stage is synthetized analogously as in the impedance case. The inductive stage is synthesized with the orthogonal transformer matrix $\Umat$, and the inductors given by the inverse of the eigenvalues of $\Dmat_0$ (i.e., the diagonal entries of $\overline{\Lmat}$). Similarly, the transformer ratios are given by $\vb{r}_\gamma = \sqrt{\lambda_\gamma}\boldsymbol{\lambda}_\gamma$ ($\vb{n}_\mu^{L,R}=\sqrt{\lambda_\mu^{L,R}}\boldsymbol{\lambda}_\mu^{L,R}$), where $\boldsymbol{\lambda}_\gamma$ ($\boldsymbol{\lambda}_\mu^{L,R}$) are the normalized eigenvectors with nonzero eigenvalue of $\Dmat_\gamma$ ($\Dmat_\mu$), and have dimension $[L]^{-1/2}$. The nonreciprocal stage $\Emat_\infty$ can be formally synthetized with a transformer matrix $\Nmat_\infty$ and a set of ideal gyrators represented by the block matrix $\Ymat_\infty^G$. However, as we do not need this synthesis for any of our results, we do not give explicit expressions for it here.

    \begin{figure}[t] 
        \centering
        \includegraphics[width=0.7\linewidth]{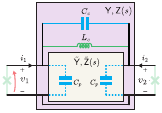}
        \caption{Generic two port circuit coupled through an admittance or impedance response plus direct inductive coupling. For such circuits  unconstrained Hamiltonian dynamics can be systematically constructed from the admittance, but not the impedance, response.}
        \label{fig:generic-2port-singular-0-poles}
    \end{figure}   
    We stress that the transformer (gyration) ratios and reactive elements for a given admittance and impedance synthesis when one is the inverse of the other are different.  
    We have chosen the same notation for both to make the duality apparent. However, in general the inverse of the impedance (admittance) is not its dual. We now make this statement more precise.
    In the context of circuit theory, electromagnetic duality refers to the invariance of the equations of motion for voltages ($\vb{v}$) and currents ($\vb{i}$) 
    under the transformation $\vb{v}\to z_0\vb{i}$ and $\vb{i}\to \vb{v}/z_0$, where $z_0$ depends on the units chosen for voltage and currents. In particular, two immittance responses $\Zmat(s)$ and $\Ymat^d(s)$ are dual if the equations of motion for voltages ($\vb{v}$) and currents ($\vb{i}$) at the ports
    are invariant under the transformation $\Zmat(s)\to \Ymat^d(s)$,  $\vb{v}\to z_0\vb{i}$ and $\vb{i}\to \vb{v}/z_0$. It follows that $\Ymat^d(s) = \Zmat(s)/z_0^2$, where the Cauer representation of $\Ymat^d(s)$ is obtained from the Cauer representation of $\Zmat(s)$, by changing capacitors to  
    inductors (and viceversa), transposing the transformer matrices, and using the LC-oscillators (gyrators) admittance instead of impedance representations. 

    % \lc{The objective of this paragraph is to motivate the fact that admittance and impedance approches have different singular cases.}
    For Josephson junction-based circuits, the shunting capacitances accompanying these JJs introduce an asymmetry in the poles between the impedance and admittance approaches, see \cref{appendix-A:reduced fractional expansion}, breaking therefore electromagnetic duality between the two approaches.
    % \aprc{For the Josephson junction-based circuits considered in this manuscript electromagnetic duality is broken mainly because Josephson Junctions (JJs) are (up to now) the sole reliable nonlinear elements in circuit quantum electrodynamics (cQED) platforms, which hinders the achievement of the complete map JJs $\rightarrow$PSs, $\Zmat\rightarrow\Ymat$, and parallel $\rightarrow$ series connection, as illustrated in \cref{fig:Duality}. Focusing only on circuits with JJs at the qubit ports, i.e., }
    % For Josephson junction-based circuits (e.g., \cref{fig:Duality}), electromagnetic duality is broken {\color{blue}because} {\color{red}for two reasons. First, in the transmon regime $E_J/E_C\gg 1$ a flux description of the Josephson junctions is required to have a weakly anharmonic potential, making it challenging to employ a dual description based on loop charges. 
    % Secondly,}
    % the shunting capacitances accompanying these JJs introduce an asymmetry in the poles between the impedance and admittance approaches, see \cref{appendix-A:reduced fractional expansion}.
    This leads to the possibility of encountering distinct singular cases for the admittance and impedance representations when constructing their respective Lagrangian. We proceed to show that this is indeed the case. Let us focus first on the singular cases of circuits in which the external ports are shunted by dipoles described by their flux degree of freedom. In this context, the kinetic matrix of the Lagrangian derived from the Cauer representation of the impedance (admittance) will be singular if $\Amat_0$ ($\Dmat_\infty$) is singular. Specifically, circuits featuring direct inductive coupling (as depicted in \cref{fig:generic-2port-singular-0-poles}) between qubit ports render $\Amat_0$ singular. This can be proven using the current-voltage relation between the inductor ports, which implies that for dc current the voltage drop across the inductor is zero and thus acts as a short circuit between the ports equalling their voltages. That is, $\lim_{s\to 0}v_1 = \lim_{s\to 0}v_2$ implies $\lim_{s\to 0}\Zmat_{1j}(s)=\lim_{s\to 0} v_1/i_j|_{i_k=0} = \lim_{s\to 0}v_2/i_j|_{i_k=0} =\lim_{s\to 0}\Zmat_{2j}(s)$ for all $j$. In other words $\Amat_0=\lim_{s\to 0}s\Zmat(s)$ has two linearly dependent rows and is singular. Notice that this statement is true even in the presence of a parallel (shunting) capacitor to the $L_c$. By the same token, if there is a shunting inductance on port $i$, then $\lim_{s\to 0} v_i = 0$ implying that $(\Amat_0)_{ij} = 0$ for all $j$ and $\Amat_0$ is singular. However, this last case can be remedied by extracting the shunting inductance out of the response and dressing the bare inductance of the dipole with it. In contrast, within the admittance framework, $\Dmat_\infty$ will always be full rank due to the shunting capacitances at the ports, hence the derived Lagrangian is not singular and derivation of the corresponding Hamiltonian and subsequent quantization, can always be done. 
    
    Finally, we note that thanks to electromagnetic duality, our results extend to weakly anharmonic qubits described by their (loop) charge degree of freedom such as phase-slip flux qubit \cite{Mooij1:2005,Mooij2:2005,Mooij:2006,Arutyunov:2008}.
    This is so because the dual circuit of an impedance shunted by Josephson junctions in parallel with capacitors, corresponds to its dual admittance with the ports shunted by phase-slips in series with inductors, see \cref{fig:Duality} (b). 
    Therefore,
    the formulas found in this work for qubits described by their flux-degree of freedom can be easily adapted to the case where the qubits are described by their loop charge-degree of freedom. 
    Singular cases for loop charge-based circuits are dual to the ones presented above.
    
    \section{Symplectic Schrieffer-Wolff transformation}\label{SSW}
    Here, we develop a form of symplectic perturbation theory in classical phase-space analogous to the standard quantum Schrieffer-Wolff perturbation theory \cite{Bravyi:2011}. Consider a Hamiltonian $H$ defined by the quadratic form $\Hmat$ as
    \begin{equation}
        H = \frac{1}{2}\vb{X}^T\Hmat\vb{X},
    \end{equation}
    where $\vb{X}$ is the vector of generalized phase-space coordinates. Without loss of generality we divide it into two sectors $A$ and $B$ such that $\vb{X}^T=(\vb{x}_A^T , \vb{p}_A^T , \vb{x}_B^T , \vb{p}_B^T ),$ where position and momenta obey the standard Poisson bracket $\qty{\vb{x}_{a},\vb{p}_{b}}=\delta_{a,b}$. With this division, the quadratic form $\Hmat$ in block form reads
    \begin{equation}
        \Hmat = \mqty(\Hmat_A & \Hmat_{AB}\\\Hmat_{AB}^T & \Hmat_B ).
    \end{equation}
    Consequently its symplectic form $\Jmat$ is also in block form, i.e.,
    \begin{equation}
        \Jmat = \mqty(\Jmat_A & \mzero \\ \mzero & \Jmat_B)
    \end{equation}
    with
    \begin{equation}
        \Jmat_A = \mqty(\mzero & \bone_{n}\\ -\bone_{n} & \mzero), \; \Jmat_B = \mqty(\mzero & \bone_{m}\\ -\bone_{m} & \mzero),\label{Appendix SW:J-def} 
    \end{equation}
    where $n$ ($m$) corresponds to the number of conjugate pair variables in $A$ ($B$). 
    The equations of motion of the system read $\dot{\vb{X}} = \Jmat \Hmat \vb{X}$ \cite{Goldstein:1950}. 
    We are seeking a symplectic transformation $\Smat$ that block diagonalizes $\Hmat$,
    \begin{equation}
        (\Smat^T)^{-1}\Hmat\Smat^{-1} \equiv \widetilde{\Hmat}= \mqty(\widetilde{\Hmat}_A & \mzero \\ \mzero & \widetilde{\Hmat}_B),\label{Appendix SW: target}
    \end{equation}    
    with $\widetilde{\vb{X}}=\Smat \vb{X}$ the transformed coordinates, such that the Hamiltonian now reads
    \begin{equation}
        H = \frac{1}{2}\vb{X}^T\Hmat\vb{X} = \frac{1}{2}\widetilde{\vb{X}}^T\widetilde{\Hmat}\widetilde{\vb{X}}.
    \end{equation}
    This symplectic transformation must satisfy $\Smat\Jmat\Smat^T = \Jmat$ and, for any symmetric matrix $\Amat$, it must hold that $\Smat = \exp(\Amat\Jmat)$ is a symplectic transformation~\cite{Goldstein:1950}. Thus, any symmetric matrix $\Amat$ is a generator of symplectic transformations, and we are interested in the generator $\Amat$ such that \cref{Appendix SW: target} is satisfied. 
    
    Because $(\Smat^T)^{-1}\Hmat\Smat^{-1} = \exp(\Jmat \Amat)\Hmat\exp(-\Amat\Jmat)$, we cannot use the well-known Baker-Cambpbell-Hausdorff expansion of $\exp(\Bmat)\Hmat\exp(-\Bmat)$ to directly evaluate the transformation. Instead, we use the general expansion 
    \begin{equation}
        \exp{\Dmat}\Hmat\exp{\Bmat} = \sum_{n=0}^\infty \frac{1}{n!}\sum_{k=0}^n\binom{n}{k} \Dmat^{n-k}\Hmat\Bmat^k.
    \end{equation}
    Substituting with $\Jmat\Amat = \Dmat$ and $-\Amat\Jmat = \Bmat$ we have
    \begin{equation}
        \widetilde{\Hmat} = \sum_{n=0}^\infty \frac{1}{n!}\sum_{k=0}^n\binom{n}{k} (-1)^k(\Jmat\Amat)^{n-k}\Hmat(\Amat\Jmat)^k.
    \end{equation}
    Furthermore, we define the transpose anticommutator of two matrices $\{\Dmat,\Bmat\}_T\equiv \Dmat\Bmat+\Bmat^T\Dmat^T$. 
    With this notation, we can show that
    \begin{equation}
    \begin{split}
        \underbrace{\{\Dmat,\dots\{\Dmat,}_{\text{n times}}\Bmat\}_T,\dots\}_T=&\sum_{k=0}^n\binom{n}{k}(\Dmat)^{n-k}\Bmat(\Dmat^T)^k  \\
        \equiv & \{\Dmat,\Bmat\}_T^{(n)}, 
    \end{split}
    \end{equation}
    which can be easily proved by induction. As $(\Jmat\Amat)^T=-\Amat\Jmat$, it follows that
    \begin{equation}
        \widetilde{\Hmat} = \sum_{n=0}^\infty \frac{1}{n!}\{\Jmat\Amat,\Hmat\}_T^{(n)}.\label{appendix SW:anticommutator-expansion}
    \end{equation}
    
    Now, we write the off-diagonal block of $\Hmat$ coupling the two sectors as $\Hmat_{AB}\equiv \lambda\Hmat_{AB}$ with $\lambda$ the customary perturbation parameter. Additionally, we express the block diagonal quadratic forms as $\Hmat_A = \Hmat_A^{(0)}+\lambda\Hmat_A'$ where $\Hmat_A^{(0)}$ and $\lambda\Hmat_A'$ correspond to its diagonal and nondiagonal entries within that sector, respectively. With similar expressions for sector $B$, we have
    \begin{equation}
    \begin{split}    
        \Hmat &= \mqty(\Hmat_A^{(0)}+\lambda\Hmat_A' & \mzero \\ \mzero & \Hmat_B'+\lambda\Hmat_B^{(1)})+\lambda\mqty(\mzero & \Hmat_{AB} \\ \Hmat_{AB}^T & \mzero )\\
        &\equiv \Hmat^{(0)} +\lambda\Hmat_D' +\lambda\Hmat_{ND}. \label{Appendix SW: H-lambda}
    \end{split}
    \end{equation}
    As is customary, we also expand $\Amat$ as a power series in $\lambda$:
    \begin{equation}
        \Amat = \sum_{n=1}^{\infty}\lambda^n\Amat^{(n)}.\label{Appendix SW: A expansion}
    \end{equation}
    Substituting \cref{Appendix SW: H-lambda,Appendix SW: A expansion} into \cref{appendix SW:anticommutator-expansion}  while also imposing condition \cref{Appendix SW: target}, we obtain the set of equations for generator $\Amat$ order by order:
    \begin{subequations}\label{Appendix SW:eq-orders-A}
    \begin{align}
        \{\Jmat\Amat^{(1)},\Hmat^{(0)}\}_T &= -\Hmat_{ND}, \label{Appendix SW:1st-order A}\\
        \{\Jmat\Amat^{(2)},\Hmat^{(0)}\}_T &= -\{\Jmat\Amat^{(1)},\Hmat_D'\}_T,\label{Appendix SW:2nd-order A}
    \end{align}
    \end{subequations}
    and so on. These expressions are analogous to those obtained for the standard quantum Schrieffer-Wolff transformation~\cite{Winkler:2003}. Using these results, we find that the transformed quadratic form of the Hamiltonian up to third-order in $\lambda$ reads
    \begin{equation}
    \begin{split}
        \widetilde{\Hmat} = &\Hmat^{(0)}+\lambda\Hmat_D'+\frac{\lambda^2}{2}\{\Jmat\Amat^{(1)},\Hmat_{ND}\}_T\\
        &+\lambda^3\left[\{\Jmat\Amat^{(2)},\Hmat_{ND}\}_T+\frac{1}{2}\{\Jmat\Amat^{(1)},\Hmat_D' \}_T^{(2)}\right]. \label{Appendix SW:H-lambda-expanded}
    \end{split}
    \end{equation}
    Thus, up to second-order in $\lambda$ only the first-order generator is needed. From \cref{Appendix SW:2nd-order A}, we have $\Amat^{(2)}=\mzero$ ($\Amat^{(2)}\neq\mzero$) if $\Hmat_D'=\mzero$ ($\Hmat_D'\neq\mzero$), and therefore taking into account only the first-order generator the expansion of $\widetilde{\Hmat}$ will be accurate up to fourth (third) order in $\lambda$. 
    
    To solve \cref{Appendix SW:1st-order A} and find $\Amat^{(1)}$ note that its diagonal (even) sector will be zero and only its nondiagonal (odd) block sector will be nonzero, and therefore we let 
    \begin{subequations}
    \begin{align}
        \Amat^{(1)} = \mqty(\mzero & \Amat_{AB}^{(1)}\\(\Amat_{AB}^{(1)})^T & \mzero),\\
        \Amat_{AB}^{(1)} = \mqty(\Amat_{x_ax_b}^{(1)} & \Amat_{x_ap_b}^{(1)} \\ \Amat_{p_ax_b}^{(1)} & \Amat_{p_ap_b}^{(1)}).
    \end{align}
    \label{Appendix SW: A(1)}
    \end{subequations}
    We also write
    \begin{subequations}
    \begin{align}
        \Hmat_{AB} &= \mqty(\Kmat_{x_ax_b} & \Kmat_{x_ap_b} \\ \Kmat_{p_ax_b} & \Kmat_{p_ap_b}), \label{Appendix SW:H_AB}\\
        \Hmat_A^{(0)} &= \text{diag}(w^a_{x1},\dots,w^a_{xn}, w^a_{p1},\dots, w^a_{pn}),
    \end{align}
    \end{subequations}
    where we have chosen the dimension of all the entries to be of frequency for reasons that will become apparent below.
    Substituting \cref{Appendix SW:J-def,Appendix SW: A(1),Appendix SW:H_AB} into \cref{Appendix SW:1st-order A} and setting the perturbation parameter $\lambda=1$ we find after some algebraic manipulations that
    \begin{subequations}\label{Appendix SW: equations-A(1)}
        \begin{align}
            (\Amat_{x_ax_b}^{(1)})_{\alpha, \beta} &= \Theta_{\alpha,\beta}^{-1}(w^a_{p\alpha}\Kmat_{x_ap_b}^{\alpha,\beta}-w^b_{p\beta}\Kmat_{p_ax_b}^{\alpha,\beta}),\\
            (\Amat_{x_ap_b}^{(1)})_{\alpha, \beta} &= \Theta_{\alpha,\beta}^{-1}(-w^a_{p\alpha}\Kmat_{x_ax_b}^{\alpha,\beta}-w^b_{x\beta}\Kmat_{p_ap_b}^{\alpha,\beta}),\\
            (\Amat_{p_ax_b}^{(1)})_{\alpha, \beta} &= \Theta_{\alpha,\beta}^{-1}(w^a_{x\alpha}\Kmat_{p_ap_b}^{\alpha,\beta}+w^b_{p\beta}\Kmat_{x_ax_b}^{\alpha,\beta}),\\
            (\Amat_{p_ap_b}^{(1)})_{\alpha, \beta} &= \Theta_{\alpha,\beta}^{-1}(-w^a_{x\alpha}\Kmat_{p_ax_b}^{\alpha,\beta}+w^b_{x\beta}\Kmat_{x_ap_b}^{\alpha,\beta}),
        \end{align}
    \end{subequations}
    where $\Theta_{\alpha,\beta}\equiv w^a_{x\alpha}w^a_{p\alpha}-w^b_{x\beta}w^b_{p\beta}=w_{\alpha}^2-\omega_{\beta}^2$ with $w_{\alpha}$ ($\omega_{\beta}$) the oscillation frequency associated with the conjugate pair $\vb{x}_{\alpha},\vb{p}_{\alpha}$ ($\vb{x}_{\beta},\vb{p}_{\beta}$) when only $\Hmat^{(0)}$ is considered. As usual, we call these frequencies the \emph{`bare'} frequencies of the problem. Here $\Kmat_{x_ax_b}^{\alpha,\beta}$ is the $\alpha,\beta$ entry of $\Kmat_{x_ax_b}$. The above expressions for the first order generator of the transformations gives us the perturbative criteria
    \begin{equation}
        {\frac{k}{\Delta_{ab}}}\ll 1,\label{Appendix SW:pert-cond}
    \end{equation}
    where $k$ is the largest entry of matrix $\Hmat-\Hmat^{(0)}$ and $\Delta_{ab}$ is the smallest frequency gap $|w^a_{\alpha}-w^b_{\beta}|$  between the bare oscillators in subsectors $A,B$.  
    Continuing with the general derivation, we substitute \cref{Appendix SW: equations-A(1)} into \cref{Appendix SW: H-lambda} and, after some algebra, we obtain the corrections up to $\mathcal{O}(k^3/\Delta_{ab}^{2})$ of the quadratic forms of both subsectors $A,B$:
    \begin{widetext}
    \begin{subequations}
        \begin{equation}
            \begin{split}
                (\widetilde{\Hmat}^A_{xx})_{\alpha,\alpha'} = \frac{1}{2}\sum_{\beta=1}^m \biggl\{&\qty(w_{p_{\beta}}\Kmat_{x_ax_b}^{\alpha,\beta}\Kmat_{x_ax_b}^{\alpha',\beta}+w_{x_{\beta}}\Kmat_{x_ap_b}^{\alpha,\beta}\Kmat_{x_ap_b}^{\alpha',\beta})\qty[\frac{1}{\Theta_{\alpha,\beta}}+\frac{1}{\Theta_{\alpha',\beta}}]\\
                & +\frac{w_{x_{\alpha}}}{\Theta_{\alpha,\beta}}\qty(\Kmat_{p_ap_b}^{\alpha,\beta}\Kmat_{x_ax_b}^{\alpha',\beta}-\Kmat_{p_ax_b}^{\alpha,\beta}\Kmat_{x_ap_b}^{\alpha',\beta})+\frac{w_{x_{\alpha'}}}{\Theta_{\alpha',\beta}}\qty(\Kmat_{p_ap_b}^{\alpha',\beta}\Kmat_{x_ax_b}^{\alpha,\beta}-\Kmat_{p_ax_b}^{\alpha',\beta}\Kmat_{x_ap_b}^{\alpha,\beta}) \biggl\},
            \end{split}
        \end{equation}
        \begin{equation}    
            \begin{split}
                (\widetilde{\Hmat}^A_{xp})_{\alpha,\alpha'}= \frac{1}{2}\sum_{\beta=1}^m \biggl\{&\qty(w_{p_{\beta}}\Kmat_{x_ax_b}^{\alpha,\beta}\Kmat_{p_ax_b}^{\alpha',\beta}+w_{x_{\beta}}\Kmat_{x_ap_b}^{\alpha,\beta}\Kmat_{p_ap_b}^{\alpha',\beta})\qty[\frac{1}{\Theta_{\alpha,\beta}}+\frac{1}{\Theta_{\alpha',\beta}}]\\
                & +\frac{w_{x_{\alpha}}}{\Theta_{\alpha,\beta}}\qty(\Kmat_{p_ap_b}^{\alpha,\beta}\Kmat_{p_ax_b}^{\alpha',\beta}-\Kmat_{p_ax_b}^{\alpha,\beta}\Kmat_{p_ap_b}^{\alpha',\beta})+\frac{w_{q_{\alpha'}}}{\Theta_{\alpha',\beta}}\qty(\Kmat_{x_ax_b}^{\alpha',\beta}\Kmat_{x_ap_b}^{\alpha,\beta}-\Kmat_{x_ap_b}^{\alpha',\beta}\Kmat_{x_ax_b}^{\alpha,\beta}) \biggl\},
            \end{split}
        \end{equation}
        \begin{equation}
            \begin{split}
                ((\widetilde{\Hmat}^A_{pp})_{\alpha,\alpha'}= \frac{1}{2}\sum_{\beta=1}^m \biggl\{&\qty(w_{p_{\beta}}\Kmat_{p_ax_b}^{\alpha,\beta}\Kmat_{p_ax_b}^{\alpha',\beta}+w_{x_{\beta}}\Kmat_{p_ap_b}^{\alpha,\beta}\Kmat_{p_ap_b}^{\alpha',\beta})\qty[\frac{1}{\Theta_{\alpha,\beta}}+\frac{1}{\Theta_{\alpha',\beta}}]\\
                & +\frac{w_{p_{\alpha}}}{\Theta_{\alpha,\beta}}\qty(\Kmat_{x_ax_b}^{\alpha,\beta}\Kmat_{p_ap_b}^{\alpha',\beta}-\Kmat_{x_ap_b}^{\alpha,\beta}\Kmat_{p_ax_b}^{\alpha',\beta})+\frac{w_{p_{\alpha'}}}{\Theta_{\alpha',\beta}}\qty(\Kmat_{x_ax_b}^{\alpha',\beta}\Kmat_{p_ap_b}^{\alpha,\beta}-\Kmat_{x_ap_b}^{\alpha',\beta}\Kmat_{p_ax_b}^{\alpha,\beta}) \biggl\}.
            \end{split}
        \end{equation}
        \label{Appendix SW:H-formulas}
    \end{subequations}
    \end{widetext}
    To obtain the corrections to $\widetilde{\Hmat}^B$, we simply exchange $\alpha$ with $\beta$, $m$ with $n$ and use $\Kmat^T$ in the above formulas. For $\widetilde{\Hmat}^A_{xp}$, the diagonal entry corrections $\alpha=\alpha'$ are equal to zero.
    
    \edit{\subsection*{Example of a symplectic SW transformation}} 
    As a simple example, take three $LC$ oscillators. Two of the oscillators are nonreciprocally coupled to the third oscillator via gyrators. Moreover, all three oscillators are capacitively coupled. Writing the Lagrangian using the flux variable on the inductances of the oscillators one readily obtains the Hamiltonian 
    \begin{subequations}
        \begin{align}
            \begin{split}
                H &= \sum_{i}\qty[\frac{{q^A_i}^2}{2C_i}+\frac{{\phi^A_i}^2}{2L_i}]+\frac{{q^B_1}^2}{2C_3}+\frac{{\phi^B_1}^2}{2L_3}+\frac{q^A_1q^A_2}{C_{k}}\\
                &+\sum_{i}\qty[\frac{q^A_iq^B_1}{C_{k}}]+\frac{q^A_1q^A_2}{C_{k}}+\sum_{i}k_{g}(\phi^A_i q^B_1-q^A_i\phi^B_1)
            \end{split}
        \end{align}
    \end{subequations}
    where $i \in\{1,2\}$. The phase space variables respect the Poisson brackets $\{\phi^a_i,q^b_j\} = \delta_{ij}\delta_{ab}$, and thus we identify $\phi\to x, q\to p$. The forms of parameters $C_i,L_i,C_k,k_{g} $ can be easily obtained following the standard approach \cite{ParraRodriguez:2019}. 
    
    Organizing the phase-space coordinate vector as $\vb{X} = (\phi^A_1,\phi^A_2, q^A_1,q^A_2, \phi^B_1, q^B_1)^T$, and rescaling them such that $\phi_i\to \phi_i/\sqrt{z_i}, q_i\to q_i\sqrt{z_i}$ with $z_i=\sqrt{L_i/C_i}$, the quadratic and symplectic forms read
    \begin{subequations}
        \begin{align}
            \Hmat &= \mqty(\Hmat_A & \Hmat_{AB} \\ \Hmat_{AB}^T & \Hmat_B), \quad \Jmat = \mqty(\Jmat_A &\mzero\\ \mzero & \Jmat_B),
        \end{align}
        with
        \begin{equation}
        \begin{split}
            \Hmat_A &= \mqty(\omega_1^A & 0 & 0 & 0 \\ 0 & \omega_2^A & 0 & 0 \\ 0 & 0 & \omega_1^A & k_q \\ 0 & 0 & k_q & \omega_2^A), \quad \Hmat_B = \omega_1^B\bone_2,\\
            \Hmat_{AB} &= \mqty( 0 & k^{1,1}_{\phi_a q_b} \\[3pt] 
             0 & k^{2,1}_{\phi_a q_b} \\[3pt]
            -k^{1,1}_{q_a\phi_b} & k^{1,1}_{q_a q_b} \\[3pt]
             -k^{2,1}_{q_a\phi_b} & k^{2,1}_{q_a q_b}), \\[3pt]
            \Jmat_A &= \mqty(\mzero & \bone_{2} \\ -\bone_{2} & \mzero), \quad \Jmat_B = \mqty(0 & 1 \\ -1 & 0).
        \end{split}
        \end{equation}
    \end{subequations}
    Here, we have defined $\omega^A_i = 1/\sqrt{L_iC_i}$, $\omega^B_1 =1\sqrt{L_3C_3}$, $k_q = 1/\sqrt{z_1z_2}C_k$, $k^{i,1}_{q_a q_b}=1/\sqrt{z_iz_3}C_k$, $k^{i,1}_{\phi_a q_b} = \sqrt{z_i/z_3}k_g$, and $k^{i,1}_{q_a\phi_b} = \sqrt{z_3/z_i}k_g$.
    
    Using these expressions, we can now obtain the perturbative corrections from \cref{Appendix SW:H-formulas} to find the effective quadratic form of sector $A$: 
    \begin{subequations}
        \begin{align}
            \begin{split}
                (\widetilde{\Hmat}^A_{\phi\phi})_{i,j} &= \frac{1}{2}\omega_1^Bk_{\phi_a q_b}^{i,1}k_{\phi_aq_b}^{j,1}\qty[\frac{1}{\Theta_{i,1}}+\frac{1}{\Theta_{j,1}}]\\
                &-\frac{\omega_i^A}{2\Theta_{i,1}}k_{q_a\phi_b}^{i,1}k_{\phi_aq_b}^{j,1}-\frac{\omega_j^A}{2\Theta_{j,1}}k_{q_a\phi_b}^{j,1}k_{\phi_aq_b}^{i,1},
            \end{split}
        \end{align}
        \begin{align}
            \begin{split}
                (\widetilde{\Hmat}^A_{\phi q})_{i,j} &= \frac{1}{2}\omega_1^Bk_{\phi_a q_b}^{i,1}k_{q_aq_b}^{j,1}\qty[\frac{1}{\Theta_{i,1}}+\frac{1}{\Theta_{j,1}}]\\
                &+\frac{\omega_i^A}{2\Theta_{i,1}}\qty(k_{q_aq_b}^{i,1}k_{q_a\phi_b}^{j,1}-k_{q_a\phi_b}^{i,1}k_{q_aq_b}^{j,1}),
            \end{split}\\
            \begin{split}
                (\widetilde{\Hmat}^A_{qq})_{i,j} &= \frac{\omega_1^B}{2}\qty(k_{q_a \phi_b}^{i,1}k_{q_a\phi_b}^{j,1}+k_{q_a q_b}^{i,1}k_{q_aq_b}^{j,1})\qty[\frac{1}{\Theta_{i,1}}+\frac{1}{\Theta_{j,1}}]\\
                &-\frac{\omega_i^A}{2\Theta_{i,1}}k_{\phi_aq_b}^{i,1}k_{q_a\phi_b}^{j,1}-\frac{\omega_j^A}{\Theta_{j,1}}k_{\phi_aq_b}^{j,1}k_{q_a\phi_b}^{i,1}\\
                &+k_q(1-\delta_{i,j}),
            \end{split}
        \end{align}
    \end{subequations}
    where $\Theta_{i,1} = [\omega_i^A)^2-(\omega^B_1)^2]$.
    For these expressions to be accurate, the perturbative criteria
    \begin{equation}
        \frac{\text{max}(k_{q_a\phi_b}^{i,1},k_{q_aq_b}^{i,1},k_{\phi_aq_b}^{i,1},k_q)}{\text{min}(|\omega^A_i-\omega^B_1|)}\ll 1
    \end{equation}
    must be satisfied.    

    For the sake of completeness, let us briefly comment on the connection of this approach to the standard quantum SW method used in \cite{Solgun:2019}. There, due to the presence of just reciprocal (capacitive) connection, only flux-flux coupling terms had to be considered. Explicitly, after arranging the phase-space coordinates as $\vb{X}^T=(\vphi_a^T , \vphi_b^T , \vb{q}_a^T , \vb{q}_b^T)$, the quadratic form has the structure 
    \begin{equation}
        \Hmat = \mqty(\Omega_{A} & \Kmat_{\phi\phi} & \mzero & \mzero \\
        \Kmat_{\phi\phi}^T & \Omega_B & \mzero & \mzero\\
        \mzero & \mzero & \Omega_A & \mzero\\
        \mzero & \mzero & \mzero & \Omega_B),
    \end{equation}
    with $\Omega_A$ ($\Omega_B$) the normal mode frequencies of sector $A$ ($B$).
    With the rescaling $\phi_i\to \phi_i/\sqrt{w_i}, q_i\to q_i\sqrt{w_i}$ the quadratic form is 
    \begin{equation}
        \Hmat = \mqty(\Omega_{A}^2 & \Mmat_{\phi\phi} & \mzero & \mzero \\
        \Mmat_{\phi\phi}^T & \Omega_B^2 & \mzero & \mzero\\
        \mzero & \mzero & \bone & \mzero\\
        \mzero & \mzero & \mzero & \bone),
    \end{equation}
    It follows that $\Smat$ will be both symplectic and orthogonal, and \cref{Appendix SW:H-formulas} reduces to
    \begin{equation}
        (\widetilde{\Hmat}_{\phi\phi}^A)_{\alpha,\alpha'}= \frac{1}{2}\sum_{\beta=1}^m \Mmat_{\phi\phi}^{\alpha,\beta}\Mmat_{\phi\phi}^{\alpha',\beta}\qty[\frac{1}{\Theta_{\alpha,\beta}}+\frac{1}{\Theta_{\alpha',\beta}}],
    \end{equation}
    which is the same as Eq. (B.15) of Ref. \cite{Winkler:2003}. Yet, in scenarios involving both flux-flux and charge-charge couplings, or flux-charge couplings, as shown here, \cref{Appendix SW:H-formulas} deviates from conventional SW formulas.

    \section{Derivation of the main results including direct capacitive, inductive, and nonreciprocal couplings} 
    \label{Derivation-main}
    In this appendix, we obtain the effective Lindbladians and Hamiltonians  for the general case with direct electrostatic and inductive coupling between the ports. In the closed system, we introduce two different approaches to include the direct capacitive (electrostatic) and inductive (magnetostatic) couplings. First, for the admittance (impedance) representation, in \cref{admittance coupler} (\cref{Impedance coupler}), we apply a numerical diagonalization of both the kinetic and inductive matrices of the junction sector and then decouple qubit modes from inner modes using SW perturbation theory. Second, in \cref{admittance coupler perturbative} (\cref{impedance coupler perturbative}) we obtain analogous coupling formulas by treating the direct couplings as a perturbation.  The main advantage of the first method  
    is that strong direct electrostatic and/or magnetostatic coupling can be accounted for exactly, but at the cost of dressing the modes, something which can
    potentially obscure the nonreciprocal coupling between the external ports. 
    
    Subsequently, our focus shifts towards the derivation of the perturbative dissipative contribution of the drive ports by deriving closed admittance and impedance formulas for the  correlated decay rates in \cref{Admitance-dissipative} and \cref{Impedance: dissipation}, respectively. Finally, we compute the drive amplitudes and classical crosstalks for the admittance (impedance) representation in \cref{driveamplitude:Y} (\cref{driveamplitude:Z}). To help the reader navigate this appendix,  \cref{tab:notation} summarizes some of the important definitions and the notation that is used.

    \renewcommand{\arraystretch}{1.0}
    \begin{table}[h]
        \centering
        \begin{tabular}{|Sc|Sc|}
            \hline 
             $\Amat_0^{-1} = \mqty(\Cmat_J & \Cmat_{JD} \\ \Cmat_{JD}^T & \Cmat_D)$ & $\Dmat_{\infty} = \mqty(\Cmat_J & \Cmat_{JD} \\ \Cmat_{JD}^T & \Cmat_D )$ \\
             \hline 
             $\Amat_{\infty} = 0$ & $\Dmat_{0} = \mqty(\Lmat_J^{-1} & \Lmat_{JD}^{-1} \\ (\Lmat_{JD}^{-1})^T & \Lmat_{D}^{-1})$ \\
             \hline
             $\Bmat_{\infty} = 0$ & $\Emat_{\infty} = \mqty(\Ymat_J^G & \Ymat_{JD}^G \\ -(\Ymat_{JD}^G)^T & \Ymat_{D}^G)$\\
             \hline
             $\Zmat^{R}(s) = \frac{\Zmat(s)+\Zmat(s)^T}{2}$ & $\Ymat^{\text{R}}(s) = \frac{\Ymat(s)+\Ymat(s)^T}{2}$\\
             \hline 
             $\Zmat^{\text{NR}}(s) = \frac{\Zmat(s)-\Zmat(s)^T}{2}$ & $\Ymat^{\text{NR}}(s) = \frac{\Ymat(s)-\Ymat(s)^T}{2}$\\
             \hline  
             $\Zmat^{ac}(s) = \sum\limits_{\beta} \frac{\Amat_{\beta}s+\Bmat_{\beta}}{\omega_{\beta}^2+s^2}$ & $\Ymat^{ac}(s) = \sum\limits_{\beta} \frac{\Dmat_{\beta}s+\Emat_{\beta}}{\omega_{\beta}^2+s^2}$\\
             \hline 
             $\Zmat^{ac,R}(s) = \frac{\Zmat^{ac}(s)+\Zmat^{ac}(s)^T}{2}$ & $\Ymat^{ac,R}(s) = \frac{\Ymat^{ac}(s)+\Ymat^{ac}(s)^T}{2}$\\
             \hline 
             $\Zmat^{ac,NR}(s) = \frac{\Zmat^{ac}(s)-\Zmat^{ac}(s)^T}{2}$ & $\Ymat^{ac,NR}(s) = \frac{\Ymat^{ac}(s)-\Ymat^{ac}(s)^T}{2}$\\
             \hline
             $\Zmat^{dc}(s) = \Amat_0/s$ & $\Ymat^{dc}(s) = \Dmat_0/s+s\Dmat_{\infty}+\Emat_{\infty}$\\
             \hline
             $\Zmat^{\text{drive}}(s) = Z_0+\Cmat_D^{-1}/s$ & $\Ymat^{\text{drive}}(s) = Z_0^{-1}+s\Cmat_D+\Lmat^{-1}_D/s$\\
             \hline
             $\Amat_{\gamma} = \frac{1}{C_{r_\gamma}}\vb{r}_{\gamma}\vb{r}_{\gamma}^T$ & $\Dmat_{\gamma} = \frac{1}{L_{r_{\gamma}}}\vb{r}_{\gamma}\vb{r}_{\gamma}^T$ \\
             \hline 
             $\Rmat^T =\mqty(\Rmat_J^T \\ \Rmat_D^T) = \mqty(\vb{r}_1,\dots,\vb{r}_{m})$ &  $\Rmat =\mqty(\Rmat_J \\ \Rmat_D) = \mqty(\vb{r}_1,\dots,\vb{r}_{m})$ \\
             \hline 
             $\Nmat^T=\mqty(\Nmat_J^T \\ \Nmat_D^T) = \mqty(\vb{n}_1^L,\dots,\vb{n}_{l}^R)$  & $\Nmat =\mqty(\Nmat_J \\ \Nmat_D) = \mqty(\vb{n}_1^L,\dots,\vb{n}_{l}^R)$  \\ 
             \hline\multicolumn{2}{|Sc|}{ $\Tmat=\mqty(\Tmat_J & \Tmat_D),\; \Tmat_J = \mqty(\Rmat_J & \Nmat_J),\; \Tmat_D = \mqty(\Rmat_D & \Nmat_D)$}\\
             \hline
             \multicolumn{2}{|Sc|}{$\Amat_{\mu} = \frac{1}{C_{g_\mu}}[\vb{n}_{\mu}^L(\vb{n}_{\mu}^L)^T+\vb{n}_{\mu}^R(\vb{n}_{\mu}^R)^T]$}\\
             \hline
             \multicolumn{2}{|Sc|}{$\Bmat_{\mu} = R_{g_\mu}\omega_{g_\mu}^2[\vb{n}_{\mu}^R(\vb{n}_{\mu}^L)^T-\vb{n}_{\mu}^L(\vb{n}_{\mu}^R)^T]$}\\
             \hline
             \multicolumn{2}{|Sc|}{$\Dmat_{\mu} = \frac{1}{L_{g_{\mu}}}[\vb{n}_{\mu}^L(\vb{n}_{\mu}^L)^T+\vb{n}_{\mu}^R(\vb{n}_{\mu}^R)^T]$}\\
             \hline
             \multicolumn{2}{|Sc|}{$\Emat_{\mu} = \frac{\omega_{g_{\mu}^2}}{R_{\mu}}[\vb{n}_{\mu}^L(\vb{n}_{\mu}^R)^T-\vb{n}_{\mu}^R(\vb{n}_{\mu}^L)^T] $}\\
             \hline
             \multicolumn{2}{|Sc|}{$(\widetilde{\Lmat}_{J})_{ij} = \delta_{ij}\widetilde{L}_{J_i},\; \widetilde{L}_{J_i}= \frac{\phi_0^2}{E_{J_i}}$}\\
             \hline
             \multicolumn{2}{|Sc|}{$\Cmat_{J_\delta}=\text{diag }\Cmat_J, \;\Lmat_{J_\delta}^{-1}=\text{diag }\Lmat_{J}^{-1}$}\\
             \hline
             \multicolumn{2}{|Sc|}{ $(\widetilde{\Wmat}_J)_{ij}=\delta_{ij}\tilde{\omega}_i,\; \tilde{\omega}_i = 1/\sqrt{\widetilde{L}_{J_i}C_{J_i}}$}\\
             \hline
        \end{tabular}
        \caption{Summary of definitions and notation.}
        \label{tab:notation}
    \end{table}
    \renewcommand{\arraystretch}{1.0}
    
    \subsection{Admittance coupling formulas with numerical diagonalization of direct coupling} \label{admittance coupler}
    
    We start by deriving the effective Hamiltonian from the admittance response. Here, and up to \cref{Admitance-dissipative}, we will not consider the drive ports. Using the Cauer circuit representation of the admittance, explained in details in \cref{Duality}, we obtain the system of equations
    
    \begin{subequations}        \label{Appendix coupling:KCL-KVL}
        \begin{align}
            -\vb{i}_J &= \Omat_C\vb{i}_C+\Omat_L\vb{i}_L+\Rmat_J\vb{i}_R+\Nmat_J\vb{i}_G+\Nmat_{\infty}\vb{i}_{G_{\infty}},\label{Appendix coupling:KCL-KVL-a}\\
            \vb{v}_C &= \Omat_C^T\vb{v}_J, \label{Appendix coupling:KCL-KVL-b}\\
            \vb{v}_L &= \Omat_L^T\vb{v}_J,\label{Appendix coupling:KCL-KVL-c}\\
            \vb{v}_R &= \Rmat_J^T\vb{v}_J\label{Appendix coupling:KCL-KVL-d}, \\
            \vb{v}_G &= \Nmat_J^T\vb{v}_J\label{Appendix coupling:KCL-KVL-e}, \\
            \vb{v}_{G_{\infty}} &= \Nmat_{\infty}^T\vb{v}_{J}, 
        \end{align}
    \end{subequations}
    where $\vb{v}_x$, $\vb{i}_x$ with $x=j,r,c,l$ correspond to the voltage and current vectors of the junctions, reciprocal resonators, capacitors and inductors respectively. The vectors $\vb{v}_G$, $\vb{i}_G$,$\vb{v}_{G_{\infty}}$, $\vb{i}_{G_{\infty}}$, 
    correspond to the currents and voltages for the left-right branches of each gyrator $\vb{v}_G=(v_{g_1}^L,v_{g_1}^R,\dots,v_{g_l}^L,v_{g_l}^R)$. 
    Moreover, $\Omat_C$, $\Omat_L$, $\Nmat_{\infty}$, $\Rmat_J=(\vb{r}_1,\dots,\vb{r}_m)$, and $\Nmat_J=(\vb{n}_1^L,\vb{n}_1^R,\dots,\vb{n}_l^L,\vb{n}_l^R)$ are the transformer matrices defined in terms of the admittance response of the junctions sector, see \cref{Appendix-A:Transformers-eqs}. Here, we changed the notation of \cref{Duality} $\Umat \to \Omat_L$ and $\Vmat\to \Omat_C$ for two reasons. First, to make more apparent their orthogonality, and second, to stress the difference with the full orthogonal decomposition of $\Dmat_{\infty}=\Vmat\overline{\Cmat}\Vmat^T$ ($\Dmat_0 = \Umat\overline{\Lmat}^{-1}\Umat^T$) when the dissipative ports are included. This will be important in \cref{Admitance-dissipative,driveamplitude:Y,Impedance: dissipation,driveamplitude:Z}.
    With an appropriate choice of flux and charge coordinates, we obtain a Lagrangian describing the dynamics of the circuit~\cite{Egusquiza:2022}. For junctions, capacitors, inductors and pure gyrators we use flux variables in writing the Lagrangian. On the other hand, for the reciprocal and nonreciprocal resonators we use charge variables. With this choice, \cref{Appendix coupling:KCL-KVL} reads
    \begin{align}
        \label{Appendix coupling:flux-charge-KCL-KVL}
        \begin{aligned}
        -\pdv{U(\vphi_J)}{\vphi_J} =&\, \Omat_C(\overline{\Cmat}_J\ddot{\vphi}_C)+\Omat_L(\overline{\Lmat}_J^{-1}\vphi_L)\\ 
                &\,+\Rmat_J\dot{\vQ}_R+\Nmat_J\dot{\vQ}_G+\Nmat_{\infty}\Ymat^G_{\infty}\dot{\vphi}_{G_{\infty}},\\
            \dot{\vphi}_C =&\; \Omat_C^T\dot{\vphi}_J,\\
            \dot{\vphi}_L =&\; \Omat_L^T\dot{\vphi}_J,\\
            \ddot{\vQ}_R+\Cmat_r^{-1}\vQ_R =&\; \Rmat_J^T\dot{\vphi}_J,\\
            \ddot{\vQ}_G+\Zmat_g\dot{\vQ}_G =&\; \Nmat_J^T\dot{\vphi}_J,\\
            \dot{\vphi}_{G_{\infty}} =&\; \Nmat_{\infty}^T\dot{\vphi}_J.
        \end{aligned}
    \end{align}
    In these expressions,  $\overline{\Cmat}_J$, $\overline{\Lmat}^{-1}_J$, and $\Cmat_r$ are diagonal matrices with entries $C_{J_i}$, $L_{J_i}$, $C_{r_{\gamma}}$ respectively. Moreover, $\Ymat^G_{\infty}$ is the admittance matrix of the ideal gyrators, and $\Zmat_g$ the impedance matrix of those forming the nonreciprocal resonators
    \begin{subequations}
        \label{Appendix coupling:Yg-Zg}
        \begin{align}
        \Ymat^G_{\infty} & = \mqty(\mqty{i\sigma_y R_1 & \\ & \ddots}  & \text{\large $\mzero$} \\
        \text{\large $\mzero$} & i\sigma_y R_k), \\
        \Zmat_g & =\mqty(\mqty{-i\sigma_y/R_1 & \\ & \ddots}  & \text{\large $\mzero$} \\
        \text{\large $\mzero$} & -i\sigma_y/R_l)=-i\Wmat_g\Sigma_y,
        \end{align}
        \begin{align}
        \Sigma_{\alpha} &= \mqty(\mqty{\sigma_{\alpha} & \\ & \ddots} & \text{\large $\mzero$}\\
        \text{\large $\mzero$} & \sigma_{\alpha}),\\
        \Wmat_g &= \mqty(\mqty{\omega_{g_1}\bone_{2}& \\ & \ddots} & \text{\large $\mzero$}\\\text{\large $\mzero$} & \omega_{g_l}\bone_{2}),
        \end{align}
    \end{subequations}
    where as above we have used $L_{g_{\mu}}=1$ implying $ \omega_{g_{\mu}}=1/R_{\mu}$, and $\sigma_{\alpha}$ corresponds to the pauli matrix $\alpha = x,y,z$.
    Substituting \cref{Appendix-A:Transformers-eqs} in \cref{Appendix coupling:flux-charge-KCL-KVL} and rearranging we obtain the set of equations
    \begin{equation}
        \begin{split}
            -\pdv{U(\vphi_J)}{\vphi_J} =&\;\Lmat_J^{-1}\vphi_J+ \Cmat_{J}\ddot{\vphi}_J+\Tmat_J\dot{\vQ}+\Ymat^G_{J}\dot{\vphi}_J,\\
            \ddot{\vQ}_R =&\; \Rmat_J^T\dot{\vphi}_J-\Cmat_r^{-1}\vQ_R,\\
            \ddot{\vQ}_G =&\; \Nmat_J^T\vphi_J-\Zmat_g\dot{\vQ}_G.
        \end{split}
    \end{equation}
    These Euler-Lagrange equations can be obtained from  the Lagrangian
    \begin{equation}
    \begin{split}
        \lgr =& \frac{1}{2}\dot{\vphi}_J^T\Cmat_J\dot{\vphi}_J-\frac{1}{2}\vphi_J^T\Lmat_J^{-1}\vphi_J-U(\vphi_J)\\
        & +\frac{1}{2}\dot{\vphi}_J^T\Ymat_J^G\vphi_J+\dot{\vphi}_J^T\Tmat_J\vQ  \\
        &+\frac{1}{2}\dot{\vQ}^T\dot{\vQ}-\frac{1}{2}\vQ^T\Cmat_I^{-1}\vQ+\frac{1}{2}\dot{\vQ}^T\Zmat_e\vQ,
    \end{split}
    \label{Appendix coupling:lgr-y}
    \end{equation}
    with $\vb{Q}^T=\mqty(\vQ_R^T & \vQ_G^T)$ the charges of the reciprocal and nonreciprocal inner modes resonators respectively, $\Tmat_J = \mqty(\Rmat_J^T & \Nmat_J^T)^T$ the transformer matrix, and
    \begin{align}
        \begin{split}
            \Cmat_I &= \mqty(\Cmat_r & \mzero \\ \mzero & \mzero), \\
            \Zmat_e &= \mqty(\mzero & \mzero \\ \mzero & \Zmat_g).
        \end{split}
    \end{align}
    Assuming that $\Cmat_J$ is not singular, we immediately obtain the classical Hamiltonian 
    \begin{equation}
    \begin{split}
        H =& \frac{1}{2}(\vq_J-\frac{1}{2}\Ymat^G_{J}\vphi_J-\Tmat_J\vQ)^T\Cmat_J^{-1}(\vq_J-\frac{1}{2}\Ymat^G_{J}\vphi_J-\Tmat_J\vQ)\\
        &+\frac{1}{2}\vphi_J^T\Lmat_J^{-1}\vphi_J+U(\vphi_J)\\
        &+\frac{1}{2}(\vPi-\frac{1}{2}\Zmat_e\vQ)^T(\vPi-\frac{1}{2}\Zmat_e\vQ)+\frac{1}{2}\vQ^T\Cmat_I^{-1}\vQ,
    \end{split}
    \label{Appendix coupling:H-y}
    \end{equation}
    where $\vq_J$, $\vPi$ are the conjugate momenta of $\vphi_J$, $\vQ$ respectively with Poisson brackets  $\{\vphi_J,\vq_J^T\}=\bone_n$, $\{\vQ,\vPi^T\}=\bone_{m+2l}$, where $n$, $m$ ($l$) are the number of qubit ports, and reciprocal (nonreciprocal) resonator modes respectively, see \cref{appendix-A:reduced fractional expansion}.
    
    Linearizing the junction potential, we write the linear part of the Hamiltonian in \cref{Appendix coupling:H-y} as $H =\vX^T\Hmat\vX/2$, with $\vX^T = (\vphi_J^T , \vq_J^T , \vQ^T , \vPi^T)$ and the quadratic form
    \begin{subequations}
    \label{Appendix coupling:H-y-quadratic-form}
    \begin{align}
        \Hmat = \mqty(\Hmat_J & \Kmat \\
        \Kmat^T & \Hmat_I),
    \end{align}
    where
    \begin{align}
        \Hmat_J &= \mqty(\Lmat_J^{-1}+\widetilde{\Lmat}_J^{-1}+(\Ymat^G_{J})^T\Cmat_J^{-1}\Ymat^G_{J}/4 & \Ymat^G_{J}\Cmat_{J}^{-1}/2 \\ \Cmat_{J}^{-1}(\Ymat^G_{J})^T/2 & \Cmat_{J}^{-1}),\\
        \Hmat_I &= \mqty(\Zmat_e^T\Zmat_e/4+\Cmat_I+\mathcal{O}(\Tmat_J^2) & \Zmat_e/2 \\ \Zmat_e^T/2 & \bone_{m+2l} ),\\
        \Kmat &= \mqty(\mathcal{O}(\Tmat_J^2) & \mzero \\ -\Cmat_{J}^{-1}\Tmat_J & \mzero).
    \end{align}
    \end{subequations}
    In these expressions, $\widetilde{\Lmat}_J$ is the diagonal matrix with entries $L_{J_{i}} = \phi_0^2/E_{J_{i}}$ obtained from the linear part of $U(\vphi_J)$, with $\phi_0=\hbar/2e$ the reduced flux quanta. We assume $||\Tmat_J||\sim ||\Ymat^G_{J}||$ and ignore the second-order terms in the coupling $\mathcal{O}(\Tmat_J^2)$, as they give rise to third and higher order corrections. 
    Furthermore, from now on, we also ignore the second-order terms in the inner modes sector as these would give fourth and higher order corrections in the effective junction sector Hamiltonian after the symplectic Schrieffer-Wolff block diagonalization, see \cref{SSW} for details. 
    
    For clarity, we separate the nonreciprocal sector from the reciprocal sector of the inner modes $H = \vX^T\Hmat\vX/2$ with $\vX^T= (\vphi_J^T , \vq_J^T , \vQ_R^T , \vPi_r^T , \vQ_G^T , \vPi_g^T)$
    \begin{equation}
        \begin{split}
            \Hmat_I &= \mqty(\Hmat_r & \mzero \\ \mzero & \Hmat_g)\\
            \Hmat_r &= \mqty(\Wmat_r^2 & \mzero \\ \mzero & \bone_{m}),\\
            \Hmat_g &= \mqty(\Zmat_g^T\Zmat_g/4 & \Zmat_g/2 \\ \Zmat^T_g/2 & \bone_{2l})\label{Appendix coupling:H-quad-I-(c)} ,
        \end{split}
    \end{equation}
    and the coupling matrix to first order then becomes
    \begin{equation}
        \begin{split}
            \Kmat &= \mqty(\mzero & \mzero &\mzero & \mzero\\
            -\Cmat_{J}^{-1}\Rmat_J & \mzero & -\Cmat_{J}^{-1}\Nmat_J & \mzero).
        \end{split}
    \end{equation}
    Since every gyrator imposes a constraint in phase-space, there is only one dynamical pair of conjugate variables (instead of two) per capacitor-gyrator-capacitor circuit. In other words, we have redundant nondynamical variables in the above Hamiltonian, see Ref.~\cite{Egusquiza:2022} for further details. In \cref{Elimination-Nondynamical} we exactly eliminate these nondynamical modes to obtain a Hamiltonian which can be immediately quantized.  Here, we take an alternative approach and approximately eliminate these nondynamical modes ignoring the second-order terms in the inner mode sector. To second-order in perturbation theory, both the exact and the approximate elimination give the same result. 
    
    Continuing our derivation, we now proceed with the approximate elimination of the nondynamical modes with the following symplectic transformation
    \begin{equation}
    \label{Appendix coupling:SI-Y}
        \begin{split}
            \Smat_I &= \mqty(\Smat_r & \mzero \\ \mzero & \Smat_g),\\
            \Smat_r &= \mqty(\Wmat_r^{1/2} & \mzero \\ \mzero & \Wmat_r^{-1/2}), \\
            \Smat_g &= \mqty(\bone_{2l}/2 & -\Sigmamat_x \\ \Sigmamat_x/2 & \bone_{2l})\mqty(\Wmat_g^{1/2} & \mzero \\ \mzero & \Wmat_g^{-1/2}).
        \end{split}
    \end{equation}
    For convenience, we have rescaled the inner modes such that every entry in the transformed quadratic form has dimensions of frequency.
    By mixing charge and fluxes of the nonreciprocal resonators, $\Smat_g$ diagonalizes the nonreciprocal sector and decouples dynamical from nondynamical modes. Indeed, we have that
    \begin{subequations}
    \label{Appendix coupling:H-g-Omega-g}
        \begin{align}
            &(\Smat_g^T)^{-1}\Hmat_g\Smat_g^{-1}\equiv \overline{\Hmat}_g = \mqty(\overline{\Wmat}_g & \mzero \\ \mzero & \overline{\Wmat}_g), \\
            &\overline{\Wmat}_g = \mqty(\mqty{\omega_{g_1}(\bone_2+\sigma_z)/2 & \\ & \ddots} & \text{\large$\mzero$}\\\text{\large$\mzero$} & \omega_{g_l}(\bone_2+\sigma_z)/2).\label{Appendix coupling:H-g-Omega-g-(b)}
        \end{align}
    \end{subequations}
    The flux-flux and charge-charge subsectors of $\Hmat_J$ are diagonalized with the additional symplectic transformation
    \begin{equation}
    \label{Appendix coupling:Sj}
        \Smat_J = \mqty(\overline{\Wmat}_J^{1/2}\Omat_{\Omega}^T\overline{\Cmat}_J^{1/2}\Omat_C^T & \mzero \\ 
        \mzero & \overline{\Wmat}_J^{-1/2}\Omat_{\Omega}^T\overline{\Cmat}_J^{-1/2}\Omat_C^T),
    \end{equation}
    where $\Omat_{\Omega}$ is the orthogonal matrix that diagonalizes $\overline{\Cmat}_J^{-1/2}\Omat_C^T(\Lmat_J^{-1}+\widetilde{\Lmat}_J^{-1})\Omat_C\overline{\Cmat}_J^{-1/2}$, such that the dressed frequencies of the junction sector due to the direct coupling between junction ports excluding the coupling from $\Ymat_{J}^G$ are 
    \begin{equation}
        \overline{\Wmat}_J^2=\Omat_{\Omega}^T\overline{\Cmat}_J^{-1/2}\Omat_C^T(\Lmat_J^{-1}+\widetilde{\Lmat}_J^{-1})\Omat_C\overline{\Cmat}_J^{-1/2}\Omat_{\Omega}.
    \end{equation} 
    After this transformation, the quadratic form in the junction sector  $(\Smat_J^T)^{-1}\Hmat_j\Smat_J^{-1} \equiv \overline{\Hmat}_j$ reads
    \begin{equation}
    \label{appendix coupling:Hmat-J-Y_JG}
        \begin{split}
            \overline{\Hmat}_j &= \mqty(\overline{\Wmat}_J+\Wmat_{\Ymat_J^G} & \overline{\Wmat}_J^{-1/2}\overline{\Ymat}^G_{J}\overline{\Wmat}_J^{1/2}/2 \\ \overline{\Wmat}_J^{1/2}(\overline{\Ymat}^G_{J})^T\overline{\Wmat}_J^{-1/2}/2 & \overline{\Wmat}_J),\\
            \overline{\Ymat}^G_{J} &\equiv \Omat_{\Omega}^T\overline{\Cmat}_J^{-1/2}\Omat_C^T\Ymat^G_{J}\Omat_C\overline{\Cmat}_J^{-1/2}\Omat_{\Omega},
        \end{split}
    \end{equation}\\[3ex]
    with $\Wmat_{\Ymat_J^G}=\overline{\Wmat}_J^{-1/2}(\overline{\Ymat}^G_{J})^T\overline{\Ymat}^G_{J}\overline{\Wmat}_J^{-1/2}/4$. Applying both transformations
    \begin{equation}
    \label{Appendix coupling:S}
        \Smat = \mqty(\Smat_J & \mzero\\ \mzero & \Smat_I),
    \end{equation}
    we get the quadratic form 
    \begin{subequations}
    \label[equations]{Appendix coupling:H-K-R-N}
        \begin{equation}
            (\Smat^T)^{-1}\Hmat\Smat^{-1}\equiv \overline{\Hmat} = \mqty(\overline{\Hmat}_j & \overline{\Kmat}_r & \overline{\Kmat}_g \\
             \overline{\Kmat}_r^T & \overline{\Hmat}_r & \mzero \\ 
             \overline{\Kmat}_g^T &\mzero & \overline{\Hmat}_g), 
        \end{equation}
    \\
        where
        \begin{equation}
        \begin{split}
            \overline{\Kmat}_r &= \mqty(\mzero& \mzero \\ \Kmat_{r}^{qQ} & \mzero), \label{Appendix coupling:H-K-R-N-(d)}\\
            \overline{\Kmat}_g &= \mqty(\mzero & \mzero \\ \Kmat_{g}^{qQ} & \Kmat_{g}^{q\Pi}),
        \end{split}
        \end{equation}
        and
        \begin{equation}
        \label{Appendix coupling:Kmats-Y}
        \begin{split}
            \Kmat_{r}^{qQ} &=  -\overline{\Wmat}_J^{1/2}\overline{\Rmat}_J\Wmat_r^{-1/2},\\
            \Kmat_{g}^{qQ} &=  -\overline{\Wmat}_J^{1/2}\overline{\Nmat}_J\Wmat_g^{-1/2},\\
            \Kmat_{g}^{q\Pi} &=  -\overline{\Wmat}_J^{1/2}\overline{\Nmat}_J\Sigmamat_x\Wmat_g^{-1/2},\\
            \overline{\Rmat}_J&=\Omat_{\Omega}^T\overline{\Cmat}_J^{-1/2}\Omat_C^T\Rmat_J,\\
            \overline{\Nmat}_J&=\Omat_{\Omega}^T\overline{\Cmat}_J^{-1/2}\Omat_C^T\Nmat_J.
        \end{split}
        \end{equation}
    \end{subequations}
    To get the effective junction sector Hamiltonian up to second-order in the couplings we can now directly apply our formulas \cref{Appendix SW:H-formulas} derived in \cref{SSW}. Doing so, we obtain the effective quadratic form of the junction sector $\widetilde{\Hmat}_J$ with entries
    \begin{widetext}
    \begin{subequations}
    \label{Appendix coupling:Hy-ssw-1}
        \begin{align}
        \begin{split}
        (\widetilde{\Hmat}_J^{qq})_{ij} &=\;\delta_{ij}\overline{\omega}_{i} +\frac{1}{2}\sum_{\gamma=1}^m \omega_{r_\gamma}(\Kmat_r^{qQ})_{i,\gamma}(\Kmat_r^{qQ})_{j,\gamma}\qty[\frac{1}{\Theta_{i,\gamma}}+\frac{1}{\Theta_{j,\gamma}}]\\
        &\;+ \frac{1}{2}\sum_{\beta=1}^{2l}\omega_{g_{\beta}}\qty[(\Kmat_g^{qQ})_{i,\beta}(\Kmat_g^{qQ})_{j,\beta}+(\Kmat_g^{q\Pi})_{i,\beta}(\Kmat_g^{q\Pi})_{j,\beta}]\qty[\frac{1}{\Theta_{i,\beta}}+\frac{1}{\Theta_{j,\beta}}],
        \end{split}\\
        (\widetilde{\Hmat}_J^{\phi q})_{ij} &=\; \frac{1}{2}\sum_{\beta=1}^{2l}\frac{\overline{\omega}_{i}}{\Theta_{i,\beta}}\qty[(\Kmat_g^{q\Pi})_{i,\beta}(\Kmat_g^{qQ})_{j,\beta}-(\Kmat_g^{qQ})_{i,\beta}(\Kmat_g^{q\Pi})_{j,\beta})]+\frac{1}{2}\sqrt{\frac{\overline{\omega}_{j}}{\overline{\omega}_{i}}}(\overline{\Ymat}^G_{J})_{ij},\\
        (\widetilde{\Hmat}_J^{\phi \phi})_{ij}&=\;\delta_{ij}\overline{\omega}_{i}+(\Wmat_{\Ymat_J^G})_{ij},\label{Appendix coupling:Hy-ssw-phiphi-1}
        \end{align}
    \end{subequations}
    \end{widetext}
    where $\overline{\omega}_{i}$ are the diagonal entries of $\overline{\Wmat}_J$, and $\Theta_{i,\gamma}=\overline{\omega}_{i}^2-\omega_{r_{\gamma}}^2$, $\Theta_{i,\beta}=\overline{\omega}_{i}^2-\omega_{g_{\beta}}^2$. Moreover, $\omega_{g_{\beta}}$ is the $\beta$th diagonal entry of $\overline{\Wmat}_g$, which for odd $\beta$ is the   nonreciprocal oscillator frequency $\omega_{g_{\beta}}= \omega_{g_{\mu}}$ with $\mu=(\beta+1)/2$ while $\omega_{g_{\beta}}=0$ for even $\beta$. 
    From \cref{Appendix coupling:H-K-R-N} we have
    \begin{subequations}
    \label{Appendix coupling:K-entries}
        \begin{align}
            (\Kmat_{r}^{qQ})_{i,\gamma} =& -\sqrt{\overline{\omega}_{i}/\omega_{r_{\gamma}}}(\overline{\vb{r}}_{\gamma})_{i}, \\
            (\Kmat_{g}^{qQ})_{i,\beta} =&
            \begin{cases}
                -\sqrt{\overline{\omega}_{i}/\omega_{g_{\mu}}}(\overline{\vb{n}}_{\mu}^{L})_{i},\text{ for odd $\beta$ } \\
                -\sqrt{\overline{\omega}_{i}/\omega_{g_{\mu}}}(\overline{\vb{n}}_{\mu}^{R})_{i},\text{ for even $\beta$ }
            \end{cases}\\
            (\Kmat_{g}^{q\Pi})_{i,\beta} =&
            \begin{cases}
                -\sqrt{\overline{\omega}_{i}/\omega_{g_{\mu}}}(\overline{\vb{n}}_{\mu}^{R})_{i},\text{ for odd $\beta$ } \\
                -\sqrt{\overline{\omega}_{i}/\omega_{g_{\mu}}}(\overline{\vb{n}}_{\mu}^{L})_{i},\text{ for even $\beta$},
            \end{cases} 
        \end{align}
    \end{subequations}
    where $\overline{\vb{r}}_{\gamma}$, $\overline{\vb{n}}_{\mu}$ are the row vectors of $\overline{\Rmat}$, $\overline{\Nmat}$, and $\mu = (\beta+1)/2$ ($\mu =\beta/2$) for odd (even) $\beta$. Substituting \cref{Appendix coupling:Hy-ssw-1} in \cref{Appendix coupling:K-entries} and rearranging the term  we obtain
    \begin{subequations}
    \label{Appendix coupling:Hy-ssw-2}
        \begin{align}
            \begin{split}
                (\widetilde{\Hmat}_J^{qq})_{ij} =\;\delta_{ij}\overline{\omega}_{i}\;&\\
                +\frac{\sqrt{\overline{\omega}_{i}\overline{\omega}_{j}}}{2}\sum_{\gamma=1}^m (\overline{\vb{r}}_{\gamma})_{i}(&\overline{\vb{r}}_{\gamma})_{j}\qty[\frac{1}{\Theta_{i,\gamma}}+\frac{1}{\Theta_{j,\gamma}}]\\
                + \frac{\sqrt{\overline{\omega}_{i}\overline{\omega}_{j}}}{2}\sum_{\mu=1}^{l}\bigg\{[(\overline{\vb{n}}^L_{\mu}&)_{i}(\overline{\vb{n}}^L_{\mu})_{j}+(\overline{\vb{n}}^R_{\mu})_{i}(\overline{\vb{n}}^R_{\mu})_{j}]\\
                &\;\; \times \qty[\frac{1}{\Theta_{i,\mu}}+\frac{1}{\Theta_{j,\mu}}]\bigg\},
        \end{split}\\
        \begin{split}
            (\widetilde{\Hmat}_J^{\phi q})_{ij} = \frac{1}{2}\sqrt{\frac{\overline{\omega}_{j}}{\overline{\omega}_{i}}}&(\overline{\Emat}_{\infty})_{ij}\\
            + \frac{\sqrt{\overline{\omega}_{i}\overline{\omega}_{j}}}{2}\sum_{\mu=1}^{l}\bigg\{[(\overline{\vb{n}}^L_{\mu}&)_{i}(\overline{\vb{n}}^R_{\mu})_{j}-(\overline{\vb{n}}^R_{\mu})_{i}(\overline{\vb{n}}^L_{\mu})_{j}]\\
                &\times \frac{1}{\omega_{g_{\mu}}}\qty[\frac{1}{\overline{\omega}_{i}}-\frac{\overline{\omega}_{i}}{\Theta_{i,\mu}}]\bigg\}.
        \end{split}
        \end{align}
    \end{subequations}
    Once again rearranging, and substituting in \cref{Appendix coupling:Hy-ssw-2} with \cref{Appendix-A:Transformers-eqs} (remembering our choice of $L_{g_{\mu}}=L_{r_{\gamma}}=1$) we get
    \begin{subequations}
    \label{Appendix coupling:Hy-ssw-final}
        \begin{align}
            \begin{split}
                (\widetilde{\Hmat}_J^{qq})_{ij} =&\;\delta_{ij}\overline{\omega}_{i}\;\\
                +\frac{i}{2}&\sqrt{\overline{\omega}_{i}\overline{\omega}_{j}}\biggl[\frac{\overline{\Ymat}^{\text{ac,R}}_{ij}(\overline{\omega}_{i})}{\overline{\omega}_{i}}+\frac{\overline{\Ymat}^{\text{ac,R}}_{ij}(\overline{\omega}_{j})}{\overline{\omega}_{j}}\biggr],
        \end{split} \\ 
        \begin{split}
            (\widetilde{\Hmat}_J^{\phi q})_{ij} =& \;\frac{1}{2}\sqrt{\frac{\overline{\omega}_{j}}{\overline{\omega}_{i}}}\overline{\Ymat}^{\text{NR}}_{ij}(\overline{\omega}_{i}), 
        \end{split}
        \end{align}
    \end{subequations}
    with
    \begin{subequations}
    \label{Appendix coupling:overline-Ymat}
        \begin{align}
             \overline{\Ymat}(s) &=\Omat_{\Omega}^T\overline{\Cmat}_J^{-1/2}\Omat_C^T \Ymat(s)\Omat_C\overline{\Cmat}_J^{-1/2}\Omat_{\Omega},\black \\
            \overline{\Ymat}^{\text{ac,R}}(s) &= \sum_{\gamma}\frac{\overline{\Dmat}_{\gamma}s}{\omega_{r_{\gamma}}^2+s^2}+\sum_{\mu}\frac{\overline{\Dmat}_{\mu}s}{\omega_{g_{\mu}}^2+s^2},\\
            \overline{\Ymat}^{\text{NR}}(s) &= \overline{\Emat}_{\infty}+\sum_{\mu}\frac{\overline{\Emat}_{\mu}}{\omega_{g_{\mu}}^2+s^2},\\
            \overline{\Dmat}_{\gamma (\mu)} &= \Omat_{\Omega}^T\overline{\Cmat}_J^{-1/2}\Omat_C^T \Dmat_{\gamma (\mu)}\Omat_C\overline{\Cmat}_J^{-1/2}\Omat_{\Omega},\\
            \overline{\Emat}_{\mu (\infty)} &= \Omat_{\Omega}^T\overline{\Cmat}_J^{-1/2}\Omat_C^T \Emat_{\mu (\infty)}\Omat_C\overline{\Cmat}_J^{-1/2}\Omat_{\Omega}. 
        \end{align}
    \end{subequations}
    Hence, the classical effective Hamiltonian for the junctions sector will be
    \begin{equation}
    \begin{split}
        \widetilde{H}_J &=\frac{1}{2}\sum_{i} \qty(\overline{\omega}_{i}(\wphi_{i}^2+\wq_{i}^2)-\Im{\overline{\Ymat}_{ij}^{ac,R}(\overline{\omega}_{i})}\wq_{i}^2)\\
        +&\sum_{i\neq j}\Bigl(\frac{(\Wmat_{\Ymat_J^G})_{ij}}{2}\widetilde{\phi}_{i}\widetilde{\phi}_{j}+ \frac{(\widetilde{\Hmat}_J^{qq})_{ij}}{2}\wq_{i}\wq_{j}\\
        &\qq{}\qq{}+(\widetilde{\Hmat}_J^{\phi q})_{ij}\wphi_{i}\wq_{j}\Bigr)+U_{\text{nl}}(\widetilde{\phi},\widetilde{q}).
    \end{split}
    \end{equation}
    Where $U_{\text{nl}}(\widetilde{\phi},\widetilde{q})$ stands for the nonlinear part of the junctions potential, now a function of both flux and charges in this new frame. 
    These final frame coordinates  are connected to the original ones through the transformation
    \begin{equation}
        \widetilde{\vb{X}} =\Smat_{sw}\Smat\vb{X}.
    \end{equation}
    How to obtain $\Smat_{sw}$ to any desired order is explained in \cref{SSW}.
    Promoting the classical variables to quantum operators ($\hbar =1$)
    \begin{subequations}
        \begin{align}
            \wphi_{i} &= (\bdag_{i}+\blow_{i})/\sqrt{2},\\
            \wq_{i} &= i(\bdag_{i}-\blow_{i})/\sqrt{2},
        \end{align}
    \end{subequations}
    rearranging, and applying the rotating wave approximation (RWA) we finally obtain the effective quantum Hamiltonian
    \begin{equation}
    \label{Appendix coupling:quantum-Hy-1}
        \begin{split}
        \hat{H}_J = & \sum_{i}\qty(\overline{\omega}_{i}-\frac{\Im[\overline{\Ymat}_{ii}^{ac,R}(\overline{\omega}_{i})]}{2})\bdag_{i}\blow_{i}\\
        +&\sum_{i\neq j}\qty(J_{ij}\blow_{i}\bdag_{j}+J_{ij}^*\bdag_{i}\blow_{j})+U_{\text{nl}}(\widetilde{\phi},\widetilde{q}),
        \end{split}
    \end{equation}
    with
    \begin{equation}
    \label{Appendix coupling:J-y}
    \begin{aligned}
        J_{ij} =&\,\frac{i}{4}\sqrt{\overline{\omega}_{i}\overline{\omega}_{j}}\qty[\frac{\overline{\Ymat}^{\Sigma}_{ij}(\overline{\omega}_{i})}{\overline{\omega}_{i}}+\frac{\overline{\Ymat}^{\Sigma}_{ij}(\overline{\omega}_{j})}{\overline{\omega}_{j}}]\\ 
        &\;+\frac{((\overline{\Ymat}^G_J)^T\overline{\Ymat}^G_J)_{ij}}{8\sqrt{\overline{\omega}_{i}\overline{\omega}_{j}}},
    \end{aligned}
    \end{equation}
    where $\overline{\Ymat}^{\Sigma}(s) = \overline{\Ymat}^{\text{ac,R}}(s)+\overline{\Ymat}^{\text{NR}}(s)$. When including the nonlinearities, the correction to the frequencies done in the main text (\cref{paper-eq:overline-w-y}) will hold to first order if the nondiagonal elements of the $\Omat_C$ ($\Omat_L$) transformations are in the same order as the transformer ratios. 
    If that is not the case,  a fully numerical treatment of the nonlinearities can be done in a systematic manner, see \cref{NL} for details.
    Finally, in the case with neither direct capacitive, inductive or nonreciprocal coupling ($\Omat_C=\Omat_L=\Omat_{\Omega}=\bone_n$, $\Ymat_J^G=\mzero$) we obtain \cref{paper:eff-J-y} of the main text.
    
    \subsection{Fully perturbative admittance coupling formulas}
    \label{admittance coupler perturbative}
    
    We now proceed to derive the effective Hamiltonian for the qubit sector, treating the direct coupling between qubit ports as a perturbation. We let $\Lmat_{J_\delta}^{-1}$ ($\Cmat^{-1}_{J_{\delta}}$) and $\Lmat_{J_{\chi}}^{-1}$ ($\Cmat^{-1}_{J_{\chi}}$) respectively be the diagonal and off-diagonal entries of $\Lmat_J^{-1}$ ($\Cmat^{-1}_J$).  We treat these off-diagonal terms 
    as first order perturbations. The derivation follows the same steps as above up to \cref{Appendix coupling:H-g-Omega-g} and, as a result,  the quadratic form of the junction sector reads
    \begin{equation}
        \begin{split}
            \Hmat_J &= \mqty(\overline{\Lmat}_J^{-1}+\Lmat_{J_\chi}^{-1} & \Ymat_J^G\Cmat_J^{-1}/2 \\ \Cmat_J^{-1}(\Ymat_J^G)^T/2 & \Cmat_{J_\delta}^{-1}+\Cmat_{J_\chi}^{-1}),
        \end{split}
    \end{equation}
    where 
    \begin{equation}
    \label{Appendix coupling:overline-Lmat-Y}
        \overline{\Lmat}_J^{-1} = \widetilde{\Lmat}_J^{-1}+\Lmat_{J_\delta}^{-1}.
    \end{equation}
    Hence, following \cref{SSW} we simply add the direct couplings at the end of our perturbative treatment to obtain the effective Hamiltonian for the junction sector. Now, instead of transforming the junction sector with \cref{Appendix coupling:Sj} we simply rescale it with
    \begin{equation}
        \overline{\Smat}_J = \mqty(\overline{\Gmat}_J^{1/2} & \mzero \\ \mzero & \overline{\Gmat}_J^{-1/2}),\label{Appendix coupling:Sj-Y-p}
    \end{equation}
    where $\overline{\Gmat}_J=(\Cmat_{J_\delta}/\overline{\Lmat}_J)^{1/2}$. Thus, for the rescaled junction sector, we have
    \begin{equation}
    \label{Appendix coupling:HmatJ-Y-p}
    \begin{split}
        \overline{\Hmat}_J &= [(\overline{\Smat}_J)^{T}]^{-1}\Hmat_J\overline{\Smat}_J^{-1}\\
        &=\mqty(\overline{\Wmat}_J+\Wmat_{L_\chi}+\Wmat_{\Ymat_J^G} & \Kmat_{\phi q}/2\\ \Kmat^T_{\phi q}/2 & \overline{\Wmat}_J+\Wmat_{C_\chi}),
    \end{split}
    \end{equation}
    with $\overline{\Wmat}_J=(\overline{\Lmat}_{J}\Cmat_{J_\delta})^{-1/2}$,
    \begin{equation}
    \label{Appendix coupling:Lmat-Cmat-Yp}
        \begin{split}
            \Wmat_{L_\chi}&=\overline{\Wmat}_J^{-1/2}\Cmat_{J_\delta}^{-1/2}\Lmat_{J_\chi}^{-1}\Cmat_{J_\delta}^{-1/2}\overline{\Wmat}_J^{-1/2}, \\
            \Wmat_{C_\chi}&=\overline{\Wmat}_J^{1/2}\Cmat_{J_\delta}^{1/2}\Cmat_{J_\chi}^{-1}\Cmat_{J_\delta}^{1/2}\overline{\Wmat}_J^{1/2},\\
            \Wmat_{\Ymat_J^G} &= \overline{\Wmat}_J^{-1/2}(\overline{\Ymat_J^G})^T\overline{\Ymat_J^G}\overline{\Wmat}_J^{-1/2}/4,
        \end{split}
    \end{equation}
     and
     \begin{equation}
         \begin{split}
             \Kmat_{\phi q} &=\overline{\Gmat}_J^{-1/2}\Ymat_J^G\Cmat_J^{-1}\overline{\Gmat}_J^{1/2}\\ 
             &=\overline{\Wmat}_J^{-1/2}\overline{\Ymat}_J^G\overline{\Wmat}_J^{1/2}+\overline{\Gmat}_J^{-1/2}\Ymat_J^G\Cmat_{J_\chi}^{-1}\overline{\Gmat}_J^{1/2},
        \end{split}
     \end{equation}
    where $\overline{\Ymat}_J^G = \Cmat_{J_\delta}^{-1/2}\Ymat_J^G\Cmat_{J_\delta}^{-1/2}$. 
    
    The symplectic transformation $\Smat$ of \cref{Appendix coupling:S} here takes the same form except for the replacement $\Smat_J\to \overline{\Smat}_J$, and \cref{Appendix coupling:H-K-R-N} are also the same under the mappings 
    \begin{subequations}
        \begin{align}
            \overline{\Wmat}_J &\to (\overline{\Lmat}_J\Cmat_{J_{\delta}})^{-1/2},\\
            \overline{\Rmat}&\to (\Cmat_{J_\delta})^{1/2}\Cmat_J^{-1}\Rmat = \Cmat_{J_{\delta}}^{-1/2}\Rmat +\Cmat_{J_{\delta}}^{-1/2}\Cmat_{J_\chi}^{-1}\Rmat , \\
            \overline{\Nmat}&\to (\Cmat_{J_\delta})^{1/2}\Cmat_J^{-1}\Nmat =\Cmat_{J_{\delta}}^{-1/2}\Nmat +\Cmat_{J_{\delta}}^{-1/2}\Cmat_{J_\chi}^{-1}\Nmat.
        \end{align}
    \end{subequations}
    Ignoring the second-order coupling terms  proportional to $\Cmat_{J_\chi}^{-1}\Rmat$, $\Cmat_{J_\chi}^{-1}\Nmat$, 
 $\Ymat_J^G\Cmat_{J_{\chi}}$ which give rise to third-order corrections and higher, it follows that \cref{Appendix coupling:Hy-ssw-phiphi-1} and \cref{Appendix coupling:Hy-ssw-final} here take the form
    \begin{subequations}
    \label{Appendix coupling:Hy-ssw-entries-2}
        \begin{align}
        \begin{split}
            (\widetilde{\Hmat}_J^{qq})_{ij} &= \delta_{ij}\overline{\omega}_{i}+ (\Wmat_{C_\chi})_{ij}\\
            +\frac{i}{2}&\sqrt{\overline{\omega}_{i}\overline{\omega}_{j}}\biggl[\frac{\overline{\Ymat}^{\text{ac,R}}_{ij}(\overline{\omega}_{i})}{\overline{\omega}_{i}}+\frac{\overline{\Ymat}^{\text{ac,R}}_{ij}(\overline{\omega}_{j})}{\overline{\omega}_{j}}\biggr],
        \end{split}\\
            (\widetilde{\Hmat}_J^{\phi q})_{ij} &= \frac{1}{2}\sqrt{\frac{\overline{\omega}_{j}}{\overline{\omega}_{i}}}\overline{\Ymat}^{\text{NR}}_{ij}(\overline{\omega}_{i}),\\
            (\widetilde{\Hmat}_J^{\phi \phi})_{ij} &= \delta_{ij}\overline{\omega}_{i}+(\Wmat_{L_\chi})_{ij}+(\Wmat_{\Ymat_J^G})_{ij},
        \end{align}
    \end{subequations}
    with $\overline{\Ymat}^{\text{ac,R}}(s),\overline{\Ymat}^{\text{NR}}(s), \overline{\Ymat}_J^G$ defined as in \cref{Appendix coupling:overline-Ymat} and \cref{appendix coupling:Hmat-J-Y_JG} with $\Omat_C=\Omat_{\Omega}=\bone_n$. 
    
    Therefore, the effective quantum Hamiltonian for the junction sector after applying the rotating-wave approximation is
    \begin{subequations}    
    \label{Appendix coupling:quantum-HY-J-2}
        \begin{equation}
            \label{Appendix coupling:quantum-Hy-2}
            \begin{split}
            \hat{H}_J = & \sum_{i}\qty(\overline{\omega}_{i}-\frac{1}{2}\Im{\overline{\Ymat}_{ii}^{ac}(\overline{\omega}_{i})})\bdag_{i}\blow_{i}\\
            +&\sum_{i\neq j}\qty(J_{ij}\blow_{i}\bdag_{j}+J_{ij}^*\bdag_{i}\blow_{j})+U_{\text{nl}}(\widetilde{\phi},\widetilde{q}),
            \end{split}
        \end{equation}
        with
       \begin{equation}
            \begin{split}
                J_{ij} =& \frac{i}{4}\sqrt{\overline{\omega}_{i}\overline{\omega}_{j}}\qty[\frac{\overline{\Ymat}^{\Sigma}_{ij}(\overline{\omega}_{i})}{\overline{\omega}_{i}}+\frac{\overline{\Ymat}^{\Sigma}_{ij}(\overline{\omega}_{j})}{\overline{\omega}_{j}}]\\
                +&\frac{1}{2}\qty[(\Wmat_{C_\chi})_{ij}+(\Wmat_{L_\chi})_{ij}]+(\Wmat_{\Ymat_J^G})_{ij}/2,
                \label{Appendix coupling:J-y-2}
            \end{split}
        \end{equation}
    \end{subequations}
    and $\overline{\Ymat}^{\Sigma}(s)= \overline{\Ymat}^{\text{ac,R}}(s)+\overline{\Ymat}^{\text{NR}}(s)$, i.e., the rescaled admittance response excluding the reciprocal dc part of the admittance ($\Dmat_0$, $\Dmat_\infty$). The frequencies in the second line are given in  \cref{Appendix coupling:Lmat-Cmat-Yp}, where $\Wmat_{C_\chi}$ ($\Wmat_{L_\chi}$) captures the correction from direct capacitive (inductive) coupling, and $\Wmat_{\Ymat_J^G}$ is a second-order correction from the direct nonreciprocal coupling between the ports. In \cref{NL}, we shift the frequencies entering \cref{Appendix coupling:Lmat-Cmat-Yp}, \cref{Appendix coupling:Hy-ssw-entries-2} and \cref{Appendix coupling:quantum-HY-J-2} to include the nonlinearities up to first order in perturbation theory. 
    
    \subsection{Impedance coupling formulas with numerical diagonalization of direct coupling} \label{Impedance coupler}
    
    We now derive the effective Hamiltonian from the Cauer representation (see \cref{fig:Duality}) of the impedance response. The derivation is similar to the one above for the admittance with some key differences that we highlight. From the transformers constitutive equations we have
    \begin{subequations}
        \begin{align}
            \vb{v}_J &= \Omat_C^T\vb{v}_C+\Rmat_J^T\vb{v}_R+\Nmat_J^T\vb{v}_G, \\
            \vb{i}_C &= -\Omat_C\vb{i}_J,\\
            \vb{i}_R &= -\Rmat_J\vb{i}_J,\\
            \vb{i}_G &= -\Nmat_J\vb{i}_J.
        \end{align}
    \end{subequations}
    We assume that $\Cmat_J$ is of full rank, and thus $\Omat_C$ is orthogonal.  Assigning flux variables for each junction and capacitive branch we have that the equations of motion of the circuit are derivable from the Lagrangian
    \begin{equation}
        \lgr = \frac{1}{2}\dot{\vphi}^T\Cmat\dot{\boldsymbol{\vphi}}-U(\vphi_J)-\frac{1}{2}\boldsymbol{\vphi}^T\Mmat_0\vphi+\frac{1}{2}\dot{\vphi}^T\Gmat\vphi,
        \label{Appendix coupling: Lgr-0}
    \end{equation}
    with $\vphi^T = (\vphi_J^T , \vphi_I^T)$, where $\vphi_J$ ($\vphi_I$) correspond to the fluxes in the external ports (inner mode resonators and gyrators). The kinetic, inductive and nonreciprocal matrices read
    \begin{subequations}\label{Appendix coupling:C-M-G}
        \begin{align}
            \Cmat &= \mqty(\Cmat_J & -\Cmat_J\Tmat_J^T\\ -\Tmat_J\Cmat_J & \bone_n + \Tmat_J\Cmat_J\Tmat_J^T),\\
            \Mmat_0 &= \mqty(\mzero & \mzero \\ \mzero & \Lmat_{R_e}^{-1}), \quad \Lmat_{R_e} = \mqty(\Lmat_R & \mzero \\ \mzero & \mzero),\\ 
            \Gmat &= \mqty(\mzero & \mzero\\ \mzero & \Ymat_e),\quad \Ymat_e = \mqty(\mzero & \mzero \\ \mzero & \Ymat),
        \end{align}
    \end{subequations}
    where $\Cmat_J=\Omat_C^T\overline{\Cmat}_J\Omat_C$. Note that in Ref.~\cite{Solgun:2019} the orthogonality of $\Omat_C$ is also assumed, and the $\Umat$ used in Eq.~(65) of that reference corresponds to our $\Omat_C^T$. 
    Moreover, $\Lmat_R$ is the diagonal inductive matrix of the inner mode resonators $L_{r_\gamma}$, $\Tmat_J = \mqty(\Rmat_J^T & \Nmat_J^T)^T$ is the transformer matrix, see \cref{Appendix-A:Transformer-eqs-impedance}
    and, $\Ymat$ is the admittance matrix of the gyrators
    \begin{subequations}
    \begin{align}
        \Ymat &= \mqty(\mqty{i\sigma_y \omega_{g_1} & \\ & \ddots}  & \text{\large$\mzero$} \\
        \text{\large$\mzero$} & i\sigma_y \omega_{g_k})= i\Sigma_y \Wmat_g,\\
        \Sigma_y &\equiv \mqty(\mqty{\sigma_y  & \\ & \ddots}  & \text{\large$\mzero$} \\
        \text{\large$\mzero$} & \sigma_y ), \quad \Wmat_g \equiv \mqty(\mqty{\omega_{g_1}\bone_2  & \\ & \ddots}  & \text{\large$\mzero$} \\
        \text{\large$\mzero$} & \omega_{g_l}\bone_2 ).
    \end{align}
    \end{subequations}
    Linearizing the nonlinear potential, the Lagrangian reads
    \begin{equation}
        \lgr = \frac{1}{2}\dot{\vphi}^T\Cmat\dot{\boldsymbol{\vphi}}-\frac{1}{2}\boldsymbol{\vphi}^T\Mmat\vphi+\frac{1}{2}\dot{\vphi}^T\Gmat\vphi+U_{\text{nl}}(\vphi_J),
        \label{Appendix coupling: Lgr-1}
    \end{equation}
    with
    \begin{equation}
        \Mmat = \mqty(\widetilde{\Lmat}_J^{-1} & \mzero \\ \mzero & \Wmat_{R_e}^{2}),\quad \Wmat_{R_e} = \mqty(\Wmat_{R} & \mzero \\ \mzero & \mzero_{2l} ),
    \end{equation}
    where $\widetilde{\Lmat}_J$ is the matrix of junction inductances $L_{j_\alpha} = \phi_0^2/E_{j_{\alpha}}$ obtained from the linear part of $U(\vphi_J)$, and $\Wmat_{R}$ is the diagonal matrix with the inner modes reciprocal resonators frequencies. We focus here on the linear sector, and treat the nonlinearities in \cref{NL}. 

    As discussed in the introduction of this appendix, we now consider the numerical diagonalization approach to include direct capacitive coupling. First, we dress the external modes with the transformation $\vphi_{co} = \Pmat_{co}\vphi$, where
    \begin{equation}
        \Pmat_{co} = \mqty(\overline{\Cmat}_J^{1/2}\Omat_C & \mzero \\ \mzero & \bone_{m+2l}),        
    \end{equation}
    after which the Lagrangian reads 
    \begin{equation}
        \lgr = \frac{1}{2}\dot{\vphi}_{co}^T\Cmat_{co}\dot{\vphi}_{co}-\frac{1}{2}\vphi_{co}^T\Mmat_{co}\vphi_{co}+\frac{1}{2}\dot{\vphi}_{co}^T\Gmat\vphi_{co},\\ \label{Appendix coupling: Lgr-2}
    \end{equation}
    with the transformed matrices 
    \begin{subequations}
    \begin{align}
        \Cmat_{co} &= \mqty(\bone_n & -\Tmat_{co}^T \\-\Tmat_{co} & \bone_{m+2l} + \Tmat_{co}\Tmat_{co}^T),\\
        \Mmat_{co} &= \mqty(\Omega'^2_J & \mzero \\ \mzero & \Wmat_{R_e}^2),
    \end{align}
    \end{subequations}
    with ${\Omega}'^2_J \equiv \overline{\Cmat}_J^{-1/2}\Omat_C\Lmat_J^{-1}\Omat_C^T\overline{\Cmat}_J^{-1/2}$ which is in general not diagonal, and $\Tmat_{co}=\Tmat_J\Omat_C^T\overline{\Cmat}_J^{1/2}$. Following Ref.~\cite{Solgun:2019} we diagonalize the kinetic matrix $\Cmat_{co}$ with the triangular point-transformation
    \begin{equation}
        \Pmat_{\sDelta} = \mqty(\bone_n & -\Tmat_{co}^T \\ \mzero & \bone_{m+2l}),
    \end{equation}
    In this new frame, the capacitive matrix is transformed into the identity $\Cmat_{co}\to \bone_{n+m+2l}$, the gyration matrix remains invariant $\Gmat_{co}\to \Gmat$, and the inductive matrix encodes the coupling between inner and weakly-dressed qubit modes 
    \begin{subequations}
        \begin{align}
            \Mmat_{co} \to \mqty(\Omega'^2_J & \Omega'^2_J\Tmat_{co}^T \\ 
            \Tmat_{co}\Omega'^2_J  & \Wmat_{R_e}^2+\Tmat_{co}\Omega'^2_J\Tmat_{co}^T).
        \end{align}
    \end{subequations}
    To obtain the normal modes of the qubit ports sector, we apply the transformation $\overline{\vphi} = \Pmat_{\Omega}\Pmat_{\sDelta}\Pmat_{co}\vphi$, with 
    \begin{equation}
        \Pmat_{\Omega} = \mqty(\Omat_{\Omega} & \mzero \\ \mzero & \bone_{m+2l}),
    \end{equation}
    where the orthogonal matrix $\Omat_{\Omega}$ diagonalizes the matrix $\Omat_\Omega\Omega'^2_J\Omat_{\Omega}^T = \overline{\Wmat}_J^2$. $\Omat_{\Omega}$ can be easily found numerically. The frequencies $\overline{\Wmat}_J$ correspond to the normal modes of the capacitively coupled external (qubits) ports. In this frame, $\overline{\Gmat}=\Gmat, \overline{\Cmat}=\bone$, and the inductive matrix reads
    \begin{subequations}
        \begin{align}
            \overline{\Mmat} = \mqty(\overline{\Wmat}_J^2 & \overline{\Wmat}_J^2\overline{\Tmat}_J^T \\ \overline{\Tmat}_J\,\overline{\Wmat}_J^2  & \Wmat_{R_e}^2+\overline{\Tmat}_J\,\overline{\Wmat}_J^2\overline{\Tmat}_J^T),
        \end{align}
    \end{subequations}
    with $\overline{\Tmat}_J=\Tmat_J\Omat_C^T\overline{\Cmat}_J^{1/2}\Omat_{\Omega}^T$. As $\overline{\Cmat}=\bone$, we directly obtain the classical Hamiltonian
    \begin{subequations}
        \begin{equation}
            H = \frac{1}{2}\qty(\overline{\vq}-\frac{\overline{\Gmat}}{2}\,\overline{\vphi})^T\qty(\overline{\vq}-\frac{\overline{\Gmat}}{2}\,\overline{\vphi})+\frac{1}{2}\overline{\vphi}^T\overline{\Mmat}\,\overline{\vphi}.
        \end{equation}
    \end{subequations}
    It is useful to express this as $H = \frac{1}{2}\vb{X}^T\Hmat\vb{X}$ with $\vb{X}^T=(\overline{\vb{\vphi}}_j , \overline{\vq}_j , \overline{\vb{\vphi}}_I , \overline{\vq}_I )$ and
        \begin{subequations}\label{Appendix coupling:H-z-quadratic-form}
    \begin{align}
        \Hmat &= \mqty(\Hmat_J & \Kmat \\
        \Kmat^T & \Hmat_I),\\
        \Hmat_J &= \mqty(\overline{\Wmat}_J^2 & \mzero \\ \mzero & \bone_n),\\
        \Hmat_I &= \mqty(\Ymat_e^T\Ymat/4+\Wmat_{R_e}^2 & \Ymat_e/2 \\ \Ymat_e^T/2 & \bone_{m+2l} ),\\
        \Kmat &= \mqty(\overline{\Wmat}_J^2\overline{\Tmat}_J^T & \mzero \\ \mzero & \mzero).
    \end{align}
    \end{subequations}
    As in the case of the admittance, we now approximately eliminate the nondynamical modes ignoring the second-order terms in the inner mode sector, as these will not change the final effective Hamiltonian of the junctions up to $\mathcal{O}((k/\Delta)^4)$. We note that the exact elimination of the nondynamical modes provides equivalent perturbative results, see \cref{Elimination-Nondynamical}. 
    
    To proceed, we now perform the symplectic transformations
    \begin{equation}
    \label{Appendix coupling:S-Sj-z}
        \begin{split}
            \Smat &= \mqty(\Smat_J & \mzero \\ \mzero & \Smat_I),\\
            \Smat_J &= \mqty(\overline{\Wmat}_J^{1/2} & \mzero \\ \mzero & \overline{\Wmat}_J^{-1/2}),
        \end{split}
    \end{equation}
    where the submatrices are
    \begin{equation}
    \label{Appendix coupling:S-SI-z}
        \begin{split}
            \Smat_I &= \mqty(\Smat_r & \mzero \\ \mzero & \Smat_g),\\
            \Smat_r &= \mqty(\Wmat_r^{1/2} & \mzero \\ \mzero & \Wmat_r^{-1/2}), \\
            \Smat_g &= \mqty(\bone_{2l}/2 & \Sigmamat_x \\ -\Sigmamat_x/2 & \bone_{2l})\mqty(\Wmat_g^{1/2} & \mzero \\ \mzero & \Wmat_g^{-1/2}).
        \end{split}
    \end{equation}
    The resulting  quadratic form is 
    \begin{align}
         \overline{\Hmat} = (\Smat^T)^{-1}\Hmat\Smat^{-1} = \mqty(\overline{\Hmat}_j & \overline{\Kmat}_r & \overline{\Kmat}_g \\\overline{\Kmat}_r^T & \overline{\Hmat}_r & \mzero \\ \overline{\Kmat}_g^T &\mzero & \overline{\Hmat}_g),
         \label{Appendix coupling:Hz-final-before-ssw}
    \end{align}
    with
    \begin{subequations}
    \label{Appendix coupling:Hz-K-R-N}
    \begin{equation}
        \begin{split}
            \overline{\Kmat}_r &= \mqty(\Kmat_{r}^{\phi\phi}& \mzero \label{Appendix coupling:Hz-K-R-N-(a)} \\ \mzero & \mzero),\\
            \overline{\Kmat}_g &= \mqty(\Kmat_{g}^{\phi\phi} & \Kmat_{g}^{\phi q}\\ \mzero & \mzero),
        \end{split}
    \end{equation}
        and
    \begin{equation}
    \label{Appendix coupling:Kmat-Z}
        \begin{split}
            \Kmat_{r}^{\phi\phi} &=  \overline{\Wmat}_J^{3/2}\overline{\Rmat}_J^T\Wmat_r^{-1/2},\\
            \Kmat_{g}^{\phi\phi} &=  \overline{\Wmat}_J^{3/2}\overline{\Nmat}_J^T\Wmat_g^{-1/2},\\
            \Kmat_{g}^{\phi q} &=  -\overline{\Wmat}_J^{3/2}\overline{\Nmat}_J^T\Sigmamat_x\Wmat_g^{-1/2},\\
            \overline{\Rmat}_J&=\Rmat_J\Omat_C^T\overline{\Cmat}_J^{1/2}\Omat_{\Omega}^T,\\
            \overline{\Nmat}_J&=\Nmat_J\Omat_C^T\overline{\Cmat}_J^{1/2}\Omat_{\Omega}^T.
        \end{split}
    \end{equation}
    \end{subequations}
    Here, $\overline{\Hmat}_j$, $\overline{\Hmat}_r$, $\overline{\Hmat}_g$ are 
    \begin{subequations}
        \begin{align}
            \overline{\Hmat}_j &= \mqty(\overline{\Wmat}_J & \mzero \\ \mzero & \overline{\Wmat}_J),\\
            \overline{\Hmat}_r &= \mqty(\Wmat_r & \mzero \\ \mzero & \Wmat_r),\\
            \overline{\Hmat}_g &= \mqty(\overline{\Wmat}_g & \mzero \\ \mzero & \overline{\Wmat}_g),
        \end{align}
    \end{subequations}
    with $\overline{\Wmat}_g$ the same as in \cref{Appendix coupling:H-g-Omega-g-(b)}. 
    
    On these expressions we perform the symplectic Schrieffer-Wolff transformation using the formulas \cref{Appendix SW:H-formulas}, and in analogous way as was done above for the admittance, we find the entries of the effective classical quadratic form 
    \begin{subequations}
    \label{Appendix coupling:Hz-ssw-final}
        \begin{align}
            \begin{split}
                (\widetilde{\Hmat}_J^{\phi\phi})_{ij} =\;&\delta_{ij}\overline{\omega}_{i}\;\\
                +\frac{i}{2}\sqrt{\overline{\omega}_{i}\overline{\omega}_{j}}&\biggl[\overline{\omega}_{j}\overline{\Zmat}^{ac,R}_{ij}(\overline{\omega}_{i})+\overline{\omega}_{i}\overline{\Zmat}^{ac,R}_{ij}(\overline{\omega}_{j})\biggr],
            \end{split} \\ 
            \begin{split}
                (\widetilde{\Hmat}_J^{\phi q})_{ij} =& \;\frac{1}{2}\sqrt{\overline{\omega}_{i}\overline{\omega}_{j}}\overline{\omega}_{i}\overline{\Zmat}^{NR}_{ij}(\overline{\omega}_{j}), 
            \end{split}\\
            (\widetilde{\Hmat}_J^{qq})_{ij}=&\;\delta_{ij}\overline{\omega}_{i},
        \end{align}
    \end{subequations}
    where 
    \begin{subequations}
    \label{Appendix coupling:Zmat-bar}
        \begin{align}
            \overline{\Zmat}(s) &= \Omat_{\Omega}\overline{\Cmat}_J^{1/2}\Omat_C\Zmat(s)\Omat_C^T\overline{\Cmat}_J^{1/2}\Omat_{\Omega}^T,\\
            \overline{\Zmat}^{ac,R}(s) &= \sum_{\gamma}\frac{\overline{\Amat}_{\gamma}s}{\omega_{r_{\gamma}}^2+s^2}+\sum_{\mu}\frac{\overline{\Amat}_{\mu}s}{\omega_{g_{\mu}}^2+s^2},\\
            \overline{\Zmat}^{NR}(s) &= \sum_{\mu}\frac{\overline{\Bmat}_{\mu}}{\omega_{g_{\mu}}^2+s^2},\\
            \overline{\Amat}_{\gamma (\mu)} &= \Omat_{\Omega}\overline{\Cmat}_J^{1/2}\Omat_C \Amat_{\gamma(\mu)}\Omat_C^T\overline{\Cmat}_J^{1/2}\Omat_{\Omega}^T,\\
            \overline{\Bmat}_{\mu} &= \Omat_{\Omega}\overline{\Cmat}_J^{1/2}\Omat_C \Bmat_{\mu}\Omat_C^T\overline{\Cmat}_J^{1/2}\Omat_{\Omega}^T.
        \end{align}
    \end{subequations}
    Quantizing and rearranging as was done with the admittance-based expressions, we obtain the effective quantum Hamiltonian for the junction sector 
    \begin{subequations}
    \label{Appendix coupling:quanutm-HZ-J-1}
        \begin{equation}
            \label{Appendix coupling:quantum-Hz-1}
            \begin{split}
            \hat{H}_J = & \sum_{i}\overline{\omega}_{i}\qty(1-\frac{\overline{\omega}_{i}}{2}\Im[\overline{\Zmat}^{ac}_{ii}(\overline{\omega}_{i})])\bdag_{i}\blow_{i}\\
            +&\sum_{i<j}\qty(J_{ij}\blow_{i}\bdag_{j}+J_{ij}^*\bdag_{i}\blow_{j})+U_{\text{nl}}(\widetilde{\phi},\widetilde{q}),
            \end{split}
        \end{equation}
    with
        \begin{equation}
        \label{Appendix coupling:J-z}
            J_{ij} = \frac{i}{4}\sqrt{\overline{\omega}_{i}\overline{\omega}_{j}}\qty[\overline{\omega}_{i}\overline{\Zmat}_{ij}(\overline{\omega}_{j})+\overline{\omega}_{j}\overline{\Zmat}_{ij}(\overline{\omega}_{i})].
        \end{equation}
    \end{subequations}
    In the case without direct capacitive or inductive coupling ($\Omat_C=\bone_n,\Omat_{\Omega}=\bone_n$) we obtain \cref{paper:eff-J-z} of the main text. Moreover, when the response is purely reciprocal, we recover the effective coupling given in Eq. (5) of Ref.~\cite{Solgun:2019}.

    \subsection{Fully perturbative impedance coupling formulas}
    \label{impedance coupler perturbative}
    
    We now consider the cased where there is direct capacitive coupling between qubit ports, which we treat as a first order perturbation. Our starting point is the Lagragian of  \cref{Appendix coupling: Lgr-1}. We first perform a triangular point-transformation 
    \begin{equation}
        \Pmat_{\sDelta}= \mqty(\bone_n & -\Tmat_J \\ \mzero & \bone_{m+2l}),
    \end{equation}
    where the new coordinates are $\vphi_{\sDelta}=\Pmat_{\sDelta}\vphi$, the  kinetic and inductive matrices are 
    \begin{subequations}
        \begin{align}
            \Cmat_{\sDelta} &= \mqty(\Cmat_J & \mzero\\ \mzero & \bone_{m+2l}),\\
            \Mmat_{\sDelta} &= \mqty(\widetilde{\Lmat}_J^{-1} & \widetilde{\Lmat}_J^{-1}\Tmat_J^T \\ \Tmat_J\widetilde{\Lmat}_J^{-1} & \Lmat_{R_e}^{-1}+\Tmat_J\widetilde{\Lmat}_J^{-1}\Tmat_J^T).
        \end{align}
    \end{subequations}
     Hence, we can immediately obtain the Hamiltonian
     \begin{equation}
         H = \frac{1}{2}\qty(\vb{q}_{\sDelta}-\frac{\Gmat}{2}\vphi_{\sDelta})^T\Cmat_{\sDelta}^{-1}\qty(\vb{q}_{\sDelta}-\frac{\Gmat}{2}\vphi_{\sDelta})+\frac{1}{2}\vphi_{\sDelta}^T\Mmat_{\sDelta}\vphi_{\sDelta}.\label{Appendix coupling:H-z-p}
     \end{equation}
    We let the matrices $\Cmat^{-1}_{J_\delta}$ and $\Cmat^{-1}_{J_\chi}$ respectively be the diagonal and offdiagonal entries of $\Cmat^{-1}_J$. We rescale the junction sector with
    \begin{equation}
        \Smat_J = \mqty((\widetilde{\Gmat}_J)^{1/2} & \mzero \\ \mzero & (\widetilde{\Gmat}_J)^{-1/2}),\label{Appendix coupling:Sj-Z-p}
    \end{equation}
    where $\widetilde{\Gmat}_J=(\Cmat_{J_\delta}/\widetilde{\Lmat}_j)^{1/2}$ such that the Hamiltonian in that sector is  
    \begin{equation}
        \overline{\Hmat}_J  = ((\Smat_J)^T)^{-1}\Hmat_j(\Smat_J)^{-1} = \mqty(\widetilde{\Wmat}_J & \mzero\\ \mzero & \widetilde{\Wmat}_J+\Wmat_{C_\chi}),\label{Appendix coupling:HmatJ-Z-p}
    \end{equation}
    with $\Wmat_{C_\chi} = \widetilde{\Gmat}_J^{1/2}\Cmat_{J_\chi}^{-1}\widetilde{\Gmat}_J^{1/2}$ and $\widetilde{\Wmat}_J = (\Cmat_{J_\delta}\widetilde{\Lmat}_J)^{-1/2}$. We now use the symplectic transformations of \cref{Appendix coupling:S-SI-z}  for the inner modes. Doing so, it is clear that the final effective classical quadratic form entries are
    \begin{subequations}
    \label{Appendix coupling:Hz-ssw-final-2}
        \begin{align}
            \begin{split}
                (\widetilde{\Hmat}_J^{\phi\phi})_{ij} =\;&\delta_{ij}\tilde{\omega}_{i}\;\\
                +\frac{i}{2}\sqrt{\tilde{\omega}_{i}\tilde{\omega}_{j}}&\biggl[\tilde{\omega}_{j}\overline{\Zmat}^{ac,R}_{ij}(\tilde{\omega}_{i})+\tilde{\omega}_{i}\overline{\Zmat}^{ac,R}_{ij}(\tilde{\omega}_{j})\biggr],
            \end{split} \\ 
            \begin{split}
                (\widetilde{\Hmat}_J^{\phi q})_{ij} =& \;\frac{1}{2}\sqrt{\tilde{\omega}_{i}\tilde{\omega}_{j}}\tilde{\omega}_{i}\overline{\Zmat}^{NR}_{ij}(\tilde{\omega}_{j}), 
            \end{split}\\
            \begin{split}   
                (\widetilde{\Hmat}_J^{qq})_{ij}=&\;\delta_{ij}\tilde{\omega}_{i}\\
                +\frac{i}{2}\sqrt{\tilde{\omega}_{i}\tilde{\omega}_{j}}&\qty[\tilde{\omega}_{i}\overline{\Zmat}^{dc}(\tilde{\omega}_{i})+\tilde{\omega}_{j}\overline{\Zmat}^{dc}(\tilde{\omega}_{j})](1-\delta_{ij}),
            \end{split}
        \end{align}
    \end{subequations}
    with $\tilde{\omega}_{i}$ the frequencies $\widetilde{\Wmat}_J$, and
    $\overline{\Zmat}^{ac,R}(s),\overline{\Zmat}^{NR}(s)$ given as in \cref{Appendix coupling:Zmat-bar} with $\Omat_C=\bone_n,\Omat_L=\bone_n$, and $\overline{\Zmat}^{dc}(s) = \Cmat_{J_{\delta}}^{1/2}\Zmat^{dc}(s)\Cmat_{J_{\delta}}^{1/2}\simeq \overline{\Cmat}_{J}^{1/2}\Zmat^{dc}(s)\overline{\Cmat}_{J}^{1/2}$. With this choice, the final effective quantum Hamiltonian reads exactly the same as in \cref{Appendix coupling:quanutm-HZ-J-1}. The above formulas hold up to third-order in perturbation theory when the direct capacitive coupling between all external ports is a first order perturbation. 
    
    \subsection{Admittance dissipative rates and Purcell decays}
    \label{Admitance-dissipative}
    
    In this section, we address dissipation resulting from the coupling of qubit ports to external drive ports via a nonreciprocal environment characterized by its admittance response, see \cref{fig:Duality}. As detailed in \cref{Duality},  the drive ports are defined as the terminals at the ends of transmission lines that connect the circuit of interest, including the junctions and inner modes, to external classical voltage sources. Furthermore, we model these external transmission lines as purely ohmic lumped elements with a characteristic impedance $Z_0$~\cite{Solgun:2019}. 
    
    Setting aside the voltage sources for the moment, the multiport synthesis of the admittance response illustrated in \cref{fig:Duality}, leads to the following dissipative classical equations of motion,
    \begin{subequations}\label{eqmotion}
    \begin{align}
    \begin{split}
        \Cmat_{J}\ddot{\vphi}_J=-\pdv{U}{\vphi_J}\;&-\Lmat_{J}^{-1}\vphi_J+\Ymat_{J}^G\dot{\vphi}_J-\Tmat_J\dot{\vQ} \\
        &-\Dissipmat^{\vphi\vphi}*\vphi_J-\Dissipmat^{\vphi \vQ}*\vQ ,
    \end{split} 
        \\[1.4ex]
         \ddot{\vQ}=-\Cmat_I^{-1}\vQ+\Tmat_J^T&\dot{\vphi}_J-\Zmat_e\dot{\vQ}-\Dissipmat^{\vQ\vphi}*\vphi_J-\Dissipmat^{\vQ \vQ}*\vQ , 
      \end{align} 
      \end{subequations}
    where $\vphi_J$ and $\vQ$ are respectively junction fluxes and inner modes loop charges. $U(\vphi_J)$ corresponds to the junctions' cosine potential. As already mentioned in \cref{admittance coupler}, the loop charge variable is the natural parametrization of the inner modes within the admittance representation. The matrices $\Lmat_J$, $\Ymat^G_J$, $\Tmat_J$, $\Cmat_I$ and $\Zmat_e$ are defined in \cref{admittance coupler} and  $*$ stands for time convolution, $f*g(t)=\int_{-\infty}^{+\infty}d\tau f(\tau)g(\tau-t).$ Moreover, the  $(n+m+2l)\times (n+m+2l)$ dissipation matrix $\Dissipmat(t)=\mqty(\Dissipmat^{\vphi\vphi} & \Dissipmat^{\vphi \vQ}\\ \Dissipmat^{\vQ\vphi} & \Dissipmat^{\vQ \vQ})$ is defined by the Fourier transform $\Dissipmat(\omega)=\int_{-\infty}^{+\infty}dt\Dissipmat(t)e^{-i\omega t}$ of its submatrices as
    \begin{subequations}
    \begin{align}
    \begin{split} \label{Dphiphi}
        \Dissipmat^{\vphi\vphi}(\omega)&=-(\Lmat_{JD}^{-1}-\omega^2 C_{JD}+i\omega \Ymat_{JD}^G)\times\\
        &\!\!\!\!\!\!\!\!\!\!(\frac{i\omega}{Z_0} \mathbbm{1}_{n_D}-\omega^2\Cmat_{D}+\Lmat_{D}^{-1})^{-1}
        (\Lmat_{JD}^{-1}-\omega^2 \Cmat_{JD}-i\omega \Ymat_{JD}^G)^T, \\
    \end{split}
    \\[2ex]
    \begin{split}\label{DphiQ}
        \Dissipmat^{\vphi\vQ}(\omega)=-&i\omega(\Lmat_{JD}^{-1}-\omega^2 \Cmat_{JD}+i\omega \Ymat_{JD}^G)\times\\
        &(\frac{i\omega}{Z_0} \mathbbm{1}_{n_D}-\omega^2\Cmat_{D}+\Lmat_{D}^{-1})^{-1} 
        \Tmat_D, \\
    \end{split}
    \\[2ex]
    \begin{split} \label{DQphi}
        \Dissipmat^{\vQ\vphi}(\omega)=i&\omega \Tmat_D^T(\frac{i\omega}{Z_0} \mathbbm{1}_{n_D}-\omega^2\Cmat_{D}+\Lmat_{D}^{-1})^{-1}\times\\ 
        &(\Lmat_{JD}^{-1}-\omega^2 \Cmat_{JD}-i\omega \Ymat_{JD}^G)^T,  \\ 
     \end{split}
    \\[2ex]
    \begin{split} \label{DQQ}
        \Dissipmat^{\vQ\vQ}(\omega)=-\omega^2 \Tmat_D^T(\frac{i\omega}{Z_0} \mathbbm{1}_{n_D}-\omega^2\Cmat_{D}+\Lmat_{D}^{-1})^{-1}\Tmat_D, 
    \end{split}
    \end{align}
      \end{subequations}
  for which we assumed that there is no direct gyration between drive ports ($\Ymat_{D}^G=0$).
  The non-zero block matrices $\Dissipmat^{\vphi\vphi}$, $\Dissipmat^{\vphi\vQ}$ and $\Dissipmat^{\vQ,\vphi}$ arise in the presence of direct coupling between qubit and drive ports characterized by $\Cmat_{JD}$, $\Lmat^{-1}_{JD}$, $\Ymat^G_{JD}$
  $\neq \mzero$. This generalizes prior studies such as Refs.~\cite{Burkard:2004,Burkard:2005, Solgun:2019,BB:2005}, where only $\Dissipmat^{\vQ,\vQ}$ was considered as the non-vanishing entry of the dissipation matrix.
  
   At this point, it is important to note some general properties of the dissipation matrix. First, $\Dissipmat(t)$ is real 
   because $\Dissipmat(-\omega)=\Dissipmat(\omega)^\star$. Additionally, the poles of $\Dissipmat(\omega)$ live in the upper half of the complex plane defined by $\mathrm{Im}(z)>0$, ensuring the causality of $\Dissipmat(t)$ (i.e., $\Dissipmat(t)=0$ for $t<0$)~\footnote{Note that if the Fourier transform is defined with opposite phase $\Dissipmat(\omega)=\int_{-\infty}^{+\infty}dt\Dissipmat(t)e^{+i\omega t}$, then the poles of $\Dissipmat(\omega)$ should live in the lower-half complex plane $\mathrm{Im}(z)<0$, to ensure causality}. Finally, unlike $\Dissipmat^{\vQ \vQ}$, $\Dissipmat^{\vphi \vphi}$ is not symmetric ($\Dissipmat^{{\vphi \vphi}^T}\neq \Dissipmat^{\vphi \vphi}$), which reflects the presence of \textit{direct} nonreciprocal interaction ($\Ymat_{ JD}^G \neq 0$) between qubit and drive ports.

   To construct a classical Lagrangian $\mathcal{L}$ that captures dissipation and reproduces classical equations of motion, we use an extended Caldeira-Leggett model \cite{CaldeiraLeggett:1981,Leggett:1984} by introducing a set of baths (collection of harmonic oscillators) that are linearly coupled (minimal coupling) to our system through \textit{all} quadratures. This Lagrangian reads
  \begin{equation} 
\mathcal{L}=\mathcal{L}_S+\mathcal{L}_B+\mathcal{L}_{SB},
  \end{equation}
    where $\mathcal{L}_S$, $\mathcal{L}_B$, $\mathcal{L}_{SB}$ are respectively system (junctions+inner modes), baths and interaction Lagrangians given by
   \begin{subequations}
   \label{Lagrange}
  \begin{align}
  \begin{split}
      \mathcal{L}_S=&\frac{1}{2}\dot{\vphi}_J^T\Cmat_{J}\dot{\vphi}_J+\frac{1}{2}\dot{\vQ}^T\dot{\vQ}+\frac{1}
      {2}\dot{\vQ}^T\Zmat_e\vQ+\frac{1}{2}\dot{\vphi}_J^T\Ymat_{J}^G\vphi_J\\
      &-\frac{1}{2}\vphi_J^T \Lmat_{J}^{-1}\vphi_J-U(\vphi_J)+\dot{\vphi}_J^T\Tmat_J\vQ, \\
  \end{split}
  \\[2ex]
  \begin{split}
   \mathcal{L}_B=\sum_i\left(\frac{1}{2}\dot{\vx}_{\alpha}^T\mmat_\alpha \dot{\vx}_\alpha-\frac{1}{2}\vx_\alpha^T\mmat_\alpha\omegamat_\alpha^2\vx_\alpha\right),
  \end{split} 
  \\[2ex]
  \begin{split}
  \mathcal{L}_{SB}=-&\vphi_J^T\sum_\alpha \mumat_\alpha \vx_\alpha-\dot{\vphi}_J^T\sum_\alpha \etamat_\alpha \dot{\vx}_\alpha-\vphi_J^T\sum_\alpha \lambdamat_\alpha \dot{\vx}_\alpha\\
  &-\dot{\vQ}^T\sum_\alpha \zetamat_\alpha \vx_\alpha, \\
  \end{split}
  \end{align}
\end{subequations}
where $\vx_\alpha=(x_{\alpha1}, \ldots, x_{\alpha B})$, $\mmat_\alpha=\text{diag}(m_{\alpha1},\ldots, m_{\alpha B} )$ and $\omegamat_\alpha=\text{diag}(\omega_{\alpha1},\ldots, \omega_{\alpha B} )$ are respectively baths flux coordinates, capacitances and eigenfrequencies with $B=\sup_{\omega}\text{rank}[\Im{\Dissipmat(\omega)}] \leq n_D$ is the number of baths needed to model dissipation, where $n_D$ is the number of drive ports~\cite{BB:2005}. The system-baths coupling matrices in $\mathcal{L}_{SB}$ correspond respectively to inductive ($\mumat_\alpha$), capacitive ($\etamat_\alpha$) and nonreciprocal $(\lambdamat_\alpha)$ couplings \cite{ParraRodriguez:2019}. While these two first type of couplings are invariant under time reversal symmetry $\vphi \to -\vphi$, $\vQ \to \vQ$, $\vx_\alpha \to -\vx_\alpha$, the charge-flux coupling arising from the gyrators breaks time reversal symmetry, leading to nonreciprocal system-baths interaction. Finally, the geometrical coupling ($\propto \dot{\vQ}\:\vx_\alpha$) that appears in the last term of $\mathcal{L}_{SB}$ emerges naturally as a consequence of using a mixed flux-charge description to parameterize the system (junctions and inner modes). The coupling matrices $\mumat $, $\etamat$, $\lambdamat$ (of size $n\times B$), $\zetamat$ (of size $(m+2l)\times B)$ and their corresponding quadratures are chosen to reproduce dissipative classical equations of motion. The resulting dissipation matrix is given by
    \begin{subequations}
    \begin{align}
        \begin{split}
          \Dissipmat^{\vphi\vphi}(\omega)&=\sum_\alpha(\mumat_\alpha+\etamat_\alpha\omega^2+i\omega\lambdamat_\alpha)\\
          &\times\mmat_\alpha^{-1}(\omega^2-\omegamat_\alpha^2)^{-1}(\mumat_\alpha+\etamat_\alpha\omega^2-i\omega\lambdamat_\alpha)^T, 
        \end{split}
        \\[2ex]
        \begin{split}
          \Dissipmat^{\vphi\vQ}(\omega)&=i\omega\sum_\alpha(\mumat_\alpha+\etamat_\alpha\omega^2+i\omega\lambdamat_\alpha)\\
          &\times\mmat_\alpha^{-1}(\omega^2-\omegamat_\alpha^2)^{-1}\zetamat_\alpha^T,  
        \end{split}
    \end{align}
    \begin{align}
    \begin{split}
      \Dissipmat^{\vQ\vphi}(\omega)=-&i\omega\sum_\alpha\zetamat_\alpha
      \mmat_\alpha^{-1}(\omega^2-\omegamat_\alpha^2)^{-1}\\&\times(\mumat_\alpha+\etamat_\alpha\omega^2-i\omega\lambdamat_\alpha)^T,\\
    \end{split}
    \\[2ex]
    \begin{split}
      \Dissipmat^{\vQ\vQ}(\omega)=\omega^2\sum_\alpha\zetamat_\alpha
      \mmat_\alpha^{-1}(\omega^2-\omegamat_\alpha^2)^{-1}\zetamat_\alpha^T. 
    \end{split}
    \end{align}
     \end{subequations}
    However, it is important to note that this matrix is not causal, since its poles are situated along the real line. Typically, to address this issue, the poles are shifted to the upper half of the complex plane following Ref.~\cite{Vool:2017}. This involves redefining $\Dissipmat(\omega)\equiv \lim_{\epsilon \to 0^{+} }\Dissipmat(\omega-i\epsilon)$ which now fulfills the requirement of dissipation matrices discussed above \footnote{The limit should be understood in the sense of distributions $\int d\omega \Dissipmat(\omega)f(\omega)=\lim_{\epsilon \to 0^{+}}\int d\omega \Dissipmat(\omega-i\epsilon)f(\omega)$ with $f$ any smooth (Schwartz) function}. Finally, it is worth noticing that in our case,  system-bath coupling via a single quadrature, as introduced in the seminal paper of  \textcite{Leggett:1984}, is insufficient to reproduce the dissipative Kirchhoff's equations. It is thus important to include all the other couplings in $\mathcal{L}_{SB}$ to match the classical dissipation matrix. 
    
    To obtain the Hamiltonian, we perform a Legendre transform up to second-order in system-bath couplings, which leads to the usual form 
    \begin{equation} \label{Y_full hamiltonian}
    H=H_S+H_B+H_{SB},
    \end{equation}
    where
    \begin{subequations}
    \begin{align}
    \begin{split}
        H_S&=\frac{1}{2}(\vq_J-\! \frac{1}{2}\Ymat_{J}^G\vphi_J-\! \Tmat_J\vQ)^T\Cmat_{J}^{-1}(\vq_J-\! \frac{1}{2}\Ymat_{J}^G\vphi_J-\! \Tmat_{J}\vQ)\\
        &+\frac{1}{2}\vphi_J^T \Lmat_J^{-1}\vphi_J+U(\vphi_J)\\
        &+\frac{1}{2}(\vPi-\frac{1}{2}\Zmat_e\vQ)^T(\vPi-\frac{1}{2}\Zmat_e\vQ)+\frac{1}{2}\vQ^T\Cmat_I^{-1}\vQ,\\
    \end{split}
    \\[1ex]
    \begin{split}
        H_B=\frac{1}{2}\sum_\alpha(\vp_\alpha^T \mmat_\alpha^{-1}\vp_\alpha+\vx_\alpha^T\mmat_\alpha\omegamat_\alpha^2\vx_\alpha), \\
     \end{split}    
    \label{H_bath}
     \\[2ex]
    \begin{split}
        H_{SB}&=\vphi_J^T\sum_\alpha \cmat_\alpha^{\vphi,\vx}\vx_\alpha+\vphi_J^T\sum_\alpha \cmat_\alpha^{\vphi,\vp}\vp_\alpha+\vq_J^T\sum_\alpha \cmat_\alpha^{\vq,\vp}\vp_\alpha\\
        &+\vQ^T\sum_\alpha \cmat_\alpha^{\vQ,\vx}\vx_\alpha+\vQ^T\sum_\alpha \cmat_\alpha^{\vQ,\vp}\vp_\alpha+\vPi^T\sum_\alpha \cmat_\alpha^{\vPi, \vx}\vx_\alpha.\\
    \end{split}
    \end{align}
      \end{subequations}
    Here, the Hamiltonian coupling matrices $\{\cmat_\alpha\}$ are expressed in terms of Lagrangian coupling matrices as
    \begin{subequations}
    \begin{align}
        \cmat_\alpha^{\vphi,\vx}&=\mumat_\alpha   &   \cmat_\alpha^{\vphi,\vp}&=\lambdamat_\alpha \mmat_\alpha^{-1}\\ 
         \cmat_\alpha^{\vq,\vp}&=\Cmat_{J}^{-1}\etamat_\alpha \mmat_\alpha^{-1}  &
         \cmat_\alpha^{\vPi,\vx}&=\zetamat_\alpha \\
         \cmat_\alpha^{\vQ,\vx}&=-\frac{\Zmat_e}{2}\zetamat_\alpha & \cmat_\alpha^{\vQ,\vp}&=-\Tmat_J^T\Cmat_{J}^{-1}\etamat_\alpha \mmat_\alpha^{-1},
    \end{align}
    \end{subequations}
    which leads to the following constraints $\cmat_\alpha^{\vQ,\vx}=-\frac{\Zmat_e}{2}\cmat_\alpha^{\vPi,\vx}$ and $\cmat_\alpha^{\vQ,\vp}=-\Tmat_J^T\cmat_\alpha^{\vq,\vp}$. Inverting the previous equations, the dissipation matrix can be expressed in terms of $\{\cmat_\alpha\}$ matrices
    \begin{subequations}
    \begin{align}
    \begin{split}
      \Dissipmat^{\vphi\vphi}(\omega)&=\sum_\alpha(\cmat_\alpha^{\vphi,\vx}+\omega^2\Cmat_{J}\cmat_\alpha^{\vq,\vp}\mmat_\alpha+i\omega \cmat_\alpha^{\vphi,\vp}\mmat_\alpha)\\
      &\times\mmat_\alpha^{-1}(\omega^2-\omegamat_\alpha^2)^{-1}\\
      &\times (\cmat_\alpha^{\vphi,\vx}+\omega^2\Cmat_{J}\cmat_\alpha^{\vq,\vp}\mmat_\alpha-i\omega \cmat_\alpha^{\vphi,\vp}\mmat_\alpha)^T,  
    \end{split}
    \\[1ex]
    \begin{split}
      \Dissipmat^{\vphi\vQ}(\omega)&=i\omega\sum_\alpha(\cmat_\alpha^{\vphi,\vx}+\omega^2\Cmat_{J}\cmat_\alpha^{\vq,\vp}\mmat_\alpha+i\omega \cmat_\alpha^{\vphi,\vp}\mmat_\alpha)\\
      &\times \mmat_\alpha^{-1}(\omega^2-\omegamat_\alpha^2)^{-1}\cmat_\alpha^{{\vPi,\vx}^T},  
    \end{split}
    \end{align}
    \begin{align}
    \begin{split}
      \Dissipmat^{\vQ\vphi}(\omega)&=-i\omega\sum_\alpha \cmat_\alpha^{\vPi,\vx}
      \mmat_\alpha^{-1}(\omega^2-\omegamat_\alpha^2)^{-1}\\& \qq{}\times(\cmat_\alpha^{\vphi,\vx}+\omega^2\Cmat_{J}\cmat_\alpha^{\vq,\vp}\mmat_\alpha-i\omega \cmat_\alpha^{\vphi,\vp}\mmat_\alpha)^T,  
    \end{split}
    \\[1ex]
    \begin{split}
    \Dissipmat^{\vQ\vQ}(\omega)&=\omega^2\sum_\alpha \cmat_\alpha^{\vPi,\vx}
      \mmat_\alpha^{-1}(\omega^2-\omegamat_\alpha^2)^{-1}\cmat_\alpha^{{\vPi,\vx}^T}. 
    \end{split}
    \end{align}
    \end{subequations}
     Using the phase-space coordinates $\vX^T = (\vphi_J^T , \vq_J^T , \vQ_R^T , \vQ_G^T  , \vPi_R^T , \vPi_G^T )$, the system-baths interaction Hamiltonian can be written compactly as
    \begin{equation}
        H_{SB}=\vX^T \sum_\alpha \cmat_\alpha \mqty(\vx_\alpha \\ \vp_\alpha).
    \end{equation}
    The coupling coefficients are regrouped as
    \begin{equation}\cmat_\alpha=\mqty(\cmat_\alpha^{\vphi,\vx} & \cmat_\alpha^{\vphi,\vp}\\
    0 & \cmat_\alpha^{\vq,\vp}\\
    0 & \cmat_\alpha^{\vQ_R, \vp}\\
    \cmat_\alpha^{\vQ_G,\vx} & \cmat_\alpha^{\vQ_G,\vp}\\
    \cmat_\alpha^{\vPi_r,\vx} & 0\\
     \cmat_\alpha^{\vPi_g,\vx} & 0).
    \end{equation}
    To analyze how the coupling coefficients transform when diagonalizing the junctions and inner modes Hamiltonian $H_S$,  we employ  the symplectic transformations computed in \cref{admittance coupler}. This process involves two sequential steps. First, we perform the symplectic transformation $\Smat=\mqty(\Smat_J & \mzero\\ \mzero & \Smat_I)$ defined in \cref{Appendix coupling:S}, allowing us to obtain normal modes for both junctions and inner modes sectors separately. Under this transformation, the coupling coefficients become
     \begin{equation}
         \cmat_\alpha \mapsto \cmat_\alpha' \equiv  (\Smat^T)^{-1} \cmat_\alpha. 
     \end{equation}
     Expanding this transformation explicitly,
     \begin{equation}\cmat_\alpha'=\mqty(\overline{\Wmat}_J^{-\frac{1}{2}}\Omat_\Omega \overline{\Cmat}_J^{-\frac{1}{2}}\Omat_C \cmat_\alpha^{\vphi,\vx} &\overline{\Omega}_j^{-\frac{1}{2}}\Omat_\Omega \overline{\Cmat}_J^{-\frac{1}{2}}\Omat_C \cmat_\alpha^{\vphi,\vp}\\
         0 & \overline{\Wmat}_J^{\frac{1}{2}}\Omat_\Omega \overline{\Cmat}_J^{\frac{1}{2}}\Omat_C \cmat_\alpha^{\vq,\vp}\\
         0 & \Omega_R^{-\frac{1}{2}} \cmat_\alpha^{\vQ_R,\vp}\\
         \Omega_G^{-\frac{1}{2}}\cmat_\alpha^{\vQ_G,\vx}-\frac{\Sigmamat_x}{2}\Omega_G^{\frac{1}{2}}\cmat_\alpha^{\vPi_G,\vx} & \Omega_G^{-\frac{1}{2}}\cmat_\alpha^{\vQ_G,\vp}\\
         \Omega_R^{\frac{1}{2}}\cmat_\alpha^{\vPi_R,\vx} & 0\\
         \Sigmamat_x\Omega_G^{-\frac{1}{2}}\cmat_\alpha^{\vQ_G,\vx}+\frac{\Omega_G^{\frac{1}{2}}}{2}\cmat_\alpha^{\vPi_G,\vx} & \Sigmamat_x\Omega_G^{-\frac{1}{2}}\cmat_\alpha^{\vQ_G,\vp} ).
     \end{equation}    
     The next step consists in eliminating the inner modes, parameterized by the classical phase-space coordinates $(\vQ_R, \vQ_G, \vPi_R,\vPi_G)$, through Schrieffer-Wolff transformation (see  \cref{Appendix SW: equations-A(1)} in \cref{SSW}), leading to effective couplings defined by $\Tilde{\cmat}_\alpha=(\Smat_{sw}^T)^{-1}\cmat_\alpha' $. They are related to the original  coupling matrices $\{\cmat_\alpha\}$ as follows 
    \begin{subequations}\label{final couplings:Y}
     \label{coupling_tilde}
    \begin{align}
    \begin{split}
         \Tilde{\cmat}_\alpha^{\vphi,\vx}&= \overline{\Wmat}_J^{-\frac{1}{2}}\Omat_\Omega \overline{\Cmat}_J^{-\frac{1}{2}}\Omat_C \cmat_\alpha^{\vphi,\vx}+\Lambdamat_x^{\vphi,\vPi_R}\cmat_\alpha^{\vPi_R,\vx}+\Lambdamat_x^{\vphi,\vQ_G}\cmat_\alpha^{\vQ_G,\vx}\\&\qq{}+\Lambdamat_x^{\vphi,\vPi_G}\cmat_\alpha^{\vPi_G,\vx},
         \end{split}\\
        \Tilde{\cmat}_\alpha^{\vphi,\vp}&= \overline{\Wmat}_J^{-\frac{1}{2}}\Omat_\Omega \overline{\Cmat}_J^{-\frac{1}{2}}\Omat_C \cmat_\alpha^{\vphi,\vp}+\Lambdamat_p^{\vphi,\vQ_G}\cmat_\alpha^{\vQ_G,\vp},
        \\[2ex]\Tilde{\cmat}_\alpha^{\vq,\vx}&=\Lambdamat_x^{\vq,\vQ_G}\cmat_\alpha^{\vQ_G,\vx}+\Lambdamat_x^{\vq,\vPi_G}\cmat_\alpha^{\vPi_G,\vx},\\[2ex]
        \Tilde{\cmat}_\alpha^{\vq,\vp}&= \overline{\Wmat}_J^{\frac{1}{2}}\Omat_\Omega \overline{\Cmat}_J^{\frac{1}{2}}\Omat_C \cmat_\alpha^{\vq,\vp}+\Lambdamat_p^{\vq,\vQ_R}\cmat_\alpha^{\vQ_R,\vp}+\Lambdamat_p^{\vq,\vQ_G}\cmat_\alpha^{\vQ_G,\vp},
    \end{align}
    \end{subequations}
    where the Schrieffer-Wolff matrices $\{\Lambdamat_x, \Lambdamat_p\}$ involved in the previous equations are defined as 
    \begin{subequations}
    \begin{align}
        {\Lambdamat_x^{\vphi,\vPi_R}}_{i\gamma}&=\frac{\overline{\Omega}_{J_i}^{\frac{3}{2}}}{\overline{\Omega}_{J_i}^2-\Omega_{R_\gamma}^2}(\overline{\Rmat}_{J})_{i\gamma},\\
          {\Lambdamat_{x}^{\vphi,\vQ_G}}_{i 2k}&=-\frac{\overline{\Omega}_{J_i}^{-\frac{1}{2}}\Omega_{G_k}}{\overline{\Omega}_{J_i}^2-\Omega_{G_k}^2}(\overline{\Nmat}_{J})_{i 2k-1},\\  
          {\Lambdamat_x^{\vphi,\vQ_G}}_{i 2k-1}&=\frac{\overline{\Omega}_{J_i}^{-\frac{1}{2}}\Omega_{G_k}}{\overline{\Omega}_{J_i}^2-\Omega_{G_k}^2}(\overline{\Nmat}_{J})_{i 2k},\\
      {\Lambdamat_{x}^{\vphi,\vPi_G}}_{i 2k}&=\frac{\overline{\Omega}_{J_i}^{-\frac{1}{2}}}{2}(1+\frac{\overline{\Omega}_{J_i}^{2}}{\overline{\Omega}_{J_i}^2-\Omega_{G_k}^2})(\overline{\Nmat}_{J})_{i 2k},\\  
      {\Lambdamat_{x}^{\vphi,\vPi_G}}_{i 2k-1}&=\frac{\overline{\Omega}_{J_i}^{-\frac{1}{2}}}{2}(1+\frac{\overline{\Omega}_{J_i}^{2}}{\overline{\Omega}_{J_i}^2-\Omega_{G_k}^2})(\overline{\Nmat}_{J})_{i 2k-1},\\  
      {\Lambdamat_p^{\vphi,\vQ_G}}_{i 2k}&=-\frac{\overline{\Omega}_{J_i}^{-\frac{1}{2}}\Omega_{G_k}}{\overline{\Omega}_{J_i}^2-\Omega_{G_k}^2}(\overline{N}_{J})_{i 2k-1},\\  
      {\Lambdamat_p^{\vphi,\vQ_G}}_{i 2k-1}&=\frac{\overline{\Omega}_{J_i}^{-\frac{1}{2}}\Omega_{G_k}}{\overline{\Omega}_{J_i}^2-\Omega_{G_k}^2}(\overline{N}_{J})_{i 2k},\\
      {\Lambdamat_x^{\vq,\vQ_G}}_{i 2k}&=-\frac{\overline{\Omega}_{J_i}^{\frac{1}{2}}}{\overline{\Omega}_{J_i}^2-\Omega_{G_k}^2}(\overline{\Nmat}_{J})_{i 2k},\\  
      {\Lambdamat_x^{\vq,\vQ_G}}_{i 2k-1}&=-\frac{\overline{\Omega}_{j_i}^{\frac{1}{2}}}{\overline{\Omega}_{J_i}^2-\Omega_{G_k}^2}(\overline{\Nmat}_{J})_{i 2k-1},\\ 
      {\Lambdamat_{x}^{\vq,\vPi_G}}_{i 2k}&=-\frac{\overline{\Omega}_{J_i}^{\frac{1}{2}}\Omega_{G_k}}{2(\overline{\Omega}_{J_i}^2-\Omega_{G_k}^2)}(\overline{\Nmat}_{J})_{i 2k-1},\\
      {\Lambdamat_{x}^{\vq,\vPi_G}}_{i 2k-1}&=\frac{\overline{\Omega}_{J_i}^{\frac{1}{2}}\Omega_{G_k}}{2(\overline{\Omega}_{J_i}^2-\Omega_{G_k}^2)}(\overline{\Nmat}_{J})_{i 2k},\\
   {\Lambdamat_p^{\vq,\vQ_R}}_{i\gamma}&=-\frac{\overline{\Omega}_{J_i}^{\frac{1}{2}}}{\overline{\Omega}_{J_i}^2-\Omega_{R_\gamma}^2}(\overline{\Rmat}_{J})_{i\gamma},
   \end{align}
    \begin{align}
       {\Lambdamat_p^{\vq,\vQ_G}}_{i 2k}&=-\frac{\overline{\Omega}_{J_i}^{\frac{1}{2}}}{\overline{\Omega}_{J_i}^2-\Omega_{R_\gamma}^2}(\overline{\Nmat}_{J})_{i 2k},\\
      {\Lambdamat_p^{\vq,\vQ_G}}_{i 2k-1}&=-\frac{\overline{\Omega}_{J_i}^{\frac{1}{2}}}{\overline{\Omega}_{J_i}^2-\Omega_{R_\gamma}^2}(\overline{\Nmat}_{J})_{i 2k-1}.
    \end{align}
    \end{subequations}
     
     In \cref{final couplings:Y}, the effective junctions-baths coupling matrices $\Tilde{\cmat}_\alpha$ incorporate two contributions. The first one arises from the dressing with inner modes as a result of SW transformations (terms proportional to $\Lambdamat_{x,p}$) while the second contribution comprises direct couplings between qubit and drive ports (terms proportional to the dressed frequency $\overline{\Wmat}_J$). The system-baths interaction Hamiltonian can be expressed in this final frame as
    \begin{equation} \label{H_SB}
    \begin{split}
     H_{SB}=&\vphi_J^T\sum_\alpha \Tilde{\cmat}_\alpha^{\vphi,\vx}\vx_\alpha+\vphi_J^T\sum_\alpha \Tilde{\cmat}_\alpha^{\vphi,\vp}\vp_\alpha\\
     &+\vq_J^T\sum_\alpha \Tilde{\cmat}_\alpha^{\vq,\vx}\vx_\alpha+\vq_J^T\sum_\alpha \Tilde{\cmat}_\alpha^{\vq,\vp}\vp_\alpha,
    \end{split}
    \end{equation}
    which captures the most general linear system-bath couplings.
    Using standard quantization method  \cite{Vool:2017} and using the results of \cref{masterequation}, the correlated decay rates are given in terms of this new effective couplings as 
    \begin{align} \label{Appendix:decay-admittance}
        \begin{split}
            \gamma_{jj'}=\frac{\pi}{2}\sum_{b=1}^{B}\sum_\alpha(s_{\alpha_{jb}}+it_{\alpha_{jb}})^\star(s_{\alpha_{j'b}}+it_{\alpha_{j'b}})\delta(\overline{\Omega}_j-\omega_{\alpha b}),
        \end{split}
    \end{align}
    where 
    \begin{align}
        \begin{split} s_{\alpha_{jb}}=\frac{\Tilde{c}_{\alpha_{jb}}^{\vphi,\vx}}{\sqrt{m_{\alpha b}\omega_{\alpha b}}}+\sqrt{m_{\alpha b}\omega_{\alpha b}}\: \Tilde{c}_{\alpha_{jb}}^{\vq,\vp}, \\ 
        \end{split}
        \\[2ex]
        \begin{split}        t_{\alpha_{jb}}=\sqrt{m_{\alpha b}\omega_{\alpha b}}\: \Tilde{c}_{\alpha_{jb}}^{\vphi,\vp}-\frac{\Tilde{c}_{\alpha_{jb}}^{\vq,\vx}}{\sqrt{m_{\alpha b}\omega_{\alpha b}}}, \\   
        \end{split}
    \end{align}
    The squared terms in \cref{Appendix:decay-admittance} can be written as 
    \begin{align}
    \label{squaredterm}
        \begin{split}
            &\frac{\pi}{2}\sum_{b=1}^{B}\sum_\alpha  s_{\alpha_{jb}} s_{\alpha_{j'b}}\delta(\overline{\Omega}_j-\omega_{\alpha b})+t_{\alpha_{jb}} t_{\alpha_{j'b}}\delta(\overline{\Omega}_j-\omega_{\alpha b})\\
            &=\frac{\pi}{2}\sum_\alpha \left[ \left(\Tilde{\cmat}_\alpha^{\vphi,\vx}+\Tilde{\cmat}_{\alpha}^{\vq,\vp}\mmat_{\alpha} \omegamat_{\alpha}\right)\mmat_\alpha^{-1}\omegamat_{\alpha}^{-1}\deltamat(\overline{\Omega}_j-\omegamat_\alpha) \right.\\
            & ( \Tilde{\cmat}_{\alpha}^{\vphi,\vx}+\Tilde{\cmat}_{\alpha}^{\vq,\vp}\mmat_{\alpha} \omegamat_{\alpha})^T 
             +(\Tilde{\cmat}_\alpha^{\vq,\vx}-\Tilde{\cmat}_{\alpha}^{\vphi,\vp}\mmat_{\alpha} \omegamat_{\alpha})\mmat_\alpha^{-1}\omegamat_\alpha^{-1}\\
             &\times\left.\deltamat(\overline{\Omega}_j-\omegamat_\alpha)( \Tilde{\cmat}_{\alpha}^{\vq,\vx}+\Tilde{\cmat}_{\alpha}^{\vphi,\vp}\mmat_{\alpha} \omegamat_\alpha)^{T}  \right]_{jj'},
        \end{split}
    \end{align}
    where $\delta(\overline{\Omega}_j-\omegamat_\alpha)\equiv\text{diag}\{\delta(\overline{\Omega}_j-\omega_{\alpha 1}),\ldots,\delta(\overline{\Omega}_j-\omega_{\alpha B})\}$. Expressing $\{\Tilde{\cmat}_\alpha\}$ using \cref{coupling_tilde}, and within the dispersive regime $\overline{\Omega}_{j}^{\frac{1}{2}}\Lambdamat_p\ll 1$, \cref{squaredterm} is equivalent to
    \begin{align}
    \begin{split}
     \frac{\pi}{2}&\left[{\overline{\Omega}_{J}}^{-\frac{1}{2}}\Omat_\Omega\overline{\Cmat}_J^{-\frac{1}{2}}\Omat_C\{ \cmat_{\alpha}^{\vphi,\vx}+\Cmat_{J}\cmat_{\alpha}^{\vq,\vp}\mmat_\alpha{\omegamat_\alpha}^2\}+\Lambdamat_{x}^{\vphi,\vPi_R} \cmat_{\alpha}^{\vPi_R,\vx} \right.\\
     &\left. +\Lambdamat_{x}^{\vphi,\vQ_G} \cmat_{\alpha}^{\vQ_G,\vx}+\Lambdamat_{x}^{\vphi,\vPi_G} \cmat_{\alpha}^{\vPi_G,\vx}\right]\mmat_{\alpha}^{-1}\omegamat_{\alpha}^{-1}\deltamat(\overline{\Wmat}_J-\omegamat_\alpha) \\
      &\times\left[{\overline{\Omega}_{J}}^{-\frac{1}{2}}\overline{\Cmat}_J^{-\frac{1}{2}}\Omat_C\{ \cmat_{\alpha}^{\vphi,\vx}+\Cmat_{J}\cmat_{\alpha}^{\vq,\vp}\mmat_\alpha{\omegamat_\alpha}^2\}+\Lambdamat_{x}^{\vphi,\vPi_R} \cmat_{\alpha}^{\vPi_R,\vx} \right.\\
     &\left. +\Lambdamat_{x}^{\vphi,\vQ_G} \cmat_{\alpha}^{\vQ_G,\vx}+\Lambdamat_{x}^{\vphi,\vPi_G} \cmat_{\alpha}^{\vPi_G,\vx}\right]^T_{jj'}.
    \end{split}
    \end{align}
    The next step  is to identify the terms present in the last equation that also appear in the dissipation matrix $\Dissipmat(\omega)$. To achieve this, we use the causality requirement $\lim_{\epsilon \to 0^{+}}\Dissipmat(\omega-i\epsilon)$ together with the Sokhotski–Plemelj formula
    \begin{equation}
        \lim_{\epsilon \to 0^{+}}\frac{1}{(\omega-i\epsilon)^2-\omega_{\alpha}^2}=\mathcal{P}\left(\frac{1}{\omega^2-\omega_{\alpha}^2}\right)+i\frac{\pi}{2}\omega_\alpha^{-1}\delta(\omega-\omega_\alpha),
    \end{equation}
   where $\mathcal{P(.)}$ corresponds to Cauchy principal value, and we obtain the following identification 
   \begin{subequations}
   \begin{align}
   \begin{split}
&\frac{1}{2}\left(\Im\left[\Dissipmat^{\vphi\vphi}\right]+\Im\left[{\Dissipmat^{\vphi \vphi}}^T\right]\right)=\frac{\pi}{2}\sum_\alpha(\cmat_{\alpha}^{\vphi,\vx}
+\Cmat_{J}\cmat_{\alpha}^{\vq,\vp}\\& \times\mmat_\alpha{\omegamat_\alpha}^2)\times\mmat_{\alpha}^{-1}\omegamat_\alpha^{-1}
\deltamat(\omega-\omegamat_\alpha)( \cmat_{\alpha}^{\vphi,\vx}
+\Cmat_{J}\cmat_{\alpha}^{\vq,\vp}\mmat_\alpha{\omegamat_\alpha}^2)^T \\ &+\frac{\pi}{2}\omega^2\sum_\alpha (\cmat_{\alpha}^{\vphi,\vp}\mmat_\alpha)\mmat_{\alpha}^{-1}\omegamat_{\alpha}^{-1}\deltamat(\omega-\omegamat_\alpha)(\cmat_{\alpha}^{\vphi,\vp}\mmat_\alpha)^T, \\
    \end{split}
    \\[3ex]
    \begin{split}
        &\frac{1}{2}\left(\Im\left[\frac{\Dissipmat^{\vphi\vQ}}{i\omega}\right]+\Im\left[\frac{{\Dissipmat^{\vQ \vphi}}^T}{-i\omega}\right]\right)=\frac{\pi}{2}\sum_\alpha(\cmat_{\alpha}^{\vphi,\vx}\\&+\Cmat_{J}\cmat_{\alpha}^{\vq,\vp}\mmat_\alpha{\omegamat_\alpha}^2)\times\mmat_{\alpha}^{-1}\omegamat_{\alpha}^{-1}
        \deltamat(\omega-\omegamat_\alpha)(\cmat_{\alpha}^{\vPi,\vx})^T, \\
    \end{split}
    \\[3ex]
    \begin{split}
        &\frac{1}{2}\left(\Re\left[\frac{{\Dissipmat^{\vQ \vphi}}^T}{-i\omega}\right]-\Re\left[\frac{\Dissipmat^{{\vphi \vQ}}}{i\omega}\right]\right)=\frac{\pi}{2}\omega\sum_\alpha( \cmat_{\alpha}^{\vphi,\vp}\mmat_\alpha)\\
        & \times\mmat_{\alpha}^{-1}\omegamat_{\alpha}^{-1}
        \deltamat(\omega-\omegamat_\alpha)(\cmat_{\alpha}^{\vPi,\vx})^T. \\
    \end{split}
    \end{align}
    \end{subequations}
On the other hand, the classical equations of motion \cref{Dphiphi,DphiQ,DQphi,DQQ} enable us to evaluate this set of equations as a function of the circuit's admittance parameters,
    \begin{subequations}
   \begin{align}  
   \begin{split}
        &\frac{1}{2}\left(\Im\left[\Dissipmat^{\vphi\vphi}\right]+\Im\left[{\Dissipmat^{{\vphi \vphi}}}^T\right]\right)=-(\Lmat_{JD}^{-1}-\omega^2 \Cmat_{JD}),\\
      &\times\Im\left[(\frac{i\omega}{Z_0} \mathbbm{1}_{n_D}-\omega^2\Cmat_{D}+\Lmat_{D}^{-1})^{-1}\right]
        (\Lmat_{JD}^{-1}-\omega^2 \Cmat_{JD})^T\\
        &-\omega^2\Ymat_{JD}^G\Im\left[(\frac{i\omega}{Z_0} \mathbbm{1}_{n_D}-\omega^2\Cmat_{D}+\Lmat_{D}^{-1})^{-1}\right]\Ymat_{JD}^G,
    \end{split}
    \end{align}
    \begin{align}
    \begin{split}
         &\frac{1}{2}\left(\Im\left[\frac{\Dissipmat^{\vphi\vQ}}{i\omega}\right]+\Im\left[\frac{{\Dissipmat^{\vQ \vphi}}^T}{-i\omega}\right]\right)=-(\Lmat_{JD}^{-1}-\omega^2 \Cmat_{JD})\\
      &\times\Im\left[(\frac{i\omega}{Z_0} \mathbbm{1}_{n_D}-\omega^2\Cmat_{D}+\Lmat_{D}^{-1})^{-1}\right]\Tmat_D,\\ 
    \end{split}
    \\[3ex]
    \begin{split}
         &\frac{1}{2}\left(\Re\left[\frac{{\Dissipmat^{\vQ \vphi}}^T}{-i\omega}\right]-\Re\left[\frac{\Dissipmat^{{\vphi \vQ}}}{i\omega}\right]\right)=-\omega \Ymat_{JD}^G
      \\
  &\times\Im\left[(\frac{i\omega}{Z_0} \mathbbm{1}_{n_D}-\omega^2\Cmat_{D}+\Lmat_{D}^{-1})^{-1}\right]\Tmat_D.\\ 
    \end{split}
   \end{align}
     \end{subequations}
    Substituting these terms in \cref{squaredterm} and using definitions of SW matrices $\{\Lambdamat_x,\Lambdamat_p\}$ yields
    \begin{align}
    \begin{split}
            \frac{\pi}{2}\sum_{b=1}^{B}&\sum_\alpha  s_{\alpha_{jb}} s_{\alpha_{j'b}}\delta(\overline{\Omega}_j-\omega_{\alpha b})+t_{\alpha_{jb}} t_{\alpha_{j'b}}\delta(\overline{\Omega}_j-\omega_{\alpha b})\\
            &=\overline{\Omega}_j\left[\overline{\Ymat}_{JD}^{R}(\overline{\Omega}_j)\mathsf{J}(\overline{\Omega}_j)\overline{\Ymat}_{JD}^R(\overline{\Omega}_j)^\dagger\right.\\
            &\left.+\overline{\Ymat}_{JD}^{NR}(\overline{\Omega}_j)\mathsf{J}(\overline{\Omega}_j)\overline{\Ymat}_{JD}^{NR}(\overline{\Omega}_j)^\dagger\right]_{jj'}.\\
    \end{split}        
    \end{align}
    Here, the Kernel $\mathsf{J}(\omega)$ is given by $n_D\times n_D$ positive semi-definite matrix,
    \begin{equation*}
        \mathsf{J}(\omega)=-\Im\left[(\frac{i\omega}{Z_0} \mathbbm{1}_{n_D}-\omega^2\Cmat_{D}+\Lmat_{D}^{-1})^{-1}\right] \: \: \text{,} \: \: \omega \geq 0
    \end{equation*}
    where $n_D$ is the number of drive ports. As defined previously in the non-dissipative case, the dressed admittance $\bar{\Ymat}$ is related to the bare $\Ymat$ via the following  similitude transformation, 
    \begin{equation}
      \bar{\Ymat}(\omega)=\Tilde{\Omat}_\Omega\Tilde{\overline{\Cmat}}_{J}^{-\frac{1}{2}}\Tilde{\Omat}_C \Ymat(\omega) \Tilde{\Omat}_C^T\Tilde{\overline{\Cmat}}_{J}^{-\frac{1}{2}}\Tilde{\Omat}_\Omega^T ,
    \end{equation}
    where 
     \begin{align}
       &\Tilde{\Omat}_\Omega=\mqty(\Omat_\Omega & 0 \\ 0 & \bone_{n_D}), \\
       &\Tilde{\Omat}_C=\mqty(\Omat_C & 0 \\ 0 & \bone_{n_D}), \\
        &\Tilde{\overline{\Cmat}}_J=\mqty(\overline{\Cmat}_J & 0 \\ 0 & \bone_{n_D }),
     \end{align}
    with the $n\times n $ matrices  $\Omat_C$, $\Omat_\Omega$, and $\overline{\Cmat}_J$  were previously defined in Appendix C1 during the diagonalization of the junction sector. As outlined in \cref{Duality}, the reciprocal $\overline{\Ymat}^{R}$ and nonreciprocal $\overline{\Ymat}^{\text{NR}}$ parts of the admittance are defined as the symmetric and antisymmetric responses
     \begin{align}
       \bar{\Ymat}^{R}&=\frac{1}{2}(\bar{\Ymat}+\bar{\Ymat}^T),\\
       \bar{\Ymat}^{NR}&=\frac{1}{2}(\bar{\Ymat}-\bar{\Ymat}^T) . 
     \end{align}    
    On the other hand, the cross terms in \cref{Appendix:decay-admittance} are computed similarly
    \begin{align}
        \begin{split}
            i\frac{\pi}{2}&\sum_{b=1}^{B}\sum_\alpha  s_{\alpha_{jb}} t_{\alpha_{j'b}}\delta(\overline{\Omega}_j-\omega_{\alpha b})-t_{\alpha_{jb}} s_{\alpha_{j'b}}\delta(\overline{\Omega}_j-\omega_{\alpha b})\\
            &=\overline{\Omega}_j\left[ \bar{\Ymat}_{JD}^R(\overline{\Omega}_j)\mathsf{J}(\overline{\Omega}_j)\bar{\Ymat}_{JD}^{NR}(\overline{\Omega}_j))^\dagger \right.\\
            &\left.+\bar{\Ymat}_{JD}^{NR}(\overline{\Omega}_j)\mathsf{J}(\overline{\Omega}_j)\bar{\Ymat}_{JD}^R(\overline{\Omega}_j)^\dagger \right]_{jj'}.\\
        \end{split}
    \end{align}
    Regrouping the square and cross terms together to finally obtain correlated dissipation rates,
    \begin{equation}
        \gamma_{jj'}=\overline{\Omega}_j\left[\overline{\Ymat}_{JD}(\overline{\Omega}_j)\mathsf{J}(\overline{\Omega}_j)\overline{\Ymat}_{JD}(\overline{\Omega}_j)^\dagger\right]_{jj'}.
    \end{equation}
with $\bar{\Ymat}=\bar{\Ymat}^R+\bar{\Ymat}^{NR}$ the total dressed admittance. Furthermore, the kernel $\mathsf{J}(\omega)$ can be written as
    \begin{equation}
        \mathsf{J}(\omega)=\frac{1}{\omega}\Re{\Ymat^{\text{drive}^{-1}}(\omega)},
    \end{equation}
    where 
    \begin{equation}\label{eq:Ydrive}
        \begin{split}
     \Ymat^{\text{drive}}(\omega)&=Z_0^{-1}\bone_{n_D}+\frac{\Lmat_{D}^{-1}}{i\omega}+i\omega \Cmat_{D}\\
     &\equiv Z_0^{-1}\bone_{n_D}+\Ymat^{dc}_{D}(\omega).
    \end{split}
    \end{equation}
    We interpret $\Ymat^{\text{drive}}$ as the external admittance seen by the inner modes, filtered by the shunting capacitances $\Cmat_D$ and inductances $\Lmat_D$ located at drive ports.
    Therefore, the correlated decay rates are fully determined by the admittance that connects qubits and drive ports $\overline{\Ymat}$ as well as the value of the characteristic impedance $Z_0$ of the external transmission lines according to
    \begin{equation} \label{gammaY}
      \gamma_{jj'}=\left[\overline{\Ymat}_{JD}(\overline{\Omega}_j)\Re{\Ymat^{{\text{drive}}^{-1}}(\overline{\Omega}_j)}\overline{\Ymat}_{JD}(\overline{\Omega}_j)^\dagger\right]_{jj'}.  
    \end{equation}
    The Purcell decays are given by diagonal elements of the dissipative rates, that is
    \begin{align} \label{T_1Y}
    \gamma_{j\kappa}=\left[\overline{\Ymat}_{JD}(\overline{\Omega}_j)\Re{\Ymat^{{\text{drive}}^{-1}}(\overline{\Omega}_j)}\overline{\Ymat}_{JD}(\overline{\Omega}_j)^\dagger\right]_{jj}. 
    \end{align}

\subsection{Impedance dissipative rates and Purcell decays} \label{Impedance: dissipation}
In this section, we derive analytical expressions for correlated decay rates and Purcell decays, akin to was done in the previous section, but here as a function of the impedance response. Applying Kirchoff's equations to Cauer circuit of the impedance representation  (see \cref{Duality,Impedance coupler}) we obtain the dissipative equation of motion
\begin{align}
    \Cmat\ddot{\vphi}=-\frac{\partial U}{\partial \vphi}-\Mmat_0\vphi-\Gmat\dot{\vphi}-\Dissipmat*\vphi,
\end{align}
where $\vphi=(\vphi_J^T,\vphi_I^T)^T$ regrouped the junctions and inner modes fluxes, $\Cmat$, $\Mmat_0$, $\Gmat$, are defined  in \cref{Impedance coupler} and $U=U(\vphi_J)$ is the junctions' cosine potential. Similarly to the admittance case, the dissipation matrix 
\begin{equation}
\Dissipmat(t)=\mqty( \Dissipmat^{JJ} & \Dissipmat^{JI} \\ \Dissipmat^{IJ} & \Dissipmat^{II})    
\end{equation}
is defined by its Fourier transform 
\begin{align}
\Dissipmat^{JJ}(\omega)&=-\omega^4\Cmat_{JD} (\frac{i\omega}{Z_0} \mathbbm{1}_{n_D}-\omega^2 \Cmat_D)^{-1}\Cmat_{JD}^T, \\
\Dissipmat^{JI}(\omega)&=\omega^4\Cmat_{JD} (\frac{i\omega}{Z_0} \mathbbm{1}_{n_D}-\omega^2 \Cmat_D)^{-1}\Cmat_{D}\Tmat_D^T,\\
\Dissipmat^{IJ}(\omega)&=\omega^4\Tmat_D\Cmat_{D} (\frac{i\omega}{Z_0} \mathbbm{1}_{n_D}-\omega^2 \Cmat_D)^{-1}\Cmat_{JD}^T ,\\
\Dissipmat^{II}(\omega)&=-\omega^4 \Tmat_D\Cmat_{D} (\frac{i\omega}{Z_0} \mathbbm{1}_{n_D}-\omega^2 \Cmat_D)^{-1}\Cmat_{D}\Tmat_D^T.
\end{align}
Here, $J$ and $I$ stand for junction and inner modes sectors respectively. Unlike the admittance case, the dissipation matrix is here symmetric, i.e., $\Dissipmat^T=\Dissipmat$. This symmetry is a consequence of two factors: First, the absence of \textit{direct} gyration in the impedance response ($\Bmat_\infty=0$) results in the symmetry of the diagonal blocks. Second, the use of the same description (flux-flux, in this case) to parameterize both the junctions and inner modes in the impedance representation implies that the off-diagonal blocks of the dissipation matrix only differ by transposition.
  
The equations of motions can be obtained from  the Caldeira-Leggett Lagrangian 
\begin{align} \label{Appendix: Lagrangian_nodrives}
\mathcal{L}&=\mathcal{L}_S+\mathcal{L}_B+\mathcal{L}_{SB},\\
\mathcal{L}_S&=\frac{1}{2}\dot{\vphi}^T\Cmat\dot{\vphi}-U(\vphi_J)-\frac{1}{2}\vphi^T\Mmat_0\vphi+\frac{1}{2}\dot{\vphi}^T \Gmat\vphi,\\
\mathcal{L}_B&=\sum_\alpha\left(\frac{1}{2}\dot{\vx}_\alpha^T\mmat_\alpha\dot{\vx}_\alpha-\frac{1}{2}\vx_\alpha^T\mmat_\alpha\omegamat_\alpha^2\vx_\alpha\right),\\
\mathcal{L}_{SB}&=-\dot{\vphi}^T\sum_\alpha \lambdamat_\alpha \dot{\vx}_\alpha, 
\end{align}
where the minimal coupling $\mathcal{L}_{SB}$ describes the direct capacitive coupling between qubits and drive ports. The couplings matrices $\{\lambdamat_\alpha\}$ are chosen to reproduce the classical dissipative equation of motion
\begin{equation}
\Dissipmat(\omega)=\lim_{\epsilon \to 0^+} \Mmat(\omega-i\epsilon),
\end{equation}
where
\begin{equation}
\qq{}\Mmat(\omega)=\omega^4\sum_\alpha\lambdamat_\alpha\mmat_\alpha^{-1}(\omega^2-\omegamat_\alpha^2)^{-1}\lambdamat_\alpha^T. 
\end{equation}

Up to correction $\mathcal{O}(\lambdamat_\alpha^2)$, the 
 classical Hamiltonian can be expressed as $H=H_S+H_B+H_{SB}$ with
\begin{align} \label{Appendix: Hamiltonian_nodrives}
\begin{split}
H_S&=\frac{1}{2}(\vq-\frac{\Gmat}{2}\vphi)^T\Cmat^{-1}(\vq-\frac{\Gmat}{2}\vphi)+U(\vphi_J)\\&+\frac{1}{2}\vphi^T\Mmat_0\vphi,
\end{split}\\
H_B&=\sum_\alpha\left(\frac{1}{2}\vp_\alpha^T\mmat_\alpha^{-1}\vp_\alpha+\frac{1}{2}\vx_\alpha^T\mmat_\alpha\omegamat_\alpha^2\vx_\alpha \right),\\
H_{SB}&=\vq^T\sum_\alpha \cmat_\alpha^{\vq,\vp}\vp_\alpha+\vphi^T\sum_\alpha \cmat_\alpha^{\vphi,\vp}\vp_\alpha,
\end{align}
where the Hamiltonian coupling matrices are now given by
\begin{align}
    \cmat_\alpha^{\vq,\vp}&\equiv\mqty(\cmat_\alpha^{\vq_J,\vp} \\ \cmat_\alpha^{\vQ_R,\vp} \\ \cmat_\alpha^{\vQ_G,\vp})=\Cmat^{-1}\lambdamat_\alpha\mmat_\alpha^{-1},\\
    \cmat_\alpha^{\vphi,\vp}&=\frac{\Gmat}{2} \cmat_\alpha^{\vq,\vp}.
\end{align}
Moreover, the dissipation matrix can be written in terms of $\{ \cmat_\alpha^{\vq,\vp}\}$ as
\begin{align}
    \Mmat(\omega)=\omega^4\Cmat\left(\sum_\alpha \cmat_\alpha^{\vq,\vp}(\omega^2-\omegamat^2)^{-1}\mmat_\alpha {\cmat_\alpha^{\vq,\vp}}^T\right)\Cmat^T.
\end{align}
In contrast to Refs \cite{Burkard:2004,Burkard:2005}, the system-bath  $H_{SB}$ in \cref{Appendix: Hamiltonian_nodrives} contains a charge-flux couplings arising from the presence of gyrators in the inner modes description ($\Gmat\neq \mzero$). 

Similar to the non-dissipative case, we apply successively the transformations
\begin{subequations}
\begin{align}
\Pmat_{cu}&=\mqty(\overline{\Cmat}_J^{1/2}\Omat_C & \mzero \\ \mzero & \bone_{m+2l}),\\
\Pmat_{\sDelta} &= \mqty(\bone_n & -\Tmat_{cu}^T \\ \mzero & \bone_{m+2l}),\\
\Pmat_o &= \mqty(\Omat_{\Omega} & \mzero \\ \mzero & \bone_n),
\end{align}
\end{subequations}
to obtain the normal modes of the junctions sectors. Hence, the couplings 
\begin{equation}
 \cmat_\alpha=\mqty(\cmat_\alpha^{\vphi_J,\vp}\\\cmat_\alpha^{\vq_J,\vp} \\\cmat_\alpha^{\vphi_R,\vp}\\ \cmat_\alpha^{\vphi_G,\vp}\\\cmat_\alpha^{\vQ_R,\vp}\\\cmat_\alpha^{\vQ_G,\vp})   
\end{equation}
transform to
 \begin{align}
   \!\!\mqty(\mzero \\ \!\!\Omat_\Omega\overline{\Cmat}_J^{1/2}\!\Omat_C\cmat_\alpha^{\vq_J,\vp}\!\!-\!\Omat_\Omega\overline{\Cmat}_J^{1/2}\!\Omat_C \Rmat_J^T\cmat_\alpha^{\vQ_R,\vp}\!\!-\!\Omat_\Omega\overline{\Cmat}_J^{1/2}\!\Omat_C \Nmat_J^T\cmat_\alpha^{\vQ_G,\vp}\!\!\!\\ \mzero\\\cmat_\alpha^{\vQ_R,\vp}\\\cmat_\alpha^{\vphi_G,\vp}\\\cmat_\alpha^{\vQ_G,\vp})
 \end{align}
 and the system-bath coupling is written compactly as $H_{SB}=\vX^T\sum_\alpha \cmat_\alpha\vp_\alpha$ with $\vX=(\vphi_J^T,\vq_J^T,\vphi_R^T,\vphi_G^T,\vQ_R^T,\vQ_G^T)^T$ the corresponding phase-space coordinate. The second step consists in diagonalizing the gyrator-inner modes (and thus eliminating the zero modes as well) via the symplectic transformations given by \cref{Appendix coupling:S-SI-z}. Consequently, the new effective couplings denoted $\cmat_\alpha'$ are given by 
 \begin{equation}
 \mqty(\mzero \\ \!\!\Omat_\Omega\overline{\Cmat}_J^{1/2}\Omat_C\cmat_\alpha^{\vq_J,\vp}\!-\!\Omat_\Omega\overline{\Cmat}_J^{1/2}\Omat_C \Rmat_J^T\cmat_\alpha^{\vQ_R,\vp}\!-\! \Omat_\Omega\overline{\Cmat}_J^{1/2}\Omat_C \Nmat_J^T\cmat_\alpha^{\vQ_G,\vp}\!\!\\ \mzero\\ \Omega_G^{-\frac{1}{2}}\cmat_\alpha^{\vphi_G,\vp}+\frac{\Sigmamat_x}{2}\Omega_G^{\frac{1}{2}}\cmat_\alpha^{\vQ_G,\vp}\\\Omega_R^{\frac{1}{2}}\cmat_\alpha^{\vQ_R,\vp}\\-\Sigmamat_x\Omega_G^{-\frac{1}{2}}\cmat_\alpha^{\vphi_G,\vp}+\frac{\Omega_G^{\frac{1}{2}}}{2}\cmat_\alpha^{\vQ_G,\vp}).
 \end{equation}

Finally, we eliminate dispersively the inner modes via the symplectic Schrieffer–Wolff transfromation $\Smat_{sw}=\exp(\Amat\Jmat)$, see \cref{SSW}.  Using \cref{Appendix SW:1st-order A} and \cref{Appendix coupling:Hz-K-R-N}, the non-vanishing matrix elements of the generator $\Amat$ are given by
\begin{subequations}\label{Appendix:Generator}
  \begin{align}
     (\Amat_{xp})_{\alpha\beta}&=-\frac{\overline{\Omega}_{J_\alpha}^{\frac{5}{2}}\Omega_{R_\beta}^{-\frac{1}{2}}}{\overline{\Omega}_{J_\alpha}^2-\Omega_{R_\beta}^2}(\overline{\Rmat}_J^T)_{\alpha\beta},\\
     (\Amat_{px})_{\alpha\beta}&=\frac{\overline{\Omega}_{J_\alpha}^{\frac{3}{2}}\Omega_{R_\beta}^{\frac{1}{2}}}{\overline{\Omega}_{J_\alpha}^2-\Omega_{R_\beta}^2}(\overline{\Rmat}_J^T)_{\alpha\beta},\\
     (\Amat_{xx})_{\alpha 2k-1}&=-\frac{\overline{\Omega}_{J_\alpha}^{\frac{5}{2}}\Omega_{G_k}^{-\frac{1}{2}}}{\overline{\Omega}_{J_\alpha}^2-\Omega_{G_k}^2}(\overline{\Nmat}_J^T)_{\alpha 2k},\\
     (\Amat_{xx})_{\alpha 2k}&=-\overline{\Omega}_{J_\alpha}^{\frac{1}{2}}\Omega_{G_k}^{-\frac{1}{2}}(\overline{\Nmat}_J^T)_{\alpha 2k-1},\\
     (\Amat_{xp})_{\alpha 2k-1}&=-\frac{\overline{\Omega}_{J_\alpha}^{\frac{5}{2}}\Omega_{G_k}^{-\frac{1}{2}}}{\overline{\Omega}_{J_\alpha}^2-\Omega_{G_k}^2}(\overline{\Nmat}_J^T)_{\alpha 2k-1},\\
     (\Amat_{xp})_{\alpha 2k}&=-\overline{\Omega}_{J_\alpha}^{\frac{1}{2}}\Omega_{G_k}^{-\frac{1}{2}}(\overline{\Nmat}_J^T)_{\alpha 2k},\\
     (\Amat_{px})_{\alpha 2k-1}&=\frac{\overline{\Omega}_{J_\alpha}^{\frac{3}{2}}\Omega_{G_k}^{\frac{1}{2}}}{\overline{\Omega}_{J_\alpha}^2-\Omega_{G_k}^2}(\overline{\Nmat}_J^T)_{\alpha 2k-1},\\
     (\Amat_{pp})_{\alpha 2k-1}&=-\frac{\overline{\Omega}_{J_\alpha}^{\frac{3}{2}}\Omega_{G_k}^{\frac{1}{2}}}{\overline{\Omega}_{J_\alpha}^2-\Omega_{G_k}^2}(\overline{\Nmat}_J^T)_{\alpha 2k}.
 \end{align}   
\end{subequations}

To first order, this leads to the following couplings,
\begin{align}
\cmat_\alpha \mapsto \Tilde{\cmat}_\alpha=(\bone+\Jmat\Amat+\mathcal{O}(\Amat^2))\cmat_\alpha'.
\end{align}
Projecting on the junctions' subspace, one obtain the final couplings as a function of the original ones
\begin{align} \label{final couplings:Z}
\Tilde{\cmat}_\alpha^{\vphi_J,\vp}&=\Lambdamat_p^{\vphi,\vphi_G}\cmat_\alpha^{\vphi_G,\vp}+\Lambdamat_p^{\vphi,\vQ_G}\cmat_\alpha^{\vQ_G,\vp},\\
\begin{split}
\Tilde{\cmat}_\alpha^{\vq_J,\vp}&=\overline{\Wmat}_J^{\frac{1}{2}}\Omat_\Omega\overline{\Cmat}_J^{\frac{1}{2}}\Omat_C\cmat_\alpha^{\vq_J,\vp}+\Lambdamat_p^{\vq,\vphi_G}\cmat_\alpha^{\vphi_G,\vp}\\
&\qq{}+\Lambdamat_p^{\vq,\vQ_R}\cmat_\alpha^{\vQ_R,\vp}+\Lambdamat_p^{\vq,\vQ_G}\cmat_\alpha^{\vQ_G,\vp},
\end{split}
\end{align}
where the Schrieffer–Wolff (SW) matrices are defined as
\begin{subequations}
    \begin{align}
         (\Lambdamat_{p}^{\vphi,\vphi_G})_{\alpha 2k-1}&=\frac{\overline{\Omega}_{J_\alpha}^{\frac{3}{2}}}{\overline{\Omega}_{J_\alpha}^2-\Omega_{G_k}^2}(\overline{\Nmat}_J^T)_{\alpha 2k-1},\\
          (\Lambdamat_{p}^{\vphi,\vphi_G})_{\alpha 2k}&=\frac{\overline{\Omega}_{J_\alpha}^{\frac{3}{2}}}{\overline{\Omega}_{J_\alpha}^2-\Omega_{G_k}^2}(\overline{\Nmat}_J^T)_{\alpha 2k},\\
           (\Lambdamat_{p}^{\vphi,\vQ_G})_{\alpha 2k-1}&=-\frac{\overline{\Omega}_{J_\alpha}^{\frac{3}{2}}\Omega_{G_k}}{2(\overline{\Omega}_{J_\alpha}^2-\Omega_{G_k}^2)}(\overline{\Nmat}_J^T)_{\alpha 2k},\\
           (\Lambdamat_{p}^{\vphi,\vQ_G})_{\alpha 2k}&=\frac{\overline{\Omega}_{J_\alpha}^{\frac{3}{2}}\Omega_{G_k}}{{2(\overline{\Omega}_{J_\alpha}^2-\Omega_{G_k}^2)}}(\overline{\Nmat}_J^T)_{\alpha 2k-1},\\
            (\Lambdamat_{p}^{\vq,\vphi_G})_{\alpha 2k-1}&=\frac{\overline{\Omega}_{J_\alpha}^{\frac{1}{2}}\Omega_{G_k}}{\overline{\Omega}_{J_\alpha}^2-\Omega_{G_k}^2}(\overline{\Nmat}_J^T)_{\alpha 2k},\\
            (\Lambdamat_{p}^{\vq,\vphi_G})_{\alpha 2k}&=-\frac{\overline{\Omega}_{J_\alpha}^{\frac{1}{2}}\Omega_{G_k}}{\overline{\Omega}_{J_\alpha}^2-\Omega_{G_k}^2}(\overline{\Nmat}_J^T)_{\alpha 2k-1},\\
             (\Lambdamat_{p}^{\vq,\vQ_R})_{\alpha \beta}&=\frac{\overline{\Omega}_{J_\alpha}^{\frac{1}{2}}\Omega_{R_\beta}^2}{\overline{\Omega}_{J_\alpha}^2-\Omega_{G_k}^2}(\overline{\Rmat}_J^T)_{\alpha \beta},\\
             (\Lambdamat_{p}^{\vq,\vQ_G})_{\alpha 2k-1}&=\frac{\overline{\Omega}_{J_\alpha}^{\frac{1}{2}}\Omega_{G_k}^2}{2(\overline{\Omega}_{J_\alpha}^2-\Omega_{G_k}^2)}(\overline{\Nmat}_J^T)_{\alpha 2k-1},\\
             (\Lambdamat_{p}^{\vq,\vQ_G})_{\alpha 2k}&=\frac{\overline{\Omega}_{J_\alpha}^{\frac{1}{2}}\Omega_{G_k}^2}{2(\overline{\Omega}_{J_\alpha}^2-\Omega_{G_k}^2)}(\overline{\Nmat}_J^T)_{\alpha 2k}.         
    \end{align}
\end{subequations}
    Similar to the admittance analysis of \cref{final couplings:Z}, the effective junctions-baths coupling matrices $\Tilde{\cmat}_\alpha$ contains two contributions, one coming from the dressing with inner modes resulting from SW transformations (terms proportional to $\Lambdamat_p$) and direct couplings between qubit and drive ports (terms proportional to the dressed frequency $\overline{\Wmat}_J$). Besides, this coupling matrices allow us to compute the quantum dissipative rates (see \cref{masterequation} for the derivation of the master equation)
    \begin{align}
    \begin{split}
        \gamma_{jj'}&=\frac{\pi}{2}\sum_\alpha \left[\left(\Tilde{\cmat}_{\alpha}^{\vq_J,\vp}+i\Tilde{\cmat}_{\alpha}^{\vphi_J,\vp}\right)^\star\mmat_\alpha\omegamat_\alpha\deltamat(\overline{\Omega}_j-\omegamat_\alpha)\right.\\
    &\left.\qq{}\qq{}\qq{}\times\left(\Tilde{\cmat}_{\alpha}^{\vq_J,\vp}+i\Tilde{\cmat}_{\alpha}^{\vphi_J,\vp}\right)^T\right]_{jj'}.
    \end{split}
    \end{align}
    Using the definitions of $\Tilde{\cmat}_\alpha$, the squared terms in the last equation can be written in terms of the dissipation matrix as
     \begin{equation}
     \begin{split}
      &\frac{\pi}{2}\sum_\alpha \left[\Tilde{\cmat}_{\alpha}^{\vq_J,\vp}\mmat_\alpha\omegamat_\alpha\deltamat(\overline{\Omega}_j-\omegamat_\alpha)\Tilde{\cmat}_{\alpha}^{{\vq_J,\vp}^T}\right]_{jj'}\\&+\frac{\pi}{2}\sum_\alpha \left[\Tilde{\cmat}_{\alpha}^{\vphi_J,\vp}\mmat_\alpha\omegamat_\alpha\deltamat(\overline{\Omega}_j-\omegamat_\alpha)\Tilde{\cmat}_{\alpha}^{{\vphi_J,\vp}^T}\right]_{jj'}\\&=\left[(\zetamat_1^J,\zetamat_1^R,\zetamat_1^G)\Cmat^{-1}\Im\left[\frac{\Dissipmat(\overline{\Omega}_j)}{\overline{\Omega}_j^2}\right]\Cmat^{-T}(\zetamat_1^J,\zetamat_1^R,\zetamat_1^G)^T\right]_{jj'}\\&+\left[(\mzero,\mzero,\zetamat_2^G)\Cmat^{-1}\Im\left[\frac{\Dissipmat(\overline{\Omega}_j)}{\overline{\Omega}_j^2}\right]\Cmat^{-T}(\mzero,\mzero,\zetamat_2^G)^T\right]_{jj'},
     \end{split}
     \end{equation}
    where
      \begin{equation}
          \begin{split}
             \zetamat_1^J&=\overline{\Wmat}_J^{\frac{1}{2}}\Omat_\Omega\overline{C}_J^\frac{1}{2}\Omat_C,\\ 
             \zetamat_1^R&=\Lambdamat_p^{\vq,\vQ_R},\\
             \zetamat_1^G&=\Lambdamat_p^{\vq,\vphi_G}\frac{\Ymat}{2}+\Lambdamat_p^{\vq,\vQ_G},\\
             \zetamat_2^G&=\Lambdamat_p^{\vphi,\vphi_G}\frac{\Ymat}{2}+\Lambdamat_p^{\vphi,\vQ_G},\\
          \end{split}
      \end{equation}
    with 
    \begin{subequations}
        \begin{align}
            \Ymat &= \mqty(\mqty{i\sigma_y \omega_{g_1} & \\ & \ddots}  & \text{\large$\mzero$} \\
            \text{\large$\mzero$} & i\sigma_y \omega_{g_k}),\\
            \Cmat^{-1} &=\mqty(\Cmat_J^{-1} & \Tmat_J^T \\ \Tmat_J & \bone_{m+2l}). 
        \end{align}
        \end{subequations}
     Using the definitions of SW matrices $\{\Lambdamat_p\}$ to evaluate the last set of equation which leads to 
     \begin{equation}
         \begin{split}
              &\frac{\pi}{2}\sum_\alpha \left[\Tilde{\cmat}_{\alpha}^{\vq_J,\vp}\mmat_\alpha\omegamat_\alpha\deltamat(\overline{\Omega}_j-\omegamat_\alpha)\Tilde{\cmat}_{\alpha}^{{\vq_J,\vp}^T}\right]_{jj'}\\&+\frac{\pi}{2}\sum_\alpha \left[\Tilde{\cmat}_{\alpha}^{\vphi_J,\vp}\mmat_\alpha\omegamat_\alpha\deltamat(\overline{\Omega}_j-\omegamat_\alpha)\Tilde{\cmat}_{\alpha}^{{\vphi_J,\vp}^T}\right]_{jj'}\\ 
              &=\overline{\Omega}_j^2\left[{\overline{\Zmat}^R}(\overline{\Omega}_j)\Re{\Zmat^{\text{drive}^{-1}}(\overline{\Omega}_j)}{\overline{\Zmat}^{R}}^\dagger(\overline{\Omega}_j)\right]_{jj'}\\
              &+\overline{\Omega}_j^2\left[\overline{\Zmat}^{NR}(\overline{\Omega}_j)\Re{\Zmat^{\text{drive}^{-1}}(\overline{\Omega}_j)}{\overline{\Zmat}^{NR}}^\dagger(\overline{\Omega}_j)\right]_{jj'}.\\
         \end{split}
     \end{equation}
     The cross terms are obtained similarly
     \begin{equation}
         \begin{split}
           &i\frac{\pi}{2}\sum_\alpha \left[\Tilde{\cmat}_{\alpha}^{\vq_J,\vp}\mmat_\alpha\omegamat_\alpha\deltamat(\overline{\Omega}_j-\omegamat_\alpha)\Tilde{\cmat}_{\alpha}^{{\vphi_J,\vp}^T}\right]_{jj'}\\ &-i\frac{\pi}{2}\sum_\alpha \left[\Tilde{\cmat}_{\alpha}^{\vphi_J,\vp}\mmat_\alpha\omegamat_\alpha\deltamat(\overline{\Omega}_j-\omegamat_\alpha)\Tilde{\cmat}_{\alpha}^{{\vq_J,\vp}^T}\right]_{jj'}\\
    &=\overline{\Omega}_j^2\left[{\overline{\Zmat}^R}(\overline{\Omega}_j)\Re{\Zmat^{\text{drive}^{-1}}(\overline{\Omega}_j)}{\overline{\Zmat}^{NR}}^\dagger(\overline{\Omega}_j)\right]_{jj'}\\
              &+\overline{\Omega}_j^2\left[\overline{\Zmat}^{NR}(\overline{\Omega}_j)\Re{\Zmat^{\text{drive}^{-1}}(\overline{\Omega}_j)}{\overline{\Zmat}^{R}}^\dagger(\overline{\Omega}_j)\right]_{jj'}.\\
         \end{split}
     \end{equation}
    Regrouping the squared and cross terms together, we finally obtain the correlated decay rates
    \begin{equation}
    \gamma_{jj'}=\overline{\Omega}_j^2\left[{\overline{\Zmat}}_{JD}(\overline{\Omega}_j)\Re{\Zmat^{\text{drive}^{-1}}(\overline{\Omega}_j)}\overline{\Zmat}_{JD}^\dagger(\overline{\Omega}_j)\right]_{jj'},
    \end{equation}
    and the Purcell decays are given by the diagonal elements,
    \begin{equation}
        \gamma_{j\kappa}=\overline{\Omega}_j^2\left[{\overline{\Zmat}}_{JD}(\overline{\Omega}_j)\Re{\Zmat^{\text{drive}^{-1}}(\overline{\Omega}_j)}\overline{\Zmat}_{JD}^\dagger(\overline{\Omega}_j)\right]_{jj},
    \end{equation}
    with $\Zmat^{\text{drive}}$  given by
    \begin{equation}
    \begin{split}
        \Zmat^{\text{drive}}(\omega)&=Z_0\bone_{n_D}+\frac{\Cmat_D^{-1}}{i\omega}\\
        &\equiv Z_0\bone_{n_D} + \Zmat^{dc}_D(\omega),
     \end{split}   
    \end{equation}
     Similarly to \cref{eq:Ydrive}, we interpret $\Zmat^{\text{drive}}$ as the external impedance seen by the inner modes, filtered by the capacitances $\Cmat_D$ connected in series with drive ports.

    \subsection{Admittance formulas for driving amplitudes} \label{driveamplitude:Y}   
    
    We now compute the qubit-drives Hamiltonian $\hat{H}_v$ that result from (time-dependent) classical voltage sources $\vV (t)=(V_1 (t),\ldots,V_{n_D}(t))^T$ present in the $n_D$ drives ports, see \cref{fig:Duality}. The classical equations of motion are given by
    \begin{subequations}
    \begin{align}
    \begin{split}
        \Cmat_{J}\ddot{\vphi}_J&=-\pdv{U}{\vphi_J}-\Lmat_{J}^{-1}\vphi_J+\Ymat_{J}^G\dot{\vphi}_J-\Tmat_J\dot{\vQ}\\
        &-\Dissipmat^{\vphi\vphi}*\vphi_J-\Dissipmat^{\vphi \vQ}*\vQ-\Voltmat^{\vphi}*\vV(t), 
    \end{split}\\[2ex]
    \begin{split}
         \ddot{\vQ}=&-\Cmat_I^{-1}\vQ+\Tmat_J^T\dot{\vphi}_J-\Zmat_e\dot{\vQ}-\Dissipmat^{\vQ\vphi}*\vphi_J\\
         &-\Dissipmat^{\vQ \vQ}*\vQ-\Voltmat^{\vQ}*\vV(t), 
       \end{split}  
      \end{align}
      \end{subequations}
    which differs from the equation of motion \cref{eqmotion} by simply adding a voltage source term $\Voltmat*\vV(t)$. Here, the $(n+m+2l)\times n_D$ voltage-source matrix $\Voltmat(t)$ is defined by its Fourier transform
    \begin{subequations}
        \begin{align}
        \begin{split}
           \Voltmat^{\vphi}(\omega)&=-(\Lmat_{JD}^{-1}-\omega^2 \Cmat_{JD}+i\omega \Ymat_{JD}^G)\\
           &\times\left(\frac{i\omega}{Z_0} \mathbbm{1}_{n_D}-\omega^2 \Cmat_{D}+\Lmat_{D}^{-1}\right)^{-1}Z_0^{-1},
        \end{split}\\
           \Voltmat^{\vQ}(\omega)&=i\omega \Tmat_D^T\left(\frac{i\omega}{Z_0} \mathbbm{1}_{n_D}-\omega^2 \Cmat_{D}+\Lmat_{D}^{-1}\right)^{-1}Z_0^{-1}.
        \end{align}
    \end{subequations}
    The existence of non-zero $\Voltmat^{\vphi}$ arises from the direct coupling between the qubit and drive ports ($\Cmat_{JD}$, $\Lmat^{-1}_{JD}$, $\Ymat^G_{JD}$
  $\neq 0 $). The Lagrangian is given by $\mathcal{L}=\mathcal{L}_{\vV=0}+\mathcal{L}_{\vV}$ where $\mathcal{L}_{\vV=0}=\mathcal{L}_S+\mathcal{L}_B+\mathcal{L}_{SB}$ is the time independent Lagrangian (i.e., without drives) given in \cref{Lagrange}. The drives-Lagrangian is given by
    \begin{equation}
        \begin{aligned}
            \mathcal{L}_{\vV}(t)=-\vphi_J^T(\Voltmat^{\vphi}*\vV(t))-\vQ^T(\Voltmat^{\vQ}*\vV(t)),
        \end{aligned}
    \end{equation}
    where $\mathcal{L}_{\vV}$ depends only on the generalized coordinates $\{\vphi_J,\vQ\}$. We note that  one could choose to drive the system via  $\{\dot{\vphi}_J,\dot{\vQ}\}$, both frames being related via a canonical transformation \cite{Goldstein:1950}. However, in the former gauge, the Hamiltonian takes a simpler form $H=H_{\vV=0}+H_{\vV}$ where $H_{\vV=0}$ was computed previously in \cref{Appendix coupling:H-y-quadratic-form} and $H_{\vV} (t)$ takes the form
    \begin{equation}
        \begin{aligned}
         H_{\vV}(t)&=\vphi_J^T(\Voltmat^{\vphi}*\vV(t))+\vQ^T(\Voltmat^{\vQ}*\vV(t))\\
         &=\vX^T(\Voltmat* \vV(t)),
        \end{aligned}
    \end{equation}
    with $\vX^T = (\vphi_J^T , \vq_J^T , \vQ_R^T , \vQ_G^T  , \vPi_R^T , \vPi_G^T )$ the phase space coordinate and $\Voltmat$ given by the $2(n+m+2l)\times n_D$ matrix 
    \begin{equation}
     \Voltmat=\mqty(\Voltmat^{\vphi}\\ 0 \\ \Voltmat^{\vQ_R}\\ \Voltmat^{\vQ_G} \\0 \\0).
    \end{equation}
    Here, we remind the reader that $n$, $m$, $l$ are the number of junction ports, LC-oscillators (reciprocal poles of the response) and gyrators (nonreciprocal poles of the response) respectively. 
    
    The voltage-source matrix transforms under a symplectic transformation $\Smat$ as 
    \begin{equation*}
         \Voltmat(t)\mapsto \widetilde{\Voltmat}(t)\equiv (\Smat^{-1})^T \Voltmat(t).
    \end{equation*}
    By linearity of the Fourier transform, it translates in frequency domain as
    \begin{equation*}
         \Voltmat(\omega)\mapsto \widetilde{\Voltmat}(\omega)\equiv(\Smat^{-1})^T \Voltmat(\omega).
    \end{equation*}
    Applying successively 
    \begin{equation}
        \mqty(\Smat_J & 0 \\ 0 & \Smat_I)
    \end{equation}
    given by \cref{Appendix coupling:S-Sj-z} and \cref{Appendix coupling:S-SI-z} to diagonalize both junctions and inner modes sectors separately, and the symplectic SW $\Smat_{sw}=\exp(\Amat\Jmat)$ to eliminate dispersively the inner modes (reciprocal and nonreciprocal poles) as detailed in \cref{SSW}, the transformed voltage-source matrix $\widetilde{\Voltmat}(\omega)$ projected on the junctions subspace is given by
    \begin{subequations}
    \begin{align}
    \begin{split}
    \widetilde{\Voltmat}^{\vphi_J}&=\overline{\Wmat}_J^{-\frac{1}{2}}\Omat_\Omega\overline{\Cmat}_J^{-\frac{1}{2}}\Omat_C \Voltmat^{\vphi}+\Amat_{px}\mqty(\Omega_R^{-\frac{1}{2}} \Voltmat^{\vQ_R}\\\Omega_G^{-\frac{1}{2}} \Voltmat^{\vQ_G} )\\
    &+\Amat_{pp}\mqty(0\\\Sigmamat_x\Omega_G^{-\frac{1}{2}} \Voltmat^{\vQ_G}),
    \end{split}\\
    \widetilde{\Voltmat}^{\vq_J}&=-\Amat_{xx}\mqty(\Omega_R^{-\frac{1}{2}} \Voltmat^{\vQ_R}\\\Omega_G^{-\frac{1}{2}} \Voltmat^{\vQ_G})-\Amat_{xp}\mqty(0\\\Sigmamat_x\Omega_G^{-\frac{1}{2}} \Voltmat^{\vQ_G} ),
    \end{align}
    \end{subequations}
    where the generators $\{\Amat_{xx},\Amat_{xp}, \Amat_{px},\Amat_{pp}\}$ are computed as in \cref{Appendix SW:1st-order A} in \cref{SSW}. This leads to 
    \begin{equation}
    \begin{aligned}
     \widetilde{\Voltmat}^{\vphi_J}(\omega)&=\overline{\Wmat}_J^{-\frac{1}{2}}\left(\overline{\Ymat}^{\text{NR}}(\overline{\Wmat}_J)+\overline{\Ymat}^{dc}(\omega)\right)\Ymat^{\text{drive}^{-1}(\omega)}/Z_0,\\
     \widetilde{\Voltmat}^{\vq_J}(\omega)&=-\overline{\Wmat}_J^{-\frac{1}{2}}\Im[\overline{\Ymat}^{\text{ac,R}}(\overline{\Wmat}_J)]\Ymat^{\text{drive}^{-1}}(\omega)/Z_0,\\
    \end{aligned}
    \end{equation}
    where $\Ymat^{\text{drive}}$ is given by:
    \begin{equation}
        \Ymat^{\text{drive}}(\omega)=Z_0^{-1}\bone_{n_D}+\frac{\Lmat_{D}^{-1}}{i\omega}+i\omega \Cmat_{D}.
    \end{equation}
    Similarly, the off-diagonal $dc$ part of the admittance $\Ymat_{JD}^{dc}$ is
    \begin{equation}
     \Ymat_{JD}^{dc}(\omega)=\frac{1}{i\omega}\Lmat_{JD}^{-1}+i\omega \Cmat_{JD}   .
    \end{equation}
    Therefore, the quantum Hamiltonian due to the external drives takes the form
    \begin{equation}
    \hat{H}_v=\sum_{j=1}^{n}\sum_{d=1}^{n_D}\left( \varepsilon_{jd}(t)\hat{b}_j+\varepsilon_{jd}^\star(t)\hat{b}_j^\dagger\right),
    \end{equation}
    where the driving amplitude contribution of  junction port $j$ from drive port $d$ is
    \begin{equation}
    \label{Appendix:drivesadmittance}
    \begin{split}
    \varepsilon_{jd}(t)&=\frac{1}{\sqrt{2}}\left(\widetilde{\Voltmat}^{\vphi_J}-i\widetilde{\Voltmat}^{\vq_J}\right)_{jd}* V_d(t) \\
    &=\frac{\overline{\Omega}_j^{-\frac{1}{2}}}{2\pi\sqrt{2}}\int_{-\infty}^{+\infty}d\omega e^{i\omega t}\\
    &\times\left(\left(\overline{\Ymat}_{JD}^{ac}(\overline{\Wmat}_J)+\overline{\Ymat}_{JD}^{dc}(\omega)\right)\Ymat^{\text{drive}^{-1}}(\omega)/Z_0\right)_{jd}V_d(\omega),
    \end{split}
    \end{equation}
    with the $ac$ part of the admittance given by $\Ymat^{ac}=\Ymat^{ac,R}+\Ymat^{\text{NR}}$. 

    For a single tone drive $V_d(t)=v_d\sin(\omega_d t)$, we obtain explicitly
    \begin{equation}\label{driveamplitudeY}
    \varepsilon_{jd}(t)=-i\frac{\overline{\Omega}_j^{-\frac{1}{2}}v_d}{2\sqrt{2}Z_0}\left(\alpha_{jd}[\omega_d]e^{i\omega_d t}-\alpha_{jd}[-\omega_d]e^{-i\omega_d t}\right), 
    \end{equation}
    with 
    \begin{equation}
        \begin{aligned}
            \alpha_{jd}[\omega_d]&=\left(\left(\overline{\Ymat}_{JD}^{ac}(\overline{\Omega}_j)+\overline{\Ymat}_{JD}^{dc}(\omega_d)\right)\Ymat^{\text{drive}^{-1}}(\omega_d)\right)_{jd}\\
        =&\sum_{d'=1}^{n_D}\left(\overline{\Ymat}_{jd'}^{ac}(\overline{\Omega}_j)+\overline{\Ymat}_{jd'}^{dc}(\omega_d)\right)\Ymat^{\text{drive}^{-1}}_{d'd}(\omega_d).
        \end{aligned}
    \end{equation}

    For realistic pulse drives with finite rise and fall times, one can either use the integral formula  \cref{Appendix:drivesadmittance} or decompose the pulse into its Fourier series $V_d(t)=\sum_{\omega_d}v_d\sin(\omega_d t)$ to account for the effect of other harmonics $\{\omega_d\}$, and the total drive amplitudes is simply the sum of single-tone drives amplitudes, i.e.,
    \begin{equation}
        \varepsilon_{jd}^{\text{tot}}(t)=\sum_{\omega_d}\varepsilon_{jd}(t),
    \end{equation}
    with $\varepsilon_{jd}(t)$ being the single-tone drives amplitude associated to frequency $\omega_d$ given by \cref{driveamplitudeY}.\\

    The analytical formula obtained in \cref{Appendix:drivesadmittance} allows us to characterize the classical crosstalk $X_{ij}$ experienced by qubit $i$ while driving from the port $d(j)$ associated to the control line of qubit $j$, taking into account not only the possible coupling between junction port $i$ and  drive port $d(j)$ but also the possible stray coupling between drives' ports $d(i)$ and $d(j)$ which occurs whenever $\Ymat^{\text{drive}}$ is not diagonal. Thus, we extend the definition of classical crosstalks $X_{ij}$, as given in Ref.~\cite{Solgun:2019}, to account for both direct and nonreciprocal couplings: 
    \begin{equation} \label{Appendix: admittance crosstalks}
        X_{ij}=20\log_{10}\abs{\frac{\alpha_{id(j)}}{\alpha_{jd(j)}}} \: \: \: \: \: (\text{dB}).
    \end{equation}
    The matrix element $\alpha_{jd}$ is obtained from an $n \times n_D$ matrix $\alphamat$, defined as $\alphamat=\{\overline{\Ymat}_{JD}^{ac}(\overline{\Wmat}_J)+\overline{\Ymat}_{JD}^{dc}(\omega_d)\}\Ymat^{\text{drive}^{-1}}(\omega_d)$, where $\omega_d$ is the drive frequency and using the notation $[\overline{\Ymat}(\overline{\Wmat}_J)]_{jd}=\overline{\Ymat}_{jd}(\overline{\Wmat}_j)$. In summary, the matrix $\alphamat$ enables the computation of drive amplitudes $\varepsilon_{jd}(t)$ (see \cref{driveamplitudeY}) and classical crosstalks between qubit ports during drive operations from different control lines (see \cref{Appendix: admittance crosstalks}).

    \subsection{Impedance formulas for driving amplitudes} \label{driveamplitude:Z}
    
    We now obtain analytical expressions for the qubit-drives Hamiltonian in terms of the impedance response. Taking into account classical voltage sources $\vV(t)=(V_1 (t),\ldots,V_{n_D}(t))^T$, the Kirchhoff's equations are given by
    \begin{equation}
        \Cmat\ddot{\vphi}=-\frac{\partial U}{\partial \vphi}-\Mmat_0\vphi-\Gmat\dot{\vphi}-\Dissipmat*\vphi-\Voltmat*\vV (t),
    \end{equation}
    where $\vphi=(\vphi_J^T,\vphi_R^T,\vphi_G^T)^T$. The $(n+m+2l)\times n_D$ voltage-matrix 
    \begin{equation}
        \Voltmat(t)=\mqty(\Voltmat^J (t) \\ \Voltmat^R (t)\\ \Voltmat^G (t) \\)
    \end{equation} 
    is defined by its  Fourier transform
    \begin{subequations}
        \begin{align}
         \Voltmat^J (\omega)&=-\omega^2\Cmat_{JD}\left(\frac{i\omega}{Z_0} \mathbbm{1}_{n_D}-\omega^2\Cmat_D\right)^{-1}Z_0^{-1},\\  
          \Voltmat^R (\omega)&=+\omega^2\Rmat_D\Cmat_{D}\left(\frac{i\omega}{Z_0} \mathbbm{1}_{n_D}-\omega^2\Cmat_D\right)^{-1}Z_0^{-1},\\
          \Voltmat^G (\omega)&=+\omega^2\Nmat_D\Cmat_{D}\left(\frac{i\omega}{Z_0} \mathbbm{1}_{n_D}-\omega^2\Cmat_D\right)^{-1}Z_0^{-1}.
        \end{align}
    \end{subequations}
    As in the admittance case, the Lagrangian can be divided in two parts, namely a time-independent Lagrangian $\mathcal{L}_{\vV=0}$ given by \cref{Appendix: Lagrangian_nodrives} and a time-dependent part $\mathcal{L}_{\vV}$
    \begin{equation}
        \mathcal{L}=\mathcal{L}_{\vV=0}+\mathcal{L}_{\vV}(t),
    \end{equation}
    where 
    \begin{equation}
    \begin{split}
     \mathcal{L}_{\vV}(t)&=-\vphi_J^T\left(\Voltmat^J*\vV(t)\right)-\vphi_R^T\left(\Voltmat^R*\vV(t)\right)\\
     &\qq{}-\vphi_G^T\left(\Voltmat^G*\vV(t)\right) .
     \end{split}
    \end{equation}
    Also as in the admittance case, we make a gauge choice in which we drive the system via the generalized coordinates $(\vphi_J,\vphi_R,\vphi_G)$. This choice is to be contrasted to Ref.~\cite{Solgun:2019} where the choice of driven quadratures is instead $(\dot{\vphi}_J,\dot{\vphi}_R)$. By definition, both pictures reproduce classical equations of motion and the corresponding lagrangians are related by a total derivative. 
    
    In this frame, the Legendre transform is simple to perform and we obtain a compact expression for the  total Hamiltonian
     \begin{equation}
      H=H_{\vV=0}+H_{\vV}(t), 
     \end{equation}
    where $H_{\vV=0}$ is given in \cref{Appendix coupling:H-z-quadratic-form}. The 
     drive Hamiltonian $H_{\vV}(t)$ is:
     \begin{equation}
      \begin{split}
        H_{\vV}(t)&=\vphi_J^T\left(\Voltmat^J*\vV(t)\right)+\vphi_R^T\left(\Voltmat^R*\vV(t)\right)\\
        &\qq{}\qq{}+\vphi_G^T\left(\Voltmat^G*\vV(t)\right)\\ 
        &=\vX^T\left(\Voltmat*\vV(t)\right),
    \end{split}
     \end{equation}
    with 
    \begin{align}
        \vX^T &=(\vphi_J^T,\vq_J^T,\vphi_R^T,\vphi_G^T,\vQ_R^T,\vQ_G^T),\\  
        \Voltmat&=\mqty(\Voltmat^J\\ \mzero\\\Voltmat^R\\\Voltmat^G\\\mzero\\ \mzero).    
    \end{align}
    After diagonalizing the junctions modes, the voltage-matrix is transformed as
    \begin{equation}
       \Voltmat \mapsto \mqty(\Omat_\Omega\overline{\Cmat}_J^{-\frac{1}{2}}\Omat_C \Voltmat^J\\ \mzero\\ \Voltmat^R+\Rmat_J\Voltmat^J\\ \Voltmat^G+\Nmat_J\Voltmat^J\\ \mzero\\ \mzero).
    \end{equation}
    Repeating the same steps of diagonalizing the inner modes and rescaling by dressed frequencies leads to
    \begin{equation}
        \Voltmat'=\mqty(\overline{\Wmat}_J^{-\frac{1}{2}}\Omat_\Omega\overline{\Cmat}_J^{-\frac{1}{2}}\Omat_C \Voltmat^J\\ \mzero\\ \overline{\Omega}_R^{-\frac{1}{2}}\Voltmat^R+\overline{\Omega}_R^{-\frac{1}{2}}\Rmat_J\Voltmat^J\\ \overline{\Omega}_G^{-\frac{1}{2}}\Voltmat^G+\overline{\Omega}_G^{-\frac{1}{2}}\Nmat_J\Voltmat^J\\ \mzero\\-\Sigmamat_x\overline{\Omega}_G^{-\frac{1}{2}}\Voltmat^G-\Sigmamat_x\overline{\Omega}_G^{-\frac{1}{2}}\Nmat_J\Voltmat^J ).
    \end{equation}
    Finally, eliminating dispersively the inner modes via a SW transformation yields the following voltage-matrices couplings
    \begin{subequations}
    \begin{align}
        \begin{split}
        \widetilde{\Voltmat}^{\vphi_J}&=\overline{\Wmat}_J^{-\frac{1}{2}}\Omat_\Omega\overline{\Cmat}_J^{-\frac{1}{2}}\Omat_C \Voltmat^J\\&+\Amat_{px}\mqty(\overline{\Omega}_R^{-\frac{1}{2}}\Voltmat^R+\overline{\Omega}_R^{-\frac{1}{2}}\Rmat_J\Voltmat^J\\ \overline{\Omega}_G^{-\frac{1}{2}}\Voltmat^G+\overline{\Omega}_G^{-\frac{1}{2}}\Nmat_J\Voltmat^J)\\
        &+\Amat_{pp}\mqty(\mzero\\-\Sigmamat_x\overline{\Omega}_G^{-\frac{1}{2}}\Voltmat^G-\Sigmamat_x\overline{\Omega}_G^{-\frac{1}{2}}\Nmat_J\Voltmat^J ),
        \end{split}
        \\
        \begin{split}
           \widetilde{\Voltmat}^{\vq_J}&=-\Amat_{xx}\mqty(\overline{\Omega}_R^{-\frac{1}{2}}\Voltmat^R+\overline{\Omega}_R^{-\frac{1}{2}}\Rmat_J\Voltmat^J\\ \overline{\Omega}_G^{-\frac{1}{2}}\Voltmat^G+\overline{\Omega}_G^{-\frac{1}{2}}\Nmat_J\Voltmat^J)\\
           &-\Amat_{xp}\mqty(\mzero \\\Sigmamat_x\overline{\Omega}_G^{-\frac{1}{2}}\Voltmat^G-\Sigmamat_x\overline{\Omega}_G^{-\frac{1}{2}}\Nmat_J\Voltmat^J ),
        \end{split}
    \end{align}
    \end{subequations}
    where $\Amat$ is the generator of the SW transformation given in \cref{Appendix:Generator}. Consequently, the previous equations can be written in terms of the impedance response
    \begin{subequations}
    \begin{align}
    \begin{split}
     (\widetilde{\Voltmat}^{\vphi_J})_{jd}&=\biggl[\left(\overline{\Omega}_j^{-\frac{1}{2}}\omega\Im\left[\overline{\Zmat}^{dc}_{JD}(\omega)\right] \right.\\ &\left.+\overline{\Omega}_j^{\frac{1}{2}}\Im\left[\overline{\Zmat}^{ac,R}_{JD}(\overline{\Omega}_j)\right]\right)\Zmat^{\text{drive}^{-1}}(\omega)\biggr]_{jd},
     \end{split}\\
    (\widetilde{\Voltmat}^{\vq_J})_{jd}&=\left[\overline{\Omega}_j^{\frac{1}{2}}\overline{\Zmat}^{NR}_{JD}(\overline{\Omega}_j)\Zmat^{\text{drive}^{-1}} (\omega)\right]_{jd}.
    \end{align}
    \end{subequations}
    
    After quantizing the junction modes  \cite{Vool:2017}, the quantum Hamiltonian accounting for the the external drives takes the form
    \begin{equation}
    \hat{H}_v=\sum_{j=1}^{n}\sum_{d=1}^{n_D}\left( \varepsilon_{jd}(t)\hat{b}_j+\varepsilon_{jd}^\star(t)\hat{b}_j^\dagger\right),
    \end{equation}
    where the drive amplitudes between junction port $j$ and drive port $d$ are
    \begin{equation}\label{Appendix:generaldrivesadmittance}
    \begin{split}
        &\varepsilon_{jd}(t)=\frac{1}{\sqrt{2}}\left(\widetilde{\Voltmat}^{\vphi_J}-i\widetilde{\Voltmat}^{\vq_J}\right)_{jd}* V_d(t) \\
        &=-\frac{i\overline{\Omega}_j^{\frac{1}{2}}}{2\pi\sqrt{2}}\int_{-\infty}^{+\infty}d\omega e^{i\omega t}\\
        &\qq{}\times\left(\left(\overline{\Zmat}_{JD}^{ac}(\overline{\Wmat}_J)+\omega\overline{\Wmat}_J^{-1}\overline{\Zmat}_{JD}^{dc}(\omega)\right)\Zmat^{\text{drive}^{-1}}(\omega)\right)_{jd}V_d(\omega),
    \end{split}
    \end{equation}
    where the $ac$ part of the impedance is given by $\Zmat^{ac}=\Zmat^{ac,R}+\Zmat^{\text{NR}}$. For a single tone drive $V_d(t)=v_d\sin(\omega_d t)$, we obtain explicitly
    \begin{equation}\label{driveamplitudeZ}
    \varepsilon_{jd}(t)=-\frac{\overline{\Omega}_j^{\frac{1}{2}}v_d}{2\sqrt{2}}\left(\alpha_{jd}[\omega_d]e^{i\omega_d t}-\alpha_{jd}[-\omega_d]e^{-i\omega_d t}\right), 
    \end{equation}
    with 
    \begin{equation}
    \begin{aligned}
        \alpha_{jd}[\omega_d]&=\left[\left(\overline{\Zmat}_{JD}^{ac}(\overline{\Omega}_j)+\omega_d \overline{\Wmat}_J^{-1}\ \overline{\Zmat}_{JD}^{dc}(\omega_d)\right)\Zmat^{\text{drive}^{-1}}(\omega_d)\right]_{jd}\\
    =&\sum_{d'=1}^{n_D}\left(\overline{\Zmat}_{jd'}^{ac}(\overline{\Omega}_j)+\omega_d \overline{\Omega}_j^{-1}\overline{\Zmat}_{jd'}^{dc}(\omega_d)\right)\Zmat^{\text{drive}^{-1}}_{d'd}(\omega_d).
    \end{aligned}
    \end{equation}
    Analogously to the admittance analysis, we can compute the classical crosstalk $X_{ij}$ experienced by a qubit $i$  while driving from the control line $d(j)$ associated to the qubit $j$ to find
    \begin{equation} \label{Appendix: impedance crosstalks}
        X_{ij}=20\log_{10}\abs{\frac{\alpha_{id(j)}}{\alpha_{jd(j)}}} \: \: \: \: \: (\text{dB}),
    \end{equation}
    with $\alpha_{jd}$ is the matrix element of $n\times n_D$ matrix $\alphamat=\left(\overline{\Zmat}_{JD}^{ac}(\overline{\Wmat}_J)+\omega\overline{\Wmat}_J^{-1}\overline{\Zmat}_{JD}^{dc}(\omega_d)\right)\Zmat^{\text{drive}^{-1}}(\omega_d)$. Here, $\omega_d$ is the drive frequency, and we use the notation $[\overline{\Zmat}(\overline{\Wmat}_J)]_{jd}=\overline{\Zmat}_{jd}(\overline{\Wmat}_j)$.

    \section{Exact elimination of nondynamical modes}
    \label{Elimination-Nondynamical}

    Because in the perturbative regime the approximate and exact elimination yield the same effective models, in  \cref{Derivation-main} we have approximately eliminated the nondynamical modes. Here, we perform an exact elimination of these nondynamical modes, extending the validity beyond the dispersive regime, partially in line with the approach outlined in Ref.~\cite{Egusquiza:2022}. We emphasize that despite its ad-hoc nature this approach is equivalent to the novel method for zero-mode elimination proposed in~\cite{ParraRodriguez:2023,ParraRodriguez:2024}.
    
    For the admittance, our starting point is \cref{Appendix coupling:H-y}. From there, we perform the symplectic triangular transformation
    \begin{equation}
    \Smat_{\sDelta_y}:
        \begin{cases}
            \vphi_J \to \vphi_J, &\; \vq_J \to \vq_J -\Tmat_J\vQ \\
            \vQ \to \vQ &\; \vPi \to \vPi-\Tmat_J^T\vphi_J.
        \end{cases}
    \end{equation}
    Following, to eliminate the nondynamical modes we apply the symplectic transformation done in the main derivation \cref{Appendix coupling:SI-Y}. After these transformations, the quadratic form of the Hamiltonian (exluding the linear terms from the junction potentials) is
    \begin{equation}
    \label{Appendix nondynamical:H-generic}
        \Hmat = \mqty(\Hmat_J & \Kmat \\ \Kmat^T & \Hmat_I),
    \end{equation}
    with
    \begin{equation}
        \begin{split}
            \Hmat_J &= \mqty(\Tmat_J\Tmat_J^T & \Ymat_J^G\Cmat_J^{-1}/2 \\
            (\Ymat_J^G\Cmat_J^{-1})^T/2 & \Cmat_J^{-1}), \\
            \Hmat_I &= \mqty(\mqty{\Wmat_r & \\ & \Wmat_r} & \mzero \\ \mzero & \mqty{\Wmat_g & \\ & \Wmat_g} ),
        \end{split}
    \end{equation}
    and 
    \begin{align}
        \Kmat &= \mqty(\mzero & -\Rmat_J\Wmat_r^{1/2} & \Nmat_J^R\Wmat_g^{1/2} & -\Nmat_J^L\Wmat_g^{1/2}\\
        \mzero & \mzero & \mzero & \mzero)\\
        \Nmat_J^{L(R)} &= \mqty(\vb{n}_1^{L(R)}, \dots \vb{n}_l^{L(R)}).
    \end{align}
    As we have not linearized the junction potential in this elimination, the full Hamiltonian is $H = \vb{X}^T\Hmat\vb{X}/2+U(\boldsymbol{\phi}_J)$ with the junctions potential $U(\boldsymbol{\phi}_J) =-\sum_n E_{J_n}\cos(\phi_{J_n}/\phi_0),$ and $\vb{X} = (\vphi_J , \vq_J , \vQ_R , \vPi_r , \vQ_G^D , \vPi_g^D)$. The superscript $D$ stands for dynamical, as all the modes in the gyrator sector now correspond to dynamical modes, and its dimension has been reduced by half, from $2l\to l$.

    For the impedance, we start directly from the Lagrangian \cref{Appendix coupling: Lgr-0}. Instead of doing a triangular transformation, we directly obtain the inverse of $\Cmat$ using known formulas for the inversion of $2\times 2$ block matrices \cite{Lu:2002}
    \begin{equation}
        \Cmat^{-1} = \mqty(\Cmat_J^{-1}+\Tmat_J^T\Tmat_J & \Tmat_J^T \\ \Tmat_J & \bone_{m+2l}).
    \end{equation}
    Then, the Hamiltonian is simply
    \begin{equation}
    \begin{split}
        H &= \frac{1}{2}(\vb{q}-\frac{\Gmat}{2}\vphi)^T\Cmat^{-1}(\vb{q}-\frac{\Gmat}{2}\vphi)\\
        &\qq{}\qq{}\;\;+\frac{1}{2}\vphi_r^T\Lmat_r^{-1}\vphi_r+U(\boldsymbol{\phi}_J).
    \end{split}
    \end{equation}
    We eliminate the nondynamical modes applying the same symplectic transformations as in the main derivation \cref{Appendix coupling:S-SI-z}. The quadratic form of the Hamiltonian (once again exluding the linear terms from the junction potentials)  finally read as  \cref{Appendix nondynamical:H-generic} now with 
    \begin{equation}
    \begin{split}
        \Hmat_J &= \mqty(\mzero & \mzero \\ \mzero & \Cmat_J^{-1}+\Tmat_J^T\Tmat_J),\\
        \Hmat_I &= \mqty(\mqty{\Wmat_r & \\ & \Wmat_r} & \mzero \\ \mzero & \mqty{\Wmat_g & \\ & \Wmat_g} ),
    \end{split}
    \end{equation}
    and
    \begin{equation}
        \Kmat = \mqty(\mzero & \mzero & \mzero & \mzero\\
        \mzero & \Rmat_J^T\Wmat_r^{1/2} & (\Nmat_J^R)^T\Wmat_g^{1/2} & (\Nmat_J^L)^T\Wmat_g^{1/2}
        ).
    \end{equation}
    The full Hamiltonian is $H = \vb{X}^T\Hmat\vb{X}/2+U(\boldsymbol{\phi}_J)$ with $\vb{X} = \mqty(\vphi_J, & \vq_J ,& \vPi_r, & \vQ_R, & \vPi_g^D, & \vQ_G^D)$, and $U(\boldsymbol{\phi}_J)$ the junctions potential. 
    
    We stress that for a proper quantization of the whole response beyond the perturbative regime, Hamiltonians where the nondynamical modes have been exactly eliminated must be used \cite{Egusquiza:2022}.
    
    \section{Josephson junctions' nonlinearities}
    \label{NL}

    We now derive the first order corrections from the weakly anharmonic nonlinear potential of the Josephson junctions. We obtain qubit anharmonicities, dispersive shifts and linear corrections from normal ordering of a Taylor expansion of the junction potential. We do so with a general derivation of the corrections from the dressing of the junctions mode, which is valid for both admittance and impedance approaches. Consider the general case in which the initial Josephson flux variables $\tilde\vphi_J$ are related to the final frame variables $\vX = (\vphi , \vq)$, expressed in dimensions of $(\text{energy}\times \text{time})^{1/2}$ via a dimensionless transformation matrix $\alpha$ of shape $(n\times t)$
    \begin{equation}
    \label{NL:eq-1-alpha}
        \tilde\vphi_J = \Zmat_J^{1/2}\alpha\vb{X}=\Zmat_J^{1/2}(\alpha^{\phi}\vphi+\alpha^{q}\vq),
    \end{equation}
    with $\Zmat_J^{1/2}$ the necessary diagonal rescaling to obtain dimensions of flux, $n$ the number of junctions and $t=(n+m+l)$ the total number of dynamical modes. We have separated the flux from charge sectors in the final frame to simplify the following calculations. 
    
    The nonlinear terms in the original frame are
    \begin{equation}
        \tilde H_{nl} = -\sum_{i}\frac{1}{24\widetilde{L}_{J_i}\phi_0^2}\tilde\phi_{J_i}^4
    \end{equation}
    such that in the final frame we have
    \begin{equation}
        H_{nl} = -\sum_{i=1}^n\frac{1}{24\widetilde{L}_{J_i}\phi_0^2}z_{J_i}^2(\sum_{k=1}^{t} \alpha^{\phi}_{ik}\phi_k+\alpha^{q}_{ik}q_k)^4.
    \end{equation}
    Introducing
    \begin{equation}
        \begin{split}
            \hat{\phi}_k &= \sqrt{\frac{\hbar}{2}}(\bdag_k+\blow_k),\\
            \hat{q}_k &= i\sqrt{\frac{\hbar}{2}}(\bdag_k-\blow_k),
        \end{split}
    \end{equation}
    substituting and rearranging we get
    \begin{equation}
    \label{appendix-NL:H_nl-fourth}
        H_{nl} = -\sum_{i}\frac{E_C^{(i)}}{12}\frac{z_{J_i}^2}{\widetilde{z}_{J_i}^2}(\sum_k \widetilde{\alpha}_{ik}\blow_k+\widetilde{\alpha}_{ik}^*\bdag_k)^4,
    \end{equation}
    with $E_{C_i}=e^2/2C_{J_i}$, $\widetilde{z}_{J_i}=\sqrt{\widetilde{L}_{J_i}/C_{J_i}}$, $\widetilde{\alpha} = \alpha^{\phi}-i\alpha^{q}$. Here, $C_{J_i}$ is simply $(\Cmat_J)_{ii}$ obtained from either $\Dmat_{\infty}$ or $\Amat_0$. 
    
    We expand the series using the general formula 
    \begin{equation}
    \label{Appendix-NL:general-multinomial-expansion}
    \begin{split}
        \big(\sum_k \alpha_{k}\blow_k+\alpha_k^*\bdag_k\big)^N &= \sum_{j,s}\Big\{ C_{j,s}^{N}\big(\sum_k \alpha_k^*\bdag_k\big)^s\\
        &\quad\;\;\;\times\big(\sum_k\alpha_k\blow_k\big)^{t}\big(\sum_k|\alpha_k|^2\big)^j\Big\},
    \end{split}
    \end{equation}
    where $t,j,s$ are nonnegative integers, $C_{j,s}^N = \boldleft N!/j!s!t!2^j\boldright$, $0\leq j\leq N$, $0\leq s \leq N-2j$, and $t=N-2j-s$. We obtained the above expression with a rearrangement of the formulas presented in Ref. \cite{Morales:1989}. Hence, expanding \cref{appendix-NL:H_nl-fourth} we have $H_{nl} = H_{\nu} + H_{\beta} + \mathcal{O}(\varphi_J^6)$ with
    \begin{equation}
    \begin{split}
        H_{\nu} &= -\sum_{rs}^{t}\qty(\nu_{rs}\blow_r\blow_s+2\nu_{\overline{r}s}\bdag_r\blow_s+\nu_{\overline{r}\overline{s}}\bdag_r\bdag_s) \\
        H_{\beta} &= -\sum_{pqrs}^{t}\Bigl(6\beta_{\overline{p}\overline{q}rs}\bdag_p\bdag_q\blow_r\blow_s\\
        &\qq{}\qq{}+(4\beta_{\overline{p}\overline{q}\overline{r}s}\bdag_p\bdag_q\bdag_r\blow_s+\text{h.c.})\\
        &\qq{}\qq{}+(\beta_{pqrs}\blow_p\blow_q\blow_r\blow_s+\text{h.c.})\Bigr),
    \end{split}
    \label{AppendixNL:expansion}
    \end{equation}
    where the coefficients are
    \begin{equation}
        \begin{split}
            \beta_{pqrs}&= \sum_i^n\frac{E_C^{(i)}}{12}\frac{z_{J_i}^2}{\widetilde{z}_{J_i}^2}\widetilde{\alpha}_{ip}\widetilde{\alpha}_{iq}\widetilde{\alpha}_{ir}\widetilde{\alpha}_{is}, \\
            \nu_{rs} &= 6\sum_p \beta_{p\overline{p}rs}.
        \end{split}
    \end{equation}
    The bar over the subindexes refer to conjugation, for example 
    \begin{equation}
        \beta_{\overline{p}\overline{q}rs}= \sum_i^n\frac{E_C^{(i)}}{12}\frac{z_{J_i}^2}{\widetilde{z}_{J_i}^2}\widetilde{\alpha}_{ip}^*\widetilde{\alpha}_{iq}^*\widetilde{\alpha}_{ir}\widetilde{\alpha}_{is}.
    \end{equation}  
    The inclusion of these conjugate coefficients is crucial because, in contrast to the expansion for the nonlinearities given in Ref.~\cite{Nigg:2012} and used in Ref.~\cite{Solgun:2019}, we are allowing $\alpha$ to mix flux and charge variables. 
    
    Now, assuming that all the nondiagonal terms of $\widetilde{\alpha}$ are $\widetilde{\alpha}_{ik}\sim \mathcal{O}(k/\Delta)$ and that $\widetilde{\alpha}_{ii}=1+\mathcal{O}(k^2/\delta^2)$, then in similar fashion as to what was done in Ref.~\cite{Solgun:2019}, to first order we obtain for the linear coefficients between qubit modes
    \begin{equation}
        \gamma_{\overline{i}j}\simeq \frac{E_c^{(i)}}{2}\qty(\frac{z_{J_i}}{\widetilde{z}_{J_i}})^2\widetilde{\alpha}_{ii}\widetilde{\alpha}_{ij}+\frac{E_c^{(j)}}{2}\qty(\frac{z_{J_j}}{\widetilde{z}_{J_j}})^2\widetilde{\alpha}_{jj}\widetilde{\alpha}_{ji}^*,
    \end{equation}
    between qubit and inner modes
    \begin{equation}
        \nu_{\overline{i}k} \simeq \frac{E_c^{(i)}}{2}\qty(\frac{z_{J_i}}{\widetilde{z}_{J_i}})^2\widetilde{\alpha}_{ii}\widetilde{\alpha}_{ik},
    \end{equation}
    and all other $\nu_{rs}\simeq 0 + \mathcal{O}(k^2/\delta^2)$. This implies that we can write the coefficients in matrix form
    \begin{equation}
        \widetilde{\nu} = \widetilde{\alpha}^{\dagger}\Zmat_J^{1/2}\Lambda\Zmat_J^{1/2}\widetilde{\alpha},
    \end{equation}
    with
    \begin{equation}
        \Lambda = \mqty(\mqty{\frac{E_c^{(1)}}{2\widetilde{z}_{J_1}\sqrt{\overline{\zeta}_1}} & & \\  & \ddots & \\ & & \frac{E_c^{(n)}}{2\widetilde{z}_{J_n}\sqrt{\overline{\zeta}_n}} } &  \mzero \\ \mzero & \mzero_{(m+l)\times(m+l)} ),
    \end{equation}
    where $\overline{\zeta}_i\equiv (\widetilde{L}_{J_i}/\overline{L}_{J_i})$, such that $\nu_{rs}\simeq (\widetilde{\nu})_{rs}$. Therefore, we obtain to first order 
    \begin{equation}
    \begin{split}
        H_{\gamma} &= -\frac{2}{\hbar}(\alpha^{\phi}\vphi+\alpha^{q}\vq)^T\Zmat_J^{1/2}\Lambda\Zmat_J^{1/2}(\alpha^{\phi}\vphi+\alpha^{q}\vq)\\
        &= -\frac{1}{2}\vphi_J^T(\Lmat_J^{nl})^{-1}\vphi_J, \label{Appendix NL:NL-addition}
    \end{split}
    \end{equation}
    where
    \begin{equation}
        \Lmat_J^{nl}
        =\widetilde{\Lmat}_J\mqty(\frac{\hbar \overline{\omega}_1}{2E_c^{(1)}} & & \\ & \ddots & \\ & & \frac{\hbar \overline{\omega}_n}{2E_c^{(n)}}),
    \end{equation}
    similar to Eq.~(138) of Ref.~\cite{Solgun:2019}. We note that $\overline{\omega}_i= 1\sqrt{\overline{L}_{J_i}C_{J_i}}$ here corresponds to the frequencies in the bare basis after including a possible frequency shift due to $\Lmat_J^{-1}$, with $\overline{L}_{J_i}$ the diagonal entries of $(\overline{\Lmat}_{J})_{ii}$ defined in \cref{Appendix coupling:overline-Lmat-Y} (\cref{Appendix coupling:HmatJ-Z-p} for the impedance), and not to the final frame frequencies. Then, if at the beginning of our treatment to Hamiltonian $H_0$ \cref{Appendix coupling:H-y} (\cref{Appendix coupling:H-z-p} for the impedance) we add and substract $H_{\gamma}$
    \begin{equation}
        H = H_0+H_{\gamma}-H_{\gamma} = H_0'-H_{\gamma}, 
    \end{equation} 
    we can account for the linear corrections coming from the nonlinearities to first order if we satisfy the condition
    \begin{equation}
        \overline{\omega}_{i}^{2} = \widetilde\omega_{J_i}^2\qty(1+\zeta_i-\frac{2E_c^{(i)}}{\hbar \overline{\omega}_{i}}),
    \end{equation}
    with $\widetilde\omega_{J_i}=1/\sqrt{C_{J_i}\widetilde{L}_{J_i}}$ and $\zeta_i = \sqrt{\widetilde{L}_{J_i}/L_{J_i}}$. Solving this expression for small anharmonicities $E_c^{(i)}/\hbar \omega_{J_i}\ll 1$ we obtain
    \begin{equation}
        \overline{\omega}_i = \widetilde\omega_{J_i}\sqrt{1+\zeta_i}\qty(1-\frac{E_c^{(i)}/\hbar \widetilde\omega_{J_i}}{(1+\zeta_i)^{3/2}-E_c^{(i)}/\hbar \widetilde\omega_{J_i}}),
    \end{equation}
    which for $\zeta_i\to 0$ (the case with no inductive poles) reduces to 
    \begin{equation}
        \overline{\omega}_i = \widetilde\omega_{J_i}-\frac{E_c^{(i)}/\hbar}{1-E_c^{(i)}/\hbar \widetilde\omega_{J_i}},
    \end{equation}
    recovering as a special case Eq.~(144) of Ref.~\cite{Solgun:2019}. These formulas are valid for both the admittance and impedance methods. 
    
    As for the self kerrs and dispersive shifts to first order in the expansion of \cref{AppendixNL:expansion} they are
    \begin{equation}
    \begin{split}
        \delta_i &= -12\beta_{\overline{i}\overline{i}ii},\\
        \chi_{ik} &= -24\beta_{\overline{i}i\overline{k}k}. 
    \end{split}
    \end{equation}
    To explicitly obtain their values, we have to obtain the entries of $\alpha$. For convenience of notation we divide $\alpha$ in submatrices
    \begin{equation}
        \alpha = \mqty(\alpha_J^{\phi} & \alpha_J^{q} & \alpha_R^{\Pi} & \alpha_R^Q & \alpha_{NR}^{\Pi} & \alpha_{NR}^Q).
    \end{equation}
    From the derivation done in \cref{Derivation-main}, for the admittance we have
    \begin{equation}
    \label{Apendix-nl:alpha}
            \alpha = (\widetilde{\Smat}_J^{\phi})^{-1}\Pmat_J^{\phi}\Smat_{sw}^{-1},
    \end{equation}
    where $\Pmat_J^{\phi}$ is the projector onto the junctions flux subspace, $(\widetilde{\Smat}_J^{\phi})^{-1} = \overline{\Zmat}_J^{-1/2}\Omat_C\overline{\Cmat}_J^{-1/2}\Omat_{\Omega}\overline{\Wmat}_J^{-1/2}$, and $\overline{\Zmat}_J = (\overline{\Lmat}_J/\overline{\Cmat}_J)^{1/2}$. In the full perturbative approach and taking $\Smat_{sw}^{-1}$ to first order we obtain
    \begin{equation}
        \begin{split}
            (\alpha_J^{\phi})^{\phi_J}_{ij} &= \delta_{ij}, \\
            (\alpha_J^{q})_{ij} &= 0, \\
            (\alpha_R^\Pi)_{i\gamma} &=-\Theta^{-1}_{i,\gamma}\sqrt{\overline{\omega}_i\omega_{r_\gamma}}(\overline{\vb{r}}_{\gamma})_i , \\
            (\alpha_R^Q)_{i\gamma} &= 0, \\
            (\alpha_{NR}^\Pi)_{i\mu} &= -\Theta_{i,\mu}^{-1}\sqrt{\overline{\omega}_i\omega_{g_\mu}}(\overline{\vb{n}}^L_{\mu})_i, \\
            (\alpha_{NR}^Q)_{i\mu} &= \Theta_{i,\mu}^{-1}\sqrt{\overline{\omega}_i\omega_{g_\mu}}(\overline{\vb{n}}^R_{\mu})_i.
        \end{split}
    \end{equation}
    with $\Theta_{i,\gamma}=\omega_i^2-\omega_{r_\gamma}^2$. Whereas for the impedance
    \begin{equation}
        \alpha= \mqty(\widetilde{\Smat}_J^{\phi} & \mzero & \widetilde{\Smat}_I^{\phi} & \widetilde{\Smat}_I^{q})\Smat_{sw}^{-1},
    \end{equation}
    with
    \begin{equation}
        \begin{split}
            \widetilde{\Smat}_J^{\phi} &= \overline{\Zmat}_J^{-1/2}\Omat_C^T\overline{\Cmat}_J^{-1/2}\Omat_{\Omega}^T\overline{\Wmat}_J^{-1/2},  \\
            \widetilde{\Smat}_I^{\phi} &= \mqty(\overline{\Zmat}_J^{-1/2}\Rmat_J^T\Omega_R^{-1/2} & \overline{\Zmat}_J^{-1/2}\Nmat_J^T\Wmat_g^{-1/2}), \\
            \widetilde{\Smat}_I^{q} &= \mqty(\mzero & -\overline{\Zmat}_J^{-1/2}\Nmat_J^T\Sigmamat_x\Wmat_g^{-1/2}).  
        \end{split}
    \end{equation}
    By simply substituting and rearranging we explicitly obtain 
    \begin{equation}
        \begin{split}
            (\alpha_J^{\phi})_{ij} &= \delta_{ij}-\frac {\Im[\Zmat_{ij}^{ac,R}(\overline{\omega}_j)]}{\sqrt{\overline{z}_{J_i}\overline{z}_{J_j}}} , \\
            (\alpha_J^{q})_{ij} &= -\frac{2\overline{\omega}_j^2}{\sqrt{\overline{z}_i\overline{z}_j}}\sum_{\mu}\frac{\Theta_{j\mu}^{-1}}{\omega_{g_\mu}}\Im\{\text{Res}\boldleft\Zmat^{\text{NR}}(\omega_{g_\mu})\boldright_{ij}\}, \\
            (\alpha_R^{\Pi})_{ij} &= \frac{1}{\sqrt{\overline{z}_{J_i}}}\frac{(\vb{r}_\gamma)_i}{\sqrt{\omega_{r_\gamma}}}\qty(\frac{\overline{\omega}_{r_\gamma}^2}{\omega_{r_\gamma}^2-\overline{\omega}_i^2}), \\
            (\alpha_R^{Q})_{ij} &= 0, \\
            (\alpha_{NR}^{\Pi})_{ij} &= \frac{1}{\sqrt{\overline{z}_{J_i}}}\frac{(\vb{n}^L_\mu)_i}{\sqrt{\omega_{g_\mu}}}\qty(\frac{\overline{\omega}_{g\mu}^2}{\omega_{g_\mu}^2-\overline{\omega}_i^2}), \\
            (\alpha_{NR}^{Q})_{ij} &= -\frac{1}{\sqrt{\overline{z}_{J_i}}}\frac{(\vb{n}^R_\mu)_i}{\sqrt{\omega_{g_\mu}}}\qty(\frac{\overline{\omega}_{g_\mu}^2}{\omega_{g_\mu}^2-\overline{\omega}_i^2}),
        \end{split}
    \end{equation}
    with the residue given by $\text{Res}\boldleft\Zmat^{\text{NR}}(\omega_{g_\mu})\boldright=\lim_{s\to i\omega_{g_\mu}}(s-i\omega_{g_\mu})\Zmat^{\text{NR}}(s) = -i\Bmat_{\mu}/2\omega_{g_\mu}$, the nonreciprocal part of the impedance residue at frequency $\omega_{g_\mu}$. Thus, we recover as a subset Eqs.~(120-122) of Ref.~\cite{Solgun:2019} with the difference steaming from the different final frame we are taking.

    It follows that the self-kerrs anharmonicities for both the admittance and the impedance approaches are 
    \begin{equation}
        \delta_i = -E_C^{(i)}(\omega_{J_i}/\overline{\omega}_i)^2.
    \end{equation}
    After rearranging, the dispersive shifts in the admittance approach are
    \begin{subequations}
    \label{Appendix-NL:dispersive-shifts-Y}
        \begin{equation}
            \begin{split}
                \chi_{i\gamma} &= 2\delta_i\qty(\frac{\omega_{r_\gamma}g_{i\gamma}^{qQ}}{\omega_{r_\gamma}^2-\overline{\omega}_i^2})^2, \\
                \chi_{i\mu} &= 2\delta_i\qty(\frac{\omega_{g_\mu}}{\omega_{g_\mu}^2-\overline{\omega}_i^2})^2\boldleft (g_{i\mu}^{qQ})^2+(g_{i\mu}^{q\Pi})^2\boldright,
            \end{split}
        \end{equation}
    with 
        \begin{equation}
            \begin{split}
                g_{i\gamma}^{qQ} &= \sqrt{\overline{\omega}_i/\omega_{r_\gamma}}(\vb{r}_{\gamma})_i/\sqrt{C_{i}},\\
                g_{i\mu}^{qQ} &= \sqrt{\overline{\omega}_i/\omega_{g_\mu}}(\vb{n}_{\mu}^L)_i/\sqrt{C_{i}}, \\
                g_{i\mu}^{q\Pi} &= \sqrt{\overline{\omega}_i/\omega_{g_\mu}}(\vb{n}_{\mu}^R)_i/\sqrt{C_{i}},
            \end{split}
            \end{equation}
    \end{subequations}
     where the last three correspond to the bare couplings between qubit and inner modes given in \cref{Appendix coupling:Kmats-Y} with $\Omat_C=\Omat_L=\Omat_{\Omega}=\bone_n$. Whereas for the impedance, the dispersive shifts are
     \begin{subequations}
    \label{Appendix-NL:dispersive-shifts-Z}
        \begin{equation}
            \begin{split}
                \chi_{i\gamma} &= 2\delta_i\qty(\frac{\omega_{r_\gamma}g_{i\gamma}^{\phi \Pi}}{\omega_{r_\gamma}^2-\overline{\omega}_i^2})^2, \\
                \chi_{i\mu} &= 2\delta_i\qty(\frac{\omega_{g_\mu}}{\omega_{g_\mu}^2-\overline{\omega}_i^2})^2\boldleft (g_{i\mu}^{\phi \Pi})^2+(g_{i\mu}^{\phi Q})^2\boldright,
            \end{split}
        \end{equation}
    with 
        \begin{equation}
            \begin{split}
                g_{i\gamma}^{\phi \Pi} &= \sqrt{\overline{\omega}_i\omega_{r_\gamma}}(\vb{r}_{\gamma})_i\sqrt{C_{i}},\\
                g_{i\mu}^{\phi \Pi} &= \sqrt{\overline{\omega}_i\omega_{g_\mu}}(\vb{n}_{\mu}^L)_i\sqrt{C_{i}}, \\
                g_{i\mu}^{\phi Q} &= \sqrt{\overline{\omega}_i\omega_{g_\mu}}(\vb{n}_{\mu}^R)_i\sqrt{C_{i}},
            \end{split}
            \end{equation}
    \end{subequations}
    where the last three correspond to the bare couplings given in \cref{Appendix coupling:Kmat-Z} with $\Omat_C=\Omat_{\Omega}=\bone_n$. For the reciprocal resonators, we recover the dispersive shift given in Eq.~(63) of Ref.~\cite{Solgun:2019}. The transformer ratios $\vb{r}_\gamma$, $\vb{n}_\mu^{L,R}$ can be obtained directly from the residues of the response as explained in \cref{Duality}. Moreover, if higher order nonlinear corrections are desired these can be straightforwardly obtained using \cref{Appendix-NL:general-multinomial-expansion}. For example, if the sixth order terms of the expansion are included the dispersive shifts are corrected to second-order in perturbation theory by the factor $\chi_{i\mu}\to (1-2E_C^{(i)}/\hbar\overline{\omega}_i)\chi_{i\mu}$. 
        
    Additionally, we note that the dispersive shifts for the impedance and admittance approaches are different. This is to be expected due to the different frames used for quantization in one approach or another, however and as we have highlighted several times throughout this work, this difference is negligible in the dispersive regime. 
    Going further, analytical formulas for qubits cross-kerrs can in principle be obtained considering higher order terms of $\Smat_{sw}^{-1}$, as was done in \cite{Solgun:2022}. Moreover, it is useful to note that a numerical treatment beyond the perturbative regime can be done using the exact Hamiltonians after elimination of the nondynamical modes given in \cref{Elimination-Nondynamical}.

    \section{Derivation of master equation} \label{masterequation}
    
In this appendix, we derive a completely positive trace preserving (CPTP) master equation for the qubits, taking into account any possible quasi-degeneracies. In the standard approach to open quantum systems, the master equation is derived from a microscopic model (system, bath, and interaction Hamiltonians) using the Born-Markov and secular approximations \cite{Davies:1974,BRE:2002,Alicki:2007,Rivas:2012}. However, the secular approximation turns out to be restrictive and fails whenever the spectral gap $\Delta\omega$ of the qubits Hamiltonian $\hat{H}_q$ is of the order of the natural linewidth $\Gamma$ dictated by dissipation, i.e.,~$\Delta \omega \lesssim \Gamma$. In this case, the rotating-wave approximation typically used to transform the Bloch-Redfield equation into a master equation in Lindblad form (CPTP) breaks down \cite{Mccauley:2020,Nathan:2020}. Recent works have introduced \emph{the partial secular approximation} to extend Lindblad's equation to systems whose energy levels are not necessarily strongly separated (i.e.,~$\Delta\omega \lesssim \Gamma$) while preserving complete positivity \cite{Mccauley:2020,Nathan:2020,Farina:2019,Trushechkin:2021}. 

The starting point of our derivation is the Caldeira-Leggett Hamiltonian $\hat{H}=\hat{H}_q+\hat{H}_B+\hat{H}_{qB}$ with interaction term of the form of \cref{H_SB} 

\begin{align}
\begin{split}
    \hat{H}_q=\sum_j\frac{\overline{\Omega}_j}{2}\left(\hat{q}_j^2+\hat{\phi}_j^2\right), 
\end{split}\\
\begin{split}
\hat{H}_B=\sum_{\alpha b}\left(\frac{\hat{p}_{\alpha b}^2}{2m_{\alpha b}}+\frac{1}{2}m_{\alpha b}\omega_{\alpha b}^2 \hat{x}_{\alpha b}^2\right),
\end{split}\\
\begin{split}
     \hat{H}_{qB}=& \sum_j \left(\hat{\phi}_j\sum_{\alpha b} \Tilde{\cmat}_\alpha^{\vphi,\vx}\hat{x}_{\alpha b}+\hat{\phi}_j\sum_{\alpha b} \Tilde{\cmat}_\alpha^{\vphi,\vp}\hat{p}_{\alpha b}\right.\\
     &\left.+\hat{q}_j\sum_{\alpha b} \Tilde{\cmat}_\alpha^{\vq,\vx}\hat{x}_{\alpha b}+\hat{q}_j\sum_{\alpha b} \Tilde{\cmat}_\alpha^{\vq,\vp}\hat{p}_{\alpha b} \right).\\
    \end{split} 
\end{align}
Here, we consider the system-bath Hamiltonian $\hat{H}_{qB}\sim \mathcal{O}(\Gamma)$ as a perturbation i.e.,~$||\hat{H}_{qB}||\ll ||\hat{H}_q||$, $||\hat{H}_{qB}||\ll ||\hat{H}_B||$ (Born approximation), which is valid when the qubits are coupled dispersively to the inner modes and when direct capacitive, inductive and nonreciprocal couplings between qubit and drive ports are weak, which is the regime of interest here~\footnote{In this section, we consider that the only impact of the nonlinearity is to define the qubit subspace. For higher-excited states, the effect of anharmonicity on the treatment of dissipation can be found in Ref~\cite{Busel:2023}}. 

The main idea of the partial secular approximation consists on separating the set of qubits (energy transitions) based on their frequency difference. With that objective, we define an equivalence relation $\mathcal{F}$ between qubit ports $j$, $j'$ such that $(j,j') \in \mathcal{F}$ if and only if their energy difference is smaller or of the order of the natural linewidth, that is  $\abs{\overline{\Omega}_j-\overline{\Omega}_{j'}}\lesssim \Gamma$. This separation of energy-scales allow us to perform the rotating wave-approximation whenever $(j,j') \notin \mathcal{F}$. However, if $(j,j') \in \mathcal{F}$, the qubits are considered to be degenerate in  first order of perturbation theory in the system-bath coupling \cite{Trushechkin:2021}. Therefore, tracing out the baths yields the master equation~\footnote{The validity of the Markovian regime is determined by the poles of the dissipation matrix, which exhibit exponential decay in time domain with a rate proportional to $Z_0^{-1}\Cmat_{D}^{-1}$. For typical values of circuit parameters, $Z_0 \Cmat_{D} \ll \Gamma^{-1}$, which justifies the validity of the Markovian approximation.}
\begin{align}
 \frac{d\hat{\rho}}{dt}=-i[\hat{H}_q,\hat{\rho}]+
 \sum_{\substack{(j,j') \in \mathcal{F} \\ \epsilon ,\epsilon'=\{ \phi,q \}}}
 \gamma_{jj'}^{\epsilon\epsilon'}(\hat{A}_{j'}^{\epsilon'}\hat{\rho} \hat{A}_{j}^{\epsilon^\dagger}-\frac{1}{2}\{ \hat{A}_{j}^{\epsilon^\dagger}\hat{A}_{j'}^{\epsilon'},\hat{\rho}\}),
\end{align}
where we have omitted the Lamb-shift Hamiltonian which introduces, in the dispersive and weak direct coupling regime, just a small renormalization on the system Hamiltonian $\hat{H}_q$. 
Notably, as shown in Ref.~\cite{Correa:2023}, in the adiabatic limit when $Z_0 \ll {\Ymat^{dc}_D}^{-1}$, the Lamb-shift  Hamiltonian is exactly cancelled by the second-order terms in the system bath couplings that we have neglected in our Hamiltonian (\cref{Y_full hamiltonian}). In typical circuit parameters, this limit is satisfied. In the previous equation, the collapse operators are defined as~\cite{BRE:2002}
\begin{align}
\begin{split}
 \hat{A}_{j}^{\phi}=\sum_{\mathcal{E'}-\mathcal{E}=\overline{\Omega}_j}\hat{\Pi}(\mathcal{E})\hat{\phi}_j\hat{\Pi}(\mathcal{E'}),
\end{split}\\
\begin{split}
 \hat{A}_{j}^{q}=\sum_{\mathcal{E'}-\mathcal{E}=\overline{\Omega}_j}\hat{\Pi}(\mathcal{E})\hat{q}_j\hat{\Pi}(\mathcal{E'}),
\end{split}
\end{align}
where $\hat{\Pi}(\mathcal{E})$ is the projector over the eigenspace of $\hat{H}_q$ with energy $\mathcal{E}$. The spectral densities $\gamma_{jj'}^{\epsilon\epsilon'}$ are defined as the Fourier transform of the bath's correlations functions
\begin{align}
\gamma_{jj'}^{\epsilon\epsilon'}=\int_{-\infty}^{+\infty}d\tau e^{i\overline{\Omega}_j\tau}\expval{\hat{B}_{j}^{\epsilon^\dagger}(\tau)\hat{B}_{j'}^{\epsilon'}(0)}.
\end{align}
 Using \cref{H_SB}, the bath's eigen-operators are defined as
 \begin{align}
    \begin{split}
    \hat{B}_{j}^{\phi}=\sum_\alpha( \Tilde{\cmat}_\alpha^{\vphi,\vx}\hat{x}_\alpha+\Tilde{\cmat}_\alpha^{\vphi,\vp}\hat{p}_\alpha),
     \end{split}\\
     \begin{split}
    \hat{B}_{j}^{q}=\sum_\alpha( \Tilde{\cmat}_\alpha^{\vq,\vx}\hat{x}_\alpha+ \Tilde{\cmat}_\alpha^{\vq,\vp}\hat{p}_\alpha).
     \end{split}
 \end{align}
Writing the normalized flux and charge operators in terms of the bosonic modes $\hat{\phi}_{j} = \frac{1}{\sqrt{2}}(\bdag_{j}+\blow_{j})$, $\hat{q}_{j} = \frac{i}{\sqrt{2}}(\bdag_{j}-\blow_{j})$, the master equation can be expressed in a compact way
\begin{align}
\frac{d\hat{\rho}}{dt}=-i[\hat{H}_q,\hat{\rho}]+\sum_{(j,j') \in \mathcal{F}}\gamma_{j j'}(\blow_{j'}\hat{\rho} \bdag_{j}-\frac{1}{2}\{ \bdag_{j}\blow_{j'},\hat{\rho}\}),  
\end{align}
with
\begin{align}
\gamma_{j j'}=\frac{1}{2}\left(\gamma_{jj'}^{\phi \phi}+\gamma_{jj'}^{qq}-i\gamma_{jj'}^{\phi q}+i\gamma_{jj'}^{q \phi}\right). 
\end{align}
At cryogenic temperatures $\beta^{-1}=k_BT\ll \overline{\Omega}_j$ such that the number of thermal photons is negligible $\overline{n}= 1/e^{\beta\hbar\overline{\Omega}_j}-1 \ll 1$ and the bath can be  approximated to be in its ground state. In this case, one can easily compute the bath correlation functions and $\gamma_{jj'}^{\epsilon,\epsilon'}$, which leads to
\begin{align}
        \begin{split}
            \gamma_{jj'}=\frac{\pi}{2}\sum_{b=1}^{B}\sum_\alpha(s_{\alpha_{jb}}+it_{\alpha_{jb}})^\star(s_{\alpha_{j'b}}+it_{\alpha_{j'b}})\delta(\overline{\Omega}_j-\omega_{\alpha b}),
        \end{split}
    \end{align}
    where 
    \begin{align}
        \begin{split} s_{\alpha_{jb}}=\frac{\Tilde{c}_{\alpha_{jb}}^{\vphi,\vx}}{\sqrt{m_{\alpha b}\omega_{\alpha b}}}+\sqrt{m_{\alpha b}\omega_{\alpha b}}\: \Tilde{c}_{\alpha_{jb}}^{\vq,\vp}, 
        \end{split}\\
        \begin{split}        t_{\alpha_{jb}}=\sqrt{m_{\alpha b}\omega_{\alpha b}}\: \Tilde{c}_{\alpha_{jb}}^{\vphi,\vp}-\frac{\Tilde{c}_{\alpha_{jb}}^{\vq,\vx}}{\sqrt{m_{\alpha b}\omega_{\alpha b}}}.   
        \end{split}
    \end{align}
    By symmetrizing the delta function, i.e.,~$\delta(\overline{\Omega}_j-\omega_{\alpha b})=\delta((\overline{\Omega}_j+\overline{\Omega}_i)/2-\omega_{\alpha b})+\mathcal{O}(\Gamma)$, one can readily verify that the matrix $\gamma_{jj'}$ is positive semi-definite within the Born approximation. This ensures that the master equation represents a completely positive trace-preserving (CPTP) map.

    \section{Circuit examples} \label{circuit-examples}

    In this appendix, we provide further  examples of circuits demonstrating the application of our results. This appendix is divided in four parts. First, we start with a comparison of the admittance and impedance approaches, revealing their equivalence in the dispersive regime. We then  focus on a specific singular circuit relying solely on the admittance response in a situation where the impedance approach cannot be applied. In the third part of this appendix, we show how to use our results to estimate Purcell decay rates, and we compare these rates to master equation simulations and previous one-port analyses. Finally, we give a detailed analysis of the nonreciprocal circuit presented in the main text. We provide here both admittance and impedance responses of this three-port device, and give a simple explanation for the conditions which achieve chiral dynamics with complete population transfer. 
    
    \subsection{Effective coupling rates: admittance vs impedance}
    \begin{figure}[h]
    \centering \includegraphics[width=1\linewidth]{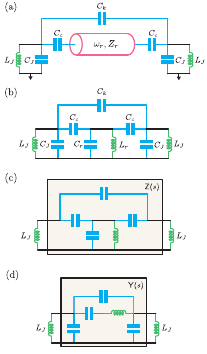}
    \caption{(a) Two (linearized) Josephson junctions coupled by a transmission line resonator  and (b) an effective circuit with only one inner-mode of the latter. Canonical circuits representions of the (c) impedance ($\msZ$) and (d) admittance ($\msY$) response matrices from the Josephson  ports.}
    \label{Appendix-example:fig-tl}
    \end{figure}

    The circuit we consider is the two-port circuit of \cref{Appendix-example:fig-tl}, already studied in Ref.~\cite{Solgun:2022}. It consist of two linearized qubits ($J$) interacting via an inner mode ($r$) which we take to be a simple LC circuit. Our objective here is to compare the admittance and the impedance methods. The impedance can be obtained with an ABCD analysis to find
    \begin{equation}
    \label{Appendix examples:Z-1}
        \begin{split}
            \Zmat(s) &= \frac{\Amat_0}{s}+\frac{\Amat_1s}{\omega_{r_z}^2+s^2},\\
            \Amat_0&= \mqty(1/\widetilde{C}_J & 0 \\ 0 & 1/\widetilde{C}_J),\\
            \Amat_1 &= r_z^2\mqty(1 & 1 \\ 1 & 1),
        \end{split}
    \end{equation}
    with $\widetilde{C}_J = (C_c+C_J)$, $\omega_{r_z} =\omega_r(r_{cj}+1)^{1/2}/(2r_{cr}+r_{cj}+1)^{1/2}$, $\omega_r=1/\sqrt{L_rC_r}$, $r_z \simeq r_{cj}/\sqrt{C_r}$, where $r_{cr}=(C_c/C_r)$ and $r_{cj}=(C_c/C_J)$. On the other hand, the admittance response is
    \begin{equation}
    \label{Appendix examples:Y-1}
        \begin{split}
           \Ymat(s) &= s\Dmat_{\infty}+\frac{\Dmat_1 s}{\omega_{r_y}^2+s^2},\\
           \Dmat_{\infty}&= \mqty(\overline{C}_J & \overline{C} \\ \overline{C} & \overline{C}_J),\\
           \Dmat_1 &= r_y^2\mqty(1 & 1 \\ 1 & 1),
        \end{split}
    \end{equation}
    with $\overline{C}_J \simeq \widetilde{C}_J(1-r_{cj}r_{cr})$, $\overline{C}\simeq C_c r_{cr} $,  $\omega_{r_y} \simeq \omega_{r_z}(1-r_{cj}r_{cr})$, and $r_y \simeq r_{cr}/\sqrt{L_r}$. These approximate values are accurate in the dispersive regime where $r_{cj},r_{cr}\ll 1$. 
    \begin{figure}[h]
        \centering
        \includegraphics[width=.48\textwidth]{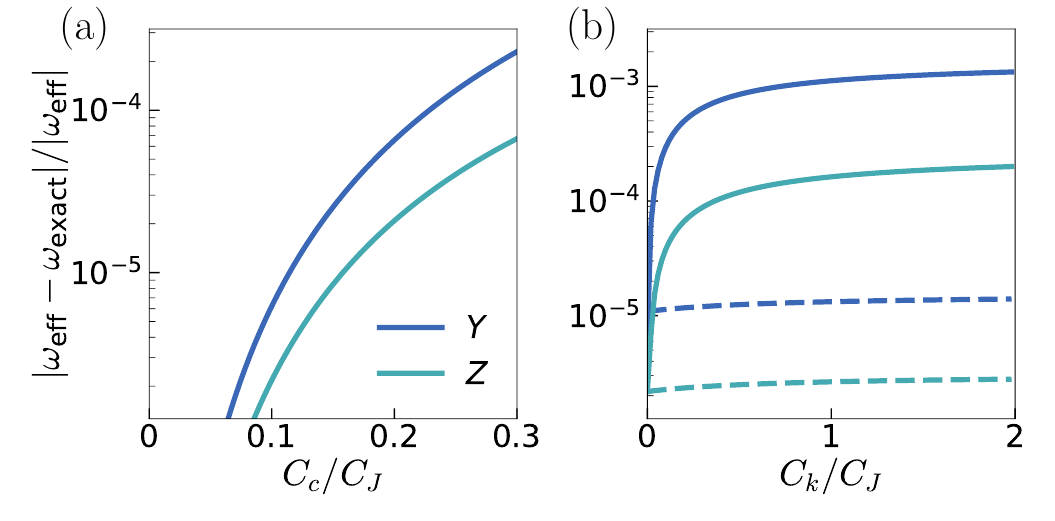}
        \caption{(a) Difference between effective and exact normal modes of the circuit in \cref{Appendix-example:fig-tl} (b) as a function of $C_c/C_J$ with $C_k=0$. Dark (light) blue lines are the normalized differences from the admittance (impedance) methods.  
        (b) Difference between effective and exact normal modes of the circuit in \cref{Appendix-example:fig-tl} (b) as a function of $C_k$ with $C_c=0.1C_J$. We plot the differences obtained from the numerical diagonalization of the direct coupling (full perturbative) method shown with dashed (solid) lines. Circuit parameters $C_J =51.0$ fF, $\omega_J/2\pi =5.57$ GHz, $\omega_r/2\pi = 7.07$ GHz, $z_r = 50$ $\Omega$.
        }
        \label{paper-fig:distances-example-1}
    \end{figure}

    To compare the two approaches, we show in \cref{paper-fig:distances-example-1}(a) the normalized difference between the normal mode frequencies obtained from the effective classical Hamiltonian derived in \cref{Derivation-main} and the exact normal modes of the circuit obtained from the standard Hamiltonian circuit analysis (dark-blue: admittance, light-blue: impedance). This is plotted as a function of $C_c/C_J$ with the qubit-resonator coupling in the dipersive limit, see parameters in the figure caption.
    We find an excellent agreement in the dispersive, where $C_c/C_J\ll 1$. As $C_c/C_J$ increases, the accuracy diminishes because the coupling between qubit and resonator modes $g=\sqrt{\omega_J \omega_r}C_c/\sqrt{C_JC_r}$ increases and we get out of the perturbative regime. It is worth noting that the larger error of the admittance approach can be attributed to the direct capacitive coupling in its dc response between the external ports (see $\overline{C}$ in \cref{Appendix examples:Y-1}), in contrast to the impedance's dc response which is directly diagonal such that the exact dressing of the capacitances is in this case automatic. As already mentioned in the main text, we stress that this difference between the two approaches arises from our perturbative derivation, which leads to different final effective frames. Of course, when an exact Hamiltonian from the impedance and the admittance response can be constructed, the normal modes of both are exactly the same~\cite{Williamson:1936,Laub:1974,ParraRodriguez:2022b,Egusquiza:2022}.
    
    In \cref{paper-fig:distances-example-1} (b) we compare the fully perturbative approach (solid) with the previous diagonalization method, where the direct coupling is treated exactly (dashed), developed in \cref{Derivation-main}. There, we plot the normalized normal mode frequency difference  for both approaches when varying $C_k/C_J$ ratios. Clearly, even in the strong coupling regime $C_k\gg C_J$, the previous diagonalization approach maintains its accuracy.

    \subsection{Direct inductive coupling and JJs' nonlinearities}
    
    \begin{figure}[h]
        \centering
        \includegraphics[width=.4\textwidth]{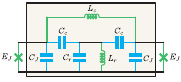}
        \caption{2-port example circuit with direct inductive coupling between qubit ports. For circuits with direct inductive coupling between inductive qubit ports (i.e. described in configuration space by their flux) $\Amat_{\infty}=0$,  $\Amat_0$ will be singular and unconstrained Hamiltonian dynamics cannot be systematically obtained from the impedance response.}
        \label{fig:Appendix-examples:2-port-reciprocal-inductive}
    \end{figure}

\begin{figure}[t]
        \centering
        \includegraphics[width=.6\linewidth]{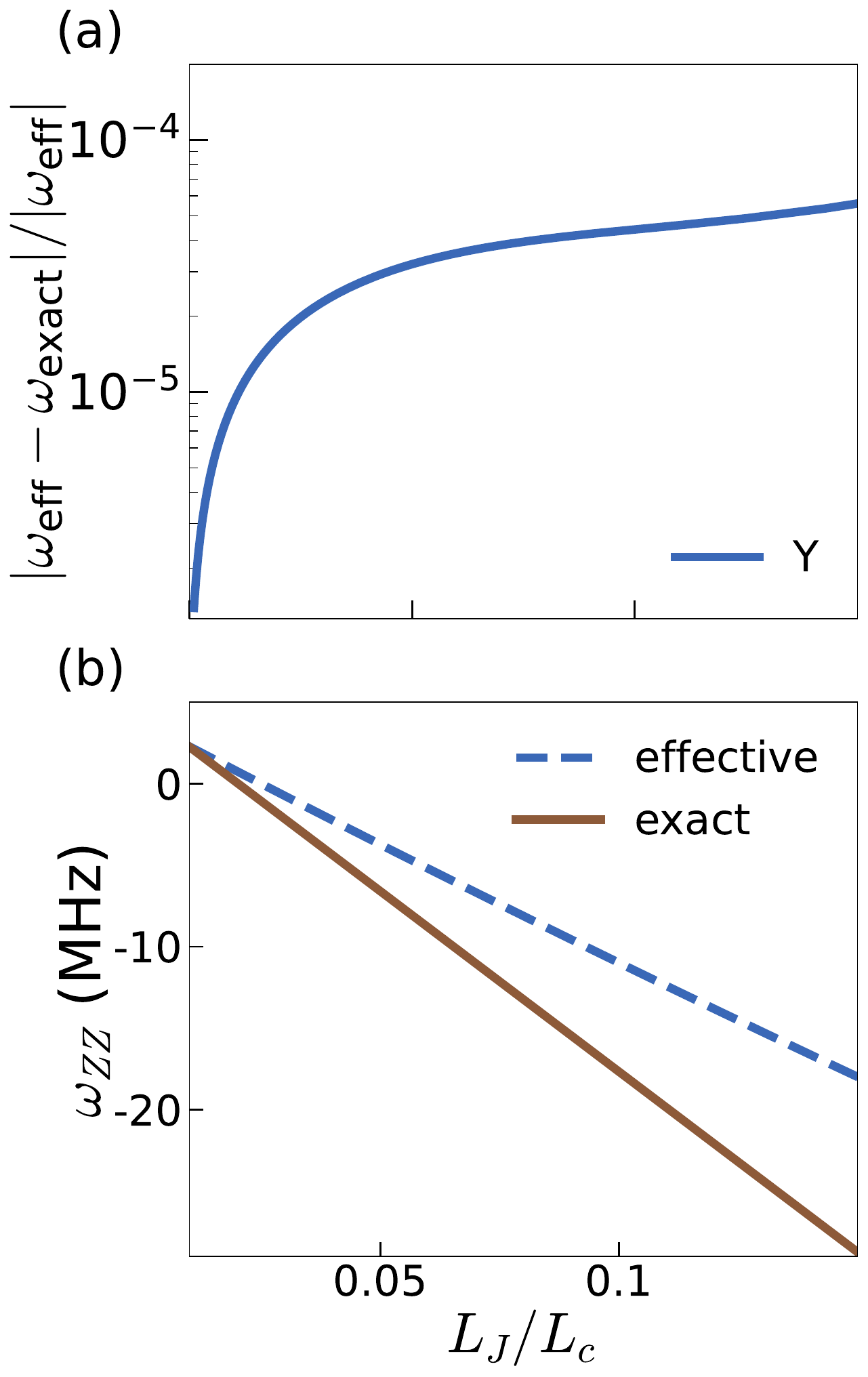}
        \caption{(a) Difference between normal modes of the linear effective and exact Hamiltonians of the circuit in \cref{fig:Appendix-examples:2-port-reciprocal-inductive}. (b) $ZZ$-interaction obtained from the diagonalization of the effective and exact quantum Hamiltonians as functions of $L_J/L_c$. Circuit parameters $C_J=100$ fF, $C_c = 0.1 C_J$, $E_{J1}/h =20.4$ GHz, $E_{J2}/h=12.6$ GHz and $\omega_r/2\pi = 7.8$ GHz.}
        \label{fig:appendix-L-nl}
    \end{figure}

    We now turn to the circuit of \cref{fig:Appendix-examples:2-port-reciprocal-inductive} where we replace the direct capacitive coupling of \cref{Appendix-example:fig-tl} between the linearized qubit modes with a direct inductive coupling. This is the simplest circuit that allows us to verify our extension of Ref.~\cite{Solgun:2019} to include direct inductive coupling between qubit ports. As discussed in \cref{Duality}, for circuits with direct inductive coupling between the qubit ports we have that $\Amat_0=\lim_{s\to 0}s\Zmat(s)$ in the impedance characterization. Due to this, the Lagrangian obtained from the Cauer representation is singular, preventing to obtain directly the Hamiltonian. In contrast, the Lagrangian for the Cauer representation of the admittance response is not singular at any pole and the circuit Hamiltonian can thus can be systematically obtained. Hence, we study only the admittance response. To do so, we note that $\Ymat(s)$ is the same as the one given in \cref{Appendix examples:Y-1} with the addition of the inductive poles $\Dmat_0/s$, where $\Dmat_0= (1/L_c)(\bone-\sigma_x)$. \Cref{fig:appendix-L-nl} (a) compares the normalized difference between the normal modes in the effective and exact linear sectors, similar to the above example. It is evident that our approach achieves a remarkably high level of precision.
    
     To benchmark the effective nonlinearities presented in the main text and derived in \cref{NL}, we compare the effective $ZZ$-interaction obtained from numerical diagonalization of the effective Hamiltonian, with the one obtained from numerical diagonalization of the Hamiltonian after full exact circuit quantization. In the qubit subspace, the diagonalized Hamiltonian takes the form 
    \begin{equation}
        \hat{H} = \frac{\omega_1}{2}\sigma_z^1+\frac{\omega_2}{2}\sigma_z^2 + \omega_{ZZ}\sigma_z^1\sigma_z^2.
    \end{equation}
    The effective and exact $ZZ$-interactions ($\omega_{ZZ}$) are shown in \cref{fig:appendix-L-nl} (b). The figure shows how a zero $ZZ$-interaction can be obtained by tuning the ratio between the junction and coupling inductances $L_J/L_c$. Importantly, the results above show that the effective $ZZ$ interaction is accurate when deep in the dispersive regime and weak direct coupling, but this accuracy decreases quickly with increasing coupling strength. This discrepancy arises from the omission of both higher order terms in the expansion of $\Smat_{sw}$ and the non-rotating terms in the nonlinear expansion \cref{AppendixNL:expansion}. For reciprocal impedances these higher-order corrections are given in Ref.~\cite{Solgun:2022}. For the most general nonreciprocal case for both impedance and admittance, analytical inclusion of these higher-order corrections remains an open-task that can be completed following the guidelines provided in \cref{SSW}. We also note that if the residues of the admittance response can be obtained, then a complete numerical treatment beyond the dispersive regime using the exact classical Hamiltonians we give in \cref{Elimination-Nondynamical} is feasible.

    % \lc{Moreover, if we take the limit of no inductive coupling $L_c\to \infty$, we can compare the ZZ-interaction rates obtained from admittance and impedance approaches. In \cref{fig:straddling} we plot the ZZ rates obtained from the effective admittance and impedance Hamiltonians, and compare them with the one obtained by numerical diagonalization of the exact circuit Hamiltonian. We fix all circuit parameters and swap the frequency of one of the qubits. We find that in the straddling regime the impedance effective Hamiltonian gives a more accurate value of ZZ, but that outside the straddling regime the ZZ obtained with the admittance effective Hamiltonian is more accurate.}

    % \begin{figure}[ht]
    %     \centering
    %     \includegraphics[width=.95\linewidth]{straddling-ZZ.pdf}
    %     \caption{\lc{ZZ-interaction rates calculated with the effective admittance (solid blue),  effective impedance (dashed green), and exact (solid brown), Hamiltonians for the circuit in \cref{fig:appendix-L-nl} in the limit of no direct inductive coupling $L_c\to \infty$, swapping the junction energy of the second qubit while keeping all other circuit parameters fixed. The circuit parameters are $C_J = 100$ fF, $C_c = 0.1 C_J$, $\omega_r/2\pi = 6.0$ GHz, $z_r = 50$ $\Omega$, $E_{J1}/h = 12.57$ GHz.} }
    %     \label{fig:straddling}
    % \end{figure} 
        
    \subsection{Readout and drive induced Purcell decays} 
    
\begin{figure}[h]
        \centering
        \includegraphics[width=1\linewidth]{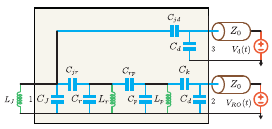}
        \caption{Example of a 3-port circuit for a typical qubit control and dispersive readout routine. It consists of a capacitively coupled qubit to its control line with characteristic impedance $Z_0= 50\:\Omega$ connected to a voltage source $V_{\text{d}}(t)$. The qubit is also coupled to its $Z_0=50\:\Omega $ readout line via two LC oscillators, corresponding to a standard scheme involving a capacitively coupled readout resonator and Purcell filter. The readout line is typically driven by  a classical voltage source $V_{RO}(t)$.}
        \label{paper-fig:R0+Drive lines}
    \end{figure}

\begin{figure}[t]
        \centering
        \includegraphics[width=0.7\linewidth]{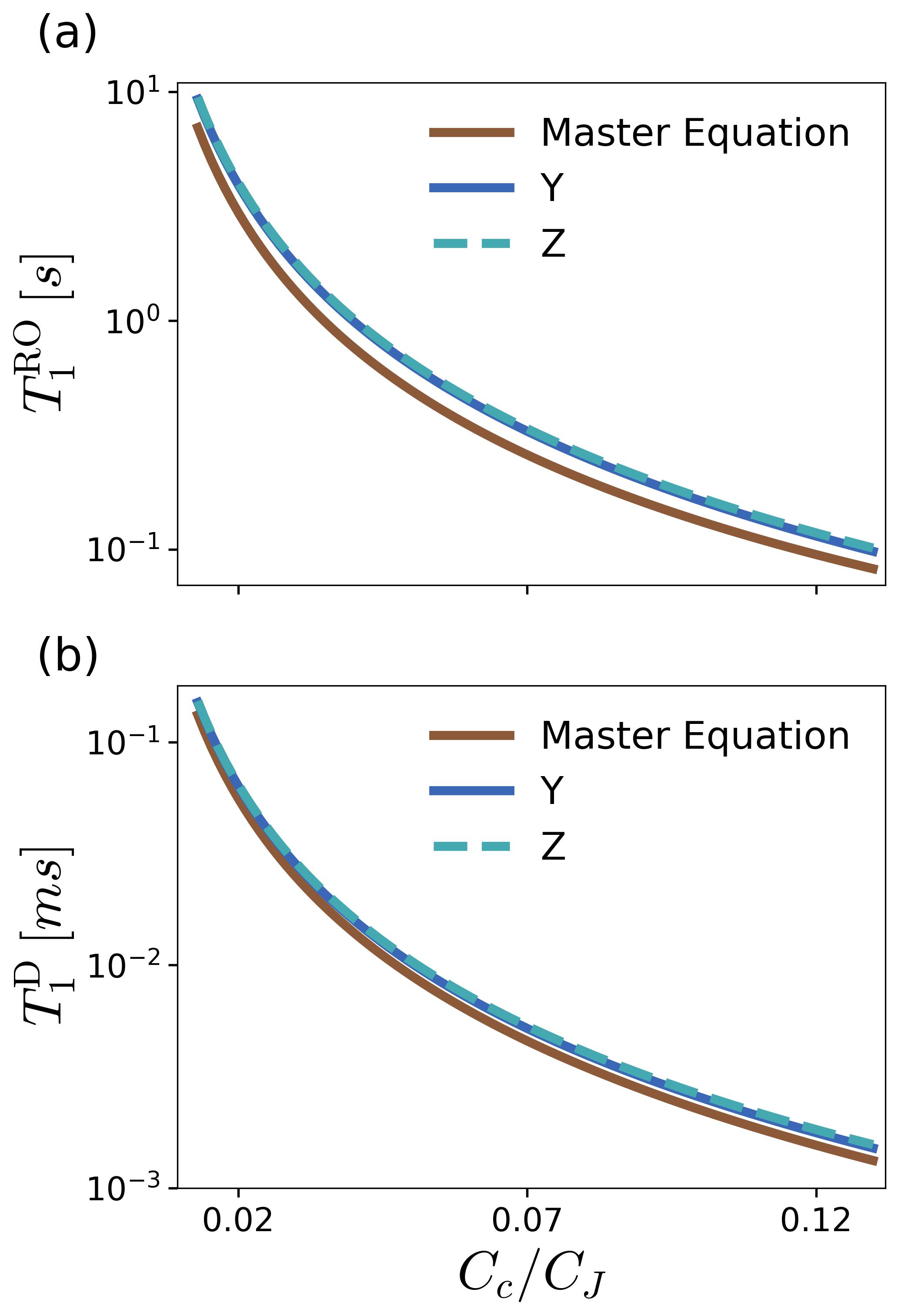}
        \caption{Analyzing circuit \cref{paper-fig:R0+Drive lines}: (a) Purcell decay  $T_1^{\text{RO}}$ from the readout line, (b) drive-induced relaxation $T_1^{\text{D}}$, as a function of the coupling capacitances $ C_{jr}=C_c$, $C_{jd}=0.1C_c$. We compare our results obtained with both impedance and admittance formulas \cref{Purcell_Y_main,Purcell_Z_main} (light and dark blue) with the ones obtained from integrating the master equation of the three modes system (brown).
        We used realistic circuit parameters $L_J=10$ nH, $C_J=77$ fF giving a plasmon frequency $\Tilde{\omega}_J/2\pi=5.73$ GHz. The frequency of readout resonator and Purcell filter are $\omega_r/2\pi=7.50$ GHz, $\omega_f/2\pi=7.51$ GHz respectively, and their impedances are $z_r=z_f=Z_0=50 \:\Omega$. The external coupling and shunting capacitance to the ground are $C_k=20$ fF, $C_d=100$ fF respectively.}
        \label{paper-fig:T_1+Drives results}
    \end{figure}

    We use the circuit in \cref{paper-fig:R0+Drive lines} to present a comparison between the analytical formulas derived for dissipation in \cref{Purcell_Y_main,Purcell_Z_main} and resulting from the coupling between the qubit and drive ports, incorporating both impedance and admittance responses. We assess the accuracy of these formulas by comparing them to master equation simulation and previous one port results presented in Ref.~\cite{Esteve:1986}, see \cref{paper-fig:T_1+Drives results}. For the circuit of \cref{paper-fig:R0+Drive lines}, the number of qubit and drive ports are $n=1$ and $n_D=2$ respectively. Once the $3\times3$ immittance response matrix is determined, which we do not present here  due to its noncompact form, we can apply our results \cref{Purcell_Y_main,Purcell_Z_main} to estimate the qubit-relaxation rate $1/T_1$. 
    To do so, we first extract the DC part of the immittance response given by $\Ymat^{dc}=\Dmat_{\infty}s$, $\Zmat^{dc}=\Amat_{0}/s$ with  $\Dmat_{\infty}=\lim_{\abs{s}\to \infty}\Ymat(s)/s$ and  $\Amat_{0}=\lim_{s\to 0}s\Zmat(s)$. This allows us to compute $\Ymat^\text{drive}$, $\Zmat^\text{drive}$ 
    given by 
    \begin{align}
    \Ymat^{\text{drive}}=Z_0^{-1}\mathbbm{1}_2+\mqty(\Ymat^{dc}_{22} & \Ymat^{dc}_{23}\\ \Ymat^{dc}_{32}& \Ymat^{dc}_{33}),    
    \end{align}
    and similarly 
    \begin{align}
    \Zmat^{\text{drive}}=Z_0 \mathbbm{1}_2+\mqty(\Zmat^{dc}_{22} & \Zmat^{dc}_{23}\\ \Zmat^{dc}_{32}& \Zmat^{dc}_{33}),
    \end{align}
    where $Z_0$ is the characteristic impedance of the external transmission lines. Here, index $1$ represents the qubit port, whereas the drive and readout ports are $2$ and $3$ respectively, see Fig.~\ref{paper-fig:R0+Drive lines}. Equipped with these matrices, we can now express the Purcell decay rates as
     \begin{equation}
        \frac{1}{T_{1,Y}} =\frac{1}{C}\sum_{d,d'=1}^{2}\Re{\Ymat_{dd'}^{\text{drive}^{-1}}(\overline{\omega})}\Ymat_{1d+1}(\overline{\omega})\Ymat_{1d'+1}^\star(\overline{\omega}),
    \end{equation}
    \begin{equation}
         \frac{1}{T_{1,Z}}=\frac{1}{\overline{L}_{J}}\sum_{d,d'=1}^{2}\Re{\Zmat_{dd'}^{\text{drive}^{-1}}(\overline{\omega})}\Zmat_{1d+1}(\overline{\omega})\Zmat_{1d'+1}^\star(\overline{\omega}),
    \end{equation}
    where $C$ is the first entry of $\Dmat_{\infty}$ and $C^{-1}$ that of $\Amat_0$, i.e., $C=(\Dmat_{\infty})_{11}$ such that for weak values of the coupling capacitances, it is approximately given by $C_j+C_{jd}+C_{jr}$. The qubit frequency is expressed as $\overline{\omega}=\Tilde{\omega}_J(1-E_c/\omega_J(1-\frac{E_c}{\omega_J}))$ where $\Tilde{\omega}_J=1/\sqrt{L_JC}$, $E_c=e^2/2C$ and $\overline{L}_J=1/(C\overline{\omega}^2)$. 
    In the case of the circuit of \cref{paper-fig:R0+Drive lines}, the drive ports are decoupled, as evidenced by the diagonal nature of $\Ymat^{\text{drive}}$, $\Zmat^{\text{drive}}$, i.e.,  $\Ymat^{dc}_{23}=\Ymat^{dc}_{32}=0$ and $\Zmat^{dc}_{23}=\Zmat^{dc}_{32}=0$. Consequently, the total relaxation rate can be decomposed into the sum of a Purcell decay rate resulting from the readout line and a drive-induced decay rate originating from the direct coupling with the drive line
    \begin{equation}
     \frac{1}{T_1}=\frac{1}{T_{1}^{\text{RO}}}+\frac{1}{T_{1}^{\text{D}}}, 
    \end{equation}
    where 
    \begin{align}
    \frac{1}{T_{1,Y}^{\text{RO}}}&=\frac{1}{C}\Re{\Ymat_{11}^{\text{drive}^{-1}}(\overline{\omega})}\abs{\Ymat_{12}(\overline{\omega})}^2,\\
    \frac{1}{T_{1,Y}^{\text{D}}}&=\frac{1}{C}\Re{\Ymat_{22}^{\text{drive}^{-1}}(\overline{\omega})}\abs{\Ymat_{13}(\overline{\omega})}^2,
    \end{align}
    and similar expressions are obtained for the impedance. In the limit of weak couplings $C_{jd},C_{jr},C_{rp},C_{k}\ll C_j,C_r,C_p$, we obtain  
    \begin{align} \label{maintext: ComparisonwithMartinis}
        \begin{split}
        \frac{1}{T_{1}^{\text{RO}}}=&\frac{Z_0C_{rp}^2C_{jr}^2C_k^2}{(1+\overline{\omega}^2(C_d+C_k)^2Z_0^2)C_jC_p^2C_r^2}\\
        &\times \frac{\overline{\omega}^{10}}{(\omega_r^2-\overline{\omega}^2)^2(\omega_f^2-\overline{\omega}^2)^2},
        \end{split}\\
        \frac{1}{T_{1}^{\text{D}}}&=\frac{Z_0\overline{\omega}^2C_{jd}^2}{C(1+Z_0^2\overline{\omega}^2(C_d+C_{jd})^2)}.
    \end{align}
    To facilitate comparison with prior studies, we rephrase the Purcell decay rate in terms of the qubit-readout resonator coupling strength $g$, the readout resonator-Purcell filter coupling $J$, and $\kappa_f$, the bare decay rate of the Purcell mode. Using our formulas \cref{paper-eq:overline-w-y,Purcell_Y_main} in the  
    weak coupling regime, we obtain $g\simeq C_{jr} (\overline{\omega}\omega_r/4C_jC_r)^{\frac{1}{2}}$, $J\simeq C_{rp} (\omega_r\omega_f/4C_rC_p)^{\frac{1}{2}}$, and $\kappa_f\simeq(\omega_f^2C_k^2Z_0)/C_p(1+\omega_f^2Z_0^2(C_d+C_k)^2)\equiv \kappa(\omega_f)$.
    Using these expressions, \cref{maintext: ComparisonwithMartinis} can be concisely written as
    \begin{equation}
    \begin{split}
    \frac{1}{T_{1}^{\text{RO}}}&=\kappa_q\left(\frac{gJ}{\Delta_{rq}\Delta_{fq}}\right)^2(1-\zeta_r)^2 (1-\zeta_f)^2\\
    &\times(\frac{1-\zeta_r}{1+\zeta_r})^2(\frac{1-\zeta_f}{1+\zeta_f}),
    \end{split}
    \end{equation}
    where $\kappa_q=(\overline{\omega}^2C_k^2Z_0)/C_p(1+\overline{\omega}^2Z_0^2(C_d+C_k)^2)=\kappa(\overline{\omega})$,  $\Delta_{rq}=\omega_r-\overline{\omega}$, $\Delta_{fq}=\omega_f-\overline{\omega}$ and $\zeta_\mu=\Delta_{\mu q}/\Sigma_{\mu q}$ with $\Sigma_{\mu q}=\omega_\mu+\overline{\omega}$ and $\mu=r,f$. From this equation, we obtain corrections to the usual Purcell protection formula given by $\left(\frac{gJ}{\Delta_{rq}\Delta_{fq}}\right)^2 \kappa_f$ of Refs.~\cite{Sete:2015, Jeffrey:2014} coming from two distinct contributions. First, the qubit probes the bath at its own frequency, rather than at the filter frequency which is reflected in $\kappa_q=\kappa(\overline{\omega})$, in accordance with Ref.~\cite{Boissonneault:2009}. This correction is significant when the bath is not entirely flat, something which cannot be captured by the standard master equation as discussed, e.g., in  Sec. IV of Ref.~\cite{Blais:2021}, see also \cref{paper-fig:T_1+Drives results}. Second, there is a correction of order $\zeta_\mu$ originating from the counter terms in the interaction between the qubit and the resonator-filter modes. Notably, this correction become particularly relevant in the deep dispersive regime (i.e., when $\Delta_{\mu q} \sim \Sigma_{\mu q}$), as highlighted in Ref.~\cite{Yan:2018}.
    
    In \cref{paper-fig:T_1+Drives results}, we present  a quantitative comparison between our immittance results (light and dark blue) and the integration of the master equation (brown) obtained after the exact quantization of the three-mode system (qubit, readout-resonator, Purcell filter). In this simulation, the values of $\kappa_f$ and $T_1^{\text{D}}$ are taken from previous one-port results, 
    in particular, here from Eqs.~(2.18) and (3.1) of Ref.~\cite{Esteve:1986} yielding to $\kappa_f=\Re{i\omega_f C_k(Z_0^{-1}+i\omega_f C_d)/C_p(Z_0^{-1}+i\omega_f(C_d+C_k))}$ and similar expression for $ 1/T_1^{\text{D}}$ by substituting $C_p$ by $C_J$, $C_k$ by $C_{jd}$ and $\omega_f$ by $\overline{\omega}$.  

\subsection{Three-port circulator circuit from the main text \edit{with capacitive filter}}
\label{Appendix main-example (a)}

We provide the calculations required to obtain the results for the example of \cref{fig:tuning-nonreciprocity-1} of the main text. We obtain the admittance of the circuit in \cref{fig:tuning-nonreciprocity-1}(a) with the $\pi$-capacitive filter of panel (b) using the admittance representation of the ideal circulator described by a scattering matrix $\Smat(\phi)$, and the combination of additional parallel and series capacitors 
    \begin{equation} \label{main: admittance tuning NR}
    \begin{split}
        \Ymat(s)(\phi) &= \Dmat_{\infty}s+\Emat_{\infty}(\phi)+\frac{\Dmat_1(\phi)s+\Emat_1(\phi)}{\omega_{y}(\phi)^2+s^2},\\
        \Dmat_{\infty} &= \overline{C}_J\bone_3,\\
        \Emat_{\infty}(\phi) &= -G(\phi)\boldsymbol{(}\Smat(2\pi/3)-\Smat(-2\pi/3)\boldsymbol{)}, \\
        \Dmat_1(\phi) &= \alpha(\phi)\boldsymbol{(}2-\Smat(-2\pi/3)-\Smat(2\pi/3)\boldsymbol{)}, \\
        \Emat_1(\phi) &= \sqrt{3}\omega_y(\phi)\alpha(\phi)\boldsymbol{(}\Smat(2\pi/3)-\Smat(-2\pi/3)\boldsymbol{)}. \\
    \end{split}
    \end{equation}
    
    In the dispersive regime, and up to second-order in $r_{cg},r_{cj}\ll 1$ with $r_{cg}=C_c/C_g$ and $r_{cj}=C_c/C_J$, we have $\overline{C}_J \simeq \boldsymbol{(}C_J(1+r_{cj})-C_gr_{cg}^2\boldsymbol{)}$, $G(\phi)\simeq \tan(\phi/2)r_{cg}^2/\sqrt{3}R$, $\omega_{y}(\phi)= \tan(\phi/2)/(C_c+C_g)R$, and $\alpha(\phi)\simeq \tan(\phi/2)^2r_{cg}^2/3R^2C_g$. The matrix $\Smat(\phi)$ corresponds to the scattering matrix defined in \cref{paper:eq-S-definition} of the main text. The inverse of the admittance gives the impedance response
    \begin{equation}
    \begin{split}
        \Zmat(s) &= \frac{\Amat_0}{s}+\frac{\Amat_1s+\Bmat_1}{\omega_{z}^2+s^2} \\
        \Amat_0 & = \widetilde{C}_J\bone_3-\widetilde{C}_k\boldsymbol{(}\Smat(2\pi/3)+\Smat(-2\pi/3)\boldsymbol{)} \\
        \Amat_1 &= \beta \boldleft 2\bone_3-\Smat(2\pi/3)-\Smat(-2\pi/3)\boldright \\
        \Bmat_1 &= \sqrt{3}\omega_z(\phi)\beta\boldleft \Smat(2\pi/3)-\Smat(-2\pi/3)\boldright,
    \end{split}
    \end{equation}
    with $\widetilde{C}_J\simeq \overline{C}_J+2\widetilde{C}_k$, $\widetilde{C}_k\simeq C_cr_{cg}/3$, $\omega_{z}(\phi)\simeq\omega_{y}(\phi)(1-r_{cg}r_{cj})$,   
    and $\beta\simeq r_{cj}^2/3C_g$. As clear from above, the frequencies of the poles are slightly different in both immittance presentations. This clearly leads to Hamiltonians with different frequencies for the inner modes after the exact circuit quantization. This difference however, is insubstantial, as the quantization is done with different frames for each response. When accounting for the exact frames within which each response is defined, all observables are predicted to be identical regardless of the chosen approach.

    Furthermore, the one third ratio needed to achieve full chiral population transfer dynamics can be readily understood by examining the exact Hamiltonian of the circuit. By introducing the quasi-momenta mode operators $\hat{B}_k=\sum_je^{i2\pi kj/3}\bdag_j/\sqrt{3}$ for the qubit modes, the linearized Hamiltonian in terms of $\phi$ reads as follows
    \vspace{1em}
    \begin{equation}
    \label{paper-eq:exact-H-example}
        \begin{split}
            \hat{H}(\phi) &= \sum_i\overline{\omega}\hat{B}^{\dagger}_i\hat{B}_i + \omega_y(\phi)\adag\alow\\
            &\;\qq{}-J(\phi)(\hat{B}_{-1}^{\dagger}a+\hat{B}_{1}^{\dagger}\adag+\text{h.c.}),
        \end{split}
    \end{equation}
    where $J(\phi) = \sqrt{\omega_J\omega_y(\phi)}\sqrt{r_{cj}r_{cg}/6}$ . By considering separately the beam-splitter-like (two-mode squeezing) interaction between the $k=-1$ ($k=1$) mode and the inner resonator mode, the elementary shifts $\lambda_{\pm 1}$ in the frequencies of the modes $\hat{B}_{\pm 1}$  are $\lambda_{-1} \simeq -2J(\phi)^2/\Delta$ and $\lambda_{+1} \simeq -2J(\phi)^2/(2\omega_y(\phi)-\Delta) $. 
    Complete chiral population transfer is obtained when the quasi-momentum modes energies are equally spaced \cite{Koch:2010}. This equal spacing is approximately obtained for a detuning $\Delta\simeq 2\omega_y(\phi)/3$, which leads to $\overline{\omega} = \omega_y(\phi)/3$. For this condition to be accurate, the perturbative criteria $k(\phi)/\Delta(\phi)\ll 1$ must be satisfied, where $k(\phi)\sim [\boldleft|\alpha(\phi)|/\overline{C_J}\boldright \boldleft \overline{\omega}/\omega_{y}(\phi)\boldright]^{1/2}$ is the coupling between the qubits and the inner mode, and $\Delta(\phi)=|\overline{\omega}-\omega_{y}(\phi)|$. In particular, when $\overline{\omega} = \omega_y(\phi)/3$, the perturbative criteria reads $k(\phi)/\Delta(\phi) \simeq (3C_g/4\overline{C}_J)^{1/2}r_{cg}/(1-r_{cg})\ll 1$. If $C_g \sim C_J$ and $r_{cg}\ll 1$, it will remain accurate for all values of $\phi$. Finally, it is noteworthy that the couplings within the Hamiltonian provide a clear explanation for the feasibility of achieving chiral dynamics for any value of $\phi$, contingent on a straightforward frequency condition. Explicitly, as $\phi$ changes, the Hamiltonian in \cref{paper-eq:exact-H-example} remains invariant, with only the strength of the coupling $J(\phi)$ and the frequency $\omega_y(\phi)$ changing, hence by setting $\overline{\omega}_i = \overline{\omega}_y(\phi)/3$ only the time scale of the resulting dynamic changes.

    \subsection{\edit{Three-port circulator circuit from the main text with resonator filter}}
    \label{Appendix main-example (b)}

    \edit{Here, we provide details of the simulation performed to obtain the qubit population transfer dynamics shown in \cref{fig:tuning-nonreciprocity-2} (b) using the $LC$ filter shown in  \cref{fig:tuning-nonreciprocity-1} (c). Additionally, we show how our simple formulas can be used to engineer the linear response of multimode nonreciprocal systems. In particular, we obtain an ideal circulator response for the circuit with the $LC$ filter in \cref{fig:tuning-nonreciprocity-1} (c) and match the effective scattering matrix of Sec. II in Ref.~\cite{Koch:2010}. }

    \begin{figure}[b]
        \centering
        \includegraphics[width=1\linewidth]{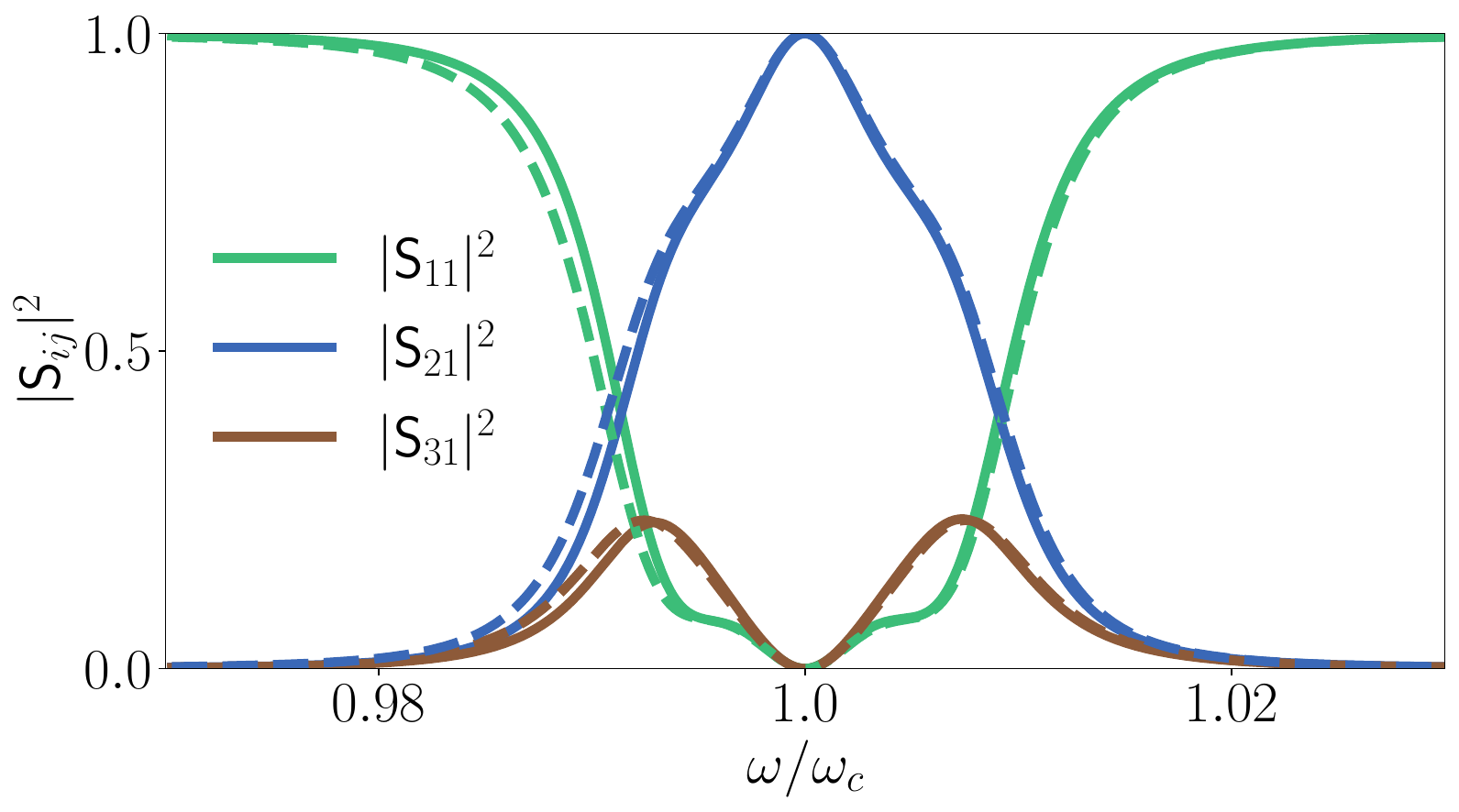}
        \caption{\edit{Comparison of the classical scattering response $\Smat$ (solid) computed from the impedance matrix $\Zmat$ of the circuit in \cref{fig:tuning-nonreciprocity-1} (a, $i$) using the $LC$ filter \cref{fig:tuning-nonreciprocity-1} (c) with reference probe transmission lines with characteristic impedance $R_0=50$ $\Omega$, with the effective one (dashed) obtained as in Sec. II of Ref.~\cite{Koch:2010}, see text for details. The frequency at which ideal circulation is achieved ($\omega_c$) is shifted from the bare frequency of the filter resonators ($\omega_{rf}$). The system parameters are $C_J=10.0$ nF, $C_{jr}=10.0$ nF, $C_{rs}=5.6$ nF, $\omega_{r_f}=\omega_y(\phi)/3 = 2\pi \times 12.08$ GHz, $z_{r_f} = 50$ $\Omega$, $R=100$ $\Omega$, $\phi\simeq 0.081 \pi$, which leads to $\omega_c \simeq 0.977 \omega_{rf}$.}}
        \label{fig:appendix-S}
    \end{figure}

    \edit{The full Hamiltonian was constructed using the impedance representation following \cref{Elimination-Nondynamical}. The admittance approach is equally valid, but the former is preferred due to the lack of poles at infinity, a feature that simplifies the construction. The effective Hamiltonian \cref{main: Hamiltonian} is constructed as prescribed in the main text. We fix the parameters $C_{jr}$, $C_{rs}$, $C_g$, and $R$, as well as the frequency and impedance of the filter resonators $\omega_{r_f}$ and $z_{r_f}$. Using \cref{paper:phaseij-Y}, we solve for $\overline{\omega}$ of \cref{paper-eq:overline-w-y} such that $\theta_{ij}=\pi/6$. In particular, the resulting expression is a second-degree polynomial in the qubit frequency $\overline{\omega}$, which admits a physical solution for any degree of nonreciprocity in the central element (i.e., $\phi\neq\{0,\pi\}$) provided one can tune the remaining circuit parameters (e.g. resonator frequencies, coupling strengths, capacitances to ground).  The quantum simulations are performed by truncating the maximum photon number to four, and convergence was checked by comparing with simulations with five photons.}

    \edit{\Cref{fig:appendix-S} shows the classical linear response (solid), and the desired quantum linear response (dashed) obtained using Eq.~(21) of Ref.~\cite{Koch:2010}.
    % \Cref{fig:appendix-S} shows the circulator response versus frequency. 
    To obtain the classical response, we first for simplicity take the limit $C_g \to 0$, where the non-reciprocal circuit becomes identical to the one studied in Sec. II of Ref.~\cite{Koch:2010}. Then, we set the filter resonator frequency using \cref{main tan-theta} such that the effective hopping phase between the resonators are $\theta_{ij}=\pi/6$. 
    % We compare the classical linear response (solid), and the quantum linear response using Eq. (21) of Ref.~\cite{Koch:2010} (dashed). 
    Finally, the response is obtained from the impedance matrix $\Zmat(\omega)$ of the circuit~\cite{Newcomb:1966,Pozar:2009}, as measured with reference probes of $R_0=50$ $\Omega$, i.e., $\Smat(\omega) = (\Zmat+R_0\bone)^{-1}(\Zmat-R_0\bone)$. The quantum mechanical response from Ref.~\cite{Koch:2010} is computed here by setting its two free parameters to $\omega_r=\omega_c$ and $\kappa = |J|$ (see Eq.~(13) of that reference together with the condition $\kappa' = 2\kappa$). Here, $\omega_c$ is the frequency where ideal circulator response is obtained, and $J$ the effective coupling between resonator modes obtained using \cref{paper:eff-J-z}. The observed small discrepancy between the two approaches arises because the targeted response (dashed line) assumes symmetric spacing between the three harmonic modes obtained after diagonalizing the effective Hamiltonian for the three resonators. Specifically, if \(\lambda_0 < \lambda_1 < \lambda_2\) are the three lowest eigenvalues of the effective Hamiltonian in the ideal case, it holds that \(\lambda_1 - \lambda_0 = \lambda_2 - \lambda_1\). However, when the full Hamiltonian is taken into account, the resulting eigenvalues exhibit a slight asymmetry, such that \(\lambda_1 - \lambda_0 = \lambda_2 - \lambda_1 + \Delta\). This small \(\Delta\) is the source of the observed minor discrepancy. This, however, does not prevent ideal circulation.

    Remarkably, the desired circulator response can be obtained for arbitrary small values of $\phi$, i.e., low nonreciprocity in the central element. This example shows how our methods can be useful not only in designing effective quantum Hamiltonian models but also in engineering desired scattering responses. All the codes to reproduce the main results are available at \cite{Labarca_Toolbox_for_nonreciprocal_2024}.}

    % \abc{Why is there a (small) discrepancy between the methods in that figure?}\lc{This is so because the ideal response (solid) assumes a symmetric spacing between the three harmonic modes obtained after diagonalization of the effective Hamiltonian for the tree resonators. In other words, if $\lambda_0<\lambda_1<\lambda_2$ are the three lowest eigenvalues of the effective Hamiltonian in the ideal case is true that $\lambda_1-\lambda0 = \lambda_2-\lambda_1$. However, when taking the full Hamiltonian the resulting eigenvalues will have a small asymmetry, i.e $\lambda_1-\lambda_0 = \lambda_2-\lambda_1+\Delta$. That small $\Delta$ causes the asymmetry in the response.} \abc{Go ahead an integrate your response to the text.}

    % \newpage
    \bibliography{bibliography.bib}

\begin{thebibliography}{83}%
\makeatletter
\providecommand \@ifxundefined [1]{%
 \@ifx{#1\undefined}
}%
\providecommand \@ifnum [1]{%
 \ifnum #1\expandafter \@firstoftwo
 \else \expandafter \@secondoftwo
 \fi
}%
\providecommand \@ifx [1]{%
 \ifx #1\expandafter \@firstoftwo
 \else \expandafter \@secondoftwo
 \fi
}%
\providecommand \natexlab [1]{#1}%
\providecommand \enquote  [1]{``#1''}%
\providecommand \bibnamefont  [1]{#1}%
\providecommand \bibfnamefont [1]{#1}%
\providecommand \citenamefont [1]{#1}%
\providecommand \href@noop [0]{\@secondoftwo}%
\providecommand \href [0]{\begingroup \@sanitize@url \@href}%
\providecommand \@href[1]{\@@startlink{#1}\@@href}%
\providecommand \@@href[1]{\endgroup#1\@@endlink}%
\providecommand \@sanitize@url [0]{\catcode `\\12\catcode `\$12\catcode
  `\&12\catcode `\#12\catcode `\^12\catcode `\_12\catcode `\%12\relax}%
\providecommand \@@startlink[1]{}%
\providecommand \@@endlink[0]{}%
\providecommand \url  [0]{\begingroup\@sanitize@url \@url }%
\providecommand \@url [1]{\endgroup\@href {#1}{\urlprefix }}%
\providecommand \urlprefix  [0]{URL }%
\providecommand \Eprint [0]{\href }%
\providecommand \doibase [0]{https://doi.org/}%
\providecommand \selectlanguage [0]{\@gobble}%
\providecommand \bibinfo  [0]{\@secondoftwo}%
\providecommand \bibfield  [0]{\@secondoftwo}%
\providecommand \translation [1]{[#1]}%
\providecommand \BibitemOpen [0]{}%
\providecommand \bibitemStop [0]{}%
\providecommand \bibitemNoStop [0]{.\EOS\space}%
\providecommand \EOS [0]{\spacefactor3000\relax}%
\providecommand \BibitemShut  [1]{\csname bibitem#1\endcsname}%
\let\auto@bib@innerbib\@empty
%</preamble>
\bibitem [{\citenamefont {{Solgun}}\ \emph {et~al.}(2019)\citenamefont
  {{Solgun}}, \citenamefont {{DiVincenzo}},\ and\ \citenamefont
  {{Gambetta}}}]{Solgun:2019}%
  \BibitemOpen
  \bibfield  {author} {\bibinfo {author} {\bibfnamefont {F.}~\bibnamefont
  {{Solgun}}}, \bibinfo {author} {\bibfnamefont {D.~P.}\ \bibnamefont
  {{DiVincenzo}}},\ and\ \bibinfo {author} {\bibfnamefont {J.~M.}\ \bibnamefont
  {{Gambetta}}},\ }\bibfield  {title} {\bibinfo {title} {Simple impedance
  response formulas for the dispersive interaction rates in the effective
  hamiltonians of low anharmonicity superconducting qubits},\ }\href
  {https://doi.org/10.1109/TMTT.2019.2893639} {\bibfield  {journal} {\bibinfo
  {journal} {IEEE Transactions on Microwave Theory and Techniques}\ }\textbf
  {\bibinfo {volume} {67}},\ \bibinfo {pages} {928} (\bibinfo {year}
  {2019})}\BibitemShut {NoStop}%
\bibitem [{\citenamefont {Devoret}\ and\ \citenamefont
  {Schoelkopf}(2013)}]{Devoret:2013}%
  \BibitemOpen
  \bibfield  {author} {\bibinfo {author} {\bibfnamefont {M.}~\bibnamefont
  {Devoret}}\ and\ \bibinfo {author} {\bibfnamefont {R.}~\bibnamefont
  {Schoelkopf}},\ }\bibfield  {title} {\bibinfo {title} {{Superconducting
  circuits for quantum information: An outlook}},\ }\href
  {https://doi.org/10.1126/science.1231930} {\bibfield  {journal} {\bibinfo
  {journal} {Science}\ }\textbf {\bibinfo {volume} {339}},\ \bibinfo {pages}
  {1169} (\bibinfo {year} {2013})}\BibitemShut {NoStop}%
\bibitem [{\citenamefont {Ofek}\ \emph {et~al.}(2016)\citenamefont {Ofek},
  \citenamefont {Petrenko}, \citenamefont {Heeres}, \citenamefont {Reinhold},
  \citenamefont {Leghtas}, \citenamefont {Vlastakis}, \citenamefont {Liu},
  \citenamefont {Frunzio}, \citenamefont {Girvin}, \citenamefont {Jiang},
  \citenamefont {Mirrahimi}, \citenamefont {Devoret},\ and\ \citenamefont
  {Schoelkopf}}]{Ofek:2016}%
  \BibitemOpen
  \bibfield  {author} {\bibinfo {author} {\bibfnamefont {N.}~\bibnamefont
  {Ofek}}, \bibinfo {author} {\bibfnamefont {A.}~\bibnamefont {Petrenko}},
  \bibinfo {author} {\bibfnamefont {R.}~\bibnamefont {Heeres}}, \bibinfo
  {author} {\bibfnamefont {P.}~\bibnamefont {Reinhold}}, \bibinfo {author}
  {\bibfnamefont {Z.}~\bibnamefont {Leghtas}}, \bibinfo {author} {\bibfnamefont
  {B.}~\bibnamefont {Vlastakis}}, \bibinfo {author} {\bibfnamefont
  {Y.}~\bibnamefont {Liu}}, \bibinfo {author} {\bibfnamefont {L.}~\bibnamefont
  {Frunzio}}, \bibinfo {author} {\bibfnamefont {S.~M.}\ \bibnamefont {Girvin}},
  \bibinfo {author} {\bibfnamefont {L.}~\bibnamefont {Jiang}}, \bibinfo
  {author} {\bibfnamefont {M.}~\bibnamefont {Mirrahimi}}, \bibinfo {author}
  {\bibfnamefont {M.~H.}\ \bibnamefont {Devoret}},\ and\ \bibinfo {author}
  {\bibfnamefont {R.~J.}\ \bibnamefont {Schoelkopf}},\ }\bibfield  {title}
  {\bibinfo {title} {Extending the lifetime of a quantum bit with error
  correction in superconducting circuits},\ }\href
  {https://doi.org/10.1038/nature18949} {\bibfield  {journal} {\bibinfo
  {journal} {Nature}\ }\textbf {\bibinfo {volume} {536}},\ \bibinfo {pages}
  {441} (\bibinfo {year} {2016})}\BibitemShut {NoStop}%
\bibitem [{\citenamefont {Arute}\ \emph {et~al.}(2019)\citenamefont {Arute},
  \citenamefont {Arya}, \citenamefont {Babbush}, \citenamefont {Bacon},
  \citenamefont {Bardin}, \citenamefont {Barends}, \citenamefont {Biswas},
  \citenamefont {Boixo}, \citenamefont {Brandao}, \citenamefont {Buell} \emph
  {et~al.}}]{Arute:2019}%
  \BibitemOpen
  \bibfield  {author} {\bibinfo {author} {\bibfnamefont {F.}~\bibnamefont
  {Arute}}, \bibinfo {author} {\bibfnamefont {K.}~\bibnamefont {Arya}},
  \bibinfo {author} {\bibfnamefont {R.}~\bibnamefont {Babbush}}, \bibinfo
  {author} {\bibfnamefont {D.}~\bibnamefont {Bacon}}, \bibinfo {author}
  {\bibfnamefont {J.~C.}\ \bibnamefont {Bardin}}, \bibinfo {author}
  {\bibfnamefont {R.}~\bibnamefont {Barends}}, \bibinfo {author} {\bibfnamefont
  {R.}~\bibnamefont {Biswas}}, \bibinfo {author} {\bibfnamefont
  {S.}~\bibnamefont {Boixo}}, \bibinfo {author} {\bibfnamefont {F.~G. S.~L.}\
  \bibnamefont {Brandao}}, \bibinfo {author} {\bibfnamefont {D.~A.}\
  \bibnamefont {Buell}}, \emph {et~al.},\ }\bibfield  {title} {\bibinfo {title}
  {Quantum supremacy using a programmable superconducting processor},\ }\href
  {https://doi.org/10.1038/s41586-019-1666-5} {\bibfield  {journal} {\bibinfo
  {journal} {Nature}\ }\textbf {\bibinfo {volume} {574}},\ \bibinfo {pages}
  {505} (\bibinfo {year} {2019})}\BibitemShut {NoStop}%
\bibitem [{\citenamefont {Campagne-Ibarcq}\ \emph {et~al.}(2020)\citenamefont
  {Campagne-Ibarcq}, \citenamefont {Eickbusch}, \citenamefont {Touzard},
  \citenamefont {Zalys-Geller}, \citenamefont {Frattini}, \citenamefont
  {Sivak}, \citenamefont {Reinhold}, \citenamefont {Puri}, \citenamefont
  {Shankar}, \citenamefont {Schoelkopf}, \citenamefont {Frunzio}, \citenamefont
  {Mirrahimi},\ and\ \citenamefont {Devoret}}]{Campagne-Ibarcq:2020}%
  \BibitemOpen
  \bibfield  {author} {\bibinfo {author} {\bibfnamefont {P.}~\bibnamefont
  {Campagne-Ibarcq}}, \bibinfo {author} {\bibfnamefont {A.}~\bibnamefont
  {Eickbusch}}, \bibinfo {author} {\bibfnamefont {S.}~\bibnamefont {Touzard}},
  \bibinfo {author} {\bibfnamefont {E.}~\bibnamefont {Zalys-Geller}}, \bibinfo
  {author} {\bibfnamefont {N.~E.}\ \bibnamefont {Frattini}}, \bibinfo {author}
  {\bibfnamefont {V.~V.}\ \bibnamefont {Sivak}}, \bibinfo {author}
  {\bibfnamefont {P.}~\bibnamefont {Reinhold}}, \bibinfo {author}
  {\bibfnamefont {S.}~\bibnamefont {Puri}}, \bibinfo {author} {\bibfnamefont
  {S.}~\bibnamefont {Shankar}}, \bibinfo {author} {\bibfnamefont {R.~J.}\
  \bibnamefont {Schoelkopf}}, \bibinfo {author} {\bibfnamefont
  {L.}~\bibnamefont {Frunzio}}, \bibinfo {author} {\bibfnamefont
  {M.}~\bibnamefont {Mirrahimi}},\ and\ \bibinfo {author} {\bibfnamefont
  {M.~H.}\ \bibnamefont {Devoret}},\ }\bibfield  {title} {\bibinfo {title}
  {Quantum error correction of a qubit encoded in grid states of an
  oscillator},\ }\href {https://doi.org/10.1038/s41586-020-2603-3} {\bibfield
  {journal} {\bibinfo  {journal} {Nature}\ }\textbf {\bibinfo {volume} {584}},\
  \bibinfo {pages} {368} (\bibinfo {year} {2020})}\BibitemShut {NoStop}%
\bibitem [{\citenamefont {Krinner}\ \emph {et~al.}(2022)\citenamefont
  {Krinner}, \citenamefont {Lacroix}, \citenamefont {Remm}, \citenamefont
  {Di~Paolo}, \citenamefont {Genois}, \citenamefont {Leroux}, \citenamefont
  {Hellings}, \citenamefont {Lazar}, \citenamefont {Swiadek}, \citenamefont
  {Herrmann}, \citenamefont {Norris}, \citenamefont {Andersen}, \citenamefont
  {Müller}, \citenamefont {Blais}, \citenamefont {Eichler},\ and\
  \citenamefont {Wallraff}}]{Krinner:2022}%
  \BibitemOpen
  \bibfield  {author} {\bibinfo {author} {\bibfnamefont {S.}~\bibnamefont
  {Krinner}}, \bibinfo {author} {\bibfnamefont {N.}~\bibnamefont {Lacroix}},
  \bibinfo {author} {\bibfnamefont {A.}~\bibnamefont {Remm}}, \bibinfo {author}
  {\bibfnamefont {A.}~\bibnamefont {Di~Paolo}}, \bibinfo {author}
  {\bibfnamefont {E.}~\bibnamefont {Genois}}, \bibinfo {author} {\bibfnamefont
  {C.}~\bibnamefont {Leroux}}, \bibinfo {author} {\bibfnamefont
  {C.}~\bibnamefont {Hellings}}, \bibinfo {author} {\bibfnamefont
  {S.}~\bibnamefont {Lazar}}, \bibinfo {author} {\bibfnamefont
  {F.}~\bibnamefont {Swiadek}}, \bibinfo {author} {\bibfnamefont
  {J.}~\bibnamefont {Herrmann}}, \bibinfo {author} {\bibfnamefont {G.~J.}\
  \bibnamefont {Norris}}, \bibinfo {author} {\bibfnamefont {C.~K.}\
  \bibnamefont {Andersen}}, \bibinfo {author} {\bibfnamefont {M.}~\bibnamefont
  {Müller}}, \bibinfo {author} {\bibfnamefont {A.}~\bibnamefont {Blais}},
  \bibinfo {author} {\bibfnamefont {C.}~\bibnamefont {Eichler}},\ and\ \bibinfo
  {author} {\bibfnamefont {A.}~\bibnamefont {Wallraff}},\ }\bibfield  {title}
  {\bibinfo {title} {Realizing repeated quantum error correction in a
  distance-three surface code},\ }\href
  {https://doi.org/10.1038/s41586-022-04566-8} {\bibfield  {journal} {\bibinfo
  {journal} {Nature}\ }\textbf {\bibinfo {volume} {605}},\ \bibinfo {pages}
  {669} (\bibinfo {year} {2022})}\BibitemShut {NoStop}%
\bibitem [{\citenamefont {Zhao}\ \emph {et~al.}(2022)\citenamefont {Zhao},
  \citenamefont {Ye}, \citenamefont {Huang}, \citenamefont {Zhang},
  \citenamefont {Wu}, \citenamefont {Guan}, \citenamefont {Zhu}, \citenamefont
  {Wei}, \citenamefont {He}, \citenamefont {Cao}, \citenamefont {Chen},
  \citenamefont {Chung}, \citenamefont {Deng}, \citenamefont {Fan},
  \citenamefont {Gong}, \citenamefont {Guo}, \citenamefont {Guo}, \citenamefont
  {Han}, \citenamefont {Li}, \citenamefont {Li}, \citenamefont {Li},
  \citenamefont {Liang}, \citenamefont {Lin}, \citenamefont {Qian},
  \citenamefont {Rong}, \citenamefont {Su}, \citenamefont {Sun}, \citenamefont
  {Wang}, \citenamefont {Wu}, \citenamefont {Xu}, \citenamefont {Ying},
  \citenamefont {Yu}, \citenamefont {Zha}, \citenamefont {Zhang}, \citenamefont
  {Huo}, \citenamefont {Lu}, \citenamefont {Peng}, \citenamefont {Zhu},\ and\
  \citenamefont {Pan}}]{Zhao:2022}%
  \BibitemOpen
  \bibfield  {author} {\bibinfo {author} {\bibfnamefont {Y.}~\bibnamefont
  {Zhao}}, \bibinfo {author} {\bibfnamefont {Y.}~\bibnamefont {Ye}}, \bibinfo
  {author} {\bibfnamefont {H.-L.}\ \bibnamefont {Huang}}, \bibinfo {author}
  {\bibfnamefont {Y.}~\bibnamefont {Zhang}}, \bibinfo {author} {\bibfnamefont
  {D.}~\bibnamefont {Wu}}, \bibinfo {author} {\bibfnamefont {H.}~\bibnamefont
  {Guan}}, \bibinfo {author} {\bibfnamefont {Q.}~\bibnamefont {Zhu}}, \bibinfo
  {author} {\bibfnamefont {Z.}~\bibnamefont {Wei}}, \bibinfo {author}
  {\bibfnamefont {T.}~\bibnamefont {He}}, \bibinfo {author} {\bibfnamefont
  {S.}~\bibnamefont {Cao}}, \bibinfo {author} {\bibfnamefont {F.}~\bibnamefont
  {Chen}}, \bibinfo {author} {\bibfnamefont {T.-H.}\ \bibnamefont {Chung}},
  \bibinfo {author} {\bibfnamefont {H.}~\bibnamefont {Deng}}, \bibinfo {author}
  {\bibfnamefont {D.}~\bibnamefont {Fan}}, \bibinfo {author} {\bibfnamefont
  {M.}~\bibnamefont {Gong}}, \bibinfo {author} {\bibfnamefont {C.}~\bibnamefont
  {Guo}}, \bibinfo {author} {\bibfnamefont {S.}~\bibnamefont {Guo}}, \bibinfo
  {author} {\bibfnamefont {L.}~\bibnamefont {Han}}, \bibinfo {author}
  {\bibfnamefont {N.}~\bibnamefont {Li}}, \bibinfo {author} {\bibfnamefont
  {S.}~\bibnamefont {Li}}, \bibinfo {author} {\bibfnamefont {Y.}~\bibnamefont
  {Li}}, \bibinfo {author} {\bibfnamefont {F.}~\bibnamefont {Liang}}, \bibinfo
  {author} {\bibfnamefont {J.}~\bibnamefont {Lin}}, \bibinfo {author}
  {\bibfnamefont {H.}~\bibnamefont {Qian}}, \bibinfo {author} {\bibfnamefont
  {H.}~\bibnamefont {Rong}}, \bibinfo {author} {\bibfnamefont {H.}~\bibnamefont
  {Su}}, \bibinfo {author} {\bibfnamefont {L.}~\bibnamefont {Sun}}, \bibinfo
  {author} {\bibfnamefont {S.}~\bibnamefont {Wang}}, \bibinfo {author}
  {\bibfnamefont {Y.}~\bibnamefont {Wu}}, \bibinfo {author} {\bibfnamefont
  {Y.}~\bibnamefont {Xu}}, \bibinfo {author} {\bibfnamefont {C.}~\bibnamefont
  {Ying}}, \bibinfo {author} {\bibfnamefont {J.}~\bibnamefont {Yu}}, \bibinfo
  {author} {\bibfnamefont {C.}~\bibnamefont {Zha}}, \bibinfo {author}
  {\bibfnamefont {K.}~\bibnamefont {Zhang}}, \bibinfo {author} {\bibfnamefont
  {Y.-H.}\ \bibnamefont {Huo}}, \bibinfo {author} {\bibfnamefont {C.-Y.}\
  \bibnamefont {Lu}}, \bibinfo {author} {\bibfnamefont {C.-Z.}\ \bibnamefont
  {Peng}}, \bibinfo {author} {\bibfnamefont {X.}~\bibnamefont {Zhu}},\ and\
  \bibinfo {author} {\bibfnamefont {J.-W.}\ \bibnamefont {Pan}},\ }\bibfield
  {title} {\bibinfo {title} {Realization of an error-correcting surface code
  with superconducting qubits},\ }\href
  {https://doi.org/10.1103/PhysRevLett.129.030501} {\bibfield  {journal}
  {\bibinfo  {journal} {Physical Review Letters}\ }\textbf {\bibinfo {volume}
  {129}},\ \bibinfo {pages} {030501} (\bibinfo {year} {2022})}\BibitemShut
  {NoStop}%
\bibitem [{\citenamefont {{Google Quantum AI}}(2023)}]{GoogleQAI:2023}%
  \BibitemOpen
  \bibfield  {author} {\bibinfo {author} {\bibnamefont {{Google Quantum AI}}},\
  }\bibfield  {title} {\bibinfo {title} {Suppressing quantum errors by scaling
  a surface code logical qubit},\ }\href
  {https://doi.org/10.1038/s41586-022-05434-1} {\bibfield  {journal} {\bibinfo
  {journal} {Nature}\ }\textbf {\bibinfo {volume} {614}},\ \bibinfo {pages}
  {676} (\bibinfo {year} {2023})}\BibitemShut {NoStop}%
\bibitem [{\citenamefont {Josephson}(1962)}]{Josephson:1962}%
  \BibitemOpen
  \bibfield  {author} {\bibinfo {author} {\bibfnamefont {B.~D.}\ \bibnamefont
  {Josephson}},\ }\bibfield  {title} {\bibinfo {title} {Possible new effects in
  superconductive tunnelling},\ }\href
  {https://doi.org/https://doi.org/10.1016/0031-9163(62)91369-0} {\bibfield
  {journal} {\bibinfo  {journal} {Physics Letters}\ }\textbf {\bibinfo {volume}
  {1}},\ \bibinfo {pages} {251} (\bibinfo {year} {1962})}\BibitemShut {NoStop}%
\bibitem [{\citenamefont {Blais}\ \emph {et~al.}(2021)\citenamefont {Blais},
  \citenamefont {Grimsmo}, \citenamefont {Girvin},\ and\ \citenamefont
  {Wallraff}}]{Blais:2021}%
  \BibitemOpen
  \bibfield  {author} {\bibinfo {author} {\bibfnamefont {A.}~\bibnamefont
  {Blais}}, \bibinfo {author} {\bibfnamefont {A.~L.}\ \bibnamefont {Grimsmo}},
  \bibinfo {author} {\bibfnamefont {S.~M.}\ \bibnamefont {Girvin}},\ and\
  \bibinfo {author} {\bibfnamefont {A.}~\bibnamefont {Wallraff}},\ }\bibfield
  {title} {\bibinfo {title} {Circuit quantum electrodynamics},\ }\href
  {https://link.aps.org/doi/10.1103/RevModPhys.93.025005} {\bibfield  {journal}
  {\bibinfo  {journal} {Review Modern Physics}\ }\textbf {\bibinfo {volume}
  {93}},\ \bibinfo {pages} {025005} (\bibinfo {year} {2021})}\BibitemShut
  {NoStop}%
\bibitem [{\citenamefont {Koch}\ \emph {et~al.}(2007)\citenamefont {Koch},
  \citenamefont {Yu}, \citenamefont {Gambetta}, \citenamefont {Houck},
  \citenamefont {Schuster}, \citenamefont {Majer}, \citenamefont {Blais},
  \citenamefont {Devoret}, \citenamefont {Girvin},\ and\ \citenamefont
  {Schoelkopf}}]{Koch:2007}%
  \BibitemOpen
  \bibfield  {author} {\bibinfo {author} {\bibfnamefont {J.}~\bibnamefont
  {Koch}}, \bibinfo {author} {\bibfnamefont {T.~M.}\ \bibnamefont {Yu}},
  \bibinfo {author} {\bibfnamefont {J.}~\bibnamefont {Gambetta}}, \bibinfo
  {author} {\bibfnamefont {A.~A.}\ \bibnamefont {Houck}}, \bibinfo {author}
  {\bibfnamefont {D.~I.}\ \bibnamefont {Schuster}}, \bibinfo {author}
  {\bibfnamefont {J.}~\bibnamefont {Majer}}, \bibinfo {author} {\bibfnamefont
  {A.}~\bibnamefont {Blais}}, \bibinfo {author} {\bibfnamefont {M.~H.}\
  \bibnamefont {Devoret}}, \bibinfo {author} {\bibfnamefont {S.~M.}\
  \bibnamefont {Girvin}},\ and\ \bibinfo {author} {\bibfnamefont {R.~J.}\
  \bibnamefont {Schoelkopf}},\ }\bibfield  {title} {\bibinfo {title}
  {Charge-insensitive qubit design derived from the cooper pair box},\ }\href
  {https://doi.org/10.1103/PhysRevA.76.042319} {\bibfield  {journal} {\bibinfo
  {journal} {Physical Review A}\ }\textbf {\bibinfo {volume} {76}},\ \bibinfo
  {pages} {042319} (\bibinfo {year} {2007})}\BibitemShut {NoStop}%
\bibitem [{\citenamefont {Wallraff}\ \emph {et~al.}(2004)\citenamefont
  {Wallraff}, \citenamefont {Schuster}, \citenamefont {Blais}, \citenamefont
  {Frunzio}, \citenamefont {Huang}, \citenamefont {Majer}, \citenamefont
  {Kumar}, \citenamefont {Girvin},\ and\ \citenamefont
  {Schoelkopf}}]{Wallraff:2004}%
  \BibitemOpen
  \bibfield  {author} {\bibinfo {author} {\bibfnamefont {A.}~\bibnamefont
  {Wallraff}}, \bibinfo {author} {\bibfnamefont {D.}~\bibnamefont {Schuster}},
  \bibinfo {author} {\bibfnamefont {A.}~\bibnamefont {Blais}}, \bibinfo
  {author} {\bibfnamefont {L.}~\bibnamefont {Frunzio}}, \bibinfo {author}
  {\bibfnamefont {R.-S.}\ \bibnamefont {Huang}}, \bibinfo {author}
  {\bibfnamefont {J.}~\bibnamefont {Majer}}, \bibinfo {author} {\bibfnamefont
  {S.}~\bibnamefont {Kumar}}, \bibinfo {author} {\bibfnamefont
  {S.}~\bibnamefont {Girvin}},\ and\ \bibinfo {author} {\bibfnamefont
  {R.}~\bibnamefont {Schoelkopf}},\ }\bibfield  {title} {\bibinfo {title}
  {Strong coupling of a single photon to a superconducting qubit using circuit
  quantum electrodynamics},\ }\href {https://doi.org/10.1038/nature02851}
  {\bibfield  {journal} {\bibinfo  {journal} {Nature}\ }\textbf {\bibinfo
  {volume} {431}},\ \bibinfo {pages} {162} (\bibinfo {year}
  {2004})}\BibitemShut {NoStop}%
\bibitem [{\citenamefont {Blais}\ \emph {et~al.}(2004)\citenamefont {Blais},
  \citenamefont {Huang}, \citenamefont {Wallraff}, \citenamefont {Girvin},\
  and\ \citenamefont {Schoelkopf}}]{Blais:2004}%
  \BibitemOpen
  \bibfield  {author} {\bibinfo {author} {\bibfnamefont {A.}~\bibnamefont
  {Blais}}, \bibinfo {author} {\bibfnamefont {R.-S.}\ \bibnamefont {Huang}},
  \bibinfo {author} {\bibfnamefont {A.}~\bibnamefont {Wallraff}}, \bibinfo
  {author} {\bibfnamefont {S.~M.}\ \bibnamefont {Girvin}},\ and\ \bibinfo
  {author} {\bibfnamefont {R.~J.}\ \bibnamefont {Schoelkopf}},\ }\bibfield
  {title} {\bibinfo {title} {Cavity quantum electrodynamics for superconducting
  electrical circuits: An architecture for quantum computation},\ }\href
  {https://doi.org/10.1103/PhysRevA.69.062320} {\bibfield  {journal} {\bibinfo
  {journal} {Physical ReviewA}\ }\textbf {\bibinfo {volume} {69}},\ \bibinfo
  {pages} {062320} (\bibinfo {year} {2004})}\BibitemShut {NoStop}%
\bibitem [{\citenamefont {Nigg}\ \emph {et~al.}(2012)\citenamefont {Nigg},
  \citenamefont {Paik}, \citenamefont {Vlastakis}, \citenamefont {Kirchmair},
  \citenamefont {Shankar}, \citenamefont {Frunzio}, \citenamefont {Devoret},
  \citenamefont {Schoelkopf},\ and\ \citenamefont {Girvin}}]{Nigg:2012}%
  \BibitemOpen
  \bibfield  {author} {\bibinfo {author} {\bibfnamefont {S.~E.}\ \bibnamefont
  {Nigg}}, \bibinfo {author} {\bibfnamefont {H.}~\bibnamefont {Paik}}, \bibinfo
  {author} {\bibfnamefont {B.}~\bibnamefont {Vlastakis}}, \bibinfo {author}
  {\bibfnamefont {G.}~\bibnamefont {Kirchmair}}, \bibinfo {author}
  {\bibfnamefont {S.}~\bibnamefont {Shankar}}, \bibinfo {author} {\bibfnamefont
  {L.}~\bibnamefont {Frunzio}}, \bibinfo {author} {\bibfnamefont {M.~H.}\
  \bibnamefont {Devoret}}, \bibinfo {author} {\bibfnamefont {R.~J.}\
  \bibnamefont {Schoelkopf}},\ and\ \bibinfo {author} {\bibfnamefont {S.~M.}\
  \bibnamefont {Girvin}},\ }\bibfield  {title} {\bibinfo {title} {Black-box
  superconducting circuit quantization},\ }\href
  {https://doi.org/10.1103/PhysRevLett.108.240502} {\bibfield  {journal}
  {\bibinfo  {journal} {Physical Review Letters}\ }\textbf {\bibinfo {volume}
  {108}},\ \bibinfo {pages} {240502} (\bibinfo {year} {2012})}\BibitemShut
  {NoStop}%
\bibitem [{\citenamefont {Solgun}\ \emph {et~al.}(2014)\citenamefont {Solgun},
  \citenamefont {Abraham},\ and\ \citenamefont {DiVincenzo}}]{Solgun:2014}%
  \BibitemOpen
  \bibfield  {author} {\bibinfo {author} {\bibfnamefont {F.}~\bibnamefont
  {Solgun}}, \bibinfo {author} {\bibfnamefont {D.~W.}\ \bibnamefont
  {Abraham}},\ and\ \bibinfo {author} {\bibfnamefont {D.~P.}\ \bibnamefont
  {DiVincenzo}},\ }\bibfield  {title} {\bibinfo {title} {Blackbox quantization
  of superconducting circuits using exact impedance synthesis},\ }\href
  {https://doi.org/10.1103/PhysRevB.90.134504} {\bibfield  {journal} {\bibinfo
  {journal} {Physical Review B}\ }\textbf {\bibinfo {volume} {90}},\ \bibinfo
  {pages} {134504} (\bibinfo {year} {2014})}\BibitemShut {NoStop}%
\bibitem [{\citenamefont {Solgun}\ and\ \citenamefont
  {DiVincenzo}(2015)}]{Solgun:2015}%
  \BibitemOpen
  \bibfield  {author} {\bibinfo {author} {\bibfnamefont {F.}~\bibnamefont
  {Solgun}}\ and\ \bibinfo {author} {\bibfnamefont {D.}~\bibnamefont
  {DiVincenzo}},\ }\bibfield  {title} {\bibinfo {title} {{Multiport impedance
  quantization}},\ }\href {https://doi.org/10.1016/j.aop.2015.07.005}
  {\bibfield  {journal} {\bibinfo  {journal} {Annals of Physics}\ }\textbf
  {\bibinfo {volume} {361}},\ \bibinfo {pages} {605} (\bibinfo {year}
  {2015})}\BibitemShut {NoStop}%
\bibitem [{\citenamefont {Minev}\ \emph {et~al.}(2021)\citenamefont {Minev},
  \citenamefont {Leghtas}, \citenamefont {Mundhada}, \citenamefont
  {Christakis}, \citenamefont {Pop},\ and\ \citenamefont
  {Devoret}}]{Minev:EPR}%
  \BibitemOpen
  \bibfield  {author} {\bibinfo {author} {\bibfnamefont {Z.~K.}\ \bibnamefont
  {Minev}}, \bibinfo {author} {\bibfnamefont {Z.}~\bibnamefont {Leghtas}},
  \bibinfo {author} {\bibfnamefont {S.~O.}\ \bibnamefont {Mundhada}}, \bibinfo
  {author} {\bibfnamefont {L.}~\bibnamefont {Christakis}}, \bibinfo {author}
  {\bibfnamefont {I.~M.}\ \bibnamefont {Pop}},\ and\ \bibinfo {author}
  {\bibfnamefont {M.~H.}\ \bibnamefont {Devoret}},\ }\bibfield  {title}
  {\bibinfo {title} {Energy-participation quantization of {Josephson}
  circuits},\ }\href {https://doi.org/10.1038/s41534-021-00461-8} {\bibfield
  {journal} {\bibinfo  {journal} {npj Quantum Information}\ }\textbf {\bibinfo
  {volume} {7}},\ \bibinfo {pages} {131} (\bibinfo {year} {2021})}\BibitemShut
  {NoStop}%
\bibitem [{\citenamefont {Newcomb}(1966)}]{Newcomb:1966}%
  \BibitemOpen
  \bibfield  {author} {\bibinfo {author} {\bibfnamefont {R.~W.}\ \bibnamefont
  {Newcomb}},\ }\href@noop {} {\emph {\bibinfo {title} {Linear Multiport
  Synthesis}}}\ (\bibinfo  {publisher} {McGraw-Hill},\ \bibinfo {address} {New
  York},\ \bibinfo {year} {1966})\BibitemShut {NoStop}%
\bibitem [{\citenamefont {Koch}\ \emph {et~al.}(2010)\citenamefont {Koch},
  \citenamefont {Houck}, \citenamefont {Hur},\ and\ \citenamefont
  {Girvin}}]{Koch:2010}%
  \BibitemOpen
  \bibfield  {author} {\bibinfo {author} {\bibfnamefont {J.}~\bibnamefont
  {Koch}}, \bibinfo {author} {\bibfnamefont {A.~A.}\ \bibnamefont {Houck}},
  \bibinfo {author} {\bibfnamefont {K.~L.}\ \bibnamefont {Hur}},\ and\ \bibinfo
  {author} {\bibfnamefont {S.~M.}\ \bibnamefont {Girvin}},\ }\bibfield  {title}
  {\bibinfo {title} {Time-reversal-symmetry breaking in circuit-qed-based
  photon lattices},\ }\href {https://doi.org/10.1103/PhysRevA.82.043811}
  {\bibfield  {journal} {\bibinfo  {journal} {Physical Review A}\ }\textbf
  {\bibinfo {volume} {82}},\ \bibinfo {pages} {043811} (\bibinfo {year}
  {2010})}\BibitemShut {NoStop}%
\bibitem [{\citenamefont {Owens}\ \emph {et~al.}(2022)\citenamefont {Owens},
  \citenamefont {Panetta}, \citenamefont {Saxberg}, \citenamefont {Roberts},
  \citenamefont {Chakram}, \citenamefont {Ma}, \citenamefont {Vrajitoarea},
  \citenamefont {Simon},\ and\ \citenamefont {Schuster}}]{Owens:2022}%
  \BibitemOpen
  \bibfield  {author} {\bibinfo {author} {\bibfnamefont {J.~C.}\ \bibnamefont
  {Owens}}, \bibinfo {author} {\bibfnamefont {M.~G.}\ \bibnamefont {Panetta}},
  \bibinfo {author} {\bibfnamefont {B.}~\bibnamefont {Saxberg}}, \bibinfo
  {author} {\bibfnamefont {G.}~\bibnamefont {Roberts}}, \bibinfo {author}
  {\bibfnamefont {S.}~\bibnamefont {Chakram}}, \bibinfo {author} {\bibfnamefont
  {R.}~\bibnamefont {Ma}}, \bibinfo {author} {\bibfnamefont {A.}~\bibnamefont
  {Vrajitoarea}}, \bibinfo {author} {\bibfnamefont {J.}~\bibnamefont {Simon}},\
  and\ \bibinfo {author} {\bibfnamefont {D.~I.}\ \bibnamefont {Schuster}},\
  }\bibfield  {title} {\bibinfo {title} {Chiral cavity quantum
  electrodynamics},\ }\href {https://doi.org/10.1038/s41567-022-01671-3}
  {\bibfield  {journal} {\bibinfo  {journal} {Nature Physics}\ }\textbf
  {\bibinfo {volume} {18}},\ \bibinfo {pages} {1048} (\bibinfo {year}
  {2022})}\BibitemShut {NoStop}%
\bibitem [{\citenamefont {Anderson}\ \emph {et~al.}(2016)\citenamefont
  {Anderson}, \citenamefont {Ma}, \citenamefont {Owens}, \citenamefont
  {Schuster},\ and\ \citenamefont {Simon}}]{Anderson:2016}%
  \BibitemOpen
  \bibfield  {author} {\bibinfo {author} {\bibfnamefont {B.~M.}\ \bibnamefont
  {Anderson}}, \bibinfo {author} {\bibfnamefont {R.}~\bibnamefont {Ma}},
  \bibinfo {author} {\bibfnamefont {C.}~\bibnamefont {Owens}}, \bibinfo
  {author} {\bibfnamefont {D.~I.}\ \bibnamefont {Schuster}},\ and\ \bibinfo
  {author} {\bibfnamefont {J.}~\bibnamefont {Simon}},\ }\bibfield  {title}
  {\bibinfo {title} {Engineering topological many-body materials in microwave
  cavity arrays},\ }\href {https://doi.org/10.1103/PhysRevX.6.041043}
  {\bibfield  {journal} {\bibinfo  {journal} {Physical Review X}\ }\textbf
  {\bibinfo {volume} {6}},\ \bibinfo {pages} {041043} (\bibinfo {year}
  {2016})}\BibitemShut {NoStop}%
\bibitem [{\citenamefont {Hassler}\ \emph {et~al.}(2019)\citenamefont
  {Hassler}, \citenamefont {Stubenrauch},\ and\ \citenamefont
  {Ciani}}]{Hassler:2019}%
  \BibitemOpen
  \bibfield  {author} {\bibinfo {author} {\bibfnamefont {F.}~\bibnamefont
  {Hassler}}, \bibinfo {author} {\bibfnamefont {J.}~\bibnamefont
  {Stubenrauch}},\ and\ \bibinfo {author} {\bibfnamefont {A.}~\bibnamefont
  {Ciani}},\ }\bibfield  {title} {\bibinfo {title} {Equation of motion approach
  to black-box quantization: Taming the multimode jaynes-cummings model},\
  }\href {https://doi.org/10.1103/PhysRevB.99.014515} {\bibfield  {journal}
  {\bibinfo  {journal} {Physical Review B}\ }\textbf {\bibinfo {volume} {99}},\
  \bibinfo {pages} {014515} (\bibinfo {year} {2019})}\BibitemShut {NoStop}%
\bibitem [{\citenamefont {Gely}\ \emph {et~al.}(2017)\citenamefont {Gely},
  \citenamefont {Parra-Rodriguez}, \citenamefont {Bothner}, \citenamefont
  {Blanter}, \citenamefont {Bosman}, \citenamefont {Solano},\ and\
  \citenamefont {Steele}}]{Gely:2017}%
  \BibitemOpen
  \bibfield  {author} {\bibinfo {author} {\bibfnamefont {M.~F.}\ \bibnamefont
  {Gely}}, \bibinfo {author} {\bibfnamefont {A.}~\bibnamefont
  {Parra-Rodriguez}}, \bibinfo {author} {\bibfnamefont {D.}~\bibnamefont
  {Bothner}}, \bibinfo {author} {\bibfnamefont {Y.~M.}\ \bibnamefont
  {Blanter}}, \bibinfo {author} {\bibfnamefont {S.~J.}\ \bibnamefont {Bosman}},
  \bibinfo {author} {\bibfnamefont {E.}~\bibnamefont {Solano}},\ and\ \bibinfo
  {author} {\bibfnamefont {G.~A.}\ \bibnamefont {Steele}},\ }\bibfield  {title}
  {\bibinfo {title} {Convergence of the multimode quantum rabi model of circuit
  quantum electrodynamics},\ }\href
  {https://doi.org/10.1103/PhysRevB.95.245115} {\bibfield  {journal} {\bibinfo
  {journal} {Physical Review B}\ }\textbf {\bibinfo {volume} {95}},\ \bibinfo
  {pages} {245115} (\bibinfo {year} {2017})}\BibitemShut {NoStop}%
\bibitem [{\citenamefont {Malekakhlagh}\ \emph {et~al.}(2017)\citenamefont
  {Malekakhlagh}, \citenamefont {Petrescu},\ and\ \citenamefont
  {T\"ureci}}]{Malekakhlagh:2017}%
  \BibitemOpen
  \bibfield  {author} {\bibinfo {author} {\bibfnamefont {M.}~\bibnamefont
  {Malekakhlagh}}, \bibinfo {author} {\bibfnamefont {A.}~\bibnamefont
  {Petrescu}},\ and\ \bibinfo {author} {\bibfnamefont {H.~E.}\ \bibnamefont
  {T\"ureci}},\ }\bibfield  {title} {\bibinfo {title} {Cutoff-free circuit
  quantum electrodynamics},\ }\href
  {https://doi.org/10.1103/PhysRevLett.119.073601} {\bibfield  {journal}
  {\bibinfo  {journal} {Physical Review Letters}\ }\textbf {\bibinfo {volume}
  {119}},\ \bibinfo {pages} {073601} (\bibinfo {year} {2017})}\BibitemShut
  {NoStop}%
\bibitem [{\citenamefont {Parra-Rodriguez}\ \emph {et~al.}(2018)\citenamefont
  {Parra-Rodriguez}, \citenamefont {Rico}, \citenamefont {Solano},\ and\
  \citenamefont {Egusquiza}}]{ParraRodriguez:2018}%
  \BibitemOpen
  \bibfield  {author} {\bibinfo {author} {\bibfnamefont {A.}~\bibnamefont
  {Parra-Rodriguez}}, \bibinfo {author} {\bibfnamefont {E.}~\bibnamefont
  {Rico}}, \bibinfo {author} {\bibfnamefont {E.}~\bibnamefont {Solano}},\ and\
  \bibinfo {author} {\bibfnamefont {I.~L.}\ \bibnamefont {Egusquiza}},\
  }\bibfield  {title} {\bibinfo {title} {Quantum networks in divergence-free
  circuit {QED}},\ }\href {https://doi.org/10.1088/2058-9565/aab1ba} {\bibfield
   {journal} {\bibinfo  {journal} {Quantum Science and Technology}\ }\textbf
  {\bibinfo {volume} {3}},\ \bibinfo {pages} {024012} (\bibinfo {year}
  {2018})}\BibitemShut {NoStop}%
\bibitem [{\citenamefont {Parra-Rodriguez}(2021)}]{ParraRodriguezPhD:2021}%
  \BibitemOpen
  \bibfield  {author} {\bibinfo {author} {\bibfnamefont {A.}~\bibnamefont
  {Parra-Rodriguez}},\ }\href {http://hdl.handle.net/10810/51132} {\emph
  {\bibinfo {title} {PhD Thesis: Canonical Quantization of Superconducting
  Circuits}}}\ (\bibinfo  {publisher} {Universidad del Pais Vasco},\ \bibinfo
  {address} {Leioa},\ \bibinfo {year} {2021})\BibitemShut {NoStop}%
\bibitem [{\citenamefont {Parra-Rodriguez}\ \emph {et~al.}(2019)\citenamefont
  {Parra-Rodriguez}, \citenamefont {Egusquiza}, \citenamefont {DiVincenzo},\
  and\ \citenamefont {Solano}}]{ParraRodriguez:2019}%
  \BibitemOpen
  \bibfield  {author} {\bibinfo {author} {\bibfnamefont {A.}~\bibnamefont
  {Parra-Rodriguez}}, \bibinfo {author} {\bibfnamefont {I.~L.}\ \bibnamefont
  {Egusquiza}}, \bibinfo {author} {\bibfnamefont {D.~P.}\ \bibnamefont
  {DiVincenzo}},\ and\ \bibinfo {author} {\bibfnamefont {E.}~\bibnamefont
  {Solano}},\ }\bibfield  {title} {\bibinfo {title} {Canonical circuit
  quantization with linear nonreciprocal devices},\ }\href
  {https://doi.org/10.1103/PhysRevB.99.014514} {\bibfield  {journal} {\bibinfo
  {journal} {Physical Review B}\ }\textbf {\bibinfo {volume} {99}},\ \bibinfo
  {pages} {014514} (\bibinfo {year} {2019})}\BibitemShut {NoStop}%
\bibitem [{\citenamefont {Egusquiza}\ and\ \citenamefont
  {Parra-Rodriguez}(2022)}]{Egusquiza:2022}%
  \BibitemOpen
  \bibfield  {author} {\bibinfo {author} {\bibfnamefont {I.~L.}\ \bibnamefont
  {Egusquiza}}\ and\ \bibinfo {author} {\bibfnamefont {A.}~\bibnamefont
  {Parra-Rodriguez}},\ }\bibfield  {title} {\bibinfo {title} {Algebraic
  canonical quantization of lumped superconducting networks},\ }\href
  {https://doi.org/10.1103/PhysRevB.106.024510} {\bibfield  {journal} {\bibinfo
   {journal} {Physical Review B}\ }\textbf {\bibinfo {volume} {106}},\ \bibinfo
  {pages} {024510} (\bibinfo {year} {2022})}\BibitemShut {NoStop}%
\bibitem [{\citenamefont {Caldeira}\ and\ \citenamefont
  {Leggett}(1981)}]{CaldeiraLeggett:1981}%
  \BibitemOpen
  \bibfield  {author} {\bibinfo {author} {\bibfnamefont {A.~O.}\ \bibnamefont
  {Caldeira}}\ and\ \bibinfo {author} {\bibfnamefont {A.~J.}\ \bibnamefont
  {Leggett}},\ }\bibfield  {title} {\bibinfo {title} {Influence of dissipation
  on quantum tunneling in macroscopic systems},\ }\href
  {https://doi.org/10.1103/PhysRevLett.46.211} {\bibfield  {journal} {\bibinfo
  {journal} {Physical Review Letters}\ }\textbf {\bibinfo {volume} {46}},\
  \bibinfo {pages} {211} (\bibinfo {year} {1981})}\BibitemShut {NoStop}%
\bibitem [{\citenamefont {Leggett}(1984)}]{Leggett:1984}%
  \BibitemOpen
  \bibfield  {author} {\bibinfo {author} {\bibfnamefont {A.~J.}\ \bibnamefont
  {Leggett}},\ }\bibfield  {title} {\bibinfo {title} {{Quantum tunneling in the
  presence of an arbitrary linear dissipation mechanism}},\ }\href
  {https://doi.org/10.1103/PhysRevB.30.1208} {\bibfield  {journal} {\bibinfo
  {journal} {Physical Review B}\ }\textbf {\bibinfo {volume} {30}},\ \bibinfo
  {pages} {1208} (\bibinfo {year} {1984})}\BibitemShut {NoStop}%
\bibitem [{Note1()}]{Note1}%
  \BibitemOpen
  \bibinfo {note} {This correction is accurate provided the qubit potential is
  still accurately described by a single-well.}\BibitemShut {Stop}%
\bibitem [{\citenamefont {Correa}\ and\ \citenamefont
  {Glatthard}(2023)}]{Correa:2023}%
  \BibitemOpen
  \bibfield  {author} {\bibinfo {author} {\bibfnamefont {L.~A.}\ \bibnamefont
  {Correa}}\ and\ \bibinfo {author} {\bibfnamefont {J.}~\bibnamefont
  {Glatthard}},\ }\bibfield  {title} {\bibinfo {title} {Potential
  renormalisation, {Lamb} shift and mean-force {Gibbs} state -- to shift or not
  to shift?},\ }\Eprint {https://arxiv.org/abs/2305.08941} {arXiv:2305.08941
  [quant-ph]}  (\bibinfo {year} {2023})\BibitemShut {NoStop}%
\bibitem [{Note2()}]{Note2}%
  \BibitemOpen
  \bibinfo {note} {Here, we assume that the inner circuit mode is always on its
  ground state. As a result, the cross-Kerr Hamiltonian $\protect \hat {H}_\chi
  $ is omitted in the master equation.}\BibitemShut {Stop}%
\bibitem [{\citenamefont {Metelmann}\ and\ \citenamefont
  {Clerk}(2015)}]{Metelmann:2015}%
  \BibitemOpen
  \bibfield  {author} {\bibinfo {author} {\bibfnamefont {A.}~\bibnamefont
  {Metelmann}}\ and\ \bibinfo {author} {\bibfnamefont {A.~A.}\ \bibnamefont
  {Clerk}},\ }\bibfield  {title} {\bibinfo {title} {Nonreciprocal photon
  transmission and amplification via reservoir engineering},\ }\href
  {https://doi.org/10.1103/PhysRevX.5.021025} {\bibfield  {journal} {\bibinfo
  {journal} {Physical Review X}\ }\textbf {\bibinfo {volume} {5}},\ \bibinfo
  {pages} {021025} (\bibinfo {year} {2015})}\BibitemShut {NoStop}%
\bibitem [{\citenamefont {Clerk}(2022)}]{Aash:2022}%
  \BibitemOpen
  \bibfield  {author} {\bibinfo {author} {\bibfnamefont {A.~A.}\ \bibnamefont
  {Clerk}},\ }\bibfield  {title} {\bibinfo {title} {{Introduction to quantum
  non-reciprocal interactions: from non-Hermitian Hamiltonians to quantum
  master equations and quantum feedforward schemes}},\ }\href
  {https://doi.org/10.21468/SciPostPhysLectNotes.44} {\bibfield  {journal}
  {\bibinfo  {journal} {SciPost Phys. Lect. Notes}\ }\textbf {\bibinfo {volume}
  {44}},\ \bibinfo {pages} {9} (\bibinfo {year} {2022})}\BibitemShut {NoStop}%
\bibitem [{\citenamefont {Miano}\ \emph {et~al.}(2023)\citenamefont {Miano},
  \citenamefont {Joshi}, \citenamefont {Liu}, \citenamefont {Dai},
  \citenamefont {Parakh}, \citenamefont {Frunzio},\ and\ \citenamefont
  {Devoret}}]{Miano:2023}%
  \BibitemOpen
  \bibfield  {author} {\bibinfo {author} {\bibfnamefont {A.}~\bibnamefont
  {Miano}}, \bibinfo {author} {\bibfnamefont {V.}~\bibnamefont {Joshi}},
  \bibinfo {author} {\bibfnamefont {G.}~\bibnamefont {Liu}}, \bibinfo {author}
  {\bibfnamefont {W.}~\bibnamefont {Dai}}, \bibinfo {author} {\bibfnamefont
  {P.}~\bibnamefont {Parakh}}, \bibinfo {author} {\bibfnamefont
  {L.}~\bibnamefont {Frunzio}},\ and\ \bibinfo {author} {\bibfnamefont
  {M.}~\bibnamefont {Devoret}},\ }\bibfield  {title} {\bibinfo {title}
  {Hamiltonian extrema of an arbitrary flux-biased josephson circuit},\ }\href
  {https://doi.org/10.1103/PRXQuantum.4.030324} {\bibfield  {journal} {\bibinfo
   {journal} {PRX Quantum}\ }\textbf {\bibinfo {volume} {4}},\ \bibinfo {pages}
  {030324} (\bibinfo {year} {2023})}\BibitemShut {NoStop}%
\bibitem [{\citenamefont {Willsch}\ \emph {et~al.}(2024)\citenamefont
  {Willsch}, \citenamefont {Rieger}, \citenamefont {Winkel}, \citenamefont
  {Willsch}, \citenamefont {Dickel}, \citenamefont {Krause}, \citenamefont
  {Ando}, \citenamefont {Lescanne}, \citenamefont {Leghtas}, \citenamefont
  {Bronn} \emph {et~al.}}]{Willsch:2024}%
  \BibitemOpen
  \bibfield  {author} {\bibinfo {author} {\bibfnamefont {D.}~\bibnamefont
  {Willsch}}, \bibinfo {author} {\bibfnamefont {D.}~\bibnamefont {Rieger}},
  \bibinfo {author} {\bibfnamefont {P.}~\bibnamefont {Winkel}}, \bibinfo
  {author} {\bibfnamefont {M.}~\bibnamefont {Willsch}}, \bibinfo {author}
  {\bibfnamefont {C.}~\bibnamefont {Dickel}}, \bibinfo {author} {\bibfnamefont
  {J.}~\bibnamefont {Krause}}, \bibinfo {author} {\bibfnamefont
  {Y.}~\bibnamefont {Ando}}, \bibinfo {author} {\bibfnamefont {R.}~\bibnamefont
  {Lescanne}}, \bibinfo {author} {\bibfnamefont {Z.}~\bibnamefont {Leghtas}},
  \bibinfo {author} {\bibfnamefont {N.~T.}\ \bibnamefont {Bronn}}, \emph
  {et~al.},\ }\bibfield  {title} {\bibinfo {title} {Observation of josephson
  harmonics in tunnel junctions},\ }\href@noop {} {\bibfield  {journal}
  {\bibinfo  {journal} {Nature Physics}\ ,\ \bibinfo {pages} {1}} (\bibinfo
  {year} {2024})}\BibitemShut {NoStop}%
\bibitem [{\citenamefont {Mooij}\ and\ \citenamefont
  {Harmans}(2005)}]{Mooij1:2005}%
  \BibitemOpen
  \bibfield  {author} {\bibinfo {author} {\bibfnamefont {J.~E.}\ \bibnamefont
  {Mooij}}\ and\ \bibinfo {author} {\bibfnamefont {C.~J. P.~M.}\ \bibnamefont
  {Harmans}},\ }\bibfield  {title} {\bibinfo {title} {Phase-slip flux qubits},\
  }\href {https://doi.org/10.1088/1367-2630/7/1/219} {\bibfield  {journal}
  {\bibinfo  {journal} {New Journal of Physics}\ }\textbf {\bibinfo {volume}
  {7}},\ \bibinfo {pages} {219} (\bibinfo {year} {2005})}\BibitemShut {NoStop}%
\bibitem [{\citenamefont {Mooij}\ and\ \citenamefont
  {Nazarov}(2006)}]{Mooij:2006}%
  \BibitemOpen
  \bibfield  {author} {\bibinfo {author} {\bibfnamefont {J.}~\bibnamefont
  {Mooij}}\ and\ \bibinfo {author} {\bibfnamefont {Y.}~\bibnamefont
  {Nazarov}},\ }\bibfield  {title} {\bibinfo {title} {{Superconducting
  nanowires as quantum phase-slip junctions}},\ }\href
  {https://doi.org/10.1038/nphys234} {\bibfield  {journal} {\bibinfo  {journal}
  {Nature Physics}\ }\textbf {\bibinfo {volume} {2}},\ \bibinfo {pages} {169}
  (\bibinfo {year} {2006})}\BibitemShut {NoStop}%
\bibitem [{\citenamefont {Arutyunov}\ \emph {et~al.}(2008)\citenamefont
  {Arutyunov}, \citenamefont {Golubev},\ and\ \citenamefont
  {Zaikin}}]{Arutyunov:2008}%
  \BibitemOpen
  \bibfield  {author} {\bibinfo {author} {\bibfnamefont {K.}~\bibnamefont
  {Arutyunov}}, \bibinfo {author} {\bibfnamefont {D.}~\bibnamefont {Golubev}},\
  and\ \bibinfo {author} {\bibfnamefont {A.}~\bibnamefont {Zaikin}},\
  }\bibfield  {title} {\bibinfo {title} {{Superconductivity in one
  dimension}},\ }\href {https://doi.org/10.1016/j.physrep.2008.04.009}
  {\bibfield  {journal} {\bibinfo  {journal} {Physics Reports}\ }\textbf
  {\bibinfo {volume} {464}},\ \bibinfo {pages} {1} (\bibinfo {year}
  {2008})}\BibitemShut {NoStop}%
\bibitem [{\citenamefont {Smith}\ \emph {et~al.}(2016)\citenamefont {Smith},
  \citenamefont {Kou}, \citenamefont {Vool}, \citenamefont {Pop}, \citenamefont
  {Frunzio}, \citenamefont {Schoelkopf},\ and\ \citenamefont
  {Devoret}}]{Smith:2016}%
  \BibitemOpen
  \bibfield  {author} {\bibinfo {author} {\bibfnamefont {W.~C.}\ \bibnamefont
  {Smith}}, \bibinfo {author} {\bibfnamefont {A.}~\bibnamefont {Kou}}, \bibinfo
  {author} {\bibfnamefont {U.}~\bibnamefont {Vool}}, \bibinfo {author}
  {\bibfnamefont {I.~M.}\ \bibnamefont {Pop}}, \bibinfo {author} {\bibfnamefont
  {L.}~\bibnamefont {Frunzio}}, \bibinfo {author} {\bibfnamefont {R.~J.}\
  \bibnamefont {Schoelkopf}},\ and\ \bibinfo {author} {\bibfnamefont {M.~H.}\
  \bibnamefont {Devoret}},\ }\bibfield  {title} {\bibinfo {title} {Quantization
  of inductively shunted superconducting circuits},\ }\href
  {https://doi.org/10.1103/PhysRevB.94.144507} {\bibfield  {journal} {\bibinfo
  {journal} {Physical Review B}\ }\textbf {\bibinfo {volume} {94}},\ \bibinfo
  {pages} {144507} (\bibinfo {year} {2016})}\BibitemShut {NoStop}%
\bibitem [{\citenamefont {Roushan}\ \emph {et~al.}(2017)\citenamefont
  {Roushan}, \citenamefont {Neill}, \citenamefont {Megrant}, \citenamefont
  {Chen}, \citenamefont {Babbush}, \citenamefont {Barends}, \citenamefont
  {Campbell}, \citenamefont {Chen}, \citenamefont {Chiaro}, \citenamefont
  {Dunsworth}, \citenamefont {Fowler}, \citenamefont {Jeffrey}, \citenamefont
  {Kelly}, \citenamefont {Lucero}, \citenamefont {Mutus}, \citenamefont
  {O’Malley}, \citenamefont {Neeley}, \citenamefont {Quintana}, \citenamefont
  {Sank}, \citenamefont {Vainsencher}, \citenamefont {Wenner}, \citenamefont
  {White}, \citenamefont {Kapit}, \citenamefont {Neven},\ and\ \citenamefont
  {Martinis}}]{Roushan:2017}%
  \BibitemOpen
  \bibfield  {author} {\bibinfo {author} {\bibfnamefont {P.}~\bibnamefont
  {Roushan}}, \bibinfo {author} {\bibfnamefont {C.}~\bibnamefont {Neill}},
  \bibinfo {author} {\bibfnamefont {A.}~\bibnamefont {Megrant}}, \bibinfo
  {author} {\bibfnamefont {Y.}~\bibnamefont {Chen}}, \bibinfo {author}
  {\bibfnamefont {R.}~\bibnamefont {Babbush}}, \bibinfo {author} {\bibfnamefont
  {R.}~\bibnamefont {Barends}}, \bibinfo {author} {\bibfnamefont
  {B.}~\bibnamefont {Campbell}}, \bibinfo {author} {\bibfnamefont
  {Z.}~\bibnamefont {Chen}}, \bibinfo {author} {\bibfnamefont {B.}~\bibnamefont
  {Chiaro}}, \bibinfo {author} {\bibfnamefont {A.}~\bibnamefont {Dunsworth}},
  \bibinfo {author} {\bibfnamefont {A.}~\bibnamefont {Fowler}}, \bibinfo
  {author} {\bibfnamefont {E.}~\bibnamefont {Jeffrey}}, \bibinfo {author}
  {\bibfnamefont {J.}~\bibnamefont {Kelly}}, \bibinfo {author} {\bibfnamefont
  {E.}~\bibnamefont {Lucero}}, \bibinfo {author} {\bibfnamefont
  {J.}~\bibnamefont {Mutus}}, \bibinfo {author} {\bibfnamefont {P.~J.~J.}\
  \bibnamefont {O’Malley}}, \bibinfo {author} {\bibfnamefont
  {M.}~\bibnamefont {Neeley}}, \bibinfo {author} {\bibfnamefont
  {C.}~\bibnamefont {Quintana}}, \bibinfo {author} {\bibfnamefont
  {D.}~\bibnamefont {Sank}}, \bibinfo {author} {\bibfnamefont {A.}~\bibnamefont
  {Vainsencher}}, \bibinfo {author} {\bibfnamefont {J.}~\bibnamefont {Wenner}},
  \bibinfo {author} {\bibfnamefont {T.}~\bibnamefont {White}}, \bibinfo
  {author} {\bibfnamefont {E.}~\bibnamefont {Kapit}}, \bibinfo {author}
  {\bibfnamefont {H.}~\bibnamefont {Neven}},\ and\ \bibinfo {author}
  {\bibfnamefont {J.}~\bibnamefont {Martinis}},\ }\bibfield  {title} {\bibinfo
  {title} {Chiral ground-state currents of interacting photons in a synthetic
  magnetic field},\ }\href {https://doi.org/10.1038/nphys3930} {\bibfield
  {journal} {\bibinfo  {journal} {Nature Physics}\ }\textbf {\bibinfo {volume}
  {13}},\ \bibinfo {pages} {146} (\bibinfo {year} {2017})}\BibitemShut
  {NoStop}%
\bibitem [{\citenamefont {Wang}\ \emph
  {et~al.}(2023{\natexlab{a}})\citenamefont {Wang}, \citenamefont {Wang},
  \citenamefont {van Geldern}, \citenamefont {Connolly}, \citenamefont
  {Clerk},\ and\ \citenamefont {Wang}}]{WangYingYing:2023}%
  \BibitemOpen
  \bibfield  {author} {\bibinfo {author} {\bibfnamefont {Y.-Y.}\ \bibnamefont
  {Wang}}, \bibinfo {author} {\bibfnamefont {Y.-X.}\ \bibnamefont {Wang}},
  \bibinfo {author} {\bibfnamefont {S.}~\bibnamefont {van Geldern}}, \bibinfo
  {author} {\bibfnamefont {T.}~\bibnamefont {Connolly}}, \bibinfo {author}
  {\bibfnamefont {A.~A.}\ \bibnamefont {Clerk}},\ and\ \bibinfo {author}
  {\bibfnamefont {C.}~\bibnamefont {Wang}},\ }\bibfield  {title} {\bibinfo
  {title} {Dispersive non-reciprocity between a qubit and a cavity},\ }\Eprint
  {https://arxiv.org/abs/2307.05298} {arXiv:2307.05298 [quant-ph]}  (\bibinfo
  {year} {2023}{\natexlab{a}})\BibitemShut {NoStop}%
\bibitem [{\citenamefont {Wang}\ \emph
  {et~al.}(2023{\natexlab{b}})\citenamefont {Wang}, \citenamefont {Wang},\ and\
  \citenamefont {Clerk}}]{WangYuXin:2023}%
  \BibitemOpen
  \bibfield  {author} {\bibinfo {author} {\bibfnamefont {Y.-X.}\ \bibnamefont
  {Wang}}, \bibinfo {author} {\bibfnamefont {C.}~\bibnamefont {Wang}},\ and\
  \bibinfo {author} {\bibfnamefont {A.~A.}\ \bibnamefont {Clerk}},\ }\bibfield
  {title} {\bibinfo {title} {Quantum nonreciprocal interactions via dissipative
  gauge symmetry},\ }\href {https://doi.org/10.1103/PRXQuantum.4.010306}
  {\bibfield  {journal} {\bibinfo  {journal} {PRX Quantum}\ }\textbf {\bibinfo
  {volume} {4}},\ \bibinfo {pages} {010306} (\bibinfo {year}
  {2023}{\natexlab{b}})}\BibitemShut {NoStop}%
\bibitem [{\citenamefont {Navarathna}\ \emph {et~al.}(2023)\citenamefont
  {Navarathna}, \citenamefont {Le}, \citenamefont {Hamann}, \citenamefont
  {Nguyen}, \citenamefont {Stace},\ and\ \citenamefont
  {Fedorov}}]{Navarathna:2023}%
  \BibitemOpen
  \bibfield  {author} {\bibinfo {author} {\bibfnamefont {R.}~\bibnamefont
  {Navarathna}}, \bibinfo {author} {\bibfnamefont {D.~T.}\ \bibnamefont {Le}},
  \bibinfo {author} {\bibfnamefont {A.~R.}\ \bibnamefont {Hamann}}, \bibinfo
  {author} {\bibfnamefont {H.~D.}\ \bibnamefont {Nguyen}}, \bibinfo {author}
  {\bibfnamefont {T.~M.}\ \bibnamefont {Stace}},\ and\ \bibinfo {author}
  {\bibfnamefont {A.}~\bibnamefont {Fedorov}},\ }\bibfield  {title} {\bibinfo
  {title} {Passive superconducting circulator on a chip},\ }\href
  {https://doi.org/10.1103/PhysRevLett.130.037001} {\bibfield  {journal}
  {\bibinfo  {journal} {Physical Review Letters}\ }\textbf {\bibinfo {volume}
  {130}},\ \bibinfo {pages} {037001} (\bibinfo {year} {2023})}\BibitemShut
  {NoStop}%
\bibitem [{\citenamefont {Ozawa}\ \emph {et~al.}(2019)\citenamefont {Ozawa},
  \citenamefont {Price}, \citenamefont {Amo}, \citenamefont {Goldman},
  \citenamefont {Hafezi}, \citenamefont {Lu}, \citenamefont {Rechtsman},
  \citenamefont {Schuster}, \citenamefont {Simon}, \citenamefont {Zilberberg},\
  and\ \citenamefont {Carusotto}}]{Ozawa:2019}%
  \BibitemOpen
  \bibfield  {author} {\bibinfo {author} {\bibfnamefont {T.}~\bibnamefont
  {Ozawa}}, \bibinfo {author} {\bibfnamefont {H.~M.}\ \bibnamefont {Price}},
  \bibinfo {author} {\bibfnamefont {A.}~\bibnamefont {Amo}}, \bibinfo {author}
  {\bibfnamefont {N.}~\bibnamefont {Goldman}}, \bibinfo {author} {\bibfnamefont
  {M.}~\bibnamefont {Hafezi}}, \bibinfo {author} {\bibfnamefont
  {L.}~\bibnamefont {Lu}}, \bibinfo {author} {\bibfnamefont {M.~C.}\
  \bibnamefont {Rechtsman}}, \bibinfo {author} {\bibfnamefont {D.}~\bibnamefont
  {Schuster}}, \bibinfo {author} {\bibfnamefont {J.}~\bibnamefont {Simon}},
  \bibinfo {author} {\bibfnamefont {O.}~\bibnamefont {Zilberberg}},\ and\
  \bibinfo {author} {\bibfnamefont {I.}~\bibnamefont {Carusotto}},\ }\bibfield
  {title} {\bibinfo {title} {Topological photonics},\ }\href
  {https://doi.org/10.1103/RevModPhys.91.015006} {\bibfield  {journal}
  {\bibinfo  {journal} {Reviews Modern Physiscs}\ }\textbf {\bibinfo {volume}
  {91}},\ \bibinfo {pages} {015006} (\bibinfo {year} {2019})}\BibitemShut
  {NoStop}%
\bibitem [{\citenamefont {Solgun}\ and\ \citenamefont
  {Srinivasan}(2022)}]{Solgun:2022}%
  \BibitemOpen
  \bibfield  {author} {\bibinfo {author} {\bibfnamefont {F.}~\bibnamefont
  {Solgun}}\ and\ \bibinfo {author} {\bibfnamefont {S.}~\bibnamefont
  {Srinivasan}},\ }\bibfield  {title} {\bibinfo {title} {Direct calculation of
  $zz$ interaction rates in multimode circuit quantum electrodynamics},\ }\href
  {https://doi.org/10.1103/PhysRevApplied.18.044025} {\bibfield  {journal}
  {\bibinfo  {journal} {Physical Review Applied}\ }\textbf {\bibinfo {volume}
  {18}},\ \bibinfo {pages} {044025} (\bibinfo {year} {2022})}\BibitemShut
  {NoStop}%
\bibitem [{\citenamefont {Labarca}\ \emph {et~al.}(2024)\citenamefont
  {Labarca}, \citenamefont {Benhayoune-Khadraoui}, \citenamefont {Blais},\ and\
  \citenamefont {Parra-Rodriguez}}]{Labarca_Toolbox_for_nonreciprocal_2024}%
  \BibitemOpen
  \bibfield  {author} {\bibinfo {author} {\bibfnamefont {L.}~\bibnamefont
  {Labarca}}, \bibinfo {author} {\bibfnamefont {O.}~\bibnamefont
  {Benhayoune-Khadraoui}}, \bibinfo {author} {\bibfnamefont {A.}~\bibnamefont
  {Blais}},\ and\ \bibinfo {author} {\bibfnamefont {A.}~\bibnamefont
  {Parra-Rodriguez}},\ }\href
  {https://github.com/LautaroLabarcaG/Toolbox-for-Nonreciprocal-Dispersive-models-in-cQED}
  {\bibinfo {title} {{Toolbox for nonreciprocal dispersive models in cQED
  simulation code}}} (\bibinfo {year} {2024})\BibitemShut {NoStop}%
\bibitem [{\citenamefont {Ciani}\ \emph {et~al.}(2023)\citenamefont {Ciani},
  \citenamefont {DiVincenzo},\ and\ \citenamefont {Terhal}}]{Ciani:2023}%
  \BibitemOpen
  \bibfield  {author} {\bibinfo {author} {\bibfnamefont {A.}~\bibnamefont
  {Ciani}}, \bibinfo {author} {\bibfnamefont {D.~P.}\ \bibnamefont
  {DiVincenzo}},\ and\ \bibinfo {author} {\bibfnamefont {B.~M.}\ \bibnamefont
  {Terhal}},\ }\bibfield  {title} {\bibinfo {title} {Lecture {Notes} on
  {Quantum} {Electrical} {Circuits}},\ }\Eprint
  {https://arxiv.org/abs/arXiv:2312.05329} {arXiv:2312.05329 [quant-ph]}
  (\bibinfo {year} {2023})\BibitemShut {NoStop}%
\bibitem [{\citenamefont {Mooij}\ and\ \citenamefont
  {Nazarov}(2005)}]{Mooij2:2005}%
  \BibitemOpen
  \bibfield  {author} {\bibinfo {author} {\bibfnamefont {J.~E.}\ \bibnamefont
  {Mooij}}\ and\ \bibinfo {author} {\bibfnamefont {Y.~V.}\ \bibnamefont
  {Nazarov}},\ }\bibfield  {title} {\bibinfo {title} {Quantum phase slip
  junctions},\ }\Eprint {https://arxiv.org/abs/cond-mat/0511535}
  {cond-mat/0511535 [cond-mat]}  (\bibinfo {year} {2005})\BibitemShut {NoStop}%
\bibitem [{\citenamefont {Bravyi}\ \emph {et~al.}(2011)\citenamefont {Bravyi},
  \citenamefont {DiVincenzo},\ and\ \citenamefont {Loss}}]{Bravyi:2011}%
  \BibitemOpen
  \bibfield  {author} {\bibinfo {author} {\bibfnamefont {S.}~\bibnamefont
  {Bravyi}}, \bibinfo {author} {\bibfnamefont {D.}~\bibnamefont {DiVincenzo}},\
  and\ \bibinfo {author} {\bibfnamefont {D.}~\bibnamefont {Loss}},\ }\bibfield
  {title} {\bibinfo {title} {Schrieffer-wolff transformation for quantum
  many-body systems},\ }\href {https://doi.org/10.1016/j.aop.2011.06.004}
  {\bibfield  {journal} {\bibinfo  {journal} {Annals of Physics}\ }\textbf
  {\bibinfo {volume} {326}},\ \bibinfo {pages} {2793} (\bibinfo {year}
  {2011})}\BibitemShut {NoStop}%
\bibitem [{\citenamefont {Goldstein}(2011)}]{Goldstein:1950}%
  \BibitemOpen
  \bibfield  {author} {\bibinfo {author} {\bibfnamefont {H.}~\bibnamefont
  {Goldstein}},\ }\href@noop {} {\emph {\bibinfo {title} {Classical
  mechanics}}}\ (\bibinfo  {publisher} {Pearson Education India},\ \bibinfo
  {year} {2011})\BibitemShut {NoStop}%
\bibitem [{\citenamefont {Winkler}(2003)}]{Winkler:2003}%
  \BibitemOpen
  \bibfield  {author} {\bibinfo {author} {\bibfnamefont {R.}~\bibnamefont
  {Winkler}},\ }\bibfield  {title} {\bibinfo {title} {Quasi-{Degenerate}
  {Perturbation} {Theory}},\ }in\ \href
  {https://doi.org/10.1007/978-3-540-36616-4_12} {\emph {\bibinfo {booktitle}
  {Spin—{Orbit} {Coupling} {Effects} in {Two}-{Dimensional} {Electron} and
  {Hole} {Systems}}}},\ \bibinfo {series and number} {Springer {Tracts} in
  {Modern} {Physics}},\ \bibinfo {editor} {edited by\ \bibinfo {editor}
  {\bibfnamefont {R.}~\bibnamefont {Winkler}}}\ (\bibinfo  {publisher}
  {Springer},\ \bibinfo {address} {Berlin, Heidelberg},\ \bibinfo {year}
  {2003})\ pp.\ \bibinfo {pages} {201--205}\BibitemShut {NoStop}%
\bibitem [{\citenamefont {Burkard}\ \emph {et~al.}(2004)\citenamefont
  {Burkard}, \citenamefont {Koch},\ and\ \citenamefont
  {DiVincenzo}}]{Burkard:2004}%
  \BibitemOpen
  \bibfield  {author} {\bibinfo {author} {\bibfnamefont {G.}~\bibnamefont
  {Burkard}}, \bibinfo {author} {\bibfnamefont {R.~H.}\ \bibnamefont {Koch}},\
  and\ \bibinfo {author} {\bibfnamefont {D.~P.}\ \bibnamefont {DiVincenzo}},\
  }\bibfield  {title} {\bibinfo {title} {Multilevel quantum description of
  decoherence in superconducting qubits},\ }\href
  {https://doi.org/10.1103/PhysRevB.69.064503} {\bibfield  {journal} {\bibinfo
  {journal} {Physical Review B}\ }\textbf {\bibinfo {volume} {69}},\ \bibinfo
  {pages} {064503} (\bibinfo {year} {2004})}\BibitemShut {NoStop}%
\bibitem [{\citenamefont {Burkard}(2005)}]{Burkard:2005}%
  \BibitemOpen
  \bibfield  {author} {\bibinfo {author} {\bibfnamefont {G.}~\bibnamefont
  {Burkard}},\ }\bibfield  {title} {\bibinfo {title} {Circuit theory for
  decoherence in superconducting charge qubits},\ }\href
  {https://doi.org/10.1103/PhysRevB.71.144511} {\bibfield  {journal} {\bibinfo
  {journal} {Physical Review B}\ }\textbf {\bibinfo {volume} {71}},\ \bibinfo
  {pages} {144511} (\bibinfo {year} {2005})}\BibitemShut {NoStop}%
\bibitem [{\citenamefont {Burkard}\ and\ \citenamefont
  {Brito}(2005)}]{BB:2005}%
  \BibitemOpen
  \bibfield  {author} {\bibinfo {author} {\bibfnamefont {G.}~\bibnamefont
  {Burkard}}\ and\ \bibinfo {author} {\bibfnamefont {F.}~\bibnamefont
  {Brito}},\ }\bibfield  {title} {\bibinfo {title} {Nonadditivity of
  decoherence rates in superconducting qubits},\ }\href
  {https://doi.org/10.1103/PhysRevB.72.054528} {\bibfield  {journal} {\bibinfo
  {journal} {Phys. Rev. B}\ }\textbf {\bibinfo {volume} {72}},\ \bibinfo
  {pages} {054528} (\bibinfo {year} {2005})}\BibitemShut {NoStop}%
\bibitem [{Note3()}]{Note3}%
  \BibitemOpen
  \bibinfo {note} {Note that if the Fourier transform is defined with opposite
  phase $\protect \mathsf {M}_D(\omega )=\DOTSI \intop \ilimits@ _{-\infty
  }^{+\infty }dt\protect \mathsf {M}_D(t)e^{+i\omega t}$, then the poles of
  $\protect \mathsf {M}_D(\omega )$ should live in the lower-half complex plane
  $\protect \mathrm {Im}(z)<0$, to ensure causality}\BibitemShut {NoStop}%
\bibitem [{\citenamefont {Vool}\ and\ \citenamefont
  {Devoret}(2017)}]{Vool:2017}%
  \BibitemOpen
  \bibfield  {author} {\bibinfo {author} {\bibfnamefont {U.}~\bibnamefont
  {Vool}}\ and\ \bibinfo {author} {\bibfnamefont {M.}~\bibnamefont {Devoret}},\
  }\bibfield  {title} {\bibinfo {title} {Introduction to quantum
  electromagnetic circuits},\ }\href
  {https://doi.org/https://doi.org/10.1002/cta.2359} {\bibfield  {journal}
  {\bibinfo  {journal} {International Journal of Circuit Theory and
  Applications}\ }\textbf {\bibinfo {volume} {45}},\ \bibinfo {pages} {897}
  (\bibinfo {year} {2017})}\BibitemShut {NoStop}%
\bibitem [{Note4()}]{Note4}%
  \BibitemOpen
  \bibinfo {note} {The limit should be understood in the sense of distributions
  $\DOTSI \intop \ilimits@ d\omega \protect \mathsf {M}_D(\omega )f(\omega
  )=\lim _{\epsilon \to 0^{+}}\DOTSI \intop \ilimits@ d\omega \protect \mathsf
  {M}_D(\omega -i\epsilon )f(\omega )$ with $f$ any smooth (Schwartz)
  function}\BibitemShut {NoStop}%
\bibitem [{\citenamefont {Parra~Rodriguez}\ and\ \citenamefont
  {Egusquiza}(2023)}]{ParraRodriguez:2023}%
  \BibitemOpen
  \bibfield  {author} {\bibinfo {author} {\bibfnamefont {A.}~\bibnamefont
  {Parra~Rodriguez}}\ and\ \bibinfo {author} {\bibfnamefont {I.~L.}\
  \bibnamefont {Egusquiza}},\ }\bibfield  {title} {\bibinfo {title}
  {Geometrical description and {Faddeev-Jackiw} quantization of electrical
  networks},\ }\Eprint {https://arxiv.org/abs/arXiv:2304.12252}
  {arXiv:2304.12252 [quant-ph]}  (\bibinfo {year} {2023})\BibitemShut {NoStop}%
\bibitem [{\citenamefont {Parra~Rodriguez}\ and\ \citenamefont
  {Egusquiza}(2024)}]{ParraRodriguez:2024}%
  \BibitemOpen
  \bibfield  {author} {\bibinfo {author} {\bibfnamefont {A.}~\bibnamefont
  {Parra~Rodriguez}}\ and\ \bibinfo {author} {\bibfnamefont {I.~L.}\
  \bibnamefont {Egusquiza}},\ }\bibfield  {title} {\bibinfo {title}
  {{Faddeev-Jackiw} quantisation of nonreciprocal quasi-lumped electrical
  networks},\ }\Eprint {https://arxiv.org/abs/arXiv:2401.09120}
  {arXiv:2401.09120 [quant-ph]}  (\bibinfo {year} {2024})\BibitemShut {NoStop}%
\bibitem [{\citenamefont {Lu}\ and\ \citenamefont {Shiou}(2002)}]{Lu:2002}%
  \BibitemOpen
  \bibfield  {author} {\bibinfo {author} {\bibfnamefont {T.-T.}\ \bibnamefont
  {Lu}}\ and\ \bibinfo {author} {\bibfnamefont {S.-H.}\ \bibnamefont {Shiou}},\
  }\bibfield  {title} {\bibinfo {title} {Inverses of 2$\times$ 2 block
  matrices},\ }\href {https://doi.org/10.1016/S0898-1221(01)00278-4} {\bibfield
   {journal} {\bibinfo  {journal} {Computers \& Mathematics with Applications}\
  }\textbf {\bibinfo {volume} {43}},\ \bibinfo {pages} {119} (\bibinfo {year}
  {2002})}\BibitemShut {NoStop}%
\bibitem [{\citenamefont {Morales}\ and\ \citenamefont
  {Flores-Riveros}(1989)}]{Morales:1989}%
  \BibitemOpen
  \bibfield  {author} {\bibinfo {author} {\bibfnamefont {J.}~\bibnamefont
  {Morales}}\ and\ \bibinfo {author} {\bibfnamefont {A.}~\bibnamefont
  {Flores-Riveros}},\ }\bibfield  {title} {\bibinfo {title} {The generalization
  of the binomial theorem},\ }\href@noop {} {\bibfield  {journal} {\bibinfo
  {journal} {Journal of mathematical physics}\ }\textbf {\bibinfo {volume}
  {30}},\ \bibinfo {pages} {393} (\bibinfo {year} {1989})}\BibitemShut
  {NoStop}%
\bibitem [{\citenamefont {Davies}(1974)}]{Davies:1974}%
  \BibitemOpen
  \bibfield  {author} {\bibinfo {author} {\bibfnamefont {E.~B.}\ \bibnamefont
  {Davies}},\ }\bibfield  {title} {\bibinfo {title} {Markovian master
  equations},\ }\href {https://doi.org/10.1007/BF01608389} {\bibfield
  {journal} {\bibinfo  {journal} {Communications in Mathematical Physics}\
  }\textbf {\bibinfo {volume} {39}},\ \bibinfo {pages} {91} (\bibinfo {year}
  {1974})}\BibitemShut {NoStop}%
\bibitem [{\citenamefont {Breuer}\ and\ \citenamefont
  {Petruccione}(2002)}]{BRE:2002}%
  \BibitemOpen
  \bibfield  {author} {\bibinfo {author} {\bibfnamefont {H.~P.}\ \bibnamefont
  {Breuer}}\ and\ \bibinfo {author} {\bibfnamefont {F.}~\bibnamefont
  {Petruccione}},\ }\href
  {https://doi.org/10.1093/acprof:oso/9780199213900.001.0001} {\emph {\bibinfo
  {title} {The theory of open quantum systems}}}\ (\bibinfo  {publisher}
  {Oxford University Press},\ \bibinfo {address} {Great Clarendon Street},\
  \bibinfo {year} {2002})\BibitemShut {NoStop}%
\bibitem [{\citenamefont {Alicki}\ and\ \citenamefont
  {Lendi}(2007)}]{Alicki:2007}%
  \BibitemOpen
  \bibfield  {author} {\bibinfo {author} {\bibfnamefont {R.}~\bibnamefont
  {Alicki}}\ and\ \bibinfo {author} {\bibfnamefont {K.}~\bibnamefont {Lendi}},\
  }\href {https://doi.org/https://doi.org/10.1007/3-540-70861-8} {\emph
  {\bibinfo {title} {Quantum Dynamical Semigroups and Applications}}}\
  (\bibinfo  {publisher} {Springer Berlin, Heidelberg},\ \bibinfo {year}
  {2007})\BibitemShut {NoStop}%
\bibitem [{\citenamefont {Rivas}\ and\ \citenamefont
  {Huelga}(2012)}]{Rivas:2012}%
  \BibitemOpen
  \bibfield  {author} {\bibinfo {author} {\bibfnamefont {{\'A}.}~\bibnamefont
  {Rivas}}\ and\ \bibinfo {author} {\bibfnamefont {S.~F.}\ \bibnamefont
  {Huelga}},\ }\href {https://doi.org/10.1007/978-3-642-23354-8} {\emph
  {\bibinfo {title} {Open quantum systems}}}\ (\bibinfo  {publisher} {Springer
  Berlin Heidelberg},\ \bibinfo {year} {2012})\BibitemShut {NoStop}%
\bibitem [{\citenamefont {McCauley}\ \emph {et~al.}(2020)\citenamefont
  {McCauley}, \citenamefont {Cruikshank}, \citenamefont {Bondar},\ and\
  \citenamefont {Jacobs}}]{Mccauley:2020}%
  \BibitemOpen
  \bibfield  {author} {\bibinfo {author} {\bibfnamefont {G.}~\bibnamefont
  {McCauley}}, \bibinfo {author} {\bibfnamefont {B.}~\bibnamefont
  {Cruikshank}}, \bibinfo {author} {\bibfnamefont {D.~I.}\ \bibnamefont
  {Bondar}},\ and\ \bibinfo {author} {\bibfnamefont {K.}~\bibnamefont
  {Jacobs}},\ }\bibfield  {title} {\bibinfo {title} {Accurate {Lindblad}-form
  master equation for weakly damped quantum systems across all regimes},\
  }\href {https://doi.org/10.1038/s41534-020-00299-6} {\bibfield  {journal}
  {\bibinfo  {journal} {npj Quantum Information}\ }\textbf {\bibinfo {volume}
  {6}},\ \bibinfo {pages} {74} (\bibinfo {year} {2020})}\BibitemShut {NoStop}%
\bibitem [{\citenamefont {Nathan}\ and\ \citenamefont
  {Rudner}(2020)}]{Nathan:2020}%
  \BibitemOpen
  \bibfield  {author} {\bibinfo {author} {\bibfnamefont {F.}~\bibnamefont
  {Nathan}}\ and\ \bibinfo {author} {\bibfnamefont {M.~S.}\ \bibnamefont
  {Rudner}},\ }\bibfield  {title} {\bibinfo {title} {Universal lindblad
  equation for open quantum systems},\ }\href
  {https://doi.org/10.1103/PhysRevB.102.115109} {\bibfield  {journal} {\bibinfo
   {journal} {Physical Review B}\ }\textbf {\bibinfo {volume} {102}},\ \bibinfo
  {pages} {115109} (\bibinfo {year} {2020})}\BibitemShut {NoStop}%
\bibitem [{\citenamefont {Farina}\ and\ \citenamefont
  {Giovannetti}(2019)}]{Farina:2019}%
  \BibitemOpen
  \bibfield  {author} {\bibinfo {author} {\bibfnamefont {D.}~\bibnamefont
  {Farina}}\ and\ \bibinfo {author} {\bibfnamefont {V.}~\bibnamefont
  {Giovannetti}},\ }\bibfield  {title} {\bibinfo {title} {Open-quantum-system
  dynamics: Recovering positivity of the redfield equation via the partial
  secular approximation},\ }\href {https://doi.org/10.1103/PhysRevA.100.012107}
  {\bibfield  {journal} {\bibinfo  {journal} {Physical Review A}\ }\textbf
  {\bibinfo {volume} {100}},\ \bibinfo {pages} {012107} (\bibinfo {year}
  {2019})}\BibitemShut {NoStop}%
\bibitem [{\citenamefont {Trushechkin}(2021)}]{Trushechkin:2021}%
  \BibitemOpen
  \bibfield  {author} {\bibinfo {author} {\bibfnamefont {A.}~\bibnamefont
  {Trushechkin}},\ }\bibfield  {title} {\bibinfo {title} {Unified
  gorini-kossakowski-lindblad-sudarshan quantum master equation beyond the
  secular approximation},\ }\href {https://doi.org/10.1103/PhysRevA.103.062226}
  {\bibfield  {journal} {\bibinfo  {journal} {Physical Review A}\ }\textbf
  {\bibinfo {volume} {103}},\ \bibinfo {pages} {062226} (\bibinfo {year}
  {2021})}\BibitemShut {NoStop}%
\bibitem [{Note5()}]{Note5}%
  \BibitemOpen
  \bibinfo {note} {In this section, we consider that the only impact of the
  nonlinearity is to define the qubit subspace. For higher-excited states, the
  effect of anharmonicity on the treatment of dissipation can be found in
  Ref~\cite {Busel:2023}}\BibitemShut {NoStop}%
\bibitem [{Note6()}]{Note6}%
  \BibitemOpen
  \bibinfo {note} {The validity of the Markovian regime is determined by the
  poles of the dissipation matrix, which exhibit exponential decay in time
  domain with a rate proportional to $Z_0^{-1}\protect \mathsf {C}_{D}^{-1}$.
  For typical values of circuit parameters, $Z_0 \protect \mathsf {C}_{D} \ll
  \Gamma ^{-1}$, which justifies the validity of the Markovian
  approximation.}\BibitemShut {Stop}%
\bibitem [{\citenamefont {Williamson}(1936)}]{Williamson:1936}%
  \BibitemOpen
  \bibfield  {author} {\bibinfo {author} {\bibfnamefont {J.}~\bibnamefont
  {Williamson}},\ }\bibfield  {title} {\bibinfo {title} {On the algebraic
  problem concerning the normal forms of linear dynamical systems},\ }\href
  {http://www.jstor.org/stable/2371062} {\bibfield  {journal} {\bibinfo
  {journal} {American Journal of Mathematics}\ }\textbf {\bibinfo {volume}
  {58}},\ \bibinfo {pages} {141} (\bibinfo {year} {1936})}\BibitemShut
  {NoStop}%
\bibitem [{\citenamefont {Laub}\ and\ \citenamefont {Meyer}(1974)}]{Laub:1974}%
  \BibitemOpen
  \bibfield  {author} {\bibinfo {author} {\bibfnamefont {A.}~\bibnamefont
  {Laub}}\ and\ \bibinfo {author} {\bibfnamefont {K.}~\bibnamefont {Meyer}},\
  }\bibfield  {title} {\bibinfo {title} {{Canonical forms for symplectic and
  Hamiltonian matrices}},\ }\href {https://doi.org/10.1007/BF01260514}
  {\bibfield  {journal} {\bibinfo  {journal} {Celestial Mechanics}\ }\textbf
  {\bibinfo {volume} {9}},\ \bibinfo {pages} {213} (\bibinfo {year}
  {1974})}\BibitemShut {NoStop}%
\bibitem [{\citenamefont {Parra-Rodriguez}\ and\ \citenamefont
  {Egusquiza}(2022)}]{ParraRodriguez:2022b}%
  \BibitemOpen
  \bibfield  {author} {\bibinfo {author} {\bibfnamefont {A.}~\bibnamefont
  {Parra-Rodriguez}}\ and\ \bibinfo {author} {\bibfnamefont {I.~L.}\
  \bibnamefont {Egusquiza}},\ }\bibfield  {title} {\bibinfo {title} {Quantum
  fluctuations in electrical multiport linear systems},\ }\href
  {https://doi.org/10.1103/PhysRevB.106.054504} {\bibfield  {journal} {\bibinfo
   {journal} {Physical Review B}\ }\textbf {\bibinfo {volume} {106}},\ \bibinfo
  {pages} {054504} (\bibinfo {year} {2022})}\BibitemShut {NoStop}%
\bibitem [{\citenamefont {Esteve}\ \emph {et~al.}(1986)\citenamefont {Esteve},
  \citenamefont {Devoret},\ and\ \citenamefont {Martinis}}]{Esteve:1986}%
  \BibitemOpen
  \bibfield  {author} {\bibinfo {author} {\bibfnamefont {D.}~\bibnamefont
  {Esteve}}, \bibinfo {author} {\bibfnamefont {M.~H.}\ \bibnamefont
  {Devoret}},\ and\ \bibinfo {author} {\bibfnamefont {J.~M.}\ \bibnamefont
  {Martinis}},\ }\bibfield  {title} {\bibinfo {title} {Effect of an arbitrary
  dissipative circuit on the quantum energy levels and tunneling of a josephson
  junction},\ }\href {https://doi.org/10.1103/PhysRevB.34.158} {\bibfield
  {journal} {\bibinfo  {journal} {Physical Review B}\ }\textbf {\bibinfo
  {volume} {34}},\ \bibinfo {pages} {158} (\bibinfo {year} {1986})}\BibitemShut
  {NoStop}%
\bibitem [{\citenamefont {Sete}\ \emph {et~al.}(2015)\citenamefont {Sete},
  \citenamefont {Martinis},\ and\ \citenamefont {Korotkov}}]{Sete:2015}%
  \BibitemOpen
  \bibfield  {author} {\bibinfo {author} {\bibfnamefont {E.~A.}\ \bibnamefont
  {Sete}}, \bibinfo {author} {\bibfnamefont {J.~M.}\ \bibnamefont {Martinis}},\
  and\ \bibinfo {author} {\bibfnamefont {A.~N.}\ \bibnamefont {Korotkov}},\
  }\bibfield  {title} {\bibinfo {title} {Quantum theory of a bandpass purcell
  filter for qubit readout},\ }\href
  {https://doi.org/10.1103/PhysRevA.92.012325} {\bibfield  {journal} {\bibinfo
  {journal} {Physical Review A}\ }\textbf {\bibinfo {volume} {92}},\ \bibinfo
  {pages} {012325} (\bibinfo {year} {2015})}\BibitemShut {NoStop}%
\bibitem [{\citenamefont {Jeffrey}\ \emph {et~al.}(2014)\citenamefont
  {Jeffrey}, \citenamefont {Sank}, \citenamefont {Mutus}, \citenamefont
  {White}, \citenamefont {Kelly}, \citenamefont {Barends}, \citenamefont
  {Chen}, \citenamefont {Chen}, \citenamefont {Chiaro}, \citenamefont
  {Dunsworth}, \citenamefont {Megrant}, \citenamefont {O'Malley}, \citenamefont
  {Neill}, \citenamefont {Roushan}, \citenamefont {Vainsencher}, \citenamefont
  {Wenner}, \citenamefont {Cleland},\ and\ \citenamefont
  {Martinis}}]{Jeffrey:2014}%
  \BibitemOpen
  \bibfield  {author} {\bibinfo {author} {\bibfnamefont {E.}~\bibnamefont
  {Jeffrey}}, \bibinfo {author} {\bibfnamefont {D.}~\bibnamefont {Sank}},
  \bibinfo {author} {\bibfnamefont {J.~Y.}\ \bibnamefont {Mutus}}, \bibinfo
  {author} {\bibfnamefont {T.~C.}\ \bibnamefont {White}}, \bibinfo {author}
  {\bibfnamefont {J.}~\bibnamefont {Kelly}}, \bibinfo {author} {\bibfnamefont
  {R.}~\bibnamefont {Barends}}, \bibinfo {author} {\bibfnamefont
  {Y.}~\bibnamefont {Chen}}, \bibinfo {author} {\bibfnamefont {Z.}~\bibnamefont
  {Chen}}, \bibinfo {author} {\bibfnamefont {B.}~\bibnamefont {Chiaro}},
  \bibinfo {author} {\bibfnamefont {A.}~\bibnamefont {Dunsworth}}, \bibinfo
  {author} {\bibfnamefont {A.}~\bibnamefont {Megrant}}, \bibinfo {author}
  {\bibfnamefont {P.~J.~J.}\ \bibnamefont {O'Malley}}, \bibinfo {author}
  {\bibfnamefont {C.}~\bibnamefont {Neill}}, \bibinfo {author} {\bibfnamefont
  {P.}~\bibnamefont {Roushan}}, \bibinfo {author} {\bibfnamefont
  {A.}~\bibnamefont {Vainsencher}}, \bibinfo {author} {\bibfnamefont
  {J.}~\bibnamefont {Wenner}}, \bibinfo {author} {\bibfnamefont {A.~N.}\
  \bibnamefont {Cleland}},\ and\ \bibinfo {author} {\bibfnamefont {J.~M.}\
  \bibnamefont {Martinis}},\ }\bibfield  {title} {\bibinfo {title} {Fast
  accurate state measurement with superconducting qubits},\ }\href
  {https://doi.org/10.1103/PhysRevLett.112.190504} {\bibfield  {journal}
  {\bibinfo  {journal} {Physical Review Letter}\ }\textbf {\bibinfo {volume}
  {112}},\ \bibinfo {pages} {190504} (\bibinfo {year} {2014})}\BibitemShut
  {NoStop}%
\bibitem [{\citenamefont {Boissonneault}\ \emph {et~al.}(2009)\citenamefont
  {Boissonneault}, \citenamefont {Gambetta},\ and\ \citenamefont
  {Blais}}]{Boissonneault:2009}%
  \BibitemOpen
  \bibfield  {author} {\bibinfo {author} {\bibfnamefont {M.}~\bibnamefont
  {Boissonneault}}, \bibinfo {author} {\bibfnamefont {J.~M.}\ \bibnamefont
  {Gambetta}},\ and\ \bibinfo {author} {\bibfnamefont {A.}~\bibnamefont
  {Blais}},\ }\bibfield  {title} {\bibinfo {title} {Dispersive regime of
  circuit qed: Photon-dependent qubit dephasing and relaxation rates},\ }\href
  {https://doi.org/10.1103/PhysRevA.79.013819} {\bibfield  {journal} {\bibinfo
  {journal} {Physical Review A}\ }\textbf {\bibinfo {volume} {79}},\ \bibinfo
  {pages} {013819} (\bibinfo {year} {2009})}\BibitemShut {NoStop}%
\bibitem [{\citenamefont {Yan}\ \emph {et~al.}(2018)\citenamefont {Yan},
  \citenamefont {Krantz}, \citenamefont {Sung}, \citenamefont {Kjaergaard},
  \citenamefont {Campbell}, \citenamefont {Orlando}, \citenamefont
  {Gustavsson},\ and\ \citenamefont {Oliver}}]{Yan:2018}%
  \BibitemOpen
  \bibfield  {author} {\bibinfo {author} {\bibfnamefont {F.}~\bibnamefont
  {Yan}}, \bibinfo {author} {\bibfnamefont {P.}~\bibnamefont {Krantz}},
  \bibinfo {author} {\bibfnamefont {Y.}~\bibnamefont {Sung}}, \bibinfo {author}
  {\bibfnamefont {M.}~\bibnamefont {Kjaergaard}}, \bibinfo {author}
  {\bibfnamefont {D.~L.}\ \bibnamefont {Campbell}}, \bibinfo {author}
  {\bibfnamefont {T.~P.}\ \bibnamefont {Orlando}}, \bibinfo {author}
  {\bibfnamefont {S.}~\bibnamefont {Gustavsson}},\ and\ \bibinfo {author}
  {\bibfnamefont {W.~D.}\ \bibnamefont {Oliver}},\ }\bibfield  {title}
  {\bibinfo {title} {Tunable {Coupling} {Scheme} for {Implementing}
  {High}-{Fidelity} {Two}-{Qubit} {Gates}},\ }\href
  {https://doi.org/10.1103/PhysRevApplied.10.054062} {\bibfield  {journal}
  {\bibinfo  {journal} {Physical Review Applied}\ }\textbf {\bibinfo {volume}
  {10}},\ \bibinfo {pages} {054062} (\bibinfo {year} {2018})}\BibitemShut
  {NoStop}%
\bibitem [{\citenamefont {Pozar}(2009)}]{Pozar:2009}%
  \BibitemOpen
  \bibfield  {author} {\bibinfo {author} {\bibfnamefont {D.~M.}\ \bibnamefont
  {Pozar}},\ }\href
  {https://www.wiley.com/en-us/Microwave+Engineering%2C+4th+Edition-p-9780470631553}
  {\emph {\bibinfo {title} {Microwave Engineering}}},\ \bibinfo {edition}
  {4th}\ ed.\ (\bibinfo  {publisher} {John Wiley \& Sons},\ \bibinfo {address}
  {Hoboken, New York},\ \bibinfo {year} {2009})\BibitemShut {NoStop}%
\bibitem [{\citenamefont {Busel}\ \emph {et~al.}(2023)\citenamefont {Busel},
  \citenamefont {Laine}, \citenamefont {Mansikkam\"aki},\ and\ \citenamefont
  {Silveri}}]{Busel:2023}%
  \BibitemOpen
  \bibfield  {author} {\bibinfo {author} {\bibfnamefont {O.}~\bibnamefont
  {Busel}}, \bibinfo {author} {\bibfnamefont {S.}~\bibnamefont {Laine}},
  \bibinfo {author} {\bibfnamefont {O.}~\bibnamefont {Mansikkam\"aki}},\ and\
  \bibinfo {author} {\bibfnamefont {M.}~\bibnamefont {Silveri}},\ }\bibfield
  {title} {\bibinfo {title} {Dissipation and dephasing of interacting photons
  in transmon arrays},\ }\href
  {https://doi.org/10.1103/PhysRevResearch.5.023121} {\bibfield  {journal}
  {\bibinfo  {journal} {Physical Review Research}\ }\textbf {\bibinfo {volume}
  {5}},\ \bibinfo {pages} {023121} (\bibinfo {year} {2023})}\BibitemShut
  {NoStop}%
\end{thebibliography}%
\end{document}